\documentclass{article}
\usepackage{amssymb,amsmath}
\usepackage[dvipdfmx]{graphicx}
\usepackage{bbm}
\usepackage{bm}
\usepackage{caption}
\usepackage{color}
\usepackage{comment}
\usepackage{comment}
\usepackage{enumerate}
\usepackage{ytableau}
\usepackage{here}
\usepackage{subfig}
\DeclareMathOperator*{\Tr}{{\rm Tr}}

\numberwithin{equation}{section}

\begin{document}

%%%%%%%%%%%%%%%%%%%%%%%%%%%%%%%%%%%%%%%%%%%%
\thispagestyle{empty}
\begin{flushright}
%%\\
%%%%%%%%%%%%%%%%%%%%%%%%%%%%%%%%%%%%%%%%%%%%%%%%%%%%%%%%%%%%%%%%%
%input \\
%%%%%%%%%%%%%%%%%%%%%%%%%%%%%%%%%%%%%%%%%%%%%%%%%%%%%%%%%%%%%%%%%

\end{flushright}
\vskip1cm
\begin{center}
{\Large \bf Abelian dualities of $\mathcal{N}=(0,4)$ boundary conditions\\
%\vskip0.3cm 
%and 
%\vskip0.5cm 
%Supersymmetric Indices
%\vskip0.25cm
}

\vskip2cm
 Tadashi Okazaki\footnote{tokazaki@perimeterinstitute.ca}

\bigskip
{\it
Perimeter Institute for Theoretical Physics,\\
Waterloo, Ontario, Canada N2L 2Y5
}

\end{center}

%%%%%%%%%%%%%%%%%%%%%%%%%%%%%%%%%%%%%%%%%%%%
\vskip1.5cm
\begin{abstract}
We propose dual pairs of $\mathcal{N}=(0,4)$ half-BPS boundary conditions for 3d $\mathcal{N}=4$ Abelian gauge theories 
related to mirror symmetry and S-duality by showing the matching of boundary 't Hooft anomalies and supersymmetric indices. 
We find simple $\mathcal{N}=(0,4)$ mirror symmetry between 2d $\mathcal{N}=(0,4)$ Abelian gauge theories and free Fermi multiplets that  
generalizes $\mathcal{N}=(0,2)$ Abelian duality. 
We also propose a prescription for computing half-index of enriched Neumann boundary condition including 2d boundary bosonic matters 
by gauging the 2d boundary flavor symmetry of Dirichlet boundary condition. 
By coupling $\mathcal{N}=(0,4)$ half-BPS boundary configurations of 3d $\mathcal{N}=4$ gauge theories 
to quarter-BPS corner configurations of 4d $\mathcal{N}=4$ Super Yang-Mills theories, 
we further obtain a new type of 4d-3d-2d duality that may involve 3d non-Abelian gauge symmetry. 
\end{abstract}
%%%%%%%%%%%%%%%%%%%%%%%%%%%%%%%%%%%%%%%%%%

%%%%%%%%%%%%%%%%%%

\newpage
\tableofcontents
%%%%%%%%%%%%%%%%%%%%%%%%%%

%%%%%%%%%%%%%%%%%%%%%%%%%%%%%%%%%%%%%%%%%%%%%
%%%%%%%%%%%%%%%%%%%%%%%%%%%%%%%%%%%%%%%%%%%%%
\section{Introduction and conclusions}
\label{sec_intro}
%%%%%%%%%%%%%%%%%%%%%%%%%%%%%%%%%%%%%%%%%%%%%
%%%%%%%%%%%%%%%%%%%%%%%%%%%%%%%%%%%%%%%%%%%%%
%backgrounds

Quantum field theories with boundaries can serve as various frameworks in studying a wide range of physical and mathematical problems. 
%2d (2,2)
Supersymmetric boundary conditions in two-dimensional $\mathcal{N}=(2,2)$ supersymmetric field theories 
can determine D-branes in the target space that play a significant role in the study of mirror symmetry \cite{Ooguri:1996ck,Hori:2000ck,Herbst:2008jq}. 
%4d
The half-BPS boundary conditions and interfaces in four-dimensional $\mathcal{N}=4$ Super Yang-Mills (SYM) theory 
are studied in \cite{Gaiotto:2008sa,Gaiotto:2008ak}. 
These boundary conditions are realized as the brane configuration in Type IIB string theory 
and admit singular half-BPS boundary condition, aka Nahm pole boundary condition.  
The action of S-duality beautifully produces dualities of these boundary conditions \cite{Gaiotto:2008ak} as well as 3d $\mathcal{N}=4$ mirror symmetry \cite{Intriligator:1996ex}. 
These half-BPS boundary conditions can also play a role in the Geometric Langlands Program 
as they can be mapped to mathematical objects in the categories relevant for the Geometric Langlands duality \cite{Henningson:2011qk,Gaiotto:2016hvd}. 
Moreover, such half-BPS interfaces in four-dimensional $\mathcal{N}=2$ gauge theories have played a central role in 
finding new duality, that is 3d-3d relation \cite{Dimofte:2011ju, Terashima:2011qi, Cecotti:2011iy}. 
While AGT relation \cite{Alday:2009aq} is related to the compactification of M5-branes on Riemann surface, 
the 3d-3d relation is concerned with the compactification of M5-branes on 3-manifold.

%3d N=2
The analysis of half-BPS boundary conditions in three-dimensional $\mathcal{N}=2$ supersymmetric field theories 
was initiated in \cite{Gadde:2013wq, Okazaki:2013kaa}. 
The half-BPS boundary conditions in three-dimensional supersymmetric field theories are more fruitful as they are classified by the chirality of preserved supercharges. 
While the $\mathcal{N}=(1,1)$ boundary conditions flow to the special Lagrangian branes, 
the $\mathcal{N}=(0,2)$ boundary conditions flow to the holomorphic branes in the IR sigma-models \cite{Okazaki:2013kaa}. 
The $\mathcal{N}=(0,2)$ boundary conditions in 3d $\mathcal{N}=2$ gauge theories 
are also important in the study of 4d-2d relation 
\cite{Gadde:2013wq, Gadde:2013sca, Gadde:2013lxa, Chun:2015gda, Gukov:2016gkn, Dedushenko:2017tdw, Dimofte:2019buf}, 
which is involved with the compactification of M5-branes on 4-manifold. 
%3d N=4
The half-BPS boundary conditions in three-dimensional $\mathcal{N}=4$ supersymmetric gauge theories 
are studied in \cite{Bullimore:2016nji, Chung:2016pgt}. 
It is argued in \cite{Bullimore:2016nji} that the $\mathcal{N}=(2,2)$ boundary conditions which flow to the holomorphic Lagrangian branes in the IR sigma-models 
can produce a certain set of modules in the categories 
so that mirror symmetry of these boundary conditions are realized as Symplectic Duality 
\cite{MR3594664,Braden:2014iea}
that is an equivalence of categories of modules. 
The $\mathcal{N}=(2,2)$ and $\mathcal{N}=(0,4)$ half-BPS boundary conditions can be realized in brane configurations \cite{Chung:2016pgt}. 
%3d N\ge4
The BPS boundary conditions in three-dimensional $\mathcal{N}\ge4$ Chern-Simons theories, 
including BLG model, ABJM model are analyzed in \cite{Berman:2009kj, Berman:2009xd, Hosomichi:2014rqa, Okazaki:2015fiq}. 
These boundary conditions can describe the M5-brane, M9-brane and M-theory wave interacting with M2-branes. 

%duality of b.c.
If one has a pair of UV gauge theories which flow to the same superconformal field theory, 
one may find a pair of UV boundary conditions 
\footnote{
%5d 
The half-BPS boundary conditions or interfaces in dimensions higher than four behave differently 
since they flow to the UV in the RG flow as discussed in \cite{Gaiotto:2015una} for the half-BPS interface in five-dimensional $\mathcal{N}=1$ gauge theories. 
} 
for these theories which flow to the same superconformal boundary condition after the RG flow. 
Recently the dualities of boundary conditions and interfaces have been confirmed by computing supersymmetric indices. 
%e.g.
In \cite{Dimofte:2017tpi} various dualities of $\mathcal{N}=(0,2)$ supersymmetric boundary conditions in 3d $\mathcal{N}=2$ gauge theories have been demonstrated. 
More recently, \cite{Gaiotto:2019jvo} have presented dualities of half-BPS interfaces and quarter-BPS corner configurations in four-dimensional $\mathcal{N}=4$ SYM theory. 
Subsequently, \cite{Okazaki:2019ony} have shown more general dualities of half-BPS boundary conditions 
and interfaces in 4d $\mathcal{N}=4$ SYM theory conjectured by Gaiotto and Witten \cite{Gaiotto:2008ak}. 
When the local operators are supported at two-dimensional boundary and corner configuration of superconformal field theories, 
one can embed the Vertex Operator Algebras (VOAs) into the algebras of local operators so that the correlation functions of certain protected operators are encoded in the VOAs \cite{Beem:2013sza,Beem:2014rza}. 
Consequently the indices reduce to characters of the VOAs in the appropriate limit. 

%%%%%%%%%%%%%%%%%%%%%%%%%%%%%%%%%%%%%%%%%%%%%
\subsection{Main results}
\label{sec_result}
%%%%%%%%%%%%%%%%%%%%%%%%%%%%%%%%%%%%%%%%%%%%%
%organization of peper
%aim of this paper 
In this paper we extend the duality network of 4d $\mathcal{N}=4$, 3d $\mathcal{N}=4$ and 2d $\mathcal{N}=(0,4)$ gauge theories 
\cite{Gaiotto:2019jvo, Okazaki:2019ony} 
by examining half-BPS $\mathcal{N}=(0,4)$ boundary conditions in 3d $\mathcal{N}=4$ gauge theories 
that can be constructed from the brane setup in Type IIB string theory \cite{Chung:2016pgt, Hanany:2018hlz, Gaiotto:2019jvo}. 
Unlike the cases studied in \cite{Gaiotto:2019jvo,Okazaki:2019ony}, 
the $\mathcal{N}=(0,4)$ configurations discussed in this paper involve  
3d $\mathcal{N}=4$ gauge theories with Neumann boundary condition $\mathcal{N}'$ or/and 2d $\mathcal{N}=(0,4)$ gauge theories. 
\footnote{In \cite{Gaiotto:2019jvo} the gauge symmetry stems from 4d gauge theories while in \cite{Okazaki:2019ony} it also results from 3d gauge theories. }
Accordingly, as we argue in the main text, they need more careful treatment in the computation of supersymmetric indices. 
We propose the appropriate prescription for evaluating the half-indices of enriched Neumann b.c. which includes 2d bosonic matter fields 
by gauging the 2d boundary flavor symmetry of Dirichlet boundary condition. 
We find $\mathcal{N}=(0,4)$ mirror symmetry of boundary conditions for 3d $\mathcal{N}=4$ Abelian gauge theories. 
Besides, we propose simple $\mathcal{N}=(0,4)$ mirror symmetry between 2d $\mathcal{N}=(0,4)$ Abelian gauge theories and free Fermi multiplets 
which is reminiscent of $\mathcal{N}=(0,2)$ Abelian duality \cite{Gadde:2013sca, Gadde:2013lxa}. 
They are descended from the action of S-duality in string theory. 
We also find that these dualities are further generalized by coupling to quarter-BPS corners of 4d $\mathcal{N}=4$ SYM theories 
studied in \cite{Gaiotto:2019jvo}. 
They can involve 3d $\mathcal{N}=4$ non-Abelian gauge theories with Neumann boundary condition $\mathcal{N}'$. 
The dual configurations are characterized by the Dirichlet/Nahm pole boundary conditions for 3d $\mathcal{N}=4$ twisted gauge theories 
with vector multiplets and twisted hypermultiplets as well as corners or boundaries of 4d $\mathcal{N}=4$ gauge theories. 
We confirm these dualities by expanding the supersymmetric indices that include half-indices of 3d $\mathcal{N}=4$ gauge theories 
and then checking several terms of series expansion. 
As a byproduct, our supersymmetric indices should reduce to the vacuum characters of the VOAs 
arising from the $\mathcal{N}=(0,4)$ supersymmetric configurations when setting the fugacities to special values. 

The paper contains many confirmed examples to gain a bottom-up understanding, 
however, the general formulae for Abelian dualities of $\mathcal{N}=(0,4)$ boundary conditions of 3d theories 
are summarized in subsection \ref{sec_sqednf_bc} 
and those for dualities of 4d-3d-2d configurations are summarized 
in subsubsection 
\ref{sec_4dun3dumNN}, 
\ref{sec_4dun3dumx}, 
\ref{sec_4duN3duMhm1}, 
\ref{sec_4d23d1hm2}, 
\ref{sec_4duL3duM4duNNN}, 
\ref{sec_4dul3dumx4dunNN}, 
and 
\ref{sec_4dul3dumhm14dunNN}.

%%%%%%%%%%%%%%%%%%%%%%%%%%%%%%%%%%%%%%%%%%%%%
\subsection{Structure}
\label{sec_organization}
%%%%%%%%%%%%%%%%%%%%%%%%%%%%%%%%%%%%%%%%%%%%%
%organization of peper
The paper is organized as follows. 
In section \ref{sec_index} we introduce main tools to test dualities of $\mathcal{N}=(0,4)$ boundary conditions. 
We define half-indices of 3d $\mathcal{N}=4$ gauge theories and full-indices of 2d $\mathcal{N}=(0,4)$ gauge theories. 
In section \ref{sec_04mirror} we study simple examples of brane box models in \cite{Hanany:2018hlz}. 
We propose the minimal $\mathcal{N}=(0,4)$ mirror symmetry by computing the indices. 
In section \ref{sec_abebc} we analyze $\mathcal{N}=(0,4)$ boundary conditions for 3d $\mathcal{N}=4$ Abelian gauge theories. 
We confirm the dualities of $\mathcal{N}=(0,4)$ boundary conditions by checking the matching of boundary anomalies and half-indices. 
We also discuss the half-indices for enriched Neumann boundary conditions which include 2d bosonic matter fields. 
In section \ref{sec_4d3d} we propose the dualities of $\mathcal{N}=(0,4)$ supersymmetric configurations 
which are composed of quarter-BPS corners of 4d $\mathcal{N}=4$ SYM theory 
and half-BPS boundaries of 3d $\mathcal{N}=4$ gauge theories. 
The configurations include non-Abelian 3d gauge theories with Neumann b.c. which map to the dual configurations with no gauge symmetry. 
We present the matching of quarter-indices as powerful evidences of proposed dualities of these corner configurations.

%%%%%%%%%%%%%%%%%%%%%%%%%%%%%%%%%%%%%%%%%%%%%
\subsection{Future works}
\label{sec_open}
%%%%%%%%%%%%%%%%%%%%%%%%%%%%%%%%%%%%%%%%%%%%%
There still remain a variety of interesting open questions which we leave for future works.

\begin{itemize}

\item {\bf Non-Abelian $\mathcal{N}=(0,4)$ configurations:} 
Most of our $\mathcal{N}=(0,4)$ configurations are limited to the cases with Abelian gauge symmetry. 
They should be promoted to more general $\mathcal{N}=(0,4)$ mirror symmetry of 2d non-Abelian gauge theories 
and dualities of $\mathcal{N}=(0,4)$ boundary conditions including enriched Neumann b.c. and Nahm pole b.c. for 3d $\mathcal{N}=4$ non-Abelian gauge theories. 
In particular, it is interesting to explore $\mathcal{N}=(0,4)$ mirror symmetry for general brane box models in \cite{Hanany:2018hlz} with several boxes of multiple D3-branes. 
We hope to report a detailed analysis in the future.

\item {\bf boundary/corner VOAs:} 
The Vertex Operator Algebras (VOAs) in 3d $\mathcal{N}=4$ gauge theories are constructed 
as the algebras of local operators obeying the holomorphic boundary conditions 
for their topologically twisted theories, which can be obtained by deforming $\mathcal{N}=(0,4)$ boundary conditions \cite{Costello:2018fnz}. 
These VOAs are motivated as powerful computational tools to study the bulk topological field theory (TFT) 
and potential applications to the gauge theory interpretation of the Geometric Langlands Program \cite{Kapustin:2006pk,Gaiotto:2008sa, Gaiotto:2008ak, Gaiotto:2016wcv}. 
Also the VOAs arising from corners of 4d $\mathcal{N}=4$ SYM theory are constructed in \cite{Gaiotto:2017euk}. 
It would be intriguing to figure out the proposed dualities 
in terms of VOAs appearing from the combined configurations of boundaries of 3d theories and corners of 4d theories studied in this paper. 

Also the relation between 3d supersymmetric field theories and 
VOAs has been recently explored in the context of 3d Modulatiry \cite{Cheng:2018vpl} and study of 4-manifolds \cite{Feigin:2018bkf}. 
It would be also interesting to study these issues related to $\mathcal{N}=(0,4)$ boundary conditions. 
\footnote{The author thanks Sergei Gukov for discussing relevant issues. }

\item {\bf Line and surface defects:} 
Our dualities should be generalized by introducing non-local operators, such as line or surface operators in $\mathcal{N}=(0,4)$ supersymmetric boundary/corner configurations. 
It would be interesting to examine the difference equations satisfied by the supersymmetric indices 
\cite{Dimofte:2011ju, Beem:2012mb, Dimofte:2011py, Dimofte:2017tpi}
which correlate with the half-BPS line or surface operators. 
From the perspective of the VOAs, including half-BPS line or surface operators can produce non-trivial modules in VOAs \cite{Costello:2018swh}. 
The analysis of these modules may be fruitfully addressed by computing 
the supersymmetric indices as they can be viewed as promotions of characters of these modules in that they reduce to these characters in the special fugacity limits.

\item {\bf Holographic duals:}
The gravitational solutions for D3-brane ending on fivebranes are argued in \cite{DHoker:2007zhm, DHoker:2007hhe, Aharony:2011yc, Assel:2011xz}. 
It would be interesting to explore the Type IIB supergravity solutions corresponding to our brane configurations 
described by quarter BPS corners of 4d $\mathcal{N}=4$ SYM theory, half-BPS boundaries of 3d $\mathcal{N}=4$ gauge theories, 
and 2d $\mathcal{N}=(0,4)$ gauge theories in the low-energy limit. 
The indices on the gravity side that count the Kaluza-Klein excitation of the massless fields in the Type IIB supergravity 
may be compared with the supersymmetric indices in the large $N$ limit. 

\item  {\bf $\mathcal{N}=(2,2)$ boundaries and corners:}
Boundaries and junctions with $\mathcal{N}=(2,2)$ supersymmetry also 
admit brane constructions \cite{Chung:2016pgt} 
which lead to conjectural dualities under the action of S-duality. 
While for $\mathcal{N}=(0,4)$ supersymmetric case one finds boundary/corner Fermi multiplets which may carry topological symmetry \cite{Hanany:2018hlz, Gaiotto:2019jvo}, 
for $\mathcal{N}=(2,2)$ supersymmetric case no Fermi multiplet can exist at the boundary/corner. 
We hope to report progress in that direction in the near future.

\item {\bf M2-M5 brane systems:}
The 3d $\mathcal{N}=4$ $U(N)$ gauge theory with an adjoint hypermultiplet 
and a fundamental hypermultiplet is known to flow to the same IR fixed point as 
the $U(N)_{1}\times U(N)_{-1}$ ABJM theory describing multiple $N$ M2-branes \cite{Bashkirov:2010kz}. 
It would be interesting to examine the $\mathcal{N}=(0,4)$ boundary conditions in this theory 
as they may describe the M5-branes on which M2-branes end on. 
In the special fugacity limit, the half-indices would reduce to the vacuum characters of the associated VOAs. 

\item {\bf Boundary conditions for $\mathcal{N}\ge 4$ Chern-Simons matter theories:} 
It would be intriguing to understand $\mathcal{N}=(0,4)$ boundary conditions for 3d $\mathcal{N}\ge 4$ Chern-Simons matter theories 
\cite{Gaiotto:2008sd, Hosomichi:2008jd, Hosomichi:2008jb}. 
The analysis of $\mathcal{N}=(0,2)$ boundary conditions for 3d $\mathcal{N}=2$ Yang Mills-Chern-Simons theories in \cite{Dimofte:2017tpi}, 
which argues the level-rank duality of boundary conditions, may be generalized by including two-dimensional chiral superalgebras 
\cite{Okazaki:2015fiq,Okazaki:2016pne}, 
as the construction of these theories is intimately related to the supergroup structure. 

\item {\bf Identities of $q$-series:} 
Although we have confirmed the matching of supersymmetric indices 
by checking several terms in the expansions, some are not still analytically proven. 
The resulting cute identities of $q$-series may have potential applications to number theory, representation theory, and combinatorics. 
In particular, we conjecture ``the minimal mirror identity'' between the half-indices (\ref{sqed1d_half1})  and (\ref{msqed1d_half1}) 
that essentially captures the $\mathcal{N}=(0,4)$ mirror symmetry 
of the boundary conditions for 3d $\mathcal{N}=4$ Abelian gauge theories. 
It would be interesting to explore the analytic proof of this identity.  

\end{itemize}

%%%%%%%%%%%%%%%%%%%%%%%%%%%%%%%%%%%%%%%%%%%%%
%%%%%%%%%%%%%%%%%%%%%%%%%%%%%%%%%%%%%%%%%%%%%
\section{Indices}
\label{sec_index}
%%%%%%%%%%%%%%%%%%%%%%%%%%%%%%%%%%%%%%%%%%%%%
%%%%%%%%%%%%%%%%%%%%%%%%%%%%%%%%%%%%%%%%%%%%%

We define the quarter-index as the trace over the cohomology of the preserved supercharges \cite{Gaiotto:2019jvo}
\begin{align}
\label{INDEX_def}
\mathbb{IV}(t,x;q)
&:={\Tr}_{\mathrm{Op}}(-1)^{F}q^{J+\frac{H+C}{4}}t^{H-C} x^{f} 
\end{align}
where $F$ is the Fermion number, 
$J$ is the generator of the $U(1)_{J}$ rotational symmetry in the two-dimensional space-time, 
$H$ and $C$ re the Cartan generators of the $SU(2)_{H}$ and $SU(2)_{C}$ R-symmetry groups respectively. 
$f$ is the Cartan generator of the global symmetry. 

When the bulk 4d theory is empty, 
the quarter-index (\ref{INDEX_def}) can be identified with the half-index of 3d theory, 
and when the bulk 4d and boundary 3d theories are absent, the quarter-index (\ref{INDEX_def}) can be viewed as the full-index of 2d theory.

We use the $q$-shifted factorial defined by 
\begin{align}
\label{qpoch_def}
(a;q)_{0}&:=1,\qquad
(a;q)_{n}:=\prod_{k=0}^{n-1}(1-aq^{k}),\qquad 
(q)_{n}:=\prod_{k=1}^{n}(1-q^{k}),\quad 
\quad  n\ge1,
\nonumber \\
(a;q)_{\infty}&:=\prod_{k=0}^{\infty}(1-aq^{k}),\qquad 
(q)_{\infty}:=\prod_{k=1}^{\infty} (1-q^k), 
\nonumber\\
(a^{\pm};q)_{\infty}&:=(a;q)_{\infty}(a^{-1};q)_{\infty}
\end{align}
where $a$ and $q$ are complex variables with $|q|<1$.

%brane
We construct the $\mathcal{N}=(0,4)$ supersymmetric configurations 
from brane configurations in Type IIB string theory whose world-volumes span the following directions:
\begin{itemize}
\item D3-branes extended along $x^{0}x^{1}x^{2}x^{6}$, 
\item NS5-branes extended along $x^{0}x^{1}x^{2}x^{3}x^{4}x^{5}$, 
\item D5-branes extended along $x^{0}x^{1}x^{2}x^{7}x^{8}x^{9}$, 
\item NS5$'$-branes extended along $x^{0}x^{1}x^{6}x^{7}x^{8}x^{9}$, 
\item D5$'$-branes extended along $x^{0}x^{1}x^{3}x^{4}x^{5}x^{6}$
\end{itemize}

The finite segments of D3-branes in the $x^6$ direction between NS5-branes 
introduce 3d $\mathcal{N}=4$ vector multiplet and 
D5-branes intersecting with the D3-branes introduce 3d $\mathcal{N}=4$ hypermultiplets \cite{Hanany:1996ie}. 
The $\mathcal{N}=(0,4)$ boundaries in 3d $\mathcal{N}=4$ gauge theories 
are realized by further adding NS5$'$- and D5$'$-branes on which the D3-branes terminate \cite{Chung:2016pgt, Hanany:2018hlz}. 
The NS5$'$-brane leads to the Neumann b.c. $\mathcal{N}'$ for 3d vector multiplet and Neumann b.c. $N'$ for 3d hypermultiplet, 
whereas the D5$'$-brane can introduce the Dirichlet b.c. $\mathcal{D}'$ and Nahm$'$ b.c. for 3d vector multiplet 
and Dirichlet $D'$ and Nahm pole boundary conditions for 3d hypermultiplet.

%Sduality
In the following, we identity the dual $\mathcal{N}=(0,4)$ configurations by applying S-duality in the brane setup. 
We then test the inspired dualities by evaluating boundary 't Hooft anomalies and supersymmetric indices.

%%%%%%%%%%%%%%%%%%%%%%%%%%%%%%%%%%%%%%%%%%%%%
\subsection{Half-indices of 3d $\mathcal{N}=4$ gauge theories}
\label{sec_3dn4_h}
%%%%%%%%%%%%%%%%%%%%%%%%%%%%%%%%%%%%%%%%%%%%%
While the half-indices for $\mathcal{N}=(0,2)$ Neumann b.c. $\mathcal{N}$ of three-dimensional gauge theories were firstly considered in \cite{Gadde:2013wq}, 
the half-indices for $\mathcal{N}=(0,2)$ Dirichlet b.c. $\mathcal{D}$ of three-dimensional gauge theories were proposed in \cite{Dimofte:2017tpi}. 
In this subsection we present the half-indices for $\mathcal{N}=(0,4)$ Neumann and Dirichlet boundary conditions which 
can be viewed as specializations of quarter-index (\ref{INDEX_def}). 

%3d N=4 hyper
The 3d $\mathcal{N}=4$ hypermultiplet includes  
a pair of complex scalars $\mathbb{H}$, $\widetilde{\mathbb{H}}$ 
forming a doublet of $SU(2)_{H}$ 
and a pair of complex fermions $\psi_{+}^{\mathbb{H}}$, $\psi_{+}^{\widetilde{\mathbb{H}}}$ 
forming a doublet of $SU(2)_{C}$. 
The charges of 3d $\mathcal{N}=4$ hypermultiplet is
\begin{align}
\label{3dn4_hm_ch}
\begin{array}{c|cccccc}
&\mathbb{H}&\widetilde{\mathbb{H}}&\psi_{+}^{\mathbb{H}}&\psi_{+}^{\widetilde{\mathbb{H}}}&\overline{\psi}_{-}^{\mathbb{H}}&\overline{\psi}_{-}^{\widetilde{\mathbb{H}}} \\ \hline
U(1)_{C}&0&0&-&-&+&+ \\
U(1)_{H}&+&+&0&0&0&0 \\
\end{array}
\end{align}

%3d N=4 vector
The 3d $\mathcal{N}=4$ Abelian vector multiplet includes 
a 3d gauge field $A_{\mu}^{\textrm{3d}}$, 
three scalars, which we denote by real and complex scalars $\sigma$, $\varphi$ forming the $SU(2)_{C}$ triplet, 
and two complex fermions $(\lambda_{\alpha}^{\textrm{3d}}, \eta_{\alpha}^{\textrm{3d}})$. 
The charges of 3d $\mathcal{N}=4$ vector multiplet is given by
\begin{align}
\label{3dn4_vm_ch}
\begin{array}{c|ccccccc}
&A_{\mu}^{\textrm{3d}}&\sigma&\varphi&\lambda_{\pm}^{\textrm{3d}}&\overline{\lambda}_{\pm}^{\textrm{3d}}
&\eta_{\pm}^{\textrm{3d}}&\overline{\eta}_{\pm}^{\textrm{3d}} \\ \hline
U(1)_{C}&0&0&2&+&-&+&- \\
U(1)_{H}&0&0&0&+&-&-&+ \\
\end{array}
\end{align}

We introduce the half-indices of $\mathcal{N}=(0,4)$ half-BPS boundary conditions for 3d $\mathcal{N}=4$ gauge theories 
by physical consideration.

%%%%%%%%%%%%%%%%%%%%%%%%%%%%%%%%%%%%%%%%%%%%%
\subsubsection{3d matter multiplets}
\label{sec_3dn4_h_m}
%%%%%%%%%%%%%%%%%%%%%%%%%%%%%%%%%%%%%%%%%%%%%

%HM b.c.
The simplest example of $\mathcal{N}=(0,4)$ half-BPS boundary conditions 
for 3d $\mathcal{N}=4$ hypermultiplet are the Neumann b.c. $N'$ and Dirichlet b.c. $D'$
\begin{align}
\begin{array}{ccc}
N':&\partial_{2}\mathbb{H}|_{\partial}=0,&\partial_{2}\widetilde{\mathbb{H}}|_{\partial}=0, \\
D':&\partial_{\mu}\mathbb{H}|_{\partial}=0,&\partial_{\mu}\widetilde{\mathbb{H}}|_{\partial}=0, \\
\end{array}
\qquad \mu=0,1
\end{align}
For Neumann b.c. $N'$ the complex scalar fields $\mathbb{H}$, $\widetilde{\mathbb{H}}$ 
and their superpartners, i.e. the right-moving fermions can fluctuate 
while for Dirichlet b.c. $D'$ 
the complex scalar fields $\mathbb{H}$, $\widetilde{\mathbb{H}}$ are fixed 
and the left-moving fermions can survive at the boundary \cite{Chung:2016pgt}.

%Neumann half index of hm
We introduce the fugacity $x$ for the flavor symmetry. 
The half-index of Neumann b.c. $N'$ for 3d $\mathcal{N}=4$ hypermultiplet is 
\begin{align}
\label{half_N_3dhm}
\mathbb{II}^{\textrm{3d HM}}_{N'}(t,x;q)
&=\frac{1}{(q^{\frac14}tx;q)_{\infty} (q^{\frac14}tx^{-1};q)_{\infty}}.
\end{align}
It has an expansion
\begin{align}
\label{half_N_3dhm1}
\mathbb{II}^{\textrm{3d HM}}_{N'}(t,x;q)
&=\sum_{n=0}^{\infty}\sum_{k=0}^{n}
\frac{1}{(q)_{k}(q)_{n-k}}x^{n-2k} q^{\frac{n}{4}}t^{n}. 
\end{align}

%H-twist limit of Neumann half-index of hm
One can deform the Neumann b.c. $N'$ for 3d hypermultiplet 
so that it is compatible with the H-twist (mirror Rozansky-Witten twist). 
The resulting deformed boundary condition supports the VOA Sb 
of symplectic bosons $X(z)$ and $Y(z)$ of conformal dimension $\frac12$ 
which obey the OPE \cite{Gaiotto:2017euk, Costello:2018fnz}
\begin{align}
\label{sb_ope}
X(z)Y(w)&\sim \frac{1}{z-w}. 
\end{align}
Correspondingly, 
the vacuum character for the symplectic boson VOA Sb can be obtained from the Neumann half-index (\ref{half_N_3dhm}) 
by taking the H-twist limit $t\rightarrow q^{\frac14}$ 
\begin{align}
\label{sb_ch}
\mathbb{II}^{\textrm{3d HM}}_{N'}(t=q^{\frac14},x;q)
&=\frac{1}{(q^{\frac12}x;q)_{\infty} (q^{\frac12}x^{-1};q)_{\infty}}
\nonumber\\
&=\chi_{\textrm{Sb}}(x;q). 
\end{align}

%Dirichlet half index of hm
The half-index of Dirichlet b.c. $D'$ for 3d $\mathcal{N}=4$ hypermultiplet takes the form
\begin{align}
\label{half_D_3dhm}
\mathbb{II}^{\textrm{3d HM}}_{D'}(t,x;q)
&=
(q^{\frac34}t^{-1}x;q)_{\infty} (q^{\frac34}t^{-1}x^{-1};q)_{\infty}. 
\end{align}
It is expanded as 
\begin{align}
\label{half_D_3dhm1}
\mathbb{II}^{\textrm{3d HM}}_{D'}
&=\frac{1}{(q)_{\infty}^2}
\sum_{n=0}^{\infty}\sum_{k=0}^{n}
(q^{1+k};q)_{\infty} (q^{1+n-k};q)_{\infty}
q^{\frac{n^2-n}{2}+k(k-n)+\frac{3n}{4}} 
(-1)^n 
x^{-n+2k}t^{-n}. 
\end{align}

%C-twist limit of Diri half-index of hm
The Dirichlet b.c. $D'$ for 3d hypermultiplet can be deformed 
in such a way that it is compatible with the C-twist. 
The deformed boundary condition supports the VOA Fc of fermionic currents 
$x(z)$ and $y(z)$ of conformal dimension $1$ with the OPE 
\begin{align}
\label{fc_ope}
x(z)y(w)&\sim \frac{1}{(z-w)^2}. 
\end{align}
In fact, in the C-twist limit $t\rightarrow q^{-\frac14}$, 
the Dirichlet half-index (\ref{half_D_3dhm}) reduces to the vacuum character for the fermionic current VOA Fc
\begin{align}
\label{fc_ch}
\mathbb{II}^{\textrm{3d HM}}_{D'}(t=q^{-\frac14},x;q)
&=(qx;q)_{\infty}(qx^{-1};q)_{\infty}
\nonumber\\
&=\chi_{\textrm{Fc}}(x;q). 
\end{align}

%half index of twisted hyper 
The half-indices for 3d twisted hypermultiplet can be obtained by setting $t\rightarrow t^{-1}$. 
The half-index of Neumann b.c. $N$ for 3d $\mathcal{N}=4$ twisted hypermultiplet is 
\begin{align}
\label{half_N_3dthm}
\mathbb{II}^{\textrm{3d tHM}}_{N}(t,x;q)
&=\frac{1}{(q^{\frac14}t^{-1}x;q)_{\infty} (q^{\frac14}t^{-1}x^{-1};q)_{\infty}}
\end{align}
and the half-index of Dirichlet b.c. $D$ for 3d $\mathcal{N}=4$ twisted hypermultiplet is 
\begin{align}
\label{half_D_3dthm}
\mathbb{II}^{\textrm{3d tHM}}_{D}(t,x;q)
&=(q^{\frac34}t x;q)_{\infty} (q^{\frac34}t x^{-1};q)_{\infty}. 
\end{align}

%%%%%%%%%%%%%%%%%%%%%%%%%%%%%%%%%%%%%%%%%%%%%
\subsubsection{3d gauge multiplets}
\label{sec_3dn4_h_g}
%%%%%%%%%%%%%%%%%%%%%%%%%%%%%%%%%%%%%%%%%%%%%

%U(1) vector (0,4) b.c.
The $\mathcal{N}=(0,4)$ half-BPS boundary conditions for 3d $\mathcal{N}=4$ $U(1)$ vector multiplet 
admit the Neumann b.c. $\mathcal{N}'$ and Dirichlet b.c. $\mathcal{D}'$
\begin{align}
\label{bc_3dvm}
\begin{array}{cccc}
\mathcal{N}':&F_{2\mu}|_{\partial}=0,&D_{\mu}\sigma|_{\partial}=0,&D_{\mu}\varphi|_{\partial}=0, \\
\mathcal{D}':&F_{\mu\nu}|_{\partial}=0,&D_{2}\sigma|_{\partial}=0,&D_{2}\varphi|_{\partial}=0, \\
\end{array}
\qquad \mu,\nu=0,1
\end{align}
For Neumann b.c. $\mathcal{N}'$, 
the 3d gauge fields and the left-moving gauginos can fluctuate at the boundary, 
while for Dirichlet b.c. $\mathcal{D}'$, 
the gauge symmetry is broken and one keeps scalar fields $\sigma$, $\varphi$ and 
their superpartners, the right-moving fermions at the boundary \cite{Chung:2016pgt}.

%3d half U(1)
The half index of Neumann b.c. $\mathcal{N}'$ for 3d $\mathcal{N}=4$ $U(1)$ vector multiplet takes the form
\begin{align}
\label{half_N_3du1}
\mathbb{II}^{\textrm{3d $U(1)$}}_{\mathcal{N}'}(t;q)&=
(q)_{\infty} (q^{\frac12}t^2;q)_{\infty}
\oint \frac{ds}{2\pi is}. 
\end{align}
When there is no charged 2d bosonic matter field at the two-dimensional boundary, 
the integration contour for variable $s$ can be taken as a unit circle. 
Except for the enriched Neumann boundary condition discussed in section \ref{sec_enrichneu_bc}, 
we choose the contour of gauge fugacity plane as a unit circle or torus in this paper. 
This allows us to expand the integrand as a series in $q$ and then to project onto the gauge invariants.  
However, as we discuss in section \ref{sec_enrichneu_bc}, 
when we try to compute half-indices of 3d gauge theories with Neumann b.c. 
coupled to 2d bosonic boundary matter, field configuration of 2d bosonic matter fields 
may support some kind of boundary monopole operator, which requires to deform the contour.

%Dirichlet b.c. 
Dirichlet b.c. $\mathcal{D}'$ for 3d $\mathcal{N}=4$ $U(1)$ vector multiplet 
leads to the following perturbative contributions to the half-index
\begin{align}
\label{half_D_3du1p}
\mathbb{II}^{\textrm{3d pert $U(1)$}}_{\mathcal{D}'}
(t;q)
&=
\frac{1}{(q)_{\infty} (q^{\frac12}t^{-2};q)_{\infty}}.
\end{align}
In addition, the Dirichlet half-index receives a non-perturbative contributions from monopole operators 
as Dirichlet b.c. $\mathcal{D}'$ allows for the boundary monopole operator \cite{Bullimore:2016nji}.   
The non-perturbative completion would be achieved by the formula proposed in \cite{Dimofte:2017tpi}
\footnote{
More precisely, the formula (\ref{half_D_3du1}) is written as 
\begin{align}
\label{half_D_3du1a}
\mathbb{II}^{\textrm{3d $U(1)$}}_{\mathcal{D}'}
(t;q)
&=
\frac{1}{(q)_{\infty} (q^{\frac12} t^{-2};q)_{\infty}}
\sum_{m\in \mathbb{Z}}
q^{\frac12 (m \mathbf{0}) k_{\textrm{eff}} \left(\begin{smallmatrix}m\\ \mathbf{0} \end{smallmatrix}\right)}
e^{(m \mathbf{0}) k_{\textrm{eff}} \left(\begin{smallmatrix}\log u\\ \log x\\ \end{smallmatrix}\right)}
\times 
\left[
\textrm{matter indices}
\right](q^{m}u)
\end{align}
where $\mathbf{0}$ is a zero vector with dimension being the number of the global fugacities $x$ 
and $k_{\textrm{eff}}$ takes the form of matrix that encodes the anomaly coefficients: 
\begin{align}
\label{k_AN}
k_{\textrm{eff}}&=
\left(
\begin{matrix}
\textrm{broken gauge anomaly coeff.}
&\textrm{mixed anomaly coeff.}\\ 
\textrm{mixed anomaly coeff.}&*
 \\
\end{matrix}
\right). 
\end{align}
Since the anomaly coefficient for the broken $U(1)_{\partial}$ gauge symmetry is only contributed from 
the 3d bosonic matter and it is always positive for the Dirichlet b.c. \cite{Hanany:2018hlz, Gaiotto:2019jvo}, 
the term $q^{\frac12 k_{\textrm{eff}}m^2}$ involves only positive powers of $q$ 
so that the formula (\ref{half_D_3du1}) gives well-defined $q$-series. 
}
\begin{align}
\label{half_D_3du1}
\mathbb{II}^{\textrm{3d $U(1)$}}_{\mathcal{D}'}
(t;q)
&=
\frac{1}{(q)_{\infty} (q^{\frac12} t^{-2};q)_{\infty}}
\sum_{m\in \mathbb{Z}}
q^{\frac{1}{2}k_{\textrm{eff}} m^2} u^{k_{\textrm{eff}}m} x^{k_{\textrm{eff}}m} 
\times 
\left[
\textrm{matter indices}
\right](q^{m}u)
.
\end{align}
Here $u$ is the fugacity for the boundary $U(1)_{\partial}$ global symmetry 
resulting from the broken $U(1)$ gauge symmetry 
and $x$ is for some other global symmetry $F$ with a boundary anomaly. 
The $\sum_{m\in \mathbb{Z}}$ stands for the summation over magnetic flux sectors. 
This prescription is distinguished from the half-index of Dirichlet b.c. for 4d gauge theories \cite{Gaiotto:2019jvo,Okazaki:2019ony}. 

The effective Chern-Simons coupling $k_{\textrm{eff}}$ encodes the total boundary $U(1)_{\partial}$ and $F$ anomalies 
in such a way that 
the boundary anomaly can be viewed as the exterior derivative of a Cherm-Simons form at level $k_{\textrm{eff}}$ \cite{Dimofte:2017tpi}.  
The terms $q^{\frac{1}{2}k_{\textrm{eff}}m^2}$ $u^{k_{\textrm{eff}}m}$ $x^{k_{\textrm{eff}}m}$ 
are the contributions from bare monopole operators with electric charge $k_{\textrm{eff}}m$, 
global charge $k_{\textrm{eff}}m$ and spin $\frac{k_{\textrm{eff}}}{2}\Tr (m^2)$. 
Since electrically charged states get effective spin in the presence of magnetic flux, 
the matter indices receive the shift of fugacity $u\rightarrow q^{m}s$.

%3d non-Abelian
For 3d $\mathcal{N}=4$ non-Abelian vector multiplet 
the $\mathcal{N}=(0,4)$ Dirichlet boundary condition $\mathcal{D}'$ would become more general boundary condition 
which allows the bulk fields to have a singular profile, that is Nahm pole boundary condition \cite{Chung:2016pgt}:
\begin{align}
\label{bc_3dvm}
\begin{array}{cccc}
\mathcal{N}':&F_{2\mu}|_{\partial}=0,&D_{\mu}\sigma|_{\partial}=0,&D_{\mu}\varphi|_{\partial}=0, \\
\textrm{Nahm}':&F_{\mu\nu}|_{\partial}=0,&D_{2}\vec{\phi}+\vec{\phi}\times \vec{\phi}|_{\partial}=0,& \\
\end{array}
\qquad \mu,\nu=0,1
\end{align}
where $\vec{\phi}$ is the $SU(2)_{C}$ triplet consisting of a real scalar $\sigma$ and a complex scalar $\varphi$. 

%3d half U(2) Neumann
The half-index of Neumann b.c. $\mathcal{N}'$ for 3d $\mathcal{N}=4$ $U(2)$ vector multiplet is 
\begin{align}
\label{half_N_3du2}
\mathbb{II}^{\textrm{3d $U(2)$}}_{\mathcal{N}'}(t;q)&=
\frac12 (q)_{\infty}^2 (q^{\frac12}t^2;q)_{\infty}^2 
\oint \frac{ds_{1}}{2\pi is_{1}} \frac{ds_{2}}{2\pi is_{2}}
\nonumber\\
&\times 
\left(\frac{s_{1}}{s_{2}};q \right)_{\infty}
\left(\frac{s_{2}}{s_{1}};q \right)_{\infty}
\left(q^{\frac12} t^2 \frac{s_{1}}{s_{2}};q \right)_{\infty}
\left(q^{\frac12} t^2 \frac{s_{2}}{s_{1}};q \right)_{\infty}. 
\end{align}
Again the integration contour for variable $s_{i}$ can be taken as a unit torus $\mathbb{T}^2$ 
when there is no 2d bosonic matter field at the two-dimensional boundary. 

Choosing the integration contour as a unit torus $\mathbb{T}^2$, 
one can evaluate the half-index (\ref{half_N_3du2}) as
\begin{align}
\label{half_N_3du2a}
\mathbb{II}^{\textrm{3d $U(2)$}}_{\mathcal{N}'}(t;q)
&=\frac12 
\frac{(q^{\frac12}t^2;q)_{\infty}^2}
{(q)_{\infty}^2} 
\sum_{n=0}^{\infty}\sum_{k=0}^{n} \sum_{l=0}^{n}
(q^{1+k};q)_{\infty}
(q^{1+n-k};q)_{\infty}
(q^{1+l};q)_{\infty}
(q^{1+n-l};q)_{\infty}
\nonumber\\
&\times 
q^{n^2-n+k(k-n)+l(l-n)+\frac12(k+l)} \cdot t^{2(k+l)}. 
\end{align}
The first term in the sum takes the form of square of half-index $\mathbb{II}_{\mathcal{N}'}^{\textrm{4d $U(1)$}}$ 
for Neumann b.c. $\mathcal{N}'$ for 3d $U(1)$ gauge theory. 
We may associate this expansion to a Higgsing precess that leads to a decomposition $U(2)\rightarrow U(1)\times U(1)$.

%3d half U(N) Neu
The half-index of Neumann b.c. $\mathcal{N}'$ for 3d $\mathcal{N}=4$ $U(N)$ vector multiplet is 
\begin{align}
\label{half_N_3duN}
\mathbb{II}^{\textrm{3d $U(N)$}}_{\mathcal{N}'}(t;q)
&=
\frac{1}{N!}(q)_{\infty}^{N}(q^{\frac12}t^{2};q)_{\infty}^{N} 
\oint \prod_{i=1}^{N}\frac{ds_{i}}{2\pi is_{i}} 
\prod_{i\neq j}\left(\frac{s_{i}}{s_{j}};q \right)_{\infty} 
\left( q^{\frac12}t^2 \frac{s_{i}}{s_{j}};q \right)_{\infty}. 
\end{align}
Again when there is no charged 2d bosonic matter field at the two-dimensional boundary, 
the integration contour for gauge fugacities $s_{i}$ can be taken as a unit torus $\mathbb{T}^{N}$.

%3d half U(N) Dirichlet
The Dirichlet b.c. $\mathcal{D}'$ for 3d $\mathcal{N}=4$ non-Abelian vector multiplet should be realized as a trivial Nahm$'$ boundary condition. 
The half-index of Dirichlet b.c. $\mathcal{D}'$ for 3d $\mathcal{N}=4$ vector multiplet of non-Abelian gauge group $G$ 
should receive the non-perturbative monopole contributions with the form
\begin{align}
\label{half_D_3duN}
\mathbb{II}^{\textrm{3d $G$}}_{\mathcal{D}'}(t,u;q)
&=
\left[
\frac{1}{(q)_{\infty} (q^{\frac12}t^{-2};q)_{\infty}}
\right]^{\mathrm{rank}(G)}
\sum_{m\in \textrm{cochar}(G)}
\left[
\prod_{\alpha\in \textrm{roots}(G)}
\frac{1}{(q^{1+m\cdot \alpha}u^{\alpha};q)_{\infty}(q^{\frac12+m\cdot \alpha}t^{-2}u^{\alpha};q)_{\infty}}
\right]
\nonumber\\
&\times 
q^{\frac12 k_{\textrm{eff}} \Tr (m^2)}
\cdot 
u^{k_{\textrm{eff}}m} 
\cdot 
x^{k_{\textrm{eff}}m}
\times 
\left[
\textrm{matter indices}
\right](q^{m}u)
\end{align}
where $u$ is the fugacity for the boundary $G_{\partial}$ global symmetry 
and $x$ is that for some other global symmetry $F$. 

Similarly to the Abelian case, the effective Chern-Simons coupling $k_{\textrm{eff}}$ encodes the boundary 't Hooft anomaly. 
$\sum_{m\in \textrm{cochar}(G)}$ is now the summation over cocharacter of $G$. 
The terms in the second line include the contributions from bare monopole and the fugacity $u$ in the matter indices are shifted due to the effective spin carried by charged states.

%%%%%%%%%%%%%%%%%%%%%%%%%%%%%%%%%%%%%%%%%%%%%
\subsection{Elliptic genera of 2d $\mathcal{N}=(0,4)$ gauge theories}
\label{sec_04_index}
%%%%%%%%%%%%%%%%%%%%%%%%%%%%%%%%%%%%%%%%%%%%%
When the both 4d theory and 3d theory are trivial, 
the quarter-index (\ref{INDEX_def}) is simply the elliptic genus of 2d $\mathcal{N}=(0,4)$ gauge theory 
\cite{Gadde:2015tra, Kim:2014dza, Putrov:2015jpa, Kim:2018gak}.

%%%%%%%%%%%%%%%%%%%%%%%%%%%%%%%%%%%%%%%%%%%%%
\subsubsection{2d matter multiplets}
\label{sec_2d04_m}
%%%%%%%%%%%%%%%%%%%%%%%%%%%%%%%%%%%%%%%%%%%%%

The full-index of 2d $\mathcal{N}=(0,2)$ chiral multiplet is 
\begin{align}
\label{cm}
C(x;q)&=\frac{1}{(x;q)_{\infty} (qx^{-1};q)_{\infty}}
\end{align}
and the full-index of 2d $\mathcal{N}=(0,2)$ Fermi multiplet is 
\begin{align}
\label{fm}
F(x;q)&=(x;q)_{\infty}(qx^{-1};q)_{\infty}. 
\end{align}

The $\mathcal{N}=(0,4)$ hypermultiplet consists of a pair of $\mathcal{N}=(0,2)$ chiral multiplets. 
Under the R-symmetry $SU(2)_{C}\times SU(2)_{H}$ 
the chiral field and the anti-chiral field transform as $({\bf 1}, {\bf 2})$. 

The full-index of 2d $\mathcal{N}=(0,4)$ hypermultiplet is given by
\begin{align}
\label{04hm}
\mathbb{I}^{\textrm{2d HM}}(t,x;q)
&=
\frac{1}
{
(q^{\frac14}tx^{\pm};q)_{\infty}
(q^{\frac34}t^{-1}x^{\pm};q)_{\infty}
}
\nonumber\\
&=
C(q^{\frac14}tx;q)\times 
C(q^{\frac14}tx^{-1};q).
\end{align}
This can be expanded as 
\begin{align}
\label{04hm1}
&
\mathbb{I}^{\textrm{2d HM}}(t,x;q)
\nonumber\\
&=
\frac{1}{(q)_{\infty}^4}
\sum_{n=0}^{\infty}
\sum_{k=0}^{n}
\sum_{l=0}^{n}
\frac{
(q^{1+k};q)_{\infty}
(q^{1+l};q)_{\infty}
(q^{1+n-k};q)_{\infty}
(q^{1+n-l};q)_{\infty}
}
{(q^{1+n};q)_{\infty}}
%\nonumber\\
%&\times 
q^{\frac{3n}{4}-\frac{k}{2}} 
\cdot 
t^{2k-n}
\cdot 
x^{2l-n}. 
\end{align}
In the H-twist limit, the hypermultiplet index (\ref{04hm}) becomes 
the square of the vacuum character (\ref{sb_ch}) for the symplectic boson VOA Sb.

Similarly, the $\mathcal{N}=(0,4)$ twisted hypermultiplet consists of a pair of $\mathcal{N}=(0,2)$ chiral multiplets. 
Under the R-symmetry $SU(2)_{C}\times SU(2)_{H}$ 
the chiral field and the anti-chiral field transform as $({\bf 2}, {\bf 1})$. 

By setting $t\rightarrow t^{-1}$, 
we can obtain the index for 2d $\mathcal{N}=(0,4)$ twisted hypermultiplet
\begin{align}
\label{04thm1}
\mathbb{I}^{\textrm{2d tHM}} (t,x;q)
&=
\frac{1}
{
(q^{\frac14}t^{-1}x^{\pm};q)_{\infty}
(q^{\frac34}t x^{\pm};q)_{\infty}
}
\nonumber\\
&=
C(q^{\frac14}t^{-1}x;q)\times 
C(q^{\frac14}t^{-1}x^{-1};q).
\end{align}
In the C-twist limit, the twisted hypermultiplet index (\ref{04thm1}) becomes 
the square of the vacuum character (\ref{sb_ch}) for the symplectic boson VOA Sb.

%flip

%%%%%%%%%%%%%%%%%%%%%%%%%%%%%%%%%%%%%%%%%%%%%
\subsubsection{2d vector multiplets}
\label{sec_2d04_g}
%%%%%%%%%%%%%%%%%%%%%%%%%%%%%%%%%%%%%%%%%%%%%
The index of 2d $\mathcal{N}=(0,4)$ $U(1)$ vector multiplet takes the form 
\begin{align}
\label{04u1vm}
\mathbb{I}^{\textrm{2d $U(1)$}}(t;q)
&=(q)_{\infty}^2 (q^{\frac12}t^2;q)_{\infty} (q^{\frac12}t^{-2};q)_{\infty}
\oint_{\mathrm{JK}} \frac{ds}{2\pi is}. 
\end{align}
Here the subscript ``JK'' in the contour integral indicates that the prescription for computing the 2d elliptic genus 
is not the naive zero charge projection given by a contour integral 
along the unit circle $|s|=1$ in the gauge fugacity plane. 

Instead, the integrand for a non-anomalous 2d symmetry is a well-defined function 
on the fugacity torus, i.e. invariant under $s\rightarrow qs$. 
The Jeffrey-Kirwan (JK) residue prescription involves a contour integral 
around poles associated to $\mathcal{N}=(0,2)$ chiral fields of, say, negative charge 
\cite{MR1318878, Benini:2013nda, Benini:2013xpa}. 
In the case of hypers or twisted hypers, 
this should pick up poles of the form $x=q^{\frac14}t^{\pm1}$. 

Notice that the  naive prescription cannot be right, as it is not modular invariant. 
On the other hand, the naive prescription clearly accounts for 
gauge-invariant operators built from elementary fields. 
The full prescription must include corrections 
due to non-local operators from twisted sectors under gauge symmetries.

%%%%%%%%%%%%%%%%%%%%%%%%%%%%%%%%%%%%%%%%%%%%%
%%%%%%%%%%%%%%%%%%%%%%%%%%%%%%%%%%%%%%%%%%%%%
\section{$\mathcal{N}=(0,4)$ box models}
\label{sec_04mirror}
%%%%%%%%%%%%%%%%%%%%%%%%%%%%%%%%%%%%%%%%%%%%%
%%%%%%%%%%%%%%%%%%%%%%%%%%%%%%%%%%%%%%%%%%%%%
In this section, we analyze simple examples of 
$\mathcal{N}=(0,4)$ brane box models in \cite{Hanany:2018hlz}. 
In particular, we propose simple $\mathcal{N}=(0,4)$ mirror symmetry 
that may be related to the 3d $\mathcal{N}=4$ mirror symmetry 
between 3d SQED$_{1}$ and free twisted hypermultiplet \cite{Kapustin:1999ha}.

%%%%%%%%%%%%%%%%%%%%%%%%%%%%%%%%%%%%%%%%%%%%%
\subsection{Abelian $\mathcal{N}=(0,4)$ theories}
\label{sec_04ab}
%%%%%%%%%%%%%%%%%%%%%%%%%%%%%%%%%%%%%%%%%%%%%

%%%%%%%%%%%%%%%%%%%%%%%%%%%%%%%%%%%%%%%%%%%%%
\subsubsection{2d $U(1)+\textrm{hyper}+\textrm{2 Fermi}$}
\label{sec_u1hm1fm2}
%%%%%%%%%%%%%%%%%%%%%%%%%%%%%%%%%%%%%%%%%%%%%
We start from a 2d $\mathcal{N}=(0,4)$ $U(1)$ gauge theory with a charged hypermultiplet and two charged Fermi multiplets. 
Let ${\bf s}$ be the field strength of $U(1)$ gauge symmetry. 
The gauge anomaly polynomial is 
\footnote{See \cite{Hanany:2018hlz,Gaiotto:2019jvo} for conventions of (boundary) anomalies. }
\begin{align}
\label{2du1hm1fm2_AN}
\mathcal{I}^{\textrm{2d $U(1)+1$HM$+2$FM}}
&=
\underbrace{-2{\bf s}^2}
_{\textrm{2d hyper}}
+
\underbrace{
2\cdot {\bf s}^2
}_{\textrm{2d Fermi}}
=0
\end{align}
so that the theory is free from gauge anomaly. 

\begin{figure}
\begin{center}
\includegraphics[width=12cm]{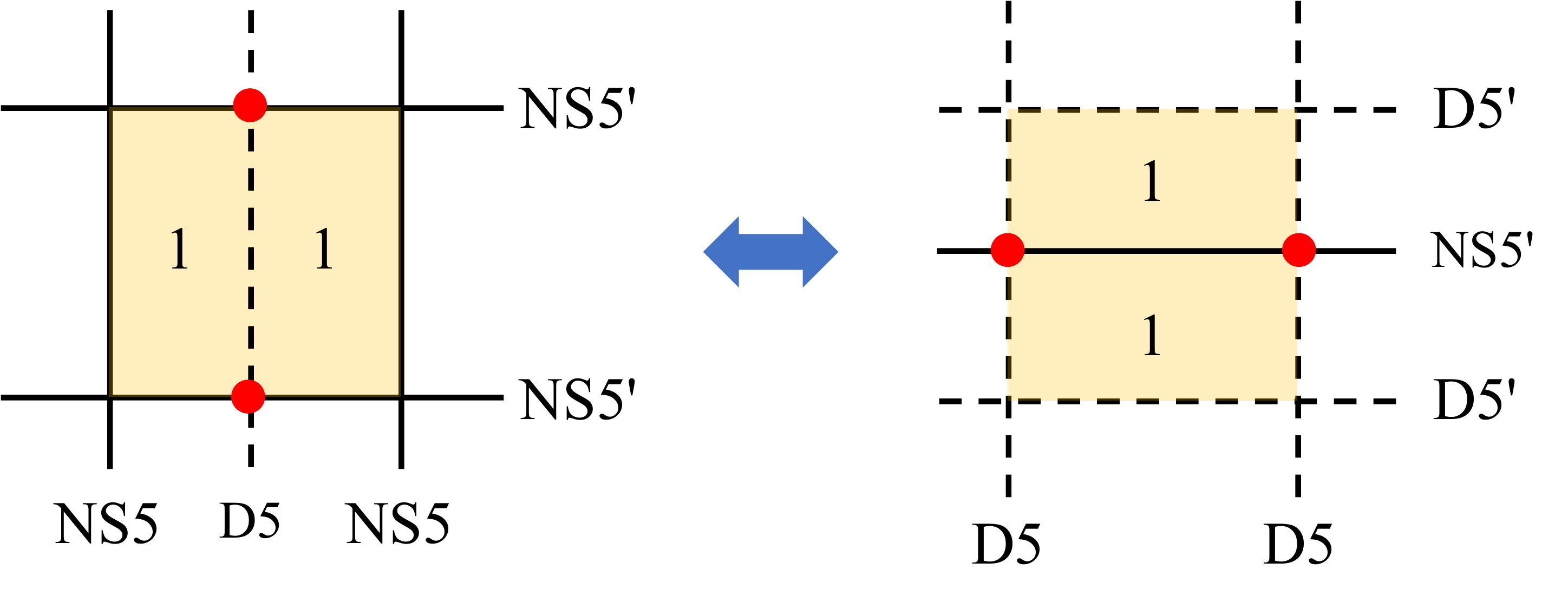}
\caption{
The brane box configurations of 
2d $\mathcal{N}=(0,4)$ $U(1)$ gauge theory with 
a charged hypermultiplet and two charged Fermi multiplets 
and of its mirror $\mathcal{N}=(0,4)$ Fermi multiplet. 
We take the horizontal and vertical directions along the $x^6$ and $x^2$ respectively.  
}
\label{fig04hm1fm2}
\end{center}
\end{figure}

The theory is constructed by the brane box configuration in Figure \ref{fig04hm1fm2}. 
Here and  hereafter we take the horizontal and vertical directions along the $x^6$ and $x^2$ respectively.  
The 2d $\mathcal{N}=(0,4)$ $U(1)$ vector multiplet appears from the D3-brane box configuration 
surrounded by NS5- and NS5$'$-branes, 
that is a 3d $\mathcal{N}=4$ $U(1)$ vector multiplet on a segment with Neumann b.c. $\mathcal{N}'$ at each end. 
Correspondingly, we have a schematic relation
\begin{align}
\label{04vm_NN}
\mathbb{I}^{\textrm{2d $U(1)$}}
&=
\frac{
\mathbb{II}_{\mathcal{N}'}^{\textrm{3d $U(1)$}}
\times 
\mathbb{II}_{\mathcal{N}'}^{\textrm{3d $U(1)$}}
}
{
\mathbb{I}^{\textrm{3d pert $U(1)$}}
}
\end{align}
where $\mathbb{I}^{\textrm{3d pert $U(1)$}}$ 
$=$ $(q^{\frac12}t^2;q)_{\infty}$ $(q^{\frac12}t^{-2};q)_{\infty}^{-1}$
is the perturbative full-index of 
3d $\mathcal{N}=4$ $U(1)$ vector multiplet \cite{Okazaki:2019ony}. 

The D5-brane gives rise to 
a 2d $\mathcal{N}=(0,4)$ fundamental hypermultiplet and 
two Fermi multiplets. 
The Fermi multiplets appear at the two intersections of D3- and D5-branes \cite{Hanany:2018hlz}, 
which are depicted as red dots in Figure \ref{fig04hm1fm2}. 

The index of the theory 2d $U(1)+\textrm{hyper}+\textrm{2 Fermi}$ 
 is evaluated as 
\begin{align}
\label{04u1hm1fm2a}
&
\mathbb{I}^{\textrm{2d $U(1)+$HM$+2$FM}}(t,z;q)
\nonumber\\
&=
\underbrace{
(q)_{\infty}^2 (q^{\frac12}t^2;q)_{\infty}(q^{\frac12}t^{-2};q)_{\infty}
\oint_{\textrm{JK}} \frac{ds}{2\pi is}
}_{\mathbb{I}^{\textrm{2d $U(1)$}}}
\cdot 
\underbrace{
\frac{
(q^{\frac12}s^{\pm}z^{\pm};q)_{\infty}
(q^{\frac12}s^{\pm}z^{\mp};q)_{\infty}
}{
(q^{\frac14}t s^{\pm};q)_{\infty}
(q^{\frac34}t^{-1}s^{\pm};q)_{\infty}
}
}_{\mathbb{I}^{\textrm{2d HM}} (s)\cdot F(q^{\frac12} sz)\cdot F(q^{\frac12}sz^{-1})}. 
\end{align}
The choice of Fermi multiplet fugacities makes the integrand modular. 
Picking the pole of hypermultiplet at $s=q^{\frac14}t$ gives cancellations 
and we find that
\begin{align}
\label{04u1hm1fm2b}
&
\mathbb{I}^{\textrm{2d $U(1)+$HM$+2$FM}}(t,z;q)
\nonumber\\
&=
\underbrace{
(q^{\frac34}tz;q)_{\infty}(q^{\frac14}t^{-1}z^{-1};q)_{\infty}
}_{F(q^{\frac34}tz)}
\cdot 
\underbrace{
(q^{\frac34}tz^{-1};q)_{\infty}(q^{\frac14}t^{-1}z;q)_{\infty}
}_{F(q^{\frac34}tz^{-1})}. 
\end{align}
This is identified with 
the index of the sort of $\mathcal{N}=(0,4)$ Fermi 
which arises from a 3d twisted hypermultiplet on a segment with Dirichlet boundary conditions $D$. 

On the other hand, the original 2d $\mathcal{N}=(0,4)$ gauge theory can be interpreted as 3d $\mathcal{N}=4$ SQED$_{1}$ 
on a segment with Neumann b.c. $\mathcal{N}'$. 
Therefore this is compatible with 3d $\mathcal{N}=4$ mirror symmetry. 
Our index computation indicates the minimal $\mathcal{N}=(0,4)$ mirror symmetry: 
\begin{align}
\label{2d04mirror1}
\begin{array}{ccc}
\textrm{2d $\mathcal{N}=(0,4)$ $U(1)$ gauge theory with $1$ hyper$+2$ Fermi}&\leftrightarrow&\textrm{2d free $\mathcal{N}=(0,4)$ Fermi} \\
\end{array}.
\end{align}
We note that the duality (\ref{2d04mirror1}) can be viewed as 
a sort of $\mathcal{N}=(0,4)$ generalization of the duality 
between 2d $\mathcal{N}=(0,2)$ SQED and free fermions proposed in \cite{Gadde:2013sca, Gadde:2013lxa}.

%%%%%%%%%%%%%%%%%%%%%%%%%%%%%%%%%%%%%%%%%%%%%
\subsubsection{2d $U(1)+\textrm{2 hypers}+\textrm{4 Fermi}$}
\label{sec_u1hm2fm4}
%%%%%%%%%%%%%%%%%%%%%%%%%%%%%%%%%%%%%%%%%%%%%
Next consider a 2d $\mathcal{N}=(0,4)$ $U(1)$ gauge theory with two charged hypermultiplets and four charged Fermi multiplets. 
The gauge anomaly polynomial is 
\begin{align}
\label{2du1hm2fm4_AN}
\mathcal{I}^{\textrm{2d $U(1)+2$HM$+4$FM}}
&=
\underbrace{-2\cdot 2{\bf s}^2}
_{\textrm{2d hypers}}
+
\underbrace{
4\cdot {\bf s}^2
}_{\textrm{2d Fermi}}
=0
\end{align}
and the theory is free from gauge anomaly.

\begin{figure}
\begin{center}
\includegraphics[width=12cm]{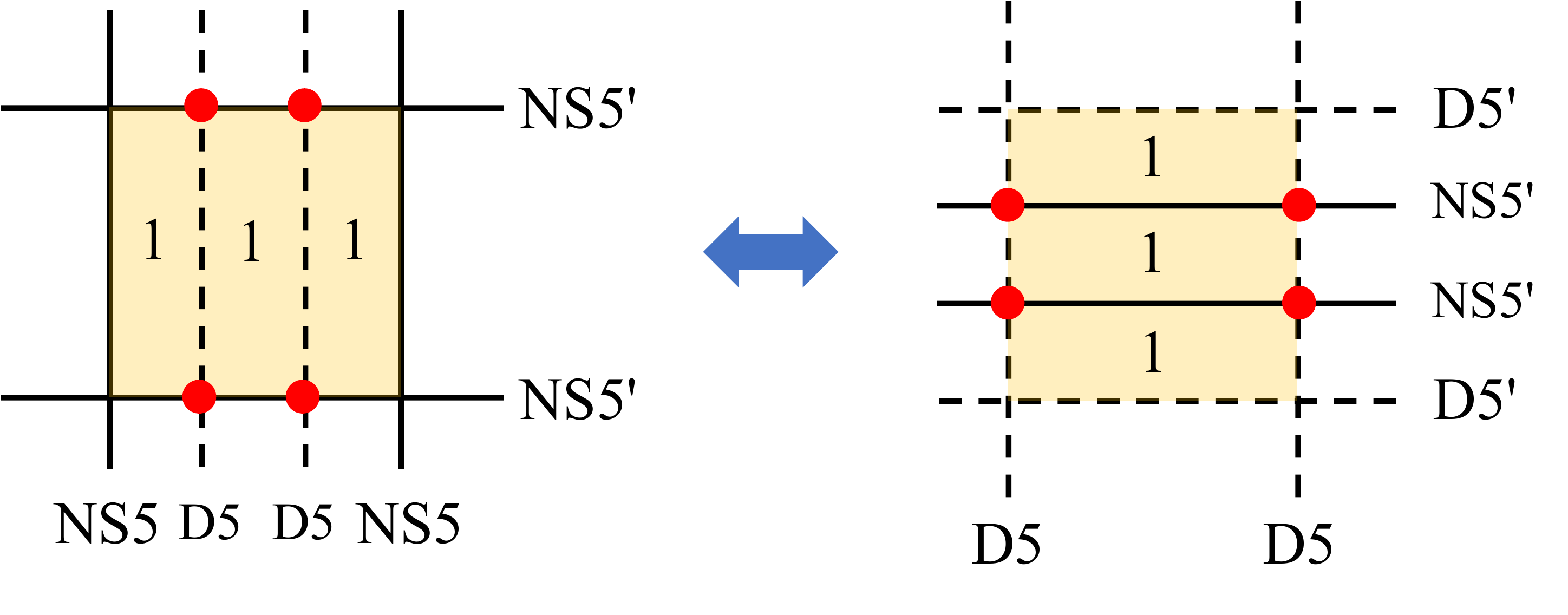}
\caption{
The brane box configuration of 
2d $\mathcal{N}=(0,4)$ $U(1)$ gauge theory with 
two charged hypermultiplets and four charged Fermi multiplets 
and the S-dual configuration. 
}
\label{fig04u1hm2fm4}
\end{center}
\end{figure}

The brane construction is depicted in Figure \ref{fig04u1hm2fm4}. 

The index takes the form 
\begin{align}
\label{04u1hm2fm4_1}
&
\mathbb{I}^{\textrm{2d $U(1)+2$HM$+4$FM}}(t,x,z_{\alpha};q)
\nonumber\\
&=
\underbrace{
(q)_{\infty}^2 (q^{\frac12}t^2;q)_{\infty} 
(q^{\frac12}t^{-2};q)_{\infty}
\oint_{\textrm{JK}} 
\frac{ds}{2\pi is}
}_{\mathbb{I}^{\textrm{2d $U(1)$}}}
\nonumber\\
&\times 
\prod_{\alpha=1}^{4} 
\underbrace{
\frac{
(q^{\frac12} s^{\pm} z_{\alpha}^{\pm};q)_{\infty}
}
{
(q^{\frac14}t s^{\pm} x^{\pm};q)_{\infty}
(q^{\frac34}t^{-1} s^{\pm} x^{\pm})_{\infty}
(q^{\frac14}t s^{\pm} x^{\mp};q)_{\infty}
(q^{\frac34}t^{-1} s^{\pm} x^{\mp};q)_{\infty}
}
}_{
\mathbb{I}^{\textrm{2d HM}}(sx)\cdot 
\mathbb{I}^{\textrm{2d HM}}(sx^{-1})\cdot 
F\left(q^{\frac12}sz_{\alpha}\right)
}
\end{align}
with $\prod_{\alpha=1}^{4}z_{\alpha}=1$. 
Here we have a total of 8 factors both in the denominator and in the numerator. 

Picking the pole for 2d hypermultiplet at $s=q^{\frac14}t x$, we get
\begin{align}
\label{04u1hm2fm4_2a}
&
\mathbb{I}^{\textrm{2d $U(1)+2$HM$+4$FM}}_{1}(t,x,z_{\alpha};q)
\nonumber\\
&=
\prod_{\alpha=1}^{4}
\underbrace{
\frac{
(q^{\frac34}txz_{\alpha};q)_{\infty}
(q^{\frac14}t^{-1}x^{-1}z_{\alpha}^{-1};q)_{\infty}
}
{
(q^{\frac12}t^2x^2;q)_{\infty} 
(q^{\frac12}t^{-2}x^{-2};q)_{\infty}
(qx^{2};q)_{\infty}
(x^{-2};q)_{\infty}
}
}_{\mathbb{I}^{\textrm{2d HM}} (q^{\frac14}tx^{2})\cdot 
F(q^{\frac34}tx z_{\alpha})}.
\end{align}
The another pole for 2d hypermultiplet at $s=q^{\frac14}tx^{-1}$ gives 
\begin{align}
\label{04u1hm2fm4_2b}
&
\mathbb{I}^{\textrm{2d $U(1)+2$HM$+4$FM}}_{2}(t,x,z_{\alpha};q)
\nonumber\\
&=
\prod_{\alpha=1}^{4}
\underbrace{
\frac{
(q^{\frac34}tx^{-1}z_{\alpha};q)_{\infty}
(q^{\frac14}t^{-1}xz_{\alpha}^{-1};q)_{\infty}
}
{
(q^{\frac12}t^2 x^{-2};q)_{\infty}
(q^{\frac12}t^{-2}x^{2};q)_{\infty}
(x^{2};q)_{\infty}
(qx^{-2};q)_{\infty}
}
}_{\mathbb{I}^{\textrm{2d HM}}(q^{\frac14}tx^{-2}) \cdot F(q^{\frac34}tx^{-1}z_{\alpha})}. 
\end{align}
The total index is the sum of the two contributions:
\begin{align}
\label{04u1hm2fm4_2}
&
\mathbb{I}^{\textrm{2d $U(1)+2$HM$+4$FM}}(t,x,z_{\alpha};q)
\nonumber\\
&=\sum_{i=1}^{2}\mathbb{I}^{\textrm{2d $U(1)+2$HM$+4$FM}}_{i}(t,x,z_{\alpha};q)
\nonumber\\
&=
1+\Biggl[
t^2 (1+x^2+x^{-2})
+t^{-2}\left( 
1+\sum_{\alpha<\beta}z_{\alpha}^{-1}z_{\beta}^{-1}
\right)
\Biggr]q^{\frac12}
+\cdots
\end{align}
The pole at $x^2=1$ cancels out 
and we get a nice fractional power series in $q$. 

Unfortunately, 
we cannot immediately make contact with a mirror symmetry, 
in the sense that the original 2d gauge theory can be interpreted as 
3d SQED$_{2}$ on a segment with Neumann b.c. $(\mathcal{N}',N')$ 
and the mirror is 3d $\widetilde{\textrm{SQED}}_{2}$ on a segment with Dirichlet b.c. $(\mathcal{D},D)$ 
and the dual photon gives a circle-valued scalar which needs to be treated with extra care. 

Still, the Neumann b.c. of 3d SQED$_{2}$ preserves the $SU(2)$ topological symmetry enhancement 
of the Coulomb branch of 3d SQED$_{2}$. 
The same enhancement should happen then in the 2d setup. 
If we look at the block-diagonal $SU(2)\times SU(2)\times U(1)$ subgroup of the $SU(4)$ rotating the Fermi multiplet, 
the $U(1)$ should enhance to a third $SU(2)$. 
This would imply that $SU(4)\simeq SO(6)$ is enhanced to $SO(7)$. 
This is indeed a symmetry of the index.

%%%%%%%%%%%%%%%%%%%%%%%%%%%%%%%%%%%%%%%%%%%%%
\subsubsection{2d $U(1)+\textrm{hyper}+\textrm{twisted hyper}+\textrm{4 Fermi}+\textrm{neutral Fermi}$}
\label{sec_u1hm1thm1fm4nfm1}
%%%%%%%%%%%%%%%%%%%%%%%%%%%%%%%%%%%%%%%%%%%%%
Let us consider a 2d $\mathcal{N}=(0,4)$ $U(1)$ gauge theory with 
a charged hypermultiplet, a charged twisted hypermultiplet, four charged Fermi multiplets and a neutral Fermi multiplet. 
The gauge anomaly polynomial is 
\begin{align}
\label{2du1hm1thm1fm4_AN}
\mathcal{I}^{\textrm{2d $U(1)+$HM$+$tHM$+4$FM}}
&=
\underbrace{-2{\bf s}^2}
_{\textrm{2d hyper}}
\underbrace{-2{\bf s}^2}
_{\textrm{2d thyper}}
+
\underbrace{
4\cdot {\bf s}^2
}_{\textrm{2d Fermi}}
=0
\end{align}
and the theory is free from gauge anomaly. 
The coexistence of hyper and twisted hyper multiplets requires 
the neutral Fermi multiplet which achieve their cubic coupling \cite{Tong:2014yna}. 

\begin{figure}
\begin{center}
\includegraphics[width=12cm]{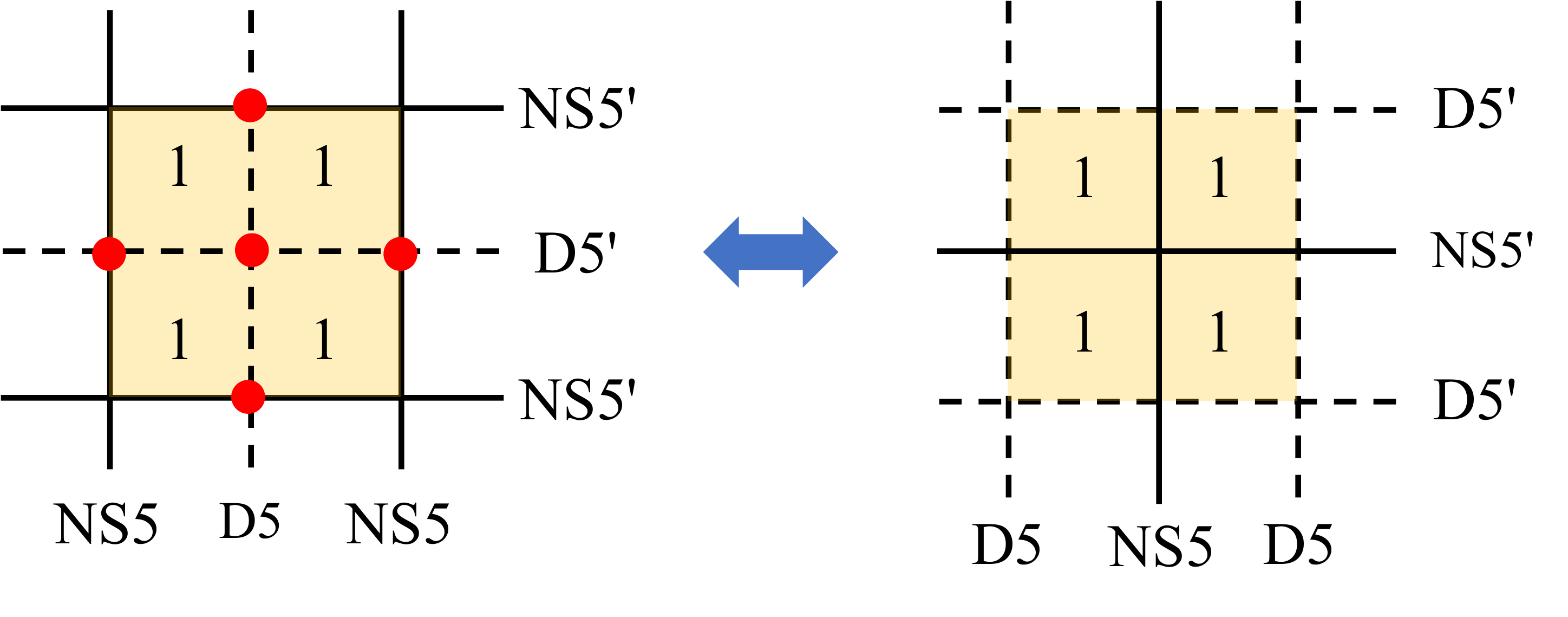}
\caption{
The brane box configuration of 
2d $\mathcal{N}=(0,4)$ $U(1)$ gauge theory with 
a charged hypermultiplet, a charged twisted hypermultiplet, four charged Fermi multiplets and a neutral Fermi multiplet 
and the S-dual configuration. 
}
\label{fig04u1hm1thm1fm4nfm1}
\end{center}
\end{figure}

The brane construction is illustrated in Figure \ref{fig04u1hm1thm1fm4nfm1}.
The $\mathcal{N}=(0,4)$ $U(1)$ vector multiplet is realized as a box of D3-brane, 
the hyper and twisted hyper multiplets are produced by the D5- and D5$'$-branes respectively. 
There are two intersections of D5- and NS$'$-branes  
and two intersections of D5$'$- and NS-branes. 
They lead to four charged Fermi multiplets. 
In addition, a neutral Fermi multiplet appears at the D5-D5$'$ junction.

The index is given by 
\begin{align}
\label{04u1hm1thm1fm4nfm1_1}
&
\mathbb{I}^{\textrm{2d $U(1)+$HM$+$tHM$+4$FM$+$nFM}}(t,x,z_{\alpha};q)
\nonumber\\
&=
\underbrace{
(q)_{\infty}^2(q^{\frac12}t^{2};q)_{\infty} (q^{\frac12}t^{-2};q)_{\infty}
\oint_{\textrm{JK}} \frac{ds}{2\pi is}
}_{\mathbb{I}^{\textrm{2d $U(1)$}}}
\nonumber\\
&\times 
\prod_{\alpha=1}^{4}
\underbrace{
\frac{
(q^{\frac12}s^{\pm}z_{\alpha}^{\pm};q)_{\infty}
\cdot 
(q^{\frac12}x^{\pm2};q)_{\infty}
}
{
(q^{\frac14}ts^{\pm}x^{\mp};q)_{\infty}
(q^{\frac34}t^{-1}s^{\pm}x^{\mp};q)_{\infty}
\cdot 
(q^{\frac14}t^{-1}s^{\pm}x^{\pm};q)_{\infty}
(q^{\frac34}ts^{\pm}x^{\pm};q)_{\infty}
}
}_
{\mathbb{I}^{\textrm{2d HM}} (sx^{-1})
\cdot \mathbb{I}^{\textrm{2d tHM}} (sx)
\cdot F(q^{\frac12}sz_{\alpha})
\cdot F(q^{\frac12}x^2)
}.
\end{align}
Here we have a total of 8 factors in the denominator and 8 factors in the numerator for the integrand. 
We need $\prod_{\alpha=1}^{4}z_{\alpha}=1$. 
We gave the neutral Fermi multiplet a fugacity needed for a cubic coupling to the hyper and twisted hyper. 

Picking the pole for 2d hypermultiplet at $s=q^{\frac14}tx$ gives 
\begin{align}
\label{04u1hm1thm1fm4nfm1_2a}
&
\mathbb{I}^{\textrm{2d $U(1)+$HM$+$tHM$+4$FM$+$nFM}}_{1}(t,x,z_{\alpha};q)
\nonumber\\
&=
\prod_{\alpha=1}^{4}
\underbrace{
\frac{
(q^{\frac34}tz_{\alpha}^{\pm}x^{\pm};q)_{\infty}
}
{
(t^{-2}x^{-2};q)_{\infty}
(qx^{2};q)_{\infty}
}
}_{C(t^{-2}x^{2})\cdot F(q^{\frac34}tz_{\alpha}x)}
\end{align}
where the Fermi multiplet cancels part of the twisted hypermultiplet contribution. 

On the other hand, 
picking the pole for 2d twisted hypermultiplet at $s=q^{\frac14}t^{-1}x^{-1}$, we obtain
\begin{align}
\label{04u1hm1thm1fm4nfm1_2b}
&
\mathbb{I}^{\textrm{2d $U(1)+$HM$+$tHM$+4$FM$+$nFM}}_{2}(t,x,z_{\alpha};q)
\nonumber\\
&=
\prod_{\alpha=1}^{4}
\underbrace{
\frac{
(q^{\frac34}t^{-1}z_{\alpha}^{\pm}x^{\mp};q)_{\infty}
}
{
(t^{2}x^{2};q)_{\infty}
(qt^{-2}x^{-2};q)_{\infty}
}
}_{C(t^2 x^2)\cdot F(q^{\frac34}t^{-1}z_{\alpha}x^{-1})}
\end{align}
with the Fermi multiplet canceling part of the hypermultiplet contribution. 

The total index is the sum of the two contributions 
\begin{align}
\label{04u1hm1thm1fm4nfm1_2}
&
\mathbb{I}^{\textrm{2d $U(1)+$HM$+$tHM$+4$FM$+$nFM}}(t,x,z_{\alpha};q)
\nonumber\\
&=
\sum_{i=1}^{2}
\mathbb{I}^{\textrm{2d $U(1)+$HM$+$tHM$+4$FM$+$nFM}}_{i}(t,x,z_{\alpha};q)
\nonumber\\
&=
1-\frac{
(z_{1}+z_{2}+z_{3}
+z_{1} z_{2}^2 z_{3}^2
+z_{1}^2 z_{2}z_{3}^2
+z_{1}^2 z_{2}^2z_{3}
)
}
{z_{1}z_{2}z_{3}}q^{\frac12}
+\cdots 
\end{align}
%The pole at $x^2 t^2=1$ cancels out and 
We see that 
the resulting index (\ref{04u1hm1thm1fm4nfm1_2}) has no dependence on the fugacities $x$ and $t$. 

From the brane picture in Figure \ref{fig04u1hm1thm1fm4nfm1}, 
the dual theory has no gauge symmetry so that the index is expected to take the form of product of full-indices. 
In fact, we find that this index matches a very simple answer:
\begin{align}
\label{04u1hm1thm1fm4nfm1_3}
&
\mathbb{I}^{\textrm{2d $U(1)+$HM$+$tHM$+4$FM$+$nFM}}(t,x,z_{\alpha};q)
\nonumber\\
&=
\underbrace{
(q^{\frac12}z_{1}^{\pm}z_{2}^{\pm};q)_{\infty}
\cdot 
(q^{\frac12}z_{2}^{\pm}z_{3}^{\pm};q)_{\infty}
\cdot 
(q^{\frac12}z_{3}^{\pm}z_{1}^{\pm};q)_{\infty}
}_
{
F(q^{\frac12}z_{1}z_{2})
\cdot 
F(q^{\frac12}z_{2}z_{3})
\cdot 
F(q^{\frac12}z_{3}z_{1})
}.
\end{align}
This is identified with 
the index of six real Fermi multiplets transforming in the fundamental representation under the $SO(6)$ flavor symmetry. 
Our index computation indicates the $\mathcal{N}=(0,4)$ mirror symmetry: 
\begin{align}
\begin{array}{ccc}
\textrm{2d $\mathcal{N}=(0,4)$ $U(1)$ with $1$ hyper$+1$ thyper
$+4$ Fermi}&\leftrightarrow&\textrm{2d free $SO(6)$ Fermi} \\
\end{array}.
\end{align}

In order to give a 3d or 4d interpretation of this, 
we would need to consider enriched Neumann boundary conditions for 3d gauge theories.

%%%%%%%%%%%%%%%%%%%%%%%%%%%%%%%%%%%%%%%%%%%%%
\subsubsection{2d $U(1)+\textrm{2 hypers}+\textrm{twisted hyper}+\textrm{6 Fermi}+\textrm{2 neutral Fermi}$}
\label{sec_u1hm1thm1fm4nfm1}
%%%%%%%%%%%%%%%%%%%%%%%%%%%%%%%%%%%%%%%%%%%%%
Let us consider a 2d $\mathcal{N}=(0,4)$ $U(1)$ gauge theory with 
two charged hypermultiplets, twisted hypermultiplet, six charged Fermi multiplets and two neutral Fermi multiplets. 
The gauge anomaly polynomial is 
\begin{align}
\label{u1hm1thm1fm4nfm1_AN}
\mathcal{I}^{\textrm{2d $U(1)+2$HM$+$tHM$+6$FM$+2$nFM}}
&=
\underbrace{
-2\cdot 2{\bf s}^2
}_{\textrm{2d hyper}}
\underbrace{
-2{\bf s}^2
}_{\textrm{2d thyper}}
+
\underbrace{
6\cdot {\bf s}^{2}
}_{\textrm{2d Fermi}}
=0
\end{align}
and the theory is gauge anomaly free. 

\begin{figure}
\begin{center}
\includegraphics[width=12cm]{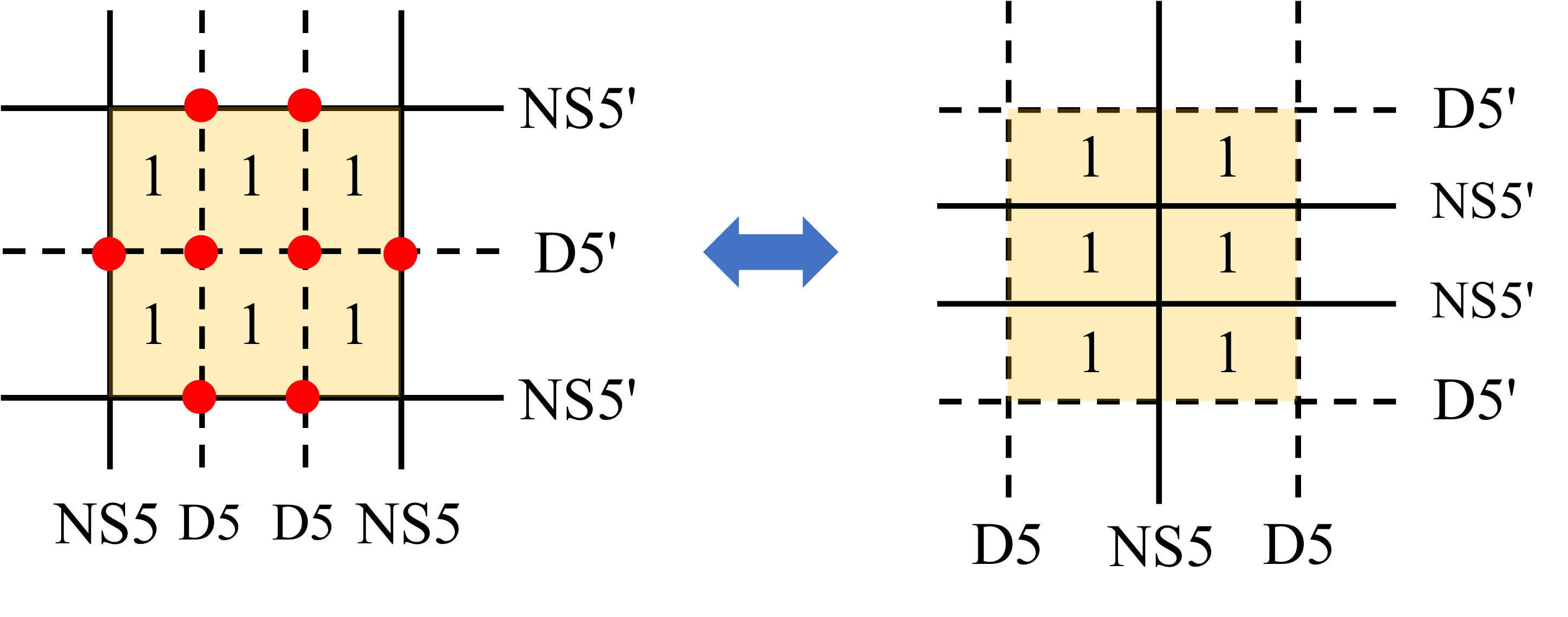}
\caption{
The brane box configuration of 
2d $\mathcal{N}=(0,4)$ $U(1)$ gauge theory with 
two charged hypermultiplets, a charged twisted hypermultiplet, six charged Fermi multiplets and two neutral Fermi multiplets 
and the S-dual configuration. 
}
\label{fig04u1hm2thm1fm6nfm2}
\end{center}
\end{figure}

The brane construction is shown in Figure \ref{fig04u1hm2thm1fm6nfm2}.
The two D5-branes and a D5$'$-brane 
lead to the two 3d charged hypermultiplets and a twisted hyper multiplet. 
Corresponding to the four NS5$'$-D5 junctions and two NS5-D5$'$ junctions, 
the effective theory has six charged Fermi multiplets. 
In addition, there are two D5-D5$'$ junction, 
which yields two neutral Fermi multiplets.

We can compute the index as 
\begin{align}
\label{04u1hm2thm1fm6nfm2_1}
&
\mathbb{I}^{\textrm{2d $U(1)+2$HM$+$tHM$+6$FM$+2$nFM}}(t,x,z_{\alpha};q)
\nonumber\\
&=
\underbrace{
(q)_{\infty}^2 (q^{\frac12}t^2;q)_{\infty} (q^{\frac12}t^{-2};q)_{\infty} 
\oint_{\textrm{JK}} \frac{ds}{2\pi is}
}_{\mathbb{I}^{\textrm{2d $U(1)$}}}
\nonumber\\
&\times 
\prod_{\alpha=1}^{6}
\underbrace{
(q^{\frac12}s^{\pm}z_{\alpha}^{\pm};q)_{\infty}
\cdot 
(q^{\frac12}x^{\pm3}u^{\mp};q)_{\infty}
\cdot 
(q^{\frac12}x^{\pm3}u^{\pm};q)_{\infty}
}_{F(q^{\frac12}sz_{\alpha})\cdot F(q^{\frac12}x^3u^{-1})\cdot F(q^{\frac12}x^3u)}
\nonumber\\
&\times 
\underbrace{
\frac{
1
}
{
(q^{\frac14}t s^{\pm}x^{\mp}u^{\pm};q)_{\infty}
(q^{\frac34}t^{-1}s^{\pm}x^{\mp}u^{\pm};q)_{\infty}
\cdot 
(q^{\frac14}t s^{\pm}x^{\mp}u^{\mp};q)_{\infty}
(q^{\frac34}t^{-1}s^{\pm}x^{\mp}u^{\mp};q)_{\infty}
}
}_{\mathbb{I}^{\textrm{2d HM}} (sx^{-1}u) \cdot \mathbb{I}^{\textrm{2d HM}} (sx^{-1}u^{-1})}
\nonumber\\
&\times 
\underbrace{
\frac{1}
{
(q^{\frac14}t^{-1}s^{\pm}x^{\pm2};q)_{\infty}
(q^{\frac34}t s^{\pm}x^{\pm2};q)_{\infty}
}
}_{\mathbb{I}^{\textrm{2d tHM}} (sx^2)}
\end{align}
where the integrand includes a total of 12 factors in the denominator 
and 12 factors in the numerator. 
Correspondingly, we have three contributions to the index. 

The first pole for 2d hypermultiplet at $s=q^{\frac14}t x u^{-1}$ gives 
\begin{align}
\label{04u1hm2thm1fm6nfm2_2a}
&
\mathbb{I}^{\textrm{2d $U(1)+2$HM$+$tHM$+6$FM$+2$nFM}}_{1}(t,x,z_{\alpha};q)
\nonumber\\
&=
\prod_{\alpha=1}^{4}
\underbrace{
\frac{
(q^{\frac34}tz_{\alpha}xu^{-1};q)_{\infty}
(q^{\frac14}t^{-1}z_{\alpha}^{-1}x^{-1}u;q)_{\infty}
\cdot 
(q^{\frac12}x^{\pm3} u^{\pm};q)_{\infty}
}
{
(q^{\frac12}t^2 u^{-2};q)_{\infty}
(q^{\frac12}t^{-2} u^{2};q)_{\infty}
\cdot 
(u^2;q)_{\infty}
(qu^{-2};q)_{\infty}
\cdot 
(t^{-2}x^{-3}u;q)_{\infty}
(qt^2x^3 u^{-1};q)_{\infty}
}
}_{
F(q^{\frac34}tz_{\alpha}xu^{-1})
\cdot 
F(q^{\frac12}x^3u)
\cdot 
C(q^{\frac12}t^2 u^{-2})
\cdot 
C(u^2)
\cdot 
C(t^{-2}x^{-3}u)
}. 
\end{align}

Picking the another pole for 2d hypermultiplet at $s=q^{\frac14}t x u$, we obtain 
\begin{align}
\label{04u1hm2thm1fm6nfm2_2b}
&
\mathbb{I}^{\textrm{2d $U(1)+2$HM$+$tHM$+6$FM$+2$nFM}}_{2}(t,x,z_{\alpha};q)
\nonumber\\
&=
\prod_{\alpha=1}^{4}
\underbrace{
\frac{
(q^{\frac34}tz_{\alpha}xu;q)_{\infty}
(q^{\frac14}t^{-1}z_{\alpha}^{-1}x^{-1}u^{-1};q)_{\infty}
\cdot 
(q^{\frac12}x^{\pm3} u^{\mp};q)_{\infty}
}
{
(q^{\frac12} t^2 u^{2};q)_{\infty}
(q^{\frac12} t^{-2} u^{-2};q)_{\infty}
\cdot 
(u^{-2};q)_{\infty}
(qu^2;q)_{\infty}
\cdot 
(t^{-2}x^{-3}u^{-1};q)_{\infty}
(qt^2x^3u;q)_{\infty}
}
}_{
F(q^{\frac34}tz_{\alpha}xu)
\cdot 
F(q^{\frac12}x^3u^{-1})
\cdot 
C(q^{\frac12}t^2u^2)
\cdot 
C(u^{-2})
\cdot 
C(t^{-2}x^{-3}u^{-1})
}. 
\end{align}

Picking the pole for twisted hypermultiplet at $s=q^{\frac14}t^{-1}x^{-2}$, we get
\begin{align}
\label{04u1hm2thm1fm6nfm2_2c}
&
\mathbb{I}^{\textrm{2d $U(1)+2$HM$+$tHM$+6$FM$+2$nFM}}_{3}(t,x,z_{\alpha};q)
\nonumber\\
&=
\prod_{\alpha=1}^{4}
\underbrace{
\frac{
(q^{\frac34} t^{-1}z_{\alpha}x^{-2};q)_{\infty}
(q^{\frac14}tz_{\alpha}^{-1}x^2;q)_{\infty}
}
{
(t^2 x^3 u;q)_{\infty}
(qt^{-2}x^{-3}u^{-1};q)_{\infty}
\cdot 
(t^2 x^3 u^{-1};q)_{\infty}
(q t^{-2}x^{-3}u;q)_{\infty}
}
}_{
F(q^{\frac34}t^{-1}z_{\alpha}x^{-2})
\cdot 
C(t^2 x^3 u)
\cdot 
C(t^2 x^3 u^{-1})
}. 
\end{align}

The sum simplifies, which is nicely compatible with the expectation. 
The total index has no $q^{\frac14}$ term but has $q^{\frac12}$ and $q^{\frac34}$ terms which do not cancel. 
However we have not understood so far the expression in terms of a simple product of free fields. 
It would be interesting to figure out the appropriate expression and identify the matter fields in the S-dual configuration of Figure \ref{fig04u1hm2thm1fm6nfm2}.

%%%%%%%%%%%%%%%%%%%%%%%%%%%%%%%%%%%%%%%%%%%%%
%%%%%%%%%%%%%%%%%%%%%%%%%%%%%%%%%%%%%%%%%%%%%
\section{$\mathcal{N}=(0,4)$ boundaries of 3d theories}
\label{sec_abebc}
%%%%%%%%%%%%%%%%%%%%%%%%%%%%%%%%%%%%%%%%%%%%%
%%%%%%%%%%%%%%%%%%%%%%%%%%%%%%%%%%%%%%%%%%%%%
The $\mathcal{N}=(0,4)$ boundary conditions for 3d $\mathcal{N}=4$ Abelian gauge theories 
can be realized in the brane setup \cite{Chung:2016pgt, Hanany:2018hlz}. 
We check the matching of boundary anomalies and half-indices for dual pairs of boundary conditions. 

%%%%%%%%%%%%%%%%%%%%%%%%%%%%%%%%%%%%%%%%%%%%%
\subsection{3d SQED$_{1}$}
\label{sec_sqed1_bc}
%%%%%%%%%%%%%%%%%%%%%%%%%%%%%%%%%%%%%%%%%%%%%
Consider 3d $\mathcal{N}=4$ $U(1)$ gauge theory with a single charged hypermultiplet, SQED$_{1}$. 
It has a $U(1)_{t}$ topological symmetry. 
The mirror of SQED$_{1}$ is the theory of a free twisted hypermultiplet \cite{Kapustin:1999ha}. 
Under mirror symmetry the topological symmetry for SQED$_{1}$ maps to a $U(1)$ flavor symmetry for a twisted hypermultiplet.

%%%%%%%%%%%%%%%%%%%%%%%%%%%%%%%%%%%%%%%%%%%%%
\subsubsection{$(\mathcal{N}',N'+\textrm{Fermi})\leftrightarrow D$}
\label{sec_sqed1nn}
\begin{figure}
\begin{center}
\includegraphics[width=11cm]{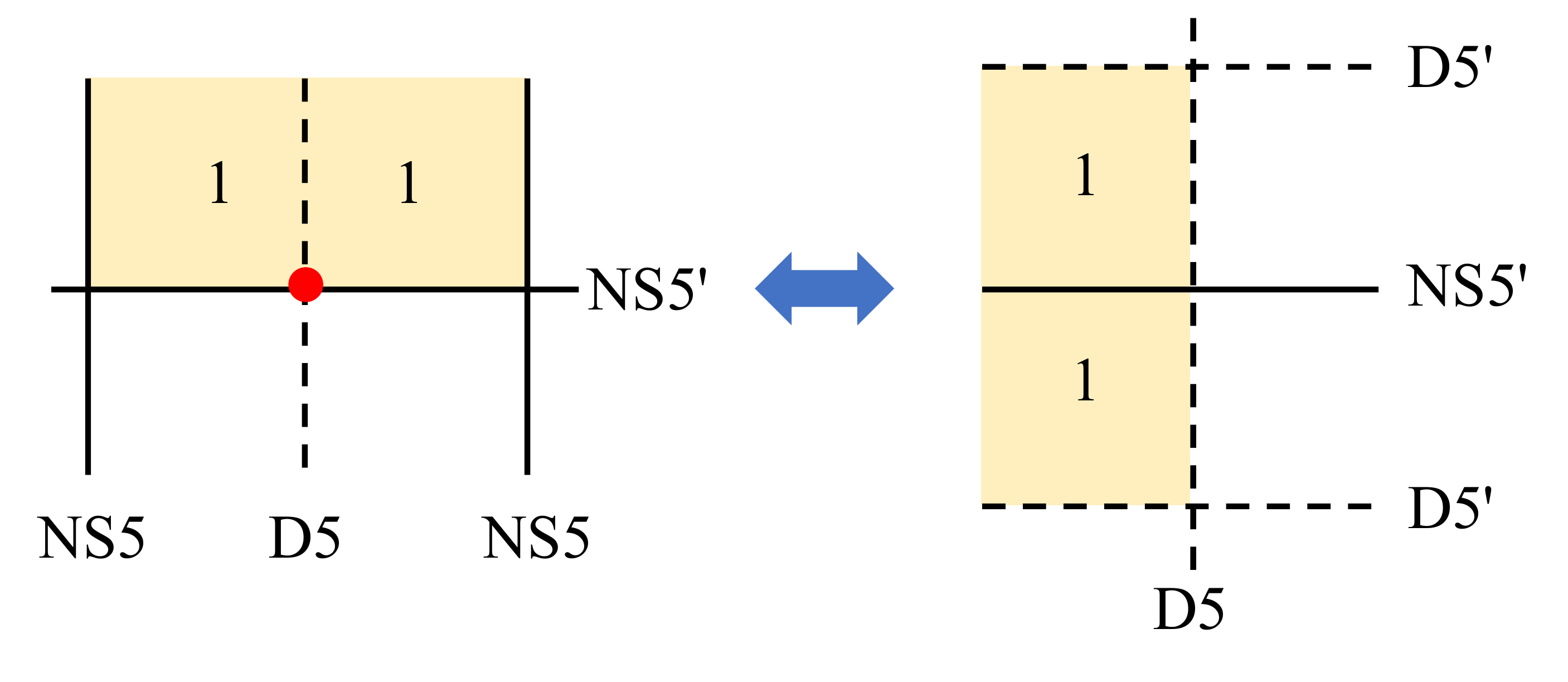}
\caption{
The brane constructions of the Neumann b.c. $(\mathcal{N}',N')+\Gamma$ for 
SQED$_{1}$ and its dual Dirichlet b.c. $D$ for twisted hypermultiplet. 
}
\label{figsqed1n}
\end{center}
\end{figure}

%%%%%%%%%%%%%%%%%%%%%%%%%%%%%%%%%%%%%%%%%%%%%
We begin with the boundary condition 
which includes Neumann b.c. $\mathcal{N}'$ for $U(1)$ vector multiplet and Neumann b.c. $N'$ for a charged hypermultiplet 
which contains a complex scalar $\mathbb{H}$. 
The corresponding brane construction is depicted in Figure \ref{figsqed1n}. 
The Neumann b.c. for gauge multiplet requires that the gauge symmetry is preserved at the boundary. 
Therefore the boundary gauge anomaly must be cancelled. 
From the brane box analysis, there would be a boundary charged Fermi multiplet \cite{Hanany:2018hlz}. 
In fact, a gauge anomaly free boundary condition can be obtained by introducing boundary charged Fermi multiplet $\Gamma$. 

From the S-dual configuration in Figure \ref{figsqed1n}, 
we expect that the b.c. $(\mathcal{N}',N'+\Gamma)$ for SQED$_{1}$ is dual to Dirichlet b.c. $D$ for twisted hypermultiplet 
which contains a complex scalar field $\mathbb{T}$. 

%charge
The charges of matter fields are given by
\begin{align}
\label{sqed1n_ch}
\begin{array}{c|c|c}
\textrm{matter}\setminus\textrm{symmetry}&G=U(1)&U(1)_{t}\\ \hline
\mathbb{H}&+&0\\
\Gamma&+&+\\ \hline 
\mathbb{T}&0&+\\
\end{array}
\end{align}

%anomaly
Let ${\bf s}$ and ${\bf z}$ be the field strengths  
of $G=U(1)$ gauge symmetry and $U(1)_{t}$ topological symmetry for SQED$_{1}$. 
The boundary anomaly polynomial for Neumann b.c. $(\mathcal{N}', N')$ 
together with boundary Fermi multiplet $\Gamma$ is 
\begin{align}
\label{sqed1N_AM}
\mathcal{I}^{\textrm{3d SQED}_{1}}_{\mathcal{N}',N'+\Gamma}
&=
\underbrace{
-{\bf s}^2
}_{N'}
+
\underbrace{
({\bf s}+{\bf z})^2
}_{\Gamma}
\underbrace{
-2{\bf s}\cdot {\bf z}
}_{\textrm{FI}}
={\bf z}^2
\end{align}
where the first term is contributed from the charged hyper with Neumann b.c. $N'$, 
the second is contributed from the boundary Fermi $\Gamma$ 
and the last is the contribution from the FI term \cite{Dimofte:2011ju}. 

The boundary anomaly polynomial for Dirichlet b.c. $D$ of free twisted hypermultiplet is 
\begin{align}
\label{msqed1N_AM}
\mathcal{I}^{\textrm{tHM}}_{D}&={\bf z}^2, 
\end{align}
which agrees with the boundary anomaly (\ref{sqed1N_AM}) for SQED$_{1}$ with b.c. $(\mathcal{N}', N',\Gamma)$.

\begin{figure}
\begin{center}
\includegraphics[width=10cm]{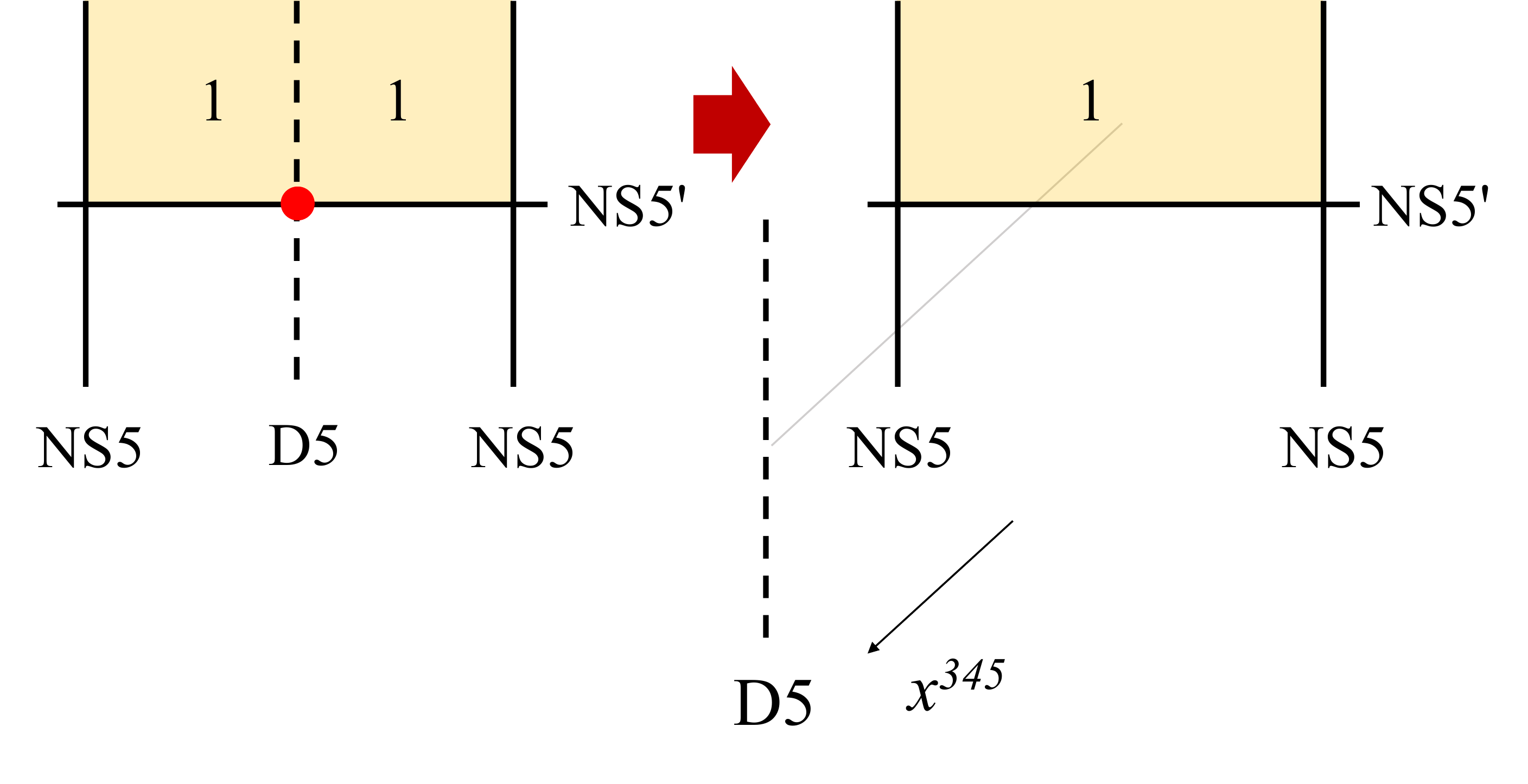}
\caption{
Higgsing precess of Neumann b.c. $(\mathcal{N}',N'+\Gamma)$ for SQED$_{1}$ 
that separates the D5-brane. }
\label{figsqed1nH}
\end{center}
\end{figure}

In addition, one can check the duality by computing half-indices. 
Let $s$ and $z$ be the fugacity for $U(1)$ gauge symmetry and $U(1)$ topological symmetry respectively.  
The half-index for the boundary condition $(\mathcal{N}',N'+\Gamma)$ for SQED$_{1}$ takes the form
\begin{align}
\label{sqed1n_half1}
&
\mathbb{II}^{\textrm{3d SQED}_{1}}_{\mathcal{N}',N'+\Gamma}
(t,z;q)
\nonumber\\
&=
\underbrace{
(q)_{\infty} (q^{\frac12}t^2;q)_{\infty}\oint \frac{ds}{2\pi is}
}_{\mathbb{II}_{\mathcal{N}'}^{\textrm{3d $U(1)$}}}
\underbrace{
\frac{
(q^{\frac12}sz;q)_{\infty} (q^{\frac12}s^{-1}z^{-1};q)_{\infty}
}
{
(q^{\frac14}ts;q)_{\infty} (q^{\frac14}ts^{-1};q)_{\infty}
}
}_{\mathbb{II}_{N'}^{\textrm{3d HM}}(s) \cdot F(q^{\frac12}sz)}. 
\end{align}
We can evaluate the half-index (\ref{sqed1n_half1}) by 
picking up the residues at poles $s=q^{\frac14+m}t$, $m=0,1,\cdots$ for the fundamental hypermultiplet
\begin{align}
\label{sqed1n_half2}
&
\mathbb{II}^{\textrm{3d SQED}_{1}}_{\mathcal{N}',N'+\Gamma}
(t,z;q)
\nonumber\\
&=
\frac{
(q^{\frac34}t z^{-1};q)_{\infty} 
(q^{\frac14}t^{-1}z;q)_{\infty} 
(q^{\frac12}t^2;q)_{\infty}
}
{
(q)_{\infty}
}
\sum_{m=0}^{\infty} 
\frac{(q^{1+m};q)_{\infty}}
{(q^{\frac12+m}t^2;q)_{\infty}}
q^{\frac{m}{4}} t^{-m} z^{m}. 
\end{align}
Making use of the $q$-binomial theorem, 
we find that the half-index (\ref{sqed1n_half2}) is equal to
\begin{align}
\label{msqed1n_half}
\mathbb{II}^{\textrm{3d tHM}}_{D}(t,z;q)
&=(q^{\frac34}tz;q)_{\infty} (q^{\frac34} tz^{-1};q)_{\infty}. 
\end{align}
As we expect, this is the half-index of Dirichlet b.c. $D$ for a free twisted hypermultiplet.

We can alternatively expand the half-index (\ref{sqed1n_half1}) as
\begin{align}
\label{sqed1n_half3}
&
\mathbb{II}^{\textrm{3d SQED}_{1}}_{\mathcal{N}',N'+\Gamma}
(t,z;q)
\nonumber\\
&=
\frac{(q^{\frac12}t^2;q)_{\infty}}
{(q)_{\infty}}
(q^{\frac14}t^{-1}z;q)_{\infty}
(q^{\frac14}t^{-1}z^{-1};q)_{\infty}
\sum_{m=0}^{\infty}
\frac{
(q^{1+m};q)_{\infty}^2
}
{
(q^{\frac14+m}t^{-1}z^{\pm}:q)_{\infty}
}. 
\end{align}
The first term in the sum (\ref{sqed1n_half3}) is identified with 
the half-index $\mathbb{II}^{\textrm{3d $U(1)$}}_{\mathcal{N}'}$ of Neumann b.c. $\mathcal{N}'$ 
for 3d $U(1)$ vector multiplet. 
The associated Higgsing precess would separate the D5-brane from the D3-NS5-NS5$'$ system 
(see Figure \ref{figsqed1nH}).

%wilson line
We can add a Wilson line operator of charge $n$ ending on a boundary. 
The half-index reads
\begin{align}
\label{sqed1nW_half}
&
\mathbb{II}^{\textrm{3d SQED$_{1}$}}_{\mathcal{N}',N'+\Gamma+\mathcal{W}_{n}}
(t,z;q)
\nonumber\\
&=
(q)_{\infty} (q^{\frac12}t^2;q)_{\infty}
\oint \frac{ds}{2\pi is} 
\frac{
(q^{\frac12}sz;q)_{\infty}
(q^{\frac12}s^{-1}z^{-1};q)_{\infty}
}
{
(q^{\frac14}ts;q)_{\infty}
(q^{\frac14}ts^{-1};q)_{\infty}
}s^{n}. 
\end{align}
By taking the sum over residues at hypermultiplet poles $s=q^{\frac14+m}t$, 
we get 
\begin{align}
\label{sqed1nW_half2}
\mathbb{II}^{\textrm{3d SQED$_{1}$}}_{\mathcal{N}',N'+\Gamma+\mathcal{W}_{n}}
&=
\frac{
(q^{\frac14}t^{-1}z^{-1};q)_{n}
}
{
(q^{\frac34}tz^{-1};q)_{n}
}
q^{\frac{n}{4}} t^{n} \times 
\mathbb{II}^{\textrm{3d tHM}}_{D}(t,z;q). 
\end{align}

%%%%%%%%%%%%%%%%%%%%%%%%%%%%%%%%%%%%%%%%%%%%%
\subsubsection{$(\mathcal{D}',D') \leftrightarrow (N+\textrm{Fermi})$}
\label{sec_sqed1dd}
\begin{figure}
\begin{center}
\includegraphics[width=11cm]{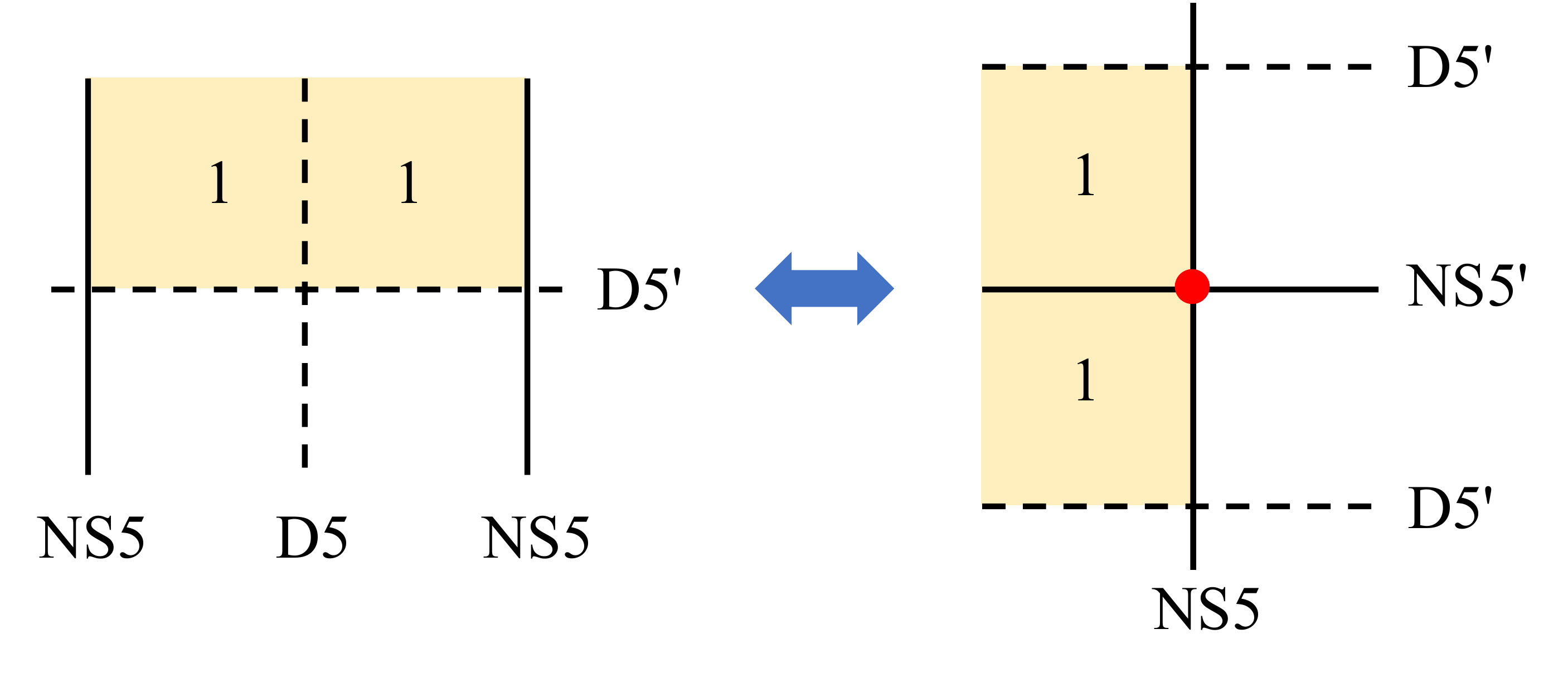}
\caption{
The brane constructions of the Dirichlet b.c. $(\mathcal{D}',D')$ for 
SQED$_{1}$ and its dual Neumann b.c. $N$ for twisted hypermultiplet with a boundary Fermi multiplet. 
}
\label{figsqed1d}
\end{center}
\end{figure}

%%%%%%%%%%%%%%%%%%%%%%%%%%%%%%%%%%%%%%%%%%%%%
Next consider the boundary condition consisting of 
Dirichlet b.c. $\mathcal{D}'$ for $U(1)$ vector multiplet 
and Dirichlet b.c. $D$ for a charged hypermultiplet. 
According to the Dirichlet b.c. $\mathcal{D}$, 
the gauge symmetry is broken to a boundary global $U(1)_{\partial}$ symmetry. 
The brane setup is illustrated in Figure \ref{figsqed1d}. 

From the brane picture, 
the dual boundary condition would include Neumann b.c. $N$ for twisted hypermultiplet 
corresponding to the NS5-brane on which the D3-brane terminate.

The boundary 't Hooft anomaly of  
the boundary condition $(\mathcal{D}',D')$ for SQED$_{1}$ is
\begin{align}
\label{sqed1d_AM}
\mathcal{I}^{\textrm{3d SQED}_{1}}_{\mathcal{D}',D'}
&=
\underbrace{
{\bf u}^2
}_{D'}
\underbrace{
-2{\bf u}\cdot {\bf z}
}_{\textrm{FI}}
\end{align}
where ${\bf u}$ is the field strength for $U(1)_{\partial}$. 

On the other hand, the boundary 't Hooft anomaly of 
the Neumann b.c. $N$ for twisted hypermultiplet is 
\begin{align}
\label{msqed1d_AM}
\mathcal{I}^{\textrm{3d tHM}}_{N}&=
-{\bf z}^2.
\end{align}
The difference 
\begin{align}
\label{sqed1d_AM1}
\mathcal{I}^{\textrm{3d SQED}_{1}}_{\mathcal{D}',D'}-
\mathcal{I}^{\textrm{3d tHM}}_{N}
&=({\bf u}-{\bf z})^2
\end{align}
of the boundary anomalies (\ref{sqed1d_AM}) and (\ref{msqed1d_AM}) 
indicates the presence of a boundary Fermi multiplet $\widetilde{\Gamma}$ 
which carries the charges $(+,-)$ under the $U(1)_{\partial}\times U(1)_{t}$. 

%charge 
The charges of matter fields are summarized as
The charges of matter fields are given by
\begin{align}
\label{sqed1d_ch}
\begin{array}{c|c|c}
\textrm{matter}\setminus\textrm{symmetry}&U(1)_{\partial}&U(1)_{t}\\ \hline
\mathbb{H}&+&0\\ \hline 
\mathbb{T}&0&+\\
\widetilde{\Gamma}&+&-\\ 
\end{array}
\end{align}

Now compute the half-indices. 
Let $u$ be the fugacity for the boundary global $U(1)_{\partial}$ symmetry. 
From the boundary 't Hooft anomaly (\ref{sqed1d_AM}), 
the half-index of boundary condition $(\mathcal{D}',D')$ for SQED$_{1}$ reads
\begin{align}
\label{sqed1d_half1}
&
\mathbb{II}^{\textrm{3d SQED}_{1}}_{\mathcal{D}',D'}(t,u;q)
\nonumber\\
&=
\frac{1}{(q)_{\infty} (q^{\frac12}t^{-2};q)_{\infty}}
\sum_{m\in \mathbb{Z}}
\underbrace{
(q^{\frac34+m} t^{-1}u;q)_{\infty} 
(q^{\frac34-m} t^{-1}u^{-1};q)_{\infty}
}_{\mathbb{II}^{\textrm{3d HM}}_{D'}(q^{m}u;q)}
\nonumber\\
&\times 
q^{\frac{m^2}{2}}
(-1)^{m}u^{m} z^{-m}. 
\end{align}

On the other hand, the half-index of boundary condition $N$ for twisted hypermultiplet 
involving the boundary Fermi multiplet $\widetilde{\Gamma}$ which takes the form 
\begin{align}
\label{msqed1d_half1}
&
\mathbb{II}^{\textrm{3d tHM}}_{N+\widetilde{\Gamma}}(t,u,z;q)
\nonumber\\
&=
\frac{
(q^{\frac12}zu^{-1};q)_{\infty}
(q^{\frac12}z^{-1}u;q)_{\infty}
}
{
(q^{\frac14}t^{-1}z;q)_{\infty}
(q^{\frac14}t^{-1}z^{-1};q)_{\infty}
}
\end{align}
where the numerator is the Fermi index $F(q^{\frac12}uz^{-1})$ 
and the denominator is the Neumann half-index $\mathbb{II}_{N}^{\textrm{3d tHM}}(z)$ for 3d twisted hypermultiplet. 
We have checked that 
the half-indices (\ref{sqed1d_half1}) and (\ref{msqed1d_half1}) agree with each other.

%%%%%%%%%%%%%%%%%%%%%%%%%%%%%%%%%%%%%%%%%%%%%
\subsection{$T[SU(2)]$}
\label{sec_tsu2_bc}
%%%%%%%%%%%%%%%%%%%%%%%%%%%%%%%%%%%%%%%%%%%%%
Consider the 3d $\mathcal{N}=4$ $U(1)$ gauge theory with two charged hypermultiplets, 
which we call $T[SU(2)]$. 
This is the self-mirror theory with $SU(2)$ flavor symmetry and $SU(2)$ topological symmetry.

%%%%%%%%%%%%%%%%%%%%%%%%%%%%%%%%%%%%%%%%%%%%%
\subsubsection{$(\mathcal{N}',N'+\textrm{Fermi})\leftrightarrow (\mathcal{D},D)$}
\label{sec_tsu2nn}
\begin{figure}
\begin{center}
\includegraphics[width=12cm]{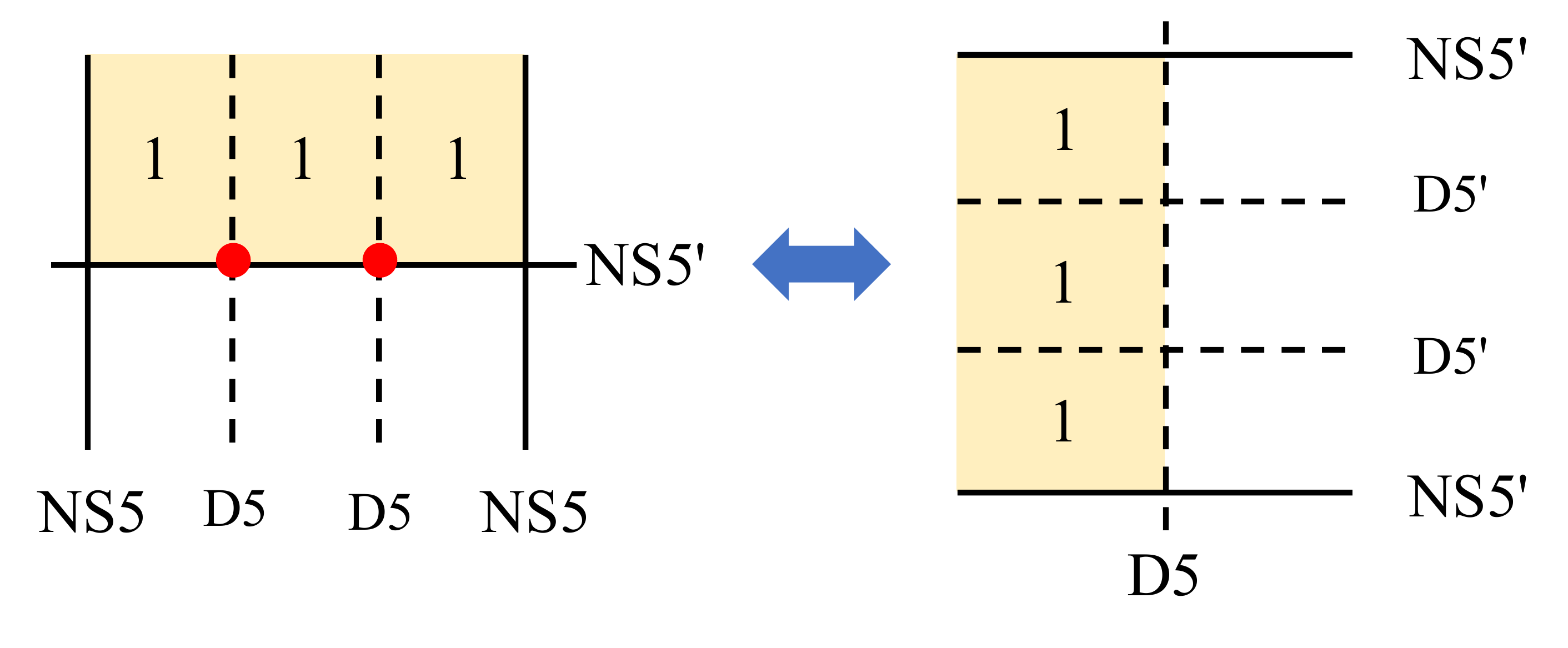}
\caption{
The brane constructions of the Neumann b.c. $(\mathcal{N}',N'+\{\Gamma^{(i)}\})$ for 
$T[SU(2)]$ and its dual Dirichlet b.c. $(\mathcal{D},D)$ for $\widetilde{T[SU(2)]}$. 
}
\label{figtsu2}
\end{center}
\end{figure}

%%%%%%%%%%%%%%%%%%%%%%%%%%%%%%%%%%%%%%%%%%%%%
One can find $\mathcal{N}=(0,4)$ boundary condition for $T[SU(2)]$ from 
the brane setup depicted in Figure \ref{figtsu2}. 
The presence of NS5$'$-brane requires the Neumann b.c. $\mathcal{N}'$ for $U(1)$ vector multiplet and Neumann b.c. $N'$ 
for two charged hypermultiplets involving two complex scalar fields $\mathbb{H}^{(1)}, \mathbb{H}^{(2)}$. 
From the brane box analysis \cite{Hanany:2018hlz}, 
there would be two boundary charged Fermi multiplets $\Gamma^{(1)}, \Gamma^{(2)}$ which cancel the gauge anomaly. 
We denote this boundary condition by $(\mathcal{N}',N'+\{\Gamma^{(i)}\}_{i=1,2})$. 

Under the action of S-duality, 
we find the dual boundary condition for $\widetilde{T[SU(2)]}$, 
which is the twisted version of $T[SU(2)]$ involving two complex scalar fields $\mathbb{T}^{(1)}, \mathbb{T}^{(2)}$, 
required from the D5-brane (see Figure \ref{figtsu2}). 
We expect that the b.c. $(\mathcal{N}',N'+\{\Gamma^{(i)}\}_{i=1,2})$ for $T[SU(2)]$ is dual to Dirichlet b.c. $(\mathcal{D},D)$ for 
$\widetilde{T[SU(2)]}$.

The charges of matter fields read
\begin{align}
\label{tsu2n_ch}
\begin{array}{c|c|c|c|c}
\textrm{matter}\setminus\textrm{symmetry}&U(1)&F&\widetilde{U(1)}_{\partial}&\widetilde{F} \\ \hline
\mathbb{H}^{(1)}&+&+&0&0 \\
\mathbb{H}^{(2)}&+&-&0&0 \\
\Gamma^{(1)}&+&+&+&+ \\ 
\Gamma^{(2)}&+&-&-&+ \\ \hline 
\mathbb{T}^{(1)}&0&0&+&+ \\
\mathbb{T}^{(2)}&0&0&+&- \\
\end{array}
\end{align}
where $F$ and $\widetilde{F}$ are the charges for $SU(2)$ flavor and topological symmetries of $T[SU(2)]$. 

Let ${\bf x}$ and ${\bf z}$ be the field strengths of the flavor and topological symmetries. 
The boundary anomaly polynomial of Neumann b.c. $(\mathcal{N}',N'+\Gamma^{(i)})$ for $T[SU(2)]$ is calculated as
 \begin{align}
 \label{tsu2_AN}
 \mathcal{I}_{\mathcal{N}',N'+\Gamma^{(i)}}^{T[SU(2)]}
 &=
 \underbrace{
% -\Tr ({\bf x}^2)
 -({\bf s}+{\bf x})^2-({\bf s}-{\bf x})^2
 }_{N'}
 \nonumber\\
 &
+
 \underbrace{
% \Tr ({\bf x}^2)+\Tr ({\bf z}^2)+
 ({\bf s}+{\bf x}+{\bf u}+{\bf z})^2
 +
 ({\bf s}-{\bf x}-{\bf u}+ {\bf z})^2
 }_{\Gamma^{(i)}}
  \underbrace{
 -2{\bf s}\cdot 2{\bf z}
 }_{\textrm{FI}}
 \nonumber\\
 &=
 2{\bf u}^2+2 {\bf z}^2+4{\bf u}\cdot {\bf x}.
 \end{align}

On the other hand, the boundary anomaly polynomial of Dirichlet b.c. $(\mathcal{D},D)$ for $\widetilde{T[SU(2)]}$ is 
\begin{align}
\label{mtsu2_AN}
\mathcal{I}^{\widetilde{T[SU(2)]}}_{\mathcal{D},D}
&=
\underbrace{
({\bf u}+{\bf z})^2
+({\bf u}-{\bf z})^2
}_{D}
+
\underbrace{
2{\bf u}\cdot 2 {\bf x}
}_{\textrm{FI}}
\nonumber\\
&=
2{\bf u}^2+2 {\bf z}^2+4{\bf u}\cdot \bf{x}.
\end{align}
As required for anomaly matching in the RG flows, 
(\ref{tsu2_AN}) and (\ref{mtsu2_AN}) agree with each other.

Let us compute the half-indices for the dual boundary conditions. 
We can evaluate the half-index of Neumann b.c. $(\mathcal{N}',N'+\Gamma^{(i)})$ for $T[SU(2)]$ as
\begin{align}
\label{tsu2_half1}
&
\mathbb{II}_{\mathcal{N},N+\Gamma^{(i)}}^{T[SU(2)]}(t,x,u,z;q)
\nonumber\\
&=
\underbrace{
(q)_{\infty} (q^{\frac12} t^2;q)_{\infty}\oint \frac{ds}{2\pi is}
}_{\mathbb{II}_{\mathcal{N}'}^{\textrm{3d $U(1)$}}}
\underbrace{
\frac{
(q^{\frac12}s^{\mp}x^{\pm}u^{\pm}z^{\pm};q)_{\infty}
}
{
(q^{\frac14}ts^{\mp}x^{\pm};q)_{\infty}
}
}_{\mathbb{II}_{N'}^{\textrm{3d HM}} (s^{-1}x)\cdot F(q^{\frac12}s^{-1}xuz)}
\cdot 
\underbrace{
\frac{
(q^{\frac12}s^{\pm}x^{\pm}u^{\pm}z^{\mp};q)_{\infty}
}
{
(q^{\frac14}ts^{\pm}x^{\pm};q)_{\infty}
}
}_{\mathbb{II}_{N'}^{\textrm{3d HM}} (sx)\cdot F(q^{\frac12}sxuz^{-1})}.
\end{align}

From the boundary 't Hooft anomaly (\ref{mtsu2_AN}), 
we obtain the half-index of 
Dirichlet b.c. $(\mathcal{D},D)$ for $\widetilde{T[SU(2)]}$
\begin{align}
\label{mtsu2_half1}
&
\mathbb{II}_{\mathcal{D},D}^{\widetilde{T[SU(2)]}}(t,x,u,z;q)
\nonumber\\
&=
\frac{1}{(q)_{\infty}(q^{\frac12} t^2;q)_{\infty}}
\sum_{m\in \mathbb{Z}}
q^{m^2}u^{2m} x^{2m}
\underbrace{
(q^{\frac34\pm m}tu^{\pm}z^{\pm};q)_{\infty}
}_{\mathbb{II}_{D}^{\textrm{3d tHM}} (q^{m}uz)}
\cdot 
\underbrace{
(q^{\frac34\pm m}tu^{\pm}z^{\mp};q)_{\infty}
}_{\mathbb{II}_{D}^{\textrm{3d tHM}} (q^{m}uz^{-1})}.
\end{align}
In fact, the half-indices (\ref{tsu2_half1}) and (\ref{mtsu2_half1}) coincide. 

%%%%%%%%%%%%%%%%%%%%%%%%%%%%%%%

%%%%%%%%%%%%%%%%%%%%%%%%%%%%%%%%%%%%%%%%%%%%%
\subsection{3d SQED$_{3}$}
\label{sec_sqed3_bc}
%%%%%%%%%%%%%%%%%%%%%%%%%%%%%%%%%%%%%%%%%%%%%
Next we turn to the 3d $\mathcal{N}=4$ $U(1)$ gauge theory with three charged hypermultiplets, SQED$_{3}$. 
The mirror theory is $\widetilde{U(1)\times U(1)}$ quiver gauge theory with three twisted hypermultiplets, 
which we denote by $\widetilde{[1]-(1)-(1)-[1]}$. 
\footnote{Here and in the following the round brackets $(N)$ represent $U(N)$ gauge symmetry 
and the square brackets $[M]$ denote $U(M)$ global symmetry. 
The dash lines are bi-fundamental (twisted)hypermultiplets. 
The tildes indicate that the corresponding gauge theories are constructed by twisted supermultiplets. }

%%%%%%%%%%%%%%%%%%%%%%%%%%%%%%%%%%%%%%%%%%%%%
\subsubsection{$(\mathcal{N}',N'+\textrm{Fermi})\leftrightarrow (\mathcal{D},D)$}
\label{sec_sqed3nn}
%%%%%%%%%%%%%%%%%%%%%%%%%%%%%%%%%%%%%%%%%%%%%

The $\mathcal{N}=(0,4)$ boundary condition can be realized in the brane setup. 
When the single D3-brane stretched between two NS5-branes 
which terminates on the NS5$'$-brane and the three D5-branes intersect with NS$'$-brane, 
one finds Neumann b.c. $\mathcal{N}'$ for $U(1)$ vector multiplet 
and Neumann b.c. $N'$ for hypermultiplets involving complex scalar fields $\{\mathbb{H}^{(i)}\}_{i=1,2,3}$. 
Besides, the configuration should have three charged Fermi multiplets that cancel the boundary gauge anomaly.

By applying S-duality, 
we find the mirror boundary condition as 
Dirichlet b.c. $\mathcal{D}$ for $\widetilde{U(1)\times U(1)}$ twisted vector multiplet 
and Dirichlet b.c. $D$ for twisted hypermultiplets containing complex scalar fields $\{\mathbb{T}^{(i)}\}_{i=1,2,3}$.

The charges of matter fields are given by
\begin{align}
\label{sqed3_ch}
\begin{array}{c|c|c|c|c}
\textrm{matter}\setminus\textrm{symmetry}&U(1)&F&\widetilde{U(1)\times U(1)}_{\partial}&\widetilde{F} \\ \hline
\mathbb{H}^{(1)}&+&(+,0,0)&(0,0)&(0,0) \\
\mathbb{H}^{(2)}&+&(0,+,0)&(0,0)&(0,0) \\
\mathbb{H}^{(3)}&+&(0,0,+)&(0,0)&(0,0) \\
\Gamma^{(1)}&+&(+,0,0)&(-,0)&(+,0) \\
\Gamma^{(2)}&+&(0,+,0)&(+,-)&(0,0) \\
\Gamma^{(3)}&+&(0,0,+)&(0,+)&(0,-) \\ \hline
\mathbb{T}^{(1)}&0&(0,0,0)&(-,0)&(+,0) \\
\mathbb{T}^{(2)}&0&(0,0,0)&(+,-)&(0,0) \\
\mathbb{T}^{(3)}&0&(0,0,0)&(0,+)&(0,-) \\
\end{array}
\end{align}
where $F$ is the charges for $SU(3)$ flavor symmetry of SQED$_{3}$ 
and $\widetilde{F}$ is the charges for $U(1)\times U(1)$ topological symmetry of SQED$_{3}$.

Let ${\bf s}$, ${\bf x}_{i}$, ${\bf u}_{i}$, ${\bf z}_{i}$ 
be the Abelian field strengths for $U(1)$, $F$, $\widetilde{U(1)\times U(1)}_{\partial}$ and $\widetilde{F}$ respectively. 
The boundary anomaly polynomial of 
Neumann b.c. $(\mathcal{N}', N'+\Gamma^{(i)})$ for SQED$_{3}$ is 
\begin{align}
\label{sqed3_AN}
\mathcal{I}_{\mathcal{N}',N'+\Gamma^{(i)}}^{\textrm{3d SQED}_{3}}
&=
\underbrace{
-({\bf s}+{\bf x}_{1})^2
-({\bf s}+{\bf x}_{2})^2
-({\bf s}+{\bf x}_{3})^2
}_{N'}
\nonumber\\
&
\underbrace{
+({\bf s}+{\bf x}_{1}-{\bf u}_{1}+{\bf z}_{1})^2
+({\bf s}+{\bf x}_{2}+{\bf u}_{1}-{\bf u}_{2})^2
+({\bf s}+{\bf x}_{3}+{\bf u}_{2}-{\bf z}_{2})^2
}_{\Gamma}
\nonumber\\
&
\underbrace{
-2{\bf s}\cdot ({\bf z}_{1}-{\bf z}_{2})
-2{\bf x}_{1}\cdot {\bf z}_{1}
+2{\bf x}_{3}\cdot {\bf z}_{2}
}_{\textrm{FI}}
\nonumber\\
&=
2{\bf u}_{1}^2+2{\bf u}_{2}^2-2{\bf u}_{1}\cdot {\bf u}_{2}
-2{\bf u}_{1}\cdot ({\bf x}_{1}-{\bf x}_{2}+{\bf z}_{1})
-2{\bf u}_{2}\cdot ({\bf x}_{2}-{\bf x}_{3}+{\bf z}_{2})
+{\bf z}_{1}^2+{\bf z}_{2}^2. 
\end{align}

For the Dirichlet b.c. $(\mathcal{D},D)$ of the mirror quiver gauge theory $\widetilde{[1]-(1)-(1)-[1]}$ 
we have the boundary 't Hooft anomaly polynomial 
\begin{align}
\label{msqed3_AN}
\mathcal{I}_{\mathcal{D},D}^{\textrm{3d $\widetilde{[1]-(1)-(1)-[1]}$}}
&=
\underbrace{
(-{\bf u}_{1}+{\bf z}_{1})^2
+({\bf u}_{1}-{\bf u}_{2})^2
+({\bf u}_{2}-{\bf z}_{2})^2
}_{D}
\nonumber\\
&
\underbrace{
-2{\bf u}_{1}\cdot ({\bf x}_{1}-{\bf x}_{2})
-2{\bf u}_{2}\cdot ({\bf x}_{2}-{\bf x}_{3})
}_{\textrm{FI}}
\nonumber\\
&=
2{\bf u}_{1}^2+2{\bf u}_{2}^2-2{\bf u}_{1}\cdot {\bf u}_{2}
-2{\bf u}_{1}\cdot ({\bf x}_{1}-{\bf x}_{2}+{\bf z}_{1})
-2{\bf u}_{2}\cdot ({\bf x}_{2}-{\bf x}_{3}+{\bf z}_{2})
+{\bf z}_{1}^2+{\bf z}_{2}^2. 
\end{align}

As expected, the anomaly polynomials (\ref{sqed3_AN}) and (\ref{msqed3_AN}) agree with each other. 

We can compute 
the half-index of Neumann b.c. $(\mathcal{N}', N'+\Gamma^{(i)})$ for SQED$_{3}$ as
\begin{align}
\label{sqed3_half1}
&
\mathbb{II}_{\mathcal{N}',N'+\Gamma^{(i)}}^{\textrm{3d SQED}_{3}}(t,x_{\alpha},u_{i},z_{\alpha};q)
\nonumber\\
&=
\underbrace{
(q)_{\infty}(q^{\frac12}t^2;q)_{\infty}\oint \frac{ds}{2\pi is}
}_{\mathbb{II}_{\mathcal{N}'}^{\textrm{3d $U(1)$}}}
\nonumber\\
&\times 
\underbrace{
\frac{(q^{\frac12} s^{\pm} x_{1}^{\pm} z_{1}^{\pm} u_{1}^{\mp};q)_{\infty}}
{(q^{\frac14}t s^{\pm} x_{1}^{\pm};q)_{\infty} }
}_{\mathbb{II}_{N'}^{\textrm{3d HM}} (sx_{1}) \cdot 
F(q^{\frac12}sx_{1}z_{1}u_{1}^{-1})}
\cdot 
\underbrace{
\frac{
(q^{\frac12} s^{\pm} x_{2}^{\pm} u_{1}^{\pm} u_{2}^{\mp};q)_{\infty}
}
{
(q^{\frac14} t s^{\pm} x_{2}^{\pm};q)_{\infty}
}
}_{\mathbb{II}_{N'}^{\textrm{3d HM}}(sx_{2})\cdot 
F(q^{\frac12}sx_{2}u_{1}u_{2}^{-1})}
\cdot 
\underbrace{
\frac{
(q^{\frac12}s^{\pm}x_{3}^{\pm}u_{2}^{\pm}z_{2}^{\mp};q)_{\infty}
}
{
(q^{\frac14}t s^{\pm}x_{3}^{\pm};q)_{\infty}
}
}_{\mathbb{II}_{N'}^{\textrm{3d HM}} (sx_{3})\cdot 
F(q^{\frac12}sx_{3}u_{2}z_{2}^{-1})}
\end{align}
where $x_{1}x_{2}x_{3}=1$.

From the boundary 't Hooft anomaly (\ref{msqed3_AN}), 
we obtain the half-index of Dirichlet b.c. $(\mathcal{D},D)$ for the mirror quiver gauge theory $\widetilde{[1]-(1)-(1)-[1]}$ 
\begin{align}
\label{msqed3_half1}
&
\mathbb{II}_{\mathcal{D},D}^{\textrm{3d $\widetilde{[1]-(1)-(1)-[1]}$}}(t,x_{\alpha},u_{i},z_{\alpha};q)
\nonumber\\
&=
\frac{1}{(q)_{\infty}(q^{\frac12}t^2;q)_{\infty}}
\sum_{m_{1}\in \mathbb{Z}}
\cdot 
\frac{1}{(q)_{\infty}(q^{\frac12}t^2;q)_{\infty}}
\sum_{m_{2}\in \mathbb{Z}}
\nonumber\\
&\times 
q^{\frac{m_{1}^2}{2}+\frac{m_{2}^2}{2}+\frac{(m_{1}-m_{2})^2}{2}}
\cdot 
u_{1}^{2m_{1}-m_{2}}
\cdot 
u_{2}^{2m_{2}-m_{1}}
z_{1}^{-m_{1}}
z_{2}^{-m_{2}}
x_{1}^{-m_{1}}
x_{2}^{m_{1}-m_{2}}
x_{3}^{m_{2}}
\nonumber\\
&\times 
\underbrace{
(q^{\frac34\pm m_{1}}t u_{1}^{\pm} z_{1}^{\mp};q)_{\infty}
}_{\mathbb{II}_{D}^{\textrm{3d tHM}} (q^{m_{1}}u_{1}z_{1}^{-1})}
\cdot 
\underbrace{
(q^{\frac34\pm m_{1}\mp m_{2}} t u_{1}^{\pm} u_{2}^{\mp};q)_{\infty}
}_{\mathbb{II}_{D}^{\textrm{3d tHM}} (q^{m_{1}-m_{2}} u_{1}u_{2}^{-1})}
\cdot 
\underbrace{
(q^{\frac34\pm m_{2}} t u_{2}^{\pm}z_{2}^{\mp};q)_{\infty}
}_{\mathbb{II}_{D}^{\textrm{3d tHM}} (q^{m_{2}} u_{2}z_{2}^{-1})}. 
\end{align}

We have confirmed that the half-indices 
(\ref{sqed3_half1}) and (\ref{msqed3_half1}) coincide to each other.

%%%%%%%%%%%%%%%%%%%%%%%%%%%%%%%%%%%%%%%%%%%%%
\subsubsection{$(\mathcal{D}',D')\leftrightarrow (\mathcal{N},N+\textrm{Fermi})$}
\label{sec_sqed3dd}
%%%%%%%%%%%%%%%%%%%%%%%%%%%%%%%%%%%%%%%%%%%%%
When the single D3-brane suspended between two NS5-branes ends on the D5$'$-brane, 
one finds Dirichlet b.c. $\mathcal{D}'$ for $U(1)$ vector multiplet 
and Dirichlet b.c. $D'$ for hypermultiplets. 

The S-dual configuration is described by 
Neumann b.c. $\mathcal{N}$ for twisted vector multiplet 
and Neumann b.c. $N$ for twisted hypermultiplet in the mirror quiver gauge theory $\widetilde{[1]-(1)-(1)-[1]}$. 
Conversely, the Fermi multiplets $\widetilde{\Gamma}^{(i)}$ appear at the boundary of the mirror quiver gauge theory 
to cancel the gauge anomaly.

The charges of matter content are 
\begin{align}
\label{sqed3_ch2}
\begin{array}{c|c|c|c|c}
\textrm{matter}\setminus\textrm{symmetry}&U(1)_{\partial}&F&\widetilde{U(1)\times U(1)}&\widetilde{F} \\ \hline
\mathbb{H}^{(1)}&+&(+,0,0)&(0,0)&(0,0) \\
\mathbb{H}^{(2)}&+&(0,+,0)&(0,0)&(0,0) \\
\mathbb{H}^{(3)}&+&(0,0,+)&(0,0)&(0,0) \\ \hline
\mathbb{T}^{(1)}&0&(0,0,0)&(-,0)&(+,0) \\
\mathbb{T}^{(2)}&0&(0,0,0)&(+,-)&(0,0) \\
\mathbb{T}^{(3)}&0&(0,0,0)&(0,+)&(0,-) \\
\widetilde{\Gamma}^{(1)}&-&(-,0,0)&(-,0)&(+,0) \\
\widetilde{\Gamma}^{(2)}&-&(0,-,0)&(+,-)&(0,0) \\
\widetilde{\Gamma}^{(3)}&-&(0,0,-)&(0,+)&(0,-) \\ 
\end{array}
\end{align}
Let ${\bf u}$, ${\bf x}_{i}$, ${\bf s_{i}}$ and ${\bf z}_{i}$ 
be the Abelian field strengths for 
$U(1)_{\partial}$, $F$, $\widetilde{U(1)\times U(1)}$ and $\widetilde{F}$ symmetries respectively.

The boundary anomaly polynomial of 
the Dirichlet b.c. $(\mathcal{D}', D')$ for SQED$_{3}$ is 
\begin{align}
\label{sqed3_AN2}
\mathcal{I}_{\mathcal{D}',D'}^{\textrm{3d SQED}_{3}}
&=
\underbrace{
({\bf u}+{\bf x}_{1})^2
+({\bf u}+{\bf x}_{2})^2
+({\bf u}+{\bf x}_{3})^2
}_{D'}
\nonumber\\
&
\underbrace{
-2{\bf u}\cdot ({\bf z}_{1}-{\bf z}_{2})
-2{\bf x}_{1}\cdot {\bf z}_{1}
+2{\bf x}_{3}\cdot {\bf z}_{2}
}_{\textrm{FI}}
\nonumber\\
&=
3{\bf u}^2
+2{\bf u}\cdot ({\bf x}_{1}+{\bf x}_{2}+{\bf x}_{3}-{\bf z}_{1}+{\bf z}_{2})
\nonumber\\
&+{\bf x}_{1}^2+{\bf x}_{2}^2+{\bf x}_{3}^2
-2{\bf x}_{1}\cdot {\bf z}_{1}+2{\bf x}_{3}\cdot {\bf z}_{2}.
\end{align}

For the Neumann b.c. $(\mathcal{N},N+\widetilde{\Gamma}^{(i)})$ of the mirror quiver gauge theory $\widetilde{[1]-(1)-(1)-[1]}$ 
we have the boundary anomaly polynomial 
\begin{align}
\label{msqed3_AN2}
\mathcal{I}_{\mathcal{N},N+\widetilde{\Gamma}^{(i)}}^{\textrm{3d $\widetilde{[1]-(1)-(1)-[1]}$}}
&=
\underbrace{
-(-{\bf s}_{1}+{\bf z}_{1})^2
-({\bf s}_{1}-{\bf s}_{2})^2
-({\bf s}_{2}-{\bf z}_{2})^2
}_{N}
\nonumber\\
&+
\underbrace{
(-{\bf s}_{1}+{\bf z}_{1}-{\bf u}-{\bf x}_{1})^2
+
({\bf s}_{1}-{\bf s}_{2}-{\bf u}-{\bf x}_{2})^2
+
({\bf s}_{2}-{\bf z}_{2}-{\bf u}-{\bf x}_{3})^2
}_{\widetilde{\Gamma}}
\nonumber\\
&
\underbrace{
-2{\bf s}_{1}({\bf x}_{1}-{\bf x}_{2})
-2{\bf s}_{2}({\bf x}_{2}-{\bf x}_{3})
}_{\textrm{FI}}
\nonumber\\
&=
3{\bf u}^2
+2{\bf u}\cdot ({\bf x}_{1}+{\bf x}_{2}+{\bf x}_{3}-{\bf z}_{1}+{\bf z}_{2})
\nonumber\\
&+{\bf x}_{1}^2+{\bf x}_{2}^2+{\bf x}_{3}^2
-2{\bf x}_{1}\cdot {\bf z}_{1}+2{\bf x}_{3}\cdot {\bf z}_{2}.
\end{align}

As expected, the anomaly polynomials (\ref{sqed3_AN2}) and (\ref{msqed3_AN2}) agree with each other.

From the anomaly polynomial (\ref{sqed3_AN2}), 
we get the half-index of Dirichlet b.c. $(\mathcal{D}', D')$ for SQED$_{3}$
\begin{align}
\label{sqed3_half2}
&
\mathbb{II}_{\mathcal{D}',D'}^{\textrm{3d SQED}_{3}}
(t,x_{\alpha},u,z_{\alpha};q)
\nonumber\\
&=
\frac{1}{(q)_{\infty} (q^{\frac12} t^{-2};q)_{\infty}}
\sum_{m\in \mathbb{Z}}
(-1)^{m} 
q^{\frac{3m^2}{2}} 
u^{3m}
z_{1}^{-m} z_{2}^{m} x_{1}^{m}x_{2}^{m} x_{3}^{m}
\nonumber\\
&\times 
\prod_{\alpha=1}^{3}
\underbrace{
(q^{\frac34\pm m}t^{-1}u^{\pm}x_{\alpha}^{\pm};q)_{\infty}
}_{\mathbb{II}_{D'}^{\textrm{3d HM}} (q^{m}ux_{\alpha})}
\end{align}

The half-index of 
he Neumann b.c. $(\mathcal{N},N+\widetilde{\Gamma}^{(i)})$ of the mirror quiver gauge theory $\widetilde{[1]-(1)-(1)-[1]}$ is given by
\begin{align}
\label{msqed3_half2}
&
\mathbb{II}_{\mathcal{N},N+\widetilde{\Gamma}^{(i)}}^{\textrm{3d $\widetilde{[1]-(1)-(1)-[1]}$}}(t,x_{\alpha},u,z_{\alpha};q)
\nonumber\\
&=
\underbrace{
(q)_{\infty}(q^{\frac12} t^{-2};q)_{\infty} \oint \frac{ds_{1}}{2\pi is_{1}}
}_{\mathbb{II}_{\mathcal{N}}^{\textrm{3d $\widetilde{U(1)}$}}}
\cdot 
\underbrace{
(q)_{\infty}(q^{\frac12} t^{-2};q)_{\infty} \oint \frac{ds_{2}}{2\pi is_{2}}
}_{\mathbb{II}_{\mathcal{N}}^{\textrm{3d $\widetilde{U(1)}$}}}
\nonumber\\
&\times 
\underbrace{
\frac{(q^{\frac12} z_{1}^{\pm}s_{1}^{\mp}u^{\mp}x_{1}^{\mp};q)_{\infty}}
{(q^{\frac14}t^{-1}z_{1}^{\pm}s_{1}^{\mp};q)_{\infty}}
}_{
\mathbb{II}_{N}^{\textrm{3d tHM}} \left(\frac{z_{1}}{s_{1}} \right)
\cdot F\left( q^{\frac12}\frac{z_{1}}{s_{1}ux_{1}} \right)
}
\cdot 
\underbrace{
\frac{(q^{\frac12}s_{1}^{\pm}s_{2}^{\mp}u^{\mp}x_{2}^{\mp};q)_{\infty}}
{
(q^{\frac14}t^{-1}s_{1}^{\pm}s_{2}^{\mp};q)_{\infty}
}
}_{\mathbb{II}_{N}^{\textrm{3d tHM}} \left(\frac{s_{1}}{s_{2}} \right)
\cdot 
F\left(q^{\frac12}\frac{s_{1}}{s_{2}ux_{2}} \right)
}
\cdot 
\underbrace{
\frac{
(q^{\frac12}s_{2}^{\pm}z_{2}^{\mp}u^{\mp}x_{3}^{\mp};q)_{\infty}
}
{
(q^{\frac14}t^{-1}s_{2}^{\pm}z_{2}^{\mp};q)_{\infty}
}
}_{\mathbb{II}_{N}^{\textrm{3d tHM}} \left(\frac{s_{2}}{z_{2}} \right)\cdot 
F\left(q^{\frac12} \frac{s_{2}}{z_{2}u x_{3}} \right)}
\end{align}

We have checked that 
the half-indices (\ref{sqed3_half2}) and (\ref{msqed3_half2}) coincide to each other.

%%%%%%%%%%%%%%%%%%%%%%%%%%%%%%%%%%%%%%%%%%%%%
\subsection{3d SQED$_{N_{f}}$}
\label{sec_sqednf_bc}
\begin{figure}
\begin{center}
\includegraphics[width=13cm]{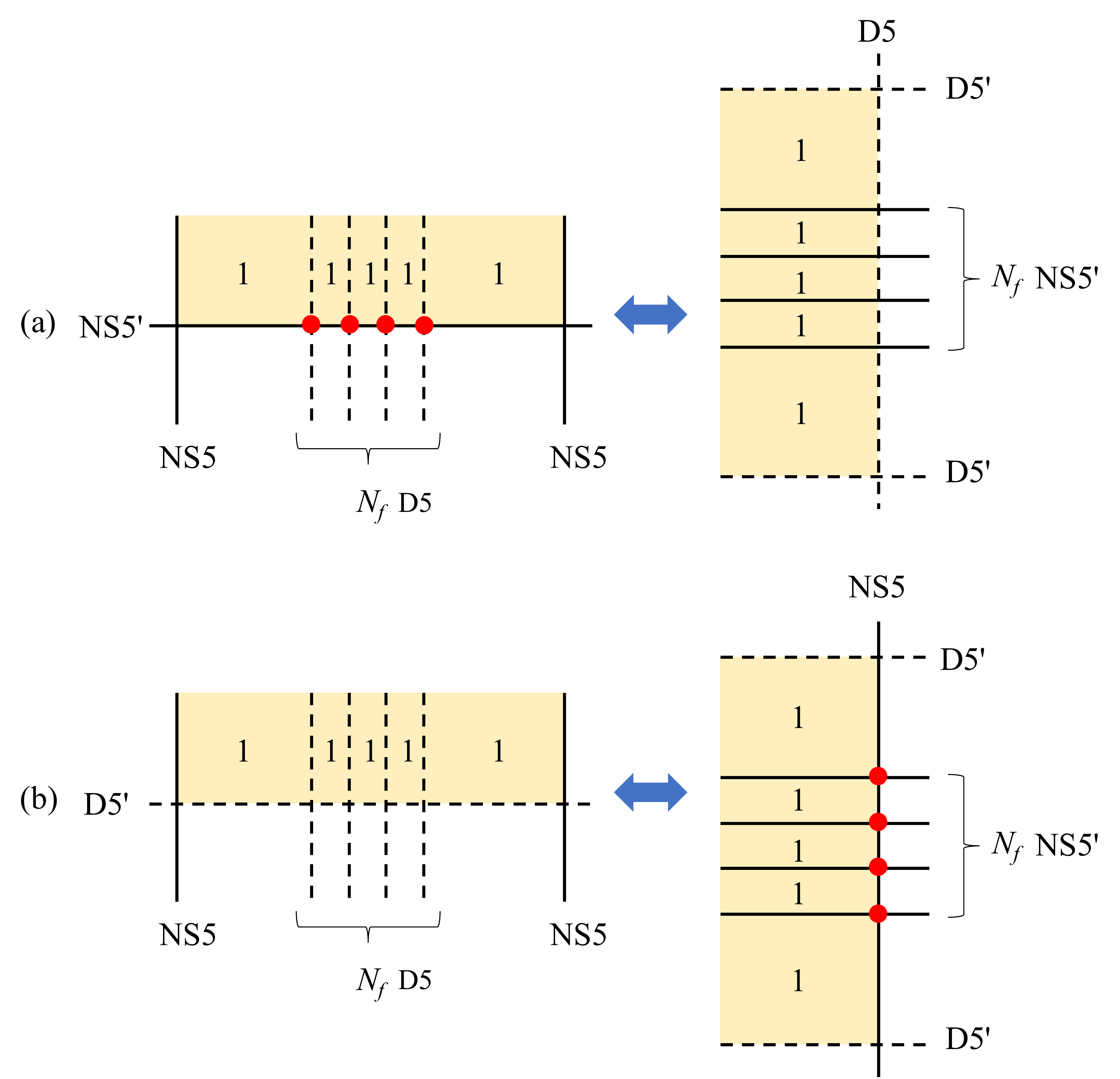}
\caption{
(a) The brane constructions of the Neumann b.c. $(\mathcal{N}',N'+\{\Gamma^{(i)}\})$ for 
SQED$_{N_{f}}$ and its dual Dirichlet b.c. $(\mathcal{D},D)$ for $\widetilde{[1]-(1)^{N_{f}-1}-[1]}$. 
(b) The brane constructions of the Dirichlet b.c. $(\mathcal{D}',D')$ for 
SQED$_{N_{f}}$ and its dual Neumann b.c. $(\mathcal{N},N+\{\widetilde{\Gamma}^{(i)}\} )$ for $\widetilde{[1]-(1)^{N_{f}-1}-[1]}$. 
}
\label{figsqednf}
\end{center}
\end{figure}

%%%%%%%%%%%%%%%%%%%%%%%%%%%%%%%%%%%%%%%%%%%%%
Let us propose the generalization of duality of $\mathcal{N}=(0,4)$ boundary conditions for 3d $\mathcal{N}=4$ SQED$_{N_{f}}$. 
One can realize the $\mathcal{N}=(0,4)$ boundary conditions by introducing a fivebrane on which the single D3-brane ends 
(see Figure \ref{figsqednf}). 
%
%
%
%
%

%%%%%%%%%%%%%%%%%%%%%%%%%%%%%%%%%%%%%%%%%%%%%
\subsubsection{$(\mathcal{N}',N'+\textrm{Fermi})\leftrightarrow (\mathcal{D},D)$}
\label{sec_sqednfnn}
%%%%%%%%%%%%%%%%%%%%%%%%%%%%%%%%%%%%%%%%%%%%%
The 3d $\mathcal{N}=4$ SQED$_{N_{f}}$ admits 
the $\mathcal{N}=(0,4)$ boundary condition 
that requires Neumann b.c. $\mathcal{N}'$ for $U(1)$ vector multiplet 
and Neumann b.c. $N'$ for the charged hypermultiplets.  
The Neumann b.c. $(\mathcal{N}',N')$ for SQED$_{N_{f}}$ is obtained when the single D3-brane 
whose world-volume theory is SQED$_{N_{f}}$ ends on the NS5$'$-brane. 
The cancellation of gauge anomaly is achieved by the boundary Fermi multiplets $\{\Gamma^{(i)}\}_{i=1,\cdots, N_{f}}$. 

Under the action of S-duality, 
the NS5$'$-brane maps to the D5-brane 
so that we find the Dirichlet b.c. $(\mathcal{D},D)$ for the mirror quiver gauge theory $\widetilde{[1]-(1)^{N_{f}-1}-[1]}$. 

Let $\{\mathbb{H}^{(i)}\}_{i=1,\cdots, N_{f}}$ and $\{\mathbb{T}^{(i)}\}_{i=1,\cdots, N_{f}}$ be the complex scalar fields 
in the hypermultiplets of SQED$_{N_{f}}$ 
and those in the twisted hypermultiplets of $\widetilde{[1]-(1)^{N_{f}-1}-[1]}$ respectively. 
The charges of the matter fields are given by
\begin{align}
\label{sqednf_ch}
\begin{array}{c|c|c|c|c}
\textrm{matter}\setminus\textrm{symmetry}&U(1)&F&\widetilde{U(1)^{N_{f}-1}}_{\partial}&\widetilde{F} \\ \hline
\mathbb{H}^{(1)}&+&(+,0,0,\cdots,0)&(0,0,0,\cdots,0)&(0.0) \\
\mathbb{H}^{(2)}&+&(0,+,0,\cdots,0)&(0,0,0,\cdots,0)&(0,0) \\
\mathbb{H}^{(3)}&+&(0,0,+,\cdots,0)&(0,0,0,\cdots,0)&(0,0) \\
\vdots&\vdots&\vdots&\vdots&\vdots \\
\mathbb{H}^{(N_{f})}&+&(0,0,0,\cdots,+)&(0,0,0,\cdots,0)&(0,0) \\
\Gamma^{(1)}&+&(+,0,0,\cdots,0)&(-,0,0,\cdots,0)&(+,0) \\
\Gamma^{(2)}&+&(0,+,0,\cdots,0)&(+,-,0,\cdots,0)&(0,0) \\
\Gamma^{(3)}&+&(0,0,+,\cdots,0)&(0,+,-,\cdots,0)&(0,0) \\
\vdots&\vdots&\vdots&\vdots&\vdots \\
\Gamma^{(N_{f})}&+&(0,0,0,\cdots,+)&(0,0,0,\cdots,+)&(0,-) \\ \hline
\mathbb{T}^{(1)}&0&(0,0,0,\cdots,0)&(-,0,0,\cdots,0)&(+,0) \\
\mathbb{T}^{(2)}&0&(0,0,0,\cdots,0)&(+,-,0,\cdots,0)&(0,0) \\
\mathbb{T}^{(3)}&0&(0,0,0,\cdots,0)&(0,+,-,\cdots,0)&(0,0) \\
\vdots&\vdots&\vdots&\vdots&\vdots \\
\mathbb{T}^{(N_{f})}&0&(0,0,0,\cdots,0)&(0,0,0,\cdots,+)&(0,-) \\
\end{array}
\end{align}

Let us denote the Abelian field strengths for 
$U(1)$, $F$, $\widetilde{U(1)^{N_{f}-1}}$ and $\widetilde{F}$ by 
${\bf s}$, ${\bf x}_{i}$, ${\bf u}_{i}$ and ${\bf z}_{i}$ respectively. 

The boundary anomaly polynomial of Neumann b.c. $(\mathcal{N}',N'+\Gamma^{(i)})$ for SQED$_{N_{f}}$ is 
\begin{align}
\label{sqednf_AN}
\mathcal{I}_{\mathcal{N}',N'+\Gamma^{(i)}}^{\textrm{3d SQED}_{N_{f}}}
&=
\underbrace{
-\sum_{i=1}^{N_{f}}({\bf s}+{\bf x}_{i})^2
}_{N'}
\nonumber\\
&+
\underbrace{
({\bf s}+{\bf x}_{1}-{\bf u}_{1}+{\bf z}_{1})^2
+\sum_{i=1}^{N_{f}-2}
({\bf s}+{\bf x}_{i+1}+{\bf u}_{i}-{\bf u}_{i+1})^2
+({\bf s}+{\bf x}_{N_{f}}+{\bf u}_{N_{f}-1}-{\bf z}_{2})^2
}_{\Gamma}
\nonumber\\
&
\underbrace{
-2{\bf s}\cdot ({\bf z}_{1}-{\bf z}_{2})
-2{\bf x}_{1}\cdot {\bf z}_{1}
+2{\bf x}_{N_{f}}\cdot {\bf z}_{2}
}_{\textrm{FI}}. 
\end{align}

The boundary anomaly polynomial of Dirichlet b.c. $(\mathcal{D},D)$ for 
$\widetilde{[1]-(1)^{N_{f}-1}-[1]}$ is computed as 
\begin{align}
\label{msqednf_AN}
\mathcal{I}_{\mathcal{D},D}^{\textrm{3d $\widetilde{[1]-(1)^{N_{f}-1}-[1]}$}}
&=
\underbrace{
(-{\bf u}_{1}+{\bf z}_{1})^2
+\sum_{i=1}^{N_{f}-2} ({\bf u}_{i}-{\bf u}_{i+1})^2
+({\bf u}_{N_{f}-1}-{\bf z}_{2})^2
}_{D}
\nonumber\\
&
\underbrace{
-2\sum_{i=1}^{N_{f}-1}{\bf u}_{i}\cdot 
({\bf x}_{i}-{\bf x}_{i+1})
}_{\textrm{FI}}. 
\end{align}
This matches with the boundary 't Hooft anomaly (\ref{sqednf_AN}) of Neumann b.c. $(\mathcal{N}',N'+\Gamma^{(i)})$ for 
SQED$_{N_{f}}$.

The half-index of Neumann b.c. $(\mathcal{N}',N'+\Gamma^{(i)})$ for 
SQED$_{N_{f}}$ is given by
\begin{align}
\label{sqednf_half1}
&
\mathbb{II}^{\textrm{3d SQED}_{N_{f}}}_{\mathcal{N}',N'+\Gamma^{(i)}}(t,x_{\alpha},u_{i},z_{\alpha};q)
\nonumber\\
&=
\underbrace{
(q)_{\infty} (q^{\frac12} t^2;q)_{\infty}\oint \frac{ds}{2\pi is}
}_{\mathbb{II}_{\mathcal{N}'}^{\textrm{3d $U(1)$}}
}
\nonumber\\
&\times 
\underbrace{
\frac{
(q^{\frac12} s^{\pm} x_{1}^{\pm}z_{1}^{\pm} u_{1}^{\mp};q)_{\infty}
}
{
(q^{\frac14}t s^{\pm}x_{1}^{\pm};q)_{\infty}
}
}_{\mathbb{II}_{N'}^{\textrm{3d HM}} (sx_{1})
\cdot F(q^{\frac12}sx_{1}z_{1}u_{1}^{-1})
}
\cdot 
\prod_{i=1}^{N_{f}-2}
\underbrace{
\frac{
(q^{\frac12}s^{\pm} x_{i+1}^{\pm} u_{i}^{\pm} u_{i+1}^{\mp};q)_{\infty}}
{(q^{\frac14}t s^{\pm} x_{i+1}^{\pm};q)_{\infty}}
}_{
\mathbb{II}_{N'}^{\textrm{3d HM}} (sx_{i+1})
\cdot F(q^{\frac12}sx_{i+1}u_{i}u_{i+1}^{-1})
}
\cdot 
\underbrace{
\frac{
(q^{\frac12} s^{\pm} x_{N_{f}}^{\pm} u_{N_{f}-1}^{\pm}z_{2}^{\mp};q)_{\infty}
}
{
(q^{\frac14}t s^{\pm} x_{N_{f}}^{\pm};q)_{\infty}
}
}_{\mathbb{II}_{N'}^{\textrm{3d HM}} (sx_{N_{f}})
\cdot F(q^{\frac12}sx_{N_{f}} u_{N_{f}-1} z_{2}^{-1})}
\end{align}

The half-index of Dirichlet b.c. $(\mathcal{D},D)$ for $\widetilde{[1]-(1)^{N_{f}-1}-[1]}$ is computed as 
\begin{align}
\label{msqednf_half1}
&
\mathbb{II}^{\textrm{3d $\widetilde{[1]-(1)^{N_{f}-1}-[1]}$}}_{\mathcal{D},D}(t,x_{\alpha},u_{i},z_{\alpha};q)
\nonumber\\
&=
\frac{1}{(q)_{\infty}^{N_{f}-1} (q^{\frac12} t^2;q)_{\infty}^{N_{f}-1}}
\sum_{m_{1},\cdots, m_{N_{f}-1}\in \mathbb{Z}}
\nonumber\\
&\times 
q^{
\frac{m_{1}^2}{2}+
\sum_{i=1}^{N_{f}-2}\frac{(m_{i}-m_{i+1})^2}{2}
+\frac{m_{N_{f}-1}^2}{2}
}
\cdot 
u_{1}^{2m_{1}-m_{2}}
\cdot
\prod_{i=1}^{N_{f}-3}u_{i+1}^{2m_{i+1}-m_{i}-m_{i+2}}
\cdot 
u_{N_{f}-1}^{2m_{N_{f}-1}-m_{N_{f}-2}}
\nonumber\\
&\times 
z_{1}^{-m_{1}}
\cdot 
z_{2}^{-m_{N_{f}-1}}
\cdot 
x_{1}^{-m_{1}}
\cdot 
\prod_{i=1}^{N_{f}-2}
x_{i+1}^{m_{i}-m_{i+1}}
\cdot 
x_{N_{f}}^{m_{N_{f}-1}}
\nonumber\\
&\times 
\underbrace{
(q^{\frac34\pm m_{1}} t u_{1}^{\pm} z_{1}^{\mp};q)_{\infty}
}_{\mathbb{II}_{D}^{\textrm{3d tHM}} (q^{m_{1}}u_{1}z_{1}^{-1})}
\cdot 
\prod_{i=1}^{N_{f}-2}
\underbrace{
(q^{\frac34\pm m_{i}\mp m_{i+1}} t u_{i}^{\pm} u_{i+1}^{\mp};q)_{\infty}
}_{\mathbb{II}_{D}^{\textrm{3d tHM}} (q^{m_{i}-m_{i+1}} u_{i}u_{i+1}^{-1})}
\cdot 
\underbrace{
(q^{\frac34\pm m_{N_{f}-1}} t u_{N_{f}-1}^{\pm} z_{2}^{\mp};q)_{\infty}
}_{\mathbb{II}_{D}^{\textrm{3d tHM}} (q^{m_{N_{f}-1}} u_{N_{f}-1} z_{2}^{-1})}.
\end{align}

Making use of the minimal mirror identity between 
(\ref{sqed1d_half1}) and (\ref{msqed1d_half1}), 
one finds that 
the half-indices (\ref{sqednf_half1}) and (\ref{msqednf_half1}) give the same answer.

%%%%%%%%%%%%%%%%%%%%%%%%%%%%%%%%%%%%%%%%%%%%%
\subsubsection{$(\mathcal{D}',D')\leftrightarrow (\mathcal{N},N+\textrm{Fermi})$}
\label{sec_sqednfdd}
%%%%%%%%%%%%%%%%%%%%%%%%%%%%%%%%%%%%%%%%%%%%%
Next consider the Dirichlet boundary condition $(\mathcal{D}',D')$ for SQED$_{N_{f}}$, 
which is realized by the D5$'$-brane on which the single D3-brane terminates. 

In the dual configuration, 
the mirror quiver gauge theory $\widetilde{[1]-(1)^{N_{f}-1}-[1]}$ 
has Neumann b.c. $\mathcal{N}$ for twisted vector multiplet and Neumann b.c. $N$ for twisted hypermultiplets. 
Again the gauge anomaly contributed from the twisted hypermultiplets is compensated 
by the boundary Fermi multiplets $\{\widetilde{\Gamma}^{(i)}\}_{i=1,\cdots, N_{f}}$.

The charges of the matter fields are 
\begin{align}
\label{sqednf_ch2}
\begin{array}{c|c|c|c|c}
\textrm{matter}\setminus\textrm{symmetry}&U(1)_{\partial}&F&\widetilde{U(1)^{N_{f}-1}}&\widetilde{F} \\ \hline
\mathbb{H}^{(1)}&+&(+,0,0,\cdots,0)&(0,0,0,\cdots,0)&(0.0) \\
\mathbb{H}^{(2)}&+&(0,+,0,\cdots,0)&(0,0,0,\cdots,0)&(0,0) \\
\mathbb{H}^{(3)}&+&(0,0,+,\cdots,0)&(0,0,0,\cdots,0)&(0,0) \\
\vdots&\vdots&\vdots&\vdots&\vdots \\
\mathbb{H}^{(N_{f})}&+&(0,0,0,\cdots,+)&(0,0,0,\cdots,0)&(0,0) \\ \hline
\mathbb{T}^{(1)}&0&(0,0,0,\cdots,0)&(-,0,0,\cdots,0)&(+,0) \\
\mathbb{T}^{(2)}&0&(0,0,0,\cdots,0)&(+,-,0,\cdots,0)&(0,0) \\
\mathbb{T}^{(3)}&0&(0,0,0,\cdots,0)&(0,+,-,\cdots,0)&(0,0) \\
\vdots&\vdots&\vdots&\vdots&\vdots \\
\mathbb{T}^{(N_{f})}&0&(0,0,0,\cdots,0)&(0,0,0,\cdots,+)&(0,-) \\
\widetilde{\Gamma}^{(1)}&-&(-,0,0,\cdots,0)&(-,0,0,\cdots,0)&(+,0) \\
\widetilde{\Gamma}^{(2)}&-&(0,-,0,\cdots,0)&(+,-,0,\cdots,0)&(0,0) \\
\widetilde{\Gamma}^{(3)}&-&(0,0,-,\cdots,0)&(0,+,-,\cdots,0)&(0,0) \\
\vdots&\vdots&\vdots&\vdots&\vdots \\
\widetilde{\Gamma}^{(N_{f})}&-&(0,0,0,\cdots,-)&(0,0,0,\cdots,+)&(0,-) \\ 
\end{array}
\end{align}

Let ${\bf u}$, ${\bf x}_{i}$, ${\bf s}_{i}$ and ${\bf z}_{i}$ be 
the Abelian field strengths for $U(1)_{\partial}$, $F$, $\widetilde{U(1)^{N_{f}-1}}$ and $\widetilde{F}$ respectively. 

The boundary 't Hooft anomaly polynomial of 
the Dirichlet b.c. $(\mathcal{D}', D')$ for SQED$_{N_{f}}$ is 
\begin{align}
\label{sqednf_AN2}
\mathcal{I}_{\mathcal{D}',D'}^{\textrm{3d SQED}_{N_{f}}}
&=
\underbrace{
\sum_{i=1}^{N_{f}}
({\bf u}-{\bf x}_{i})^2
}_{D'}
\underbrace{
-2{\bf u}\cdot ({\bf z}_{1}-{\bf z}_{2})
-2{\bf x}_{1}\cdot {\bf z}_{1}
+2{\bf x}_{N_{f}}\cdot {\bf z}_{2}
}_{\textrm{FI}}.
\end{align}
This matches with 
the boundary anomaly polynomial of the Neumann b.c. $(\mathcal{N},N+\widetilde{\Gamma}^{(i)})$ 
for the mirror quiver gauge theory $\widetilde{[1]-(1)^{N_{f}-1}-[1]}$
\begin{align}
\label{msqednf_AN2}
\mathcal{I}_{\mathcal{N},N+\widetilde{\Gamma}^{(i)}}^{\textrm{3d $\widetilde{[1]-(1)^{N_{f}-1}-[1]}$}}
&=
\underbrace{
-(-{\bf s}_{1}+{\bf z}_{1})^2
-\sum_{i=1}^{N_{f}-2}({\bf s}_{i}-{\bf s}_{i+1})^2
-({\bf s}_{N_{f}-1}-{\bf z}_{2})^2
}_{N}
\nonumber\\
&+
\underbrace{
(-{\bf s}_{1}+{\bf z}_{1}-{\bf u}-{\bf x}_{1})^2
+\sum_{i=1}^{N_{f}-2}
({\bf s}_{i}-{\bf s}_{i+1}-{\bf u}-{\bf x}_{i+1})^2
+({\bf s}_{N_{f}-1}-{\bf z}_{2}-{\bf u}-{\bf x}_{N_{f}})^2
}_{\widetilde{\Gamma}}
\nonumber\\
&
\underbrace{
-2\sum_{i=1}^{N_{f}-1}{\bf s}_{i}\cdot ({\bf x}_{i}-{\bf x}_{i+1})
}_{\textrm{FI}}.
\end{align}

From the  boundary 't Hooft anomaly (\ref{sqednf_AN2}) 
we obtain the half-index of the Dirichlet b.c. $(\mathcal{D}', D')$ for SQED$_{N_{f}}$
\begin{align}
\label{sqednf_half2}
&
\mathbb{II}_{\mathcal{D}',D'}^{\textrm{3d SQED}_{N_{f}}}(t,x_{\alpha},u,z_{\alpha};q)
\nonumber\\
&=
\frac{1}{(q)_{\infty} (q^{\frac12}t^{-2};q)_{\infty}}
\sum_{m\in \mathbb{Z}}
(-1)^{N_{f}m} 
q^{\frac{N_{f}m^2}{2}}
\cdot 
u^{N_{f}m}
\cdot 
z_{1}^{-m}
\cdot 
z_{2}^{m}
\cdot 
\prod_{\alpha=1}^{N_{f}}x_{\alpha}^{m}
\cdot 
\underbrace{
(q^{\frac34\pm m} t^{-1}u^{\pm} x_{\alpha}^{\pm};q)_{\infty}
}_{\mathbb{II}_{D'}^{\textrm{3d HM}} (q^{m}ux_{\alpha})}.
\end{align}

On the other hand, 
the half-index of the Neumann b.c. $(\mathcal{N},N+\widetilde{\Gamma}^{(i)})$ 
for the mirror quiver gauge theory $\widetilde{[1]-(1)^{N_{f}-1}-[1]}$ is calculated as 
\begin{align}
\label{msqednf_half2}
&
\mathbb{II}_{\mathcal{N},N+\widetilde{\Gamma}^{(i)}}^{\textrm{3d $\widetilde{[1]-(1)^{N_{f}-1}-(1)}$}}(t,x_{\alpha},u,z_{\alpha};q)
\nonumber\\
&=
\underbrace{
(q)_{\infty}^{N_{f}-1} (q^{\frac12}t^{-2};q)_{\infty}^{N_{f}-1}
\oint \prod_{i=1}^{N_{f}-1}\frac{ds_{i}}{2\pi is_{i}}
}_{\mathbb{II}_{\mathcal{N}}^{\textrm{3d $\widetilde{U(1)^{N_{f}-1}}$}}}
\nonumber\\
&\times 
\underbrace{
\frac{(q^{\frac12}z_{1}^{\pm}s_{1}^{\mp}u^{\mp}x_{1}^{\mp};q)_{\infty}}
{(q^{\frac14}t^{-1}z_{1}^{\pm}s_{1}^{\mp};q)_{\infty}}
}_{\mathbb{II}_{N}^{\textrm{3d tHM}} \left(\frac{z_{1}}{s_{1}} \right)
\cdot 
F\left(q^{\frac12} \frac{z_{1}}{s_{1}u x_{1}} \right)
}
\cdot 
\prod_{i=1}^{N_{f}-2}
\underbrace{
\frac{(q^{\frac12}s_{i}^{\pm}s_{i+1}^{\mp}u^{\mp}x_{i+1}^{\mp};q)_{\infty}}
{(q^{\frac14}t^{-1}s_{i}^{\pm}s_{i+1}^{\mp};q)_{\infty}}
}_{\mathbb{II}_{N}^{\textrm{3d tHM}} \left(\frac{s_{i}}{s_{i+1}} \right)
\cdot F\left(q^{\frac12} \frac{s_{i}}{s_{i+1} ux_{i+1}} \right)
}
\cdot 
\underbrace{
\frac{
(q^{\frac12}s_{N_{f}-1}^{\pm}z_{2}^{\mp}u^{\mp}x_{N_{f}}^{\mp};q)_{\infty}
}
{
(q^{\frac14}t^{-1}s_{N_{f}-1}^{\pm}z_{2}^{\mp};q)_{\infty}
}
}_{\mathbb{II}_{N}^{\textrm{3d tHM}} \left(\frac{s_{N_{f}-1}}{z_{2}} \right)
\cdot F\left(q^{\frac12} \frac{s_{N_{f}-1}}{z_{2}u x_{N_{f}}} \right)}. 
\end{align}

The half-indices (\ref{sqednf_half2}) and (\ref{msqednf_half2}) would agree with each other. 
In fact, given that the minimal mirror identity holds, one can show that they lead to the same answer.

We therefore propose the following dualities of $\mathcal{N}=(0,4)$ boundary conditions 
for 3d $\mathcal{N}=4$ Abelian gauge theories:
\begin{align}
\label{sqednf_dual_bc}
\begin{array}{ccc}
\textrm{$(\mathcal{N}', N'+\{\Gamma^{(i)}\})$ for SQED$_{N_{f}}$}
&\leftrightarrow&\textrm{$(\mathcal{D},D)$ for $\widetilde{[1]-(1)^{N_{f}-1}-[1]}$} \\
\textrm{$(\mathcal{D}', D')$ for SQED$_{N_{f}}$}
&\leftrightarrow&\textrm{$(\mathcal{N},N+\{\widetilde{\Gamma}^{(i)}\})$ for $\widetilde{[1]-(1)^{N_{f}-1}-[1]}$} \\
\end{array}.
\end{align}
In general, a pair of $\mathcal{N}=(0,4)$ Neumann b.c. for 3d $\mathcal{N}=4$ $U(1)$ vector multiplet and hypermultiplets 
should be accompanied with boundary Fermi multiplets that cancel the gauge anomaly. 
We have confirmed that 
it is dual to a pair of $\mathcal{N}=(0,4)$ Dirichlet b.c. for 3d $\mathcal{N}=4$ $U(1)$ twisted vector multiplet and twisted hypermultiplets in the mirror theory.

%%%%%%%%%%%%%%%%%%%%%%%%%%%%%%%%%%%%%%%%%%%%%
\subsection{Enriched Neumann from Dirichlet}
\label{sec_enrichneu_bc}
%%%%%%%%%%%%%%%%%%%%%%%%%%%%%%%%%%%%%%%%%%%%%
To compute the half-index of 3d gauge theories with enriched Neumann b.c. coupled to 2d boundary matter other than Fermi multiplets, 
we expect that monopole operators exist at the boundary 
in such a way that the boundary condition $*F=J_{\textrm{2d}}$ admits a boundary monopole 
supported by an appropriate field configuration of the 2d bosonic fields. 

We propose the prescription 
to derive the half-index of 3d gauge theories with enriched Neumann b.c. 
by adding 2d matter to Dirichlet half-index 
and gauging 2d boundary system in the Jeffrey-Kirwan residue method.

%%%%%%%%%%%%%%%%%%%%%%%%%%%%%%%%%%%%%%%%%%%%%
\subsubsection{3d SQED$_{1}$}
\label{sec_enrichsqed1}
%%%%%%%%%%%%%%%%%%%%%%%%%%%%%%%%%%%%%%%%%%%%%
Let us reconsider the half-index for Neumann b.c. $(\mathcal{N}',N')$ of SQED$_{1}$
\begin{align}
\label{sqed1EN_half1}
&
\mathbb{II}^{\textrm{3d SQED}_{1}}_{\mathcal{N}',N'+\Gamma}
(t,z;q)
\nonumber\\
&=
\underbrace{
(q)_{\infty} (q^{\frac12}t^2;q)_{\infty}\oint \frac{ds}{2\pi is}
}_{\mathbb{II}_{\mathcal{N}'}^{\textrm{3d $U(1)$}}}
\underbrace{
\frac{
(q^{\frac12}sz;q)_{\infty} (q^{\frac12}s^{-1}z^{-1};q)_{\infty}
}
{
(q^{\frac14}ts;q)_{\infty} (q^{\frac14}ts^{-1};q)_{\infty}
}
}_{\mathbb{II}_{N'}^{\textrm{3d HM}}(s) \cdot F(q^{\frac12}sz)}. 
\end{align}
This is the same as the Dirichlet half-index $\mathbb{II}_{D}^{\textrm{3d tHM}}(z)$ for twisted hyper, 
however, we would like to manipulate in a different way. 
We can tentatively rewrite the contour integral as a sum over residues of poles 
\begin{align}
\label{sqed1EN_half2}
&
\mathbb{II}_{\mathcal{N}',N'+\Gamma}^{\textrm{3d SQED}_{1}}(t,z;q)
\nonumber\\
&=(q)_{\infty} (q^{\frac12} t^2;q)_{\infty}
\sum_{n=0}^{\infty}
\oint_{s=q^{\frac14+n}t} 
\frac{ds}{2\pi is}
\frac{(q^{\frac12}sz;q)_{\infty} (q^{\frac12}s^{-1}z^{-1};q)_{\infty}}
{(q^{\frac14}t s;q)_{\infty} (q^{\frac14}ts^{-1};q)_{\infty}}.
\end{align}
We can now shift the integration variables to rewrite
\begin{align}
\label{sqed1EN_half3}
&
\mathbb{II}_{\mathcal{N}',N'+\Gamma}^{\textrm{3d SQED}_{1}}(t,z;q)
\nonumber\\
&=
\underbrace{
(q)_{\infty}^2 (q^{\frac12} t^2;q)_{\infty} (q^{\frac12} t^{-2};q)_{\infty}\oint_{s=q^{\frac14}t}
\frac{ds}{2\pi is}
}_{\mathbb{I}^{\textrm{2d $U(1)$}}}
\nonumber\\
&\times 
\left[
\underbrace{
\frac{1}{(q)_{\infty}(q^{\frac12}t^{-2};q)_{\infty}}
}_{\mathbb{II}_{\mathcal{D}'}^{\textrm{3d $U(1)$}}}
\sum_{n\in \mathbb{Z}}
\frac{
(q^{\frac12+n}sz;q)_{\infty}
(q^{\frac12-n}s^{-1}z^{-1};q)_{\infty}
}
{(q^{\frac14+n}ts;q)_{\infty} (q^{\frac14-n}ts^{-1};q)_{\infty}}
\right].
\end{align}
Here we have extended the summation to negative $n$ 
because the integrand has no residue there. 
This can be interpreted as 2d gauging of the boundary global symmetry of 
Dirichlet boundary conditions for the gauge field.

Making use of the relation 
\begin{align}
\label{fm_shift}
F(q^{\frac12+n} s)&=(-1)^{n} q^{-\frac{n^2}{2}}s^{-n} F(q^{\frac12}s), 
\end{align}
the Fermi index can be also brought out of the sum and we get
\begin{align}
\label{sqed1EN_half4}
&
\mathbb{II}_{\mathcal{N}',N'+\Gamma}^{\textrm{3d SQED}_{1}}(t,z;q)
\nonumber\\
&=
\underbrace{
(q)_{\infty}^2 (q^{\frac12} t^2;q)_{\infty} (q^{\frac12} t^{-2};q)_{\infty}\oint_{s=q^{\frac14}t}
\frac{ds}{2\pi is}
}_{\mathbb{I}^{\textrm{2d $U(1)$}}} 
\cdot 
\underbrace{
(q^{\frac12}sx;q)_{\infty} 
(q^{\frac12}s^{-1}x^{-1};q)_{\infty}
}_{F(q^{\frac12}sx)}
\nonumber\\
&\times 
\left[
\underbrace{
\frac{1}{(q)_{\infty}(q^{\frac12}t^{-2};q)_{\infty}}
}_{\mathbb{II}_{\mathcal{D}'}^{\textrm{3d $U(1)$}}}
\sum_{n\in \mathbb{Z}}
(-1)^n q^{-\frac{n^2}{2}} s^{-n} z^{-n}
\underbrace{
\frac{1}{(q^{\frac14+n} ts;q)_{\infty} (q^{\frac14-n}t s^{-1};q)_{\infty}}
}_{\mathbb{II}_{N'}^{\textrm{3d HM}} (q^n s)}
\right].
\end{align}

We can now propose the correct form for 
the half-index of enriched Neumann b.c. in such a way that 
one firstly adds the 2d matter to the Dirichlet half-index and 
then gauge the 2d boundary symmetry in a standard JK-like way.

%%%%%%%%%%%%%%%%%%%%%%%%%%%%%%%%%%%%%%%%%%%%%
\subsubsection{3d SQED$_{1}+$2d twisted hyper$+$3 Fermi$+$neutral Fermi}
\label{sec_enrichsqed1}
%%%%%%%%%%%%%%%%%%%%%%%%%%%%%%%%%%%%%%%%%%%%%

For example, 
consider adding a 2d twisted hypermultiplet at the boundary, 
together with a total of three Fermi multiplets of gauge charge 1 
and one neutral to couple to the twisted hyper and the boundary value of the 3d hyper. 

\begin{figure}
\begin{center}
\includegraphics[width=14cm]{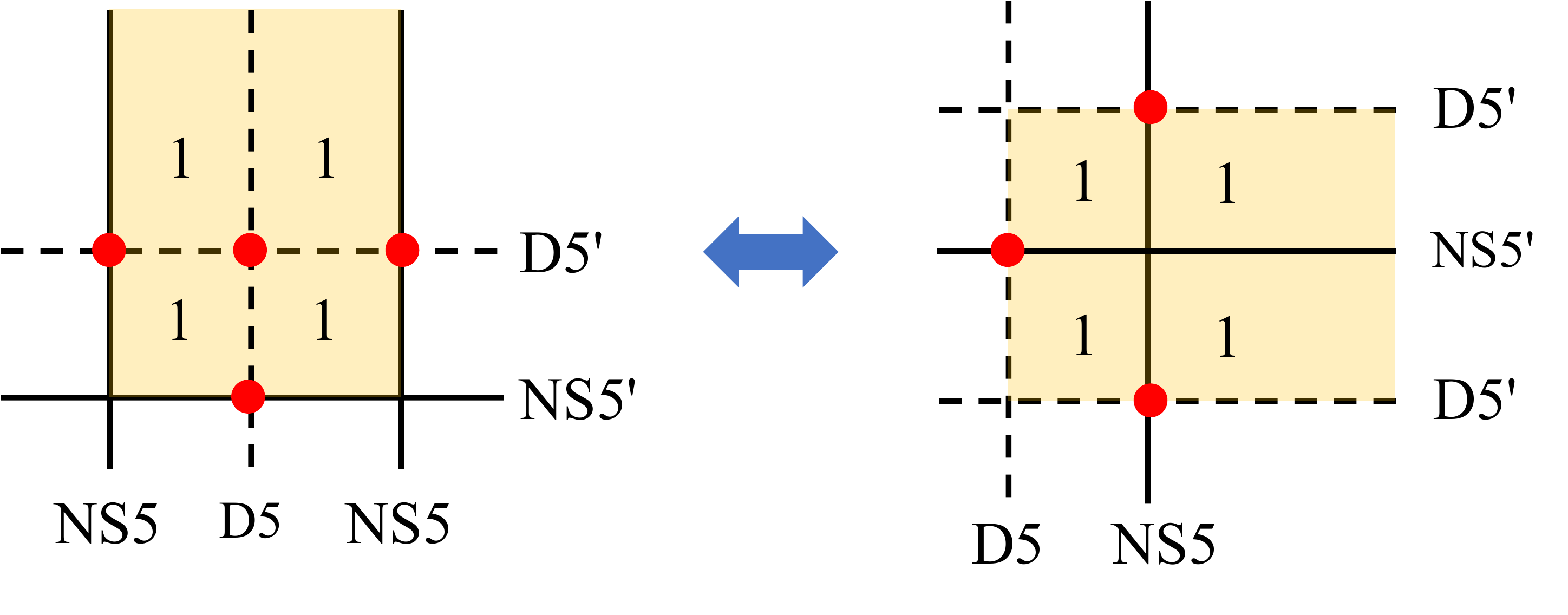}
\caption{
The brane constructions of the enriched Neumann b.c. 
involving 2d twisted hyper, 3 charged Fermi and a neutral Fermi for SQED$_{1}$ and its dual configuration. 
The dual configuration would be described by Neumann b.c. for 3d twisted hyper 
as well as three Fermi multiplets. 
}
\label{figsqed1nEN}
\end{center}
\end{figure}

The brane construction is illustrated in Figure \ref{figsqed1nEN}. 
The defect of D5$'$-brane would couple a 2d twisted hypermultipelt to 3d $U(1)$ vector multiplet. 
At the NS5-D5$'$-junctions and NS5$'$-D5 junction, three charged Fermi multiplets would appear 
while at the D5-D5$'$ junction a neutral Fermi multiplet would show up. 

The half-index reads
\begin{align}
\label{sqed1ENthm1}
&
\mathbb{II}_{\mathcal{N}',N'+\Gamma+T}^{\textrm{3d SQED$_{1}$}}
(t,z,u,x_{\alpha};q)
\nonumber\\
&=
\underbrace{
(q)_{\infty}^2 (q^{\frac12}t^2;q)_{\infty} (q^{\frac12}t^{-2};q)_{\infty}
\oint_{\textrm{JK}} \frac{ds}{2\pi is}
}_{\mathbb{I}^{\textrm{2d $U(1)$}}}
\cdot 
\nonumber\\
&\times 
\underbrace{
(q^{\frac12}u^{\pm3}z^{\pm};q)_{\infty}
}_{F(q^{\frac12} u^{3}z)}
\underbrace{
\frac{1}{
(q^{\frac14}t^{-1} s^{\pm}u^{\pm3}z^{\pm};q)_{\infty}
(q^{\frac34}t s^{\pm}u^{\pm 3}z^{\pm};q)_{\infty}
}
}_{\mathbb{I}^{\textrm{2d tHM}} (su^3 z)}
\cdot 
\prod_{\alpha=1}^{3}
\underbrace{
(q^{\frac12}s^{\pm} z^{\pm} u^{\pm 2}x_{\alpha}^{\pm};q)_{\infty}
}_{F(q^{\frac12}s zu^2 x_{\alpha})}
\nonumber\\
&\times 
\left[
\frac{1}{(q)_{\infty} (q^{\frac12} t^{-2};q)_{\infty}}
\sum_{n\in \mathbb{Z}}
(-1)^{n}q^{-\frac{n^2}{2}} s^{-n}z^{-n}
\underbrace{
\frac{1}{(q^{\frac14\pm n}ts^{\pm};q)_{\infty}}
}_{\mathbb{II}_{N'}^{\textrm{3d HM}} (q^{n}s)}
\right]
\end{align}
with $\prod_{\alpha=1}^{3}x_{\alpha}=1$. 

The JK prescription should pick two poles. 
One is the pole of 3d hyper at $s=q^{\frac14}t$ 
and the other is the pole for 2d twisted hyper at $s=q^{\frac14}t^{-1}z^{-1}u^{-3}$. 

The contributions from the first pole of 3d hyper can be shifted back to 
\begin{align}
\label{sqed1ENthm2a}
&
\mathbb{II}_{1,\ \mathcal{N}',N'+\Gamma+T}^{\textrm{3d SQED$_{1}$}}
(t,z,u,x_{\alpha};q)
\nonumber\\
&=
(q)_{\infty}(q^{\frac12} t^2;q)_{\infty}
\cdot 
\underbrace{
(q^{\frac12}u^{\pm3}z^{\pm};q)_{\infty}
}_{F(q^{\frac12} u^3z)}
\sum_{n=0}^{\infty}\oint_{s=q^{\frac14+n}t} \frac{ds}{2\pi is}
\nonumber\\
&\times 
\prod_{\alpha=1}^{3} 
\underbrace{
\frac{ (q^{\frac12} s^{\pm} z^{\pm} u^{\pm2} x_{\alpha}^{\pm};q )_{\infty} }
{ (q^{\frac14} t s^{\pm};q)_{\infty} \cdot 
(q^{\frac14}t^{-1} s^{\pm} u^{\pm3} z^{\pm};q)_{\infty} 
(q^{\frac34}t s^{\pm} u^{\pm3}z^{\pm};q )_{\infty}
 }
 }_{
 \mathbb{II}_{N'}^{\textrm{3d HM}}(s)
 \cdot 
 \mathbb{I}^{\textrm{2d tHM}}(su^{3}z)
 \cdot 
 F(q^{\frac12}szu^2 x_{\alpha})
 }
 \nonumber\\
 &=
 \frac{(q^{\frac12}t^2;q)_{\infty}}
 {(q)_{\infty}}
 \prod_{\alpha=1}^{3}
 \underbrace{
\frac{
(q^{\frac34}tzu^2 x_{\alpha};q)_{\infty}
(q^{\frac14}t^{-1}z^{-1}u^{-2}x_{\alpha}^{-1};q)_{\infty}
}
{
(q t^2 u^3 z;q)_{\infty} 
(t^{-2}u^{-3}z^{-1};q)_{\infty}
}
}_{
F(q^{\frac34}tzu^{2}x_{\alpha})
\cdot 
C(qt^2 u^3 z)
}
\nonumber\\
&\times 
\sum_{n=0}^{\infty}
\frac{(q^{1+n};q)_{\infty}}
{(q^{\frac12+n}t^2;q)_{\infty}}
q^{\frac{n}{4}}
\cdot 
t^{-n}
\cdot 
z^{-n}
\cdot 
\prod_{\alpha=1}^{3} x_{\alpha}^{-n}.
\end{align}

The contributions from the second pole of 2d twisted hyper become
\begin{align}
\label{sqed1ENthm2b}
&
\mathbb{II}_{2,\ \mathcal{N}',N'+\Gamma+T}^{\textrm{3d SQED$_{1}$}}
(t,z,u,x_{\alpha};q)
\nonumber\\
&=
(q)_{\infty}(q^{\frac12} t^2;q)_{\infty}
\cdot 
\underbrace{
(q^{\frac12}u^{\pm3}z^{\pm};q)_{\infty}
}_{F(q^{\frac12} u^3z)}
\sum_{n=-\infty}^{\infty}\oint_{s=q^{\frac14+n}t^{-1}z^{-1}u^{-3}} \frac{ds}{2\pi is}
\nonumber\\
&\times 
\prod_{\alpha=1}^{3} 
\underbrace{
\frac{ (q^{\frac12} s^{\pm} z^{\pm} u^{\pm2} x_{\alpha}^{\pm};q )_{\infty} }
{ (q^{\frac14} t s^{\pm};q)_{\infty} \cdot 
(q^{\frac14}t^{-1} s^{\pm} u^{\pm3} z^{\pm};q)_{\infty} 
(q^{\frac34}t s^{\pm} u^{\pm3}z^{\pm};q )_{\infty}
 }
 }_{
 \mathbb{II}_{N'}^{\textrm{3d HM}}(s)
 \cdot 
 \mathbb{I}^{\textrm{2d tHM}}(su^{3}z)
 \cdot 
 F(q^{\frac12}szu^2 x_{\alpha})
 }
 \nonumber\\
 &=
 \frac{1}{(q)_{\infty} (q^{\frac12}t^{-2};q)_{\infty} }
 \prod_{\alpha=1}^{3}
 \underbrace{
 \frac{
(q^{\frac34}t^{-1}u^{-1}x_{\alpha};q)_{\infty}
(q^{\frac14}tux_{\alpha}^{-1};q)_{\infty}
\cdot 
(q^{\frac12}u^3z;q)_{\infty}
(q^{\frac12}u^{-3}z^{-1};q)_{\infty}
 }
 {
 (t^2u^3z;q)_{\infty}
 (qt^{-2}u^{-3}z^{-1};q)_{\infty}
 }
 }_{
 F(q^{\frac34}t^{-1}u^{-1}x_{\alpha}) 
\cdot 
F(q^{\frac12}u^3z)
\cdot 
C(t^2u^3z)
 }
 \nonumber\\
 &\times 
 \sum_{n=-\infty}^{\infty}
 \frac{
 (q^{1+n}t^{-2}u^{-3}z^{-1};q)_{\infty}
 }
 {
 (q^{\frac12+n}u^{-3}z^{-1};q)_{\infty}
 }
 q^{\frac{n}{4}}
 \cdot 
 t^{-n}
 \cdot 
 z^{-n}
 \cdot 
 \prod_{\alpha=1}^{3}
 x_{\alpha}^{-n}. 
\end{align}
The above argument means that 
the correct prescription for enriched Neumann half-index would be to pick the standard semi-infinite sequence of poles for 3d matter fields obeying Neumann b.c. 
However, one needs a full infinite sequence of poles of only one charge/type for 2d bosonic matter fields. 

The final answer is surprisingly simple:
\begin{align}
\label{sqed1ENthm3}
&
\mathbb{II}_{\mathcal{N}',N'+\Gamma+T}^{\textrm{3d SQED$_{1}$}}
(t,z,u,x_{\alpha};q)
\nonumber\\
&=
\sum_{i=1}^{2}
\mathbb{II}_{i,\ \mathcal{N}',N'+\Gamma+T}^{\textrm{3d SQED$_{1}$}}
(t,z,u,x_{\alpha};q)
\nonumber\\
&=
\underbrace{
\frac{1}{(q^{\frac14}t^{-1}z^{\pm};q)_{\infty}}
}_{\mathbb{II}_{N}^{\textrm{3d tHM}} (z)}
\cdot 
\prod_{\alpha=1}^{3}
\underbrace{
(q^{\frac12}u^{\pm} z^{\pm} x_{\alpha}^{\pm};q)_{\infty}
}_{F(q^{\frac12}uz x_{\alpha}^{-1})}.
\end{align}
It includes the half-index of Neumann b.c. $N$ for 3d twisted hypermultiplet of fugacity $z$ 
and three full-indices of Fermi multiplets of fugacities $uzx_{\alpha}^{-1}$ with $\alpha=1,2,3$. 
This beautifully agrees with the brane box picture! 

The result supports the duality:
\begin{align}
\label{EN_duality1}
&
(\mathcal{N}',N'+\textrm{2d twisted hyper$+$3 Fermi$+$neutral Ferm})
\textrm{ for 3d SQED$_{1}$}
\nonumber\\
&\leftrightarrow 
\textrm{$N$ for 3d twisted hyper $+$ 3 Fermi}. 
\end{align}

%%%%%%%%%%%%%%%%%%%%%%%%%%%%%%%%%%%%%%%%%%%%%
%%%%%%%%%%%%%%%%%%%%%%%%%%%%%%%%%%%%%%%%%%%%%
\section{$\mathcal{N}=(0,4)$ boundaries of 3d theories and corners of 4d theories}
\label{sec_4d3d}
%%%%%%%%%%%%%%%%%%%%%%%%%%%%%%%%%%%%%%%%%%%%%
%%%%%%%%%%%%%%%%%%%%%%%%%%%%%%%%%%%%%%%%%%%%%
We examine the configurations which involve 
the quarter-BPS corners of 4d $\mathcal{N}=4$ SYM theory 
and the half-BPS boundary of 3d $\mathcal{N}=4$ gauge theories. 
We show the dualities between the configurations of gauge theories, 
which involve local operators in 4d, 3d and 2d theories, by computing the quarter-indices. 
\footnote{See \cite{Gaiotto:2019jvo} for the half-indices and quarter-indices of 4d $\mathcal{N}=4$ gauge theory. }

%%%%%%%%%%%%%%%%%%%%%%%%%%%%%%%%%%%%%%%%%%%%%
%%%%%%%%%%%%%%%%%%%%%%%%%%%%%%%%%%%%%%%%%%%%%
\subsection{4d $U(N)|$3d $U(M)$ with $\mathcal{N}'$}
\label{sec_4d3dNN}
\begin{figure}
\begin{center}
\includegraphics[width=11cm]{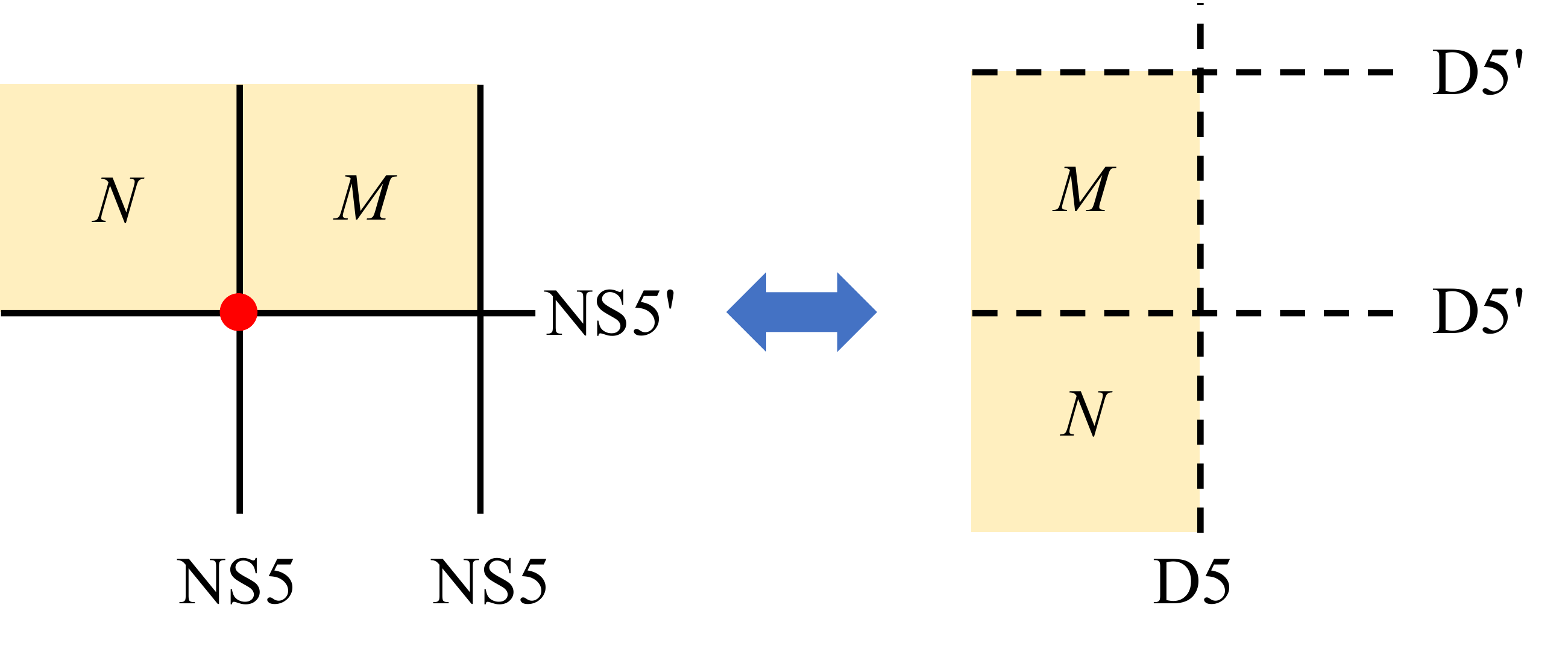}
\caption{
The brane constructions of the configuration 4d $U(N)|$3d $U(M)$ with Neumann b.c. $\mathcal{N}'$ 
and its mirror corner configuration that has no gauge symmetry. 
}
\label{fig4dun3dum}
\end{center}
\end{figure}

%%%%%%%%%%%%%%%%%%%%%%%%%%%%%%%%%%%%%%%%%%%%%
%%%%%%%%%%%%%%%%%%%%%%%%%%%%%%%%%%%%%%%%%%%%%
Consider the configuration 
where the corner of 4d $\mathcal{N}=4$ $U(N)$ gauge theory 
obeying a pair of Neumann b.c. $(\mathcal{N}, \mathcal{N}')$ 
coupled to the boundary of 3d $\mathcal{N}=4$ $U(M)$ gauge theory 
satisfying Neumann b.c. $\mathcal{N}'$ 
through the 3d hypermultiplet with Neumann b.c. $N'$ 
transforming as $({\bf N}, \overline{\bf M})$ $\oplus$ $(\overline{\bf N}, {\bf M})$ 
under the $U(N)\times U(M)$ gauge symmetry. 
The corresponding brane configuration is shown in Figure \ref{fig4dun3dum}. 
From the brane box analysis \cite{Hanany:2018hlz}, 
we expect that the Abelian part of gauge anomaly is cancelled by 
the 2d cross-determinant Fermi multiplet living at the junction. 
We denote this configuration by 4d $U(N)|$3d $U(M)$ with Neumann b.c. $\mathcal{N}'$. 
The dual configuration can be obtained by applying S-duality to the brane configuration in Figure \ref{fig4dun3dum}. 
As it does not have gauge symmetry, we find that the corresponding quarter-index is expressed as a product of 
indices of free fields. 
These dual configurations are the generalizations 
of NS5-NS5$'$ and D5-D5$'$ junctions studied in \cite{Gaiotto:2019jvo}.

%%%%%%%%%%%%%%%%%%%%%%%%%%%%%%%%%%%%%%%%%%%%%
\subsubsection{4d $U(1)|$3d $U(1)$ with $\mathcal{N}'$}
\label{sec_4du13du1NN}
%%%%%%%%%%%%%%%%%%%%%%%%%%%%%%%%%%%%%%%%%%%%%
Let us begin with the configuration 4d $U(1)|$3d $U(1)$ with $\mathcal{N}'$. 
The junction keeps 4d $U(1)$ gauge symmetry and 3d $U(1)$ gauge symmetry 
and involves the bi-fundamental 3d hypermultiplet and the 2d cross-determinant Fermi multiplet.

The quarter-index reads
\begin{align}
\label{4du1_3du1NN1}
&
\mathbb{IV}^{\textrm{4d $U(1)|(1)|$}}_{\mathcal{N}\mathcal{N}'}
\nonumber\\
&=
\underbrace{
(q)_{\infty} \oint \frac{ds_{1}}{2\pi is_{1}}
}_{\mathbb{IV}_{\mathcal{N}\mathcal{N}'}^{\textrm{4d $U(1)$}}}
\cdot 
\underbrace{
(q)_{\infty} (q^{\frac12} t^2;q)_{\infty} 
\oint \frac{ds_{2}}{2\pi is_{2}}
}_{\mathbb{II}_{\mathcal{N}'}^{\textrm{3d $U(1)$}}}
\cdot 
\underbrace{
\frac{
(q^{\frac12}s_{1}^{\pm} s_{2}^{\mp} z^{\pm};q)_{\infty}
}
{
(q^{\frac14}t s_{1}^{\pm} s_{2}^{\mp};q)_{\infty}
}
}_{\mathbb{II}_{N'}^{\textrm{3d HM}}\left( \frac{s_{1}}{s_{2}} \right)
\cdot F\left(q^{\frac12}\frac{s_{1}}{s_{2}}z\right)
}.
\end{align}

In the S-dual brane configuration there exists a single D3-brane for each side of one of the D5$'$-branes 
so that the 3d twisted hypermultiplet arising from D3-D5$'$ string. 
According to the presence of D5-brane on which D3-brane ends, 
the 3d twisted hypermultiplet should obey the Dirichlet b.c. $D$. 

In fact, the quarter-index (\ref{4du1_3du1NN1}) coincides with 
\begin{align}
\label{4du1_3du1NN2}
\mathbb{IV}^{1|[1]|}_{\mathcal{D}\mathcal{D}'}
&=
\underbrace{
(q)_{\infty}}_{\mathbb{IV}_{\mathcal{D}\mathcal{D}'}^{\textrm{4d $U(1)$}}}
 \cdot 
\underbrace{
(q^{\frac34}t z;q)_{\infty} (q^{\frac34}tz^{-1};q)_{\infty}
}_{\mathbb{II}_{D}^{\textrm{3d tHM}}(z)}.
\end{align}

%%%%%%%%%%%%%%%%%%%%%%%%%%%%%%%%%%%%%%%%%%%%%
\subsubsection{4d $U(2)|$3d $U(1)$ with $\mathcal{N}'$}
\label{sec_4du23du1NN}
%%%%%%%%%%%%%%%%%%%%%%%%%%%%%%%%%%%%%%%%%%%%%
Next consider the junction 4d $U(2)|$3d $U(1)$ with $\mathcal{N}'$. 
The configuration has 4d $U(2)$ gauge symmetry and 3d $U(1)$ gauge symmetry. 

One can evaluate the quarter-index as
\begin{align}
\label{4du2_3du1NN1}
&
\mathbb{IV}^{\textrm{4d $U(2)|(1)|$}}_{\mathcal{N}\mathcal{N}'}
\nonumber\\
&=
\underbrace{
\frac12 (q)_{\infty}^2 \oint \frac{ds_{1}}{2\pi is_{1}} \frac{ds_{2}}{2\pi is_{2}}
\left( \frac{s_{1}}{s_{2}};q \right)_{\infty}
\left( \frac{s_{2}}{s_{1}};q \right)_{\infty}
}_{\mathbb{IV}_{\mathcal{N}\mathcal{N}'}^{\textrm{4d $U(2)$}}}
\cdot 
\underbrace{
(q)_{\infty} (q^{\frac12} t^2;q)_{\infty} 
\oint \frac{ds_{3}}{2\pi is_{3}}
}_{\mathbb{II}_{\mathcal{N}'}^{\textrm{3d $U(1)$}}}
\nonumber\\
&\times 
\prod_{i=1}^{2}
\underbrace{
\frac{
1
}
{
(q^{\frac14}t s_{i}^{\pm} s_{3}^{\mp};q)_{\infty}
}
}_{\mathbb{II}^{\textrm{3d HM}}_{N'}\left(\frac{s_{i}}{s_{3}} \right)}
\cdot 
\underbrace{
(q^{\frac12}s_{1}^{\pm} s_{2}^{\pm} s_{3}^{\mp}z^{\pm};q)_{\infty}
}_{F\left(q^{\frac12} \frac{s_{1}s_{2}}{s_{3}} z \right)}
.
\end{align}

Unlike the previous case, the number of D3-branes jumps across the D5$'$-brane in the dual configuration 
and therefore there is no twisted hypermultiplet. 

We find that the quarter-index (\ref{4du2_3du1NN1}) agrees with 
\begin{align}
\label{4du2_3du1NN2}
\mathbb{IV}^{2|[1]|}_{\mathcal{D}\mathcal{D}'}
&=
\underbrace{
(q)_{\infty}
}_{\mathbb{IV}_{\mathcal{D}\mathcal{D}'}^{\textrm{4d $U(1)$}}}.
\end{align}
This is identified with the quarter-index of 4d $U(1)$ gauge theory 
obeying a pair of Dirichlet b.c. $(\mathcal{D},\mathcal{D}')$.

%%%%%%%%%%%%%%%%%%%%%%%%%%%%%%%%%%%%%%%%%%%%%
\subsubsection{4d $U(3)|$3d $U(1)$ with $\mathcal{N}'$}
\label{sec_4du33du1NN}
%%%%%%%%%%%%%%%%%%%%%%%%%%%%%%%%%%%%%%%%%%%%%

The quarter-index is 
\begin{align}
\label{4du3_3du1NN1}
&
\mathbb{IV}^{\textrm{4d $U(3)|(1)|$}}_{\mathcal{N}\mathcal{N}'}
\nonumber\\
&=
\underbrace{
\frac{1}{3!} (q)_{\infty}^3 \oint \prod_{i=1}^{3} \frac{ds_{i}}{2\pi is_{i}} 
\prod_{i\neq j}
\left( \frac{s_{i}}{s_{j}};q \right)_{\infty}
}_{\mathbb{IV}_{\mathcal{N}\mathcal{N}'}^{\textrm{4d $U(3)$}}}
\cdot 
\underbrace{
(q)_{\infty} (q^{\frac12} t^2;q)_{\infty} 
\oint \frac{ds_{4}}{2\pi is_{4}}
}_{\mathbb{II}_{\mathcal{N}'}^{\textrm{3d $U(1)$}}}
\nonumber\\
&\times 
\prod_{i=1}^{3}
\underbrace{
\frac{
1
}
{
(q^{\frac14}t s_{i}^{\pm} s_{4}^{\mp};q)_{\infty}
}
}_{\mathbb{II}^{\textrm{3d HM}}_{N'}\left(\frac{s_{i}}{s_{4}} \right)}
\cdot 
\underbrace{
(q^{\frac12}s_{1}^{\pm} s_{2}^{\pm} s_{3}^{\pm}s_{4}^{\mp}z^{\pm};q)_{\infty}
}_{F\left(q^{\frac12} \frac{s_{1}s_{2}s_{3}}{s_{4}}z \right)}
.
\end{align}

The quarter-index (\ref{4du3_3du1NN1}) agrees with 
\begin{align}
\label{4du3_3du1NN2}
\mathbb{IV}^{2|[1]|}_{\mathcal{D}\mathcal{D}'}
&=
\underbrace{
(q)_{\infty}
}_{\mathbb{IV}_{\mathcal{D}\mathcal{D}'}^{\textrm{4d $U(1)$}}}.
\end{align}

%%%%%%%%%%%%%%%%%%%%%%%%%%%%%%%%%%%%%%%%%%%%%
\subsubsection{4d $U(N)|$3d $U(1)$ with $\mathcal{N}'$}
\label{sec_4duN3du1NN}
%%%%%%%%%%%%%%%%%%%%%%%%%%%%%%%%%%%%%%%%%%%%%
Now we are led to propose the generalization of quarter-index for 
the junction 4d $U(N)|$3d $U(1)$ with $\mathcal{N}'$ for $ N>1$. 
The junction preserves 4d $U(N)$ gauge symmetry and 3d $U(1)$ gauge symmetry. 
It also contains 3d bi-fundamental hyper satisfying Neumann b.c. $N'$ and 2d cross-determinant Fermi multiplet. 

The quarter-index is given by
\begin{align}
\label{4duN_3du1NN1}
&
\mathbb{IV}^{\textrm{4d $U(N)|(1)|$}}_{\mathcal{N}\mathcal{N}'}
\nonumber\\
&=
\underbrace{
\frac{1}{N!} (q)_{\infty}^N \oint \prod_{i=1}^{N} \frac{ds_{i}}{2\pi is_{i}} 
\prod_{i\neq j}
\left( \frac{s_{i}}{s_{j}};q \right)_{\infty}
}_{\mathbb{IV}_{\mathcal{N}\mathcal{N}'}^{\textrm{4d $U(N)$}}}
\cdot 
\underbrace{
(q)_{\infty} (q^{\frac12} t^2;q)_{\infty} 
\oint \frac{ds_{N+1}}{2\pi is_{N+1}}
}_{\mathbb{II}_{\mathcal{N}'}^{\textrm{3d $U(1)$}}}
\nonumber\\
&\times 
\prod_{i=1}^{N}
\underbrace{
\frac{
1
}
{
(q^{\frac14}t s_{i}^{\pm} s_{N+1}^{\mp};q)_{\infty}
}
}_{\mathbb{II}^{\textrm{3d HM}}_{N'}\left(\frac{s_{i}}{s_{N+1}} \right)}
\cdot 
\underbrace{
(q^{\frac12} \prod_{i=1}^{N} s_{i}^{\pm} s_{N+1}^{\mp}z^{\pm};q)_{\infty}
}_{F\left( q^{\frac12} \frac{\prod_{i=1}^{N}s_{i}}{s_{N+1}} z \right)}. 
\end{align}

The quarter-index (\ref{4duN_3du1NN1}) would be equal to 
the quarter-index of 4d $U(1)$ gauge theory with a pair of Dirichlet b.c. $(\mathcal{D}, \mathcal{D}')$
\begin{align}
\label{4duN_3du1NN2}
\mathbb{IV}^{N|[1]|}_{\mathcal{D}\mathcal{D}'}
&=
\underbrace{
(q)_{\infty}
}_{\mathbb{IV}_{\mathcal{D}\mathcal{D}'}^{\textrm{4d $U(1)$}}}. 
\end{align}

%%%%%%%%%%%%%%%%%%%%%%%%%%%%%%%%%%%%%%%%%%%%%
\subsubsection{4d $U(2)|$3d $U(2)$ with $\mathcal{N}'$}
\label{sec_4du23du2NN}
%%%%%%%%%%%%%%%%%%%%%%%%%%%%%%%%%%%%%%%%%%%%%
Consider the case with both 4d and 3d gauge symmetries are non-Abelian. 
For the configuration 4d $U(2)|$3d $U(2)$ with $\mathcal{N}'$, 
we have 4d $U(2)$ gauge symmetry and 3d $U(2)$ gauge symmetry. 
At the junction there are 3d bi-fundamental hyper obeying Neumann b.c. $N'$ and 2d cross-determinant Fermi multiplet.

The quarter-index is evaluated as 
\begin{align}
\label{4du2_3du2NN1}
&
\mathbb{IV}^{\textrm{4d $U(2)|(2)|$}}_{\mathcal{N}\mathcal{N}'}
\nonumber\\
&=
\underbrace{
\frac12 (q)_{\infty}^2 \oint \prod_{i=1}^{2} 
\frac{ds_{i}}{2\pi is_{i}}
\prod_{i\neq j}\left( \frac{s_{i}}{s_{j}};q \right)_{\infty}
}_{\mathbb{IV}_{\mathcal{N}\mathcal{N}'}^{\textrm{4d $U(2)$}}}
\nonumber\\
&\times 
\underbrace{
\frac12 (q)_{\infty}^2 (q^{\frac12} t^2;q)_{\infty}^2 
\oint \prod_{i=3}^{4} \frac{ds_{i}}{2\pi is_{i}} 
\prod_{i\neq j}
\left( \frac{s_{i}}{s_{j}};q \right)_{\infty} 
\left( q^{\frac12} t^2 \frac{s_{i}}{s_{j}};q \right)_{\infty}
}_{\mathbb{II}_{\mathcal{N}'}^{\textrm{3d $U(2)$}}}
\nonumber\\
&\times 
\prod_{i=1}^{2}\prod_{j=3}^{4} 
\underbrace{
\frac{
1
}
{(q^{\frac14}t s_{i}^{\pm} s_{j}^{\mp};q)_{\infty}}
}_{\mathbb{II}_{N'}^{\textrm{3d HM}} \left(\frac{s_{i}}{s_{j}} \right) }
\cdot 
\underbrace{
(q^{\frac12} s_{1}^{\pm} s_{2}^{\pm} s_{3}^{\mp} s_{4}^{\mp}z^{\pm};q)_{\infty}
}_{F\left( q^{\frac12} \frac{s_{1}s_{2}}{s_{3}s_{4}}z \right) }. 
\end{align}

In the S-dual configuration, 
there are two D3-branes in each side of one of the D5$'$-branes, 
which would lead to the 3d $\mathcal{N}=4$ fundamental twisted hypermultiplets. 
They should obey the boundary condition corresponding to the D5-brane 
which is characterized by the Nahm pole with 
and embedding $\rho:$ $\mathfrak{su}(2)$ $\rightarrow$ $\mathfrak{u}(2)$. 

We have confirmed that 
the quarter-index (\ref{4du2_3du2NN1}) agrees with 
\begin{align}
\label{4du2_3du2NN2}
\mathbb{IV}^{2|[2]|}_{\mathcal{D}\mathcal{D}'}
&=
\underbrace{
(q)_{\infty}}_{\mathbb{IV}_{\mathcal{D}\mathcal{D}'}^{\textrm{4d $U(1)$}}} 
\cdot 
(q t^2 z;q)_{\infty} (q t^2 z^{-1};q)_{\infty}.
\end{align}
As discussed in \cite{Gaiotto:2019jvo}, 
the contributions $(qt^2z^{\pm};q)_{\infty}$ from 3d $\mathcal{N}=4$ fundamental twisted hyper can be obtained by the Higgsing procedure 
that produces a Nahm pole. 
Identifying the fugacities $z_{1}$, $z_{2}$ for the $U(2)$ global symmetry with $q^{-\frac14}z$, $q^{\frac14}z$ 
gives $(q^{\frac12}z^{\pm};q)_{\infty} (q t^{2}z^{\pm};q)_{\infty}$. 
Then stripping off the Fermi index, we get the terms in (\ref{4du2_3du2NN2}).

%%%%%%%%%%%%%%%%%%%%%%%%%%%%%%%%%%%%%%%%%%%%%
\subsubsection{4d $U(3)|$3d $U(2)$ with $\mathcal{N}'$}
\label{sec_4du33du2NN}
%%%%%%%%%%%%%%%%%%%%%%%%%%%%%%%%%%%%%%%%%%%%%
For the junction 4d $U(3)|$3d $U(2)$ with $\mathcal{N}'$, 
we have 4d $U(3)$ gauge symmetry and 3d $U(2)$ gauge symmetry. 
There are 3d $\mathcal{N}=4$ bi-fundamental hyper with Neumann b.c. $N'$ 
and 2d cross-determinant Fermi multiplet.

We have the quarter-index 
\begin{align}
\label{4du3_3du2NN1}
&
\mathbb{IV}^{\textrm{4d $U(3)|(2)|$}}_{\mathcal{N}\mathcal{N}'}
\nonumber\\
&=
\underbrace{
\frac{1}{3!} (q)_{\infty}^3 
\oint \prod_{i=1}^{3} \frac{ds_{i}}{2\pi is_{i}}
\prod_{i\neq j}
\left( \frac{s_{i}}{s_{j}};q \right)_{\infty}
}_{\mathbb{IV}_{\mathcal{N}\mathcal{N}'}^{\textrm{4d $U(3)$}}} 
\nonumber\\
&\times 
\underbrace{
\frac12 (q)_{\infty}^2 (q^{\frac12}t^2;q)_{\infty}^2 
\oint \prod_{i=4}^{5} \frac{ds_{i}}{2\pi is_{i}} 
\prod_{i\neq j} \left( \frac{s_{i}}{s_{j}};q \right)_{\infty} 
\left( q^{\frac12}t^2 \frac{s_{i}}{s_{j}};q \right)_{\infty}
}_{\mathbb{II}_{\mathcal{N}'}^{\textrm{3d $U(2)$}}}
\nonumber\\
&\times 
\prod_{i=1}^{3} 
\prod_{j=4}^{5}
\underbrace{
\frac{1}{(q^{\frac14}t s_{i}^{\pm} s_{j}^{\mp};q)_{\infty}}
}_{\mathbb{II}_{N'}^{\textrm{3d HM}} \left( \frac{s_{i}}{s_{j}} \right)}
\cdot 
\underbrace{
(q^{\frac12} s_{1}^{\pm} s_{2}^{\pm} s_{3}^{\pm} s_{4}^{\mp} s_{5}^{\mp} z^{\pm};q)_{\infty}
}_{F \left(q^{\frac12}\frac{s_{1}s_{2}s_{3}}{s_{4}s_{5}}z \right)}. 
\end{align}

In the dual configuration, 
according to the unequal numbers of D3-branes in each side of D5$'$-branes, 
there is no 3d fundamental twisted hypermultiplet. 

In fact, the quarter-index (\ref{4du3_3du2NN1}) coincides with 
the quarter-index of 4d $\mathcal{N}=4$ $U(1)$ gauge theory 
with a pair of Dirichlet b.c. $(\mathcal{D},\mathcal{D}')$ 
\begin{align}
\label{4du3_3du2NN2}
\mathbb{IV}^{3|(2)|}_{\mathcal{D}\mathcal{D}'}
&=
\underbrace{
(q)_{\infty}}_{\mathbb{IV}_{\mathcal{D}\mathcal{D}'}^{\textrm{4d $U(1)$}}}.
\end{align}

%%%%%%%%%%%%%%%%%%%%%%%%%%%%%%%%%%%%%%%%%%%%%
\subsubsection{4d $U(3)|$3d $U(3)$ with $\mathcal{N}'$}
\label{sec_4du33du3NN}
%%%%%%%%%%%%%%%%%%%%%%%%%%%%%%%%%%%%%%%%%%%%%

Similarly, we can compute the quarter-index for 
the configuration 4d $U(3)|$3d $U(3)$ with $\mathcal{N}'$ as
\begin{align}
\label{4du3_3du3NN1}
&
\mathbb{IV}^{\textrm{4d $U(3)|(3)|$}}_{\mathcal{N}\mathcal{N}'}
\nonumber\\
&=
\underbrace{
\frac{1}{3!} (q)_{\infty}^3 
\oint \prod_{i=1}^3 \frac{ds_{i}}{2\pi is_{i}}
\prod_{i\neq j}
\left( \frac{s_{i}}{s_{j}};q \right)_{\infty}
}_{\mathbb{IV}_{\mathcal{N}\mathcal{N}'}^{\textrm{3d $U(3)$}}}
\nonumber\\
&\times 
\underbrace{
\frac{1}{3!}
(q)_{\infty}^3 (q^{\frac12}t^2;q)_{\infty}^3 
\oint \prod_{i=4}^{6}\frac{ds_{i}}{2\pi is_{i}}
\prod_{i\neq j}
\left( \frac{s_{i}}{s_{j}};q \right)_{\infty}
\left(q^{\frac12}t^2 \frac{s_{i}}{s_{j}};q \right)_{\infty}
}_{\mathbb{II}_{\mathcal{N}'}^{\textrm{3d $U(3)$}}}
\nonumber\\
&\times 
\prod_{i=1}^{3}
\prod_{j=4}^{6}
\underbrace{
\frac{1}{(q^{\frac14}ts_{i}^{\pm} s_{j}^{\mp};q)_{\infty}}
}_{\mathbb{II}_{N'}^{\textrm{3d HM}} \left( \frac{s_{i}}{s_{j}} \right) }
\cdot 
\underbrace{
(q^{\frac12}s_{1}^{\pm}s_{2}^{\pm}s_{3}^{\pm} s_{4}^{\mp}s_{5}^{\mp} s_{6}^{\mp}z^{\pm};q)_{\infty}
}_{F \left(q^{\frac12} \frac{s_{1}s_{2}s_{3}}{s_{4}s_{5}s_{6}} z\right)}. 
\end{align}

Although the dual junction has no gauge symmetry, 
it includes 3d $\mathcal{N}=4$ fundamental twisted hyper 
since the equal numbers of D3-branes exist in both sides of D5$'$-brane. 

We find that 
the quarter-index (\ref{4du3_3du3NN1}) coincides with
\begin{align}
\label{4du3_3du3NN2}
\mathbb{IV}^{3|(3)|}_{\mathcal{D}\mathcal{D}'}
&=
\underbrace{
(q)_{\infty}
}_{\mathbb{IV}_{\mathcal{D}\mathcal{D}'}^{\textrm{4d $U(1)$}}}
\cdot 
(q^{\frac54}t^3 z;q)_{\infty}(q^{\frac54}t^3 z^{-1};q)_{\infty}. 
\end{align}
Here the contributions from 3d twisted hyper are not Dirichlet b.c. $D$ 
but rather the Nahm pole boundary condition with an embedding 
$\rho:$ $\mathfrak{su}(2)$ $\rightarrow$ $\mathfrak{u}(3)$. 
To achieve the RG flow from Dirichlet to a Nahm pole, 
the corresponding Higgsing procedure fixes  the 
fugacities $z_{1}$, $z_{2}$, $z_{3}$ for $U(3)$ global symmetry 
to $q^{\frac12}t^2 z$, $q^{\frac12}t^2 z^{-1}$, $z$, 
which yields 
$(q^{\frac{5}{4}} t^3z^{\pm};q)_{\infty}$
$(q^{\frac{1}{4}} t^{-1}z^{\pm};q)_{\infty}$
$(q^{\frac34}tz^{\pm};q)_{\infty}$. 
Stripping off the Fermi indices 
$F(q^{\frac14}t^{-1}z)$ $F(q^{\frac14}t^{-1}z^{-1})$, 
we get the correct contributions in (\ref{4du3_3du3NN2}).

%%%%%%%%%%%%%%%%%%%%%%%%%%%%%%%%%%%%%%%%%%%%%
\subsubsection{4d $U(N)|$3d $U(M)$ with $\mathcal{N}'$}
\label{sec_4dun3dumNN}
%%%%%%%%%%%%%%%%%%%%%%%%%%%%%%%%%%%%%%%%%%%%%
Now we would like to propose the generalization. 
The configuration keeps 4d $U(N)$ gauge symmetry and 3d $U(M)$ gauge symmetry. 
At the junction, there is a 3d bi-fundamental hypermultiplet transforming as 
$({\bf N}, \overline{\bf M})$ $\oplus$ $(\overline{\bf N},{\bf M})$ with Neumann  b.c. $N'$. 
In addition, there is a cross-determinant Fermi multiplet 
which cancels the Abelian part of gauge anomaly.

We have the quarter-index 
\begin{align}
\label{4dun_3dumNN1}
&
\mathbb{IV}^{\textrm{4d $U(N)|(M)|$}}_{\mathcal{N}\mathcal{N}'}
\nonumber\\
&=
\underbrace{
\frac{1}{N!} (q)_{\infty}^{N} 
\oint \prod_{i=1}^{N} \frac{ds_{i}}{2\pi is_{i}} 
\prod_{i\neq j}\left( \frac{s_{i}}{s_{j}};q \right)_{\infty}
}_{\mathbb{IV}_{\mathcal{N}\mathcal{N}'}^{\textrm{4d $U(N)$}}}
\nonumber\\
&\times 
\underbrace{
\frac{1}{M!}(q)_{\infty}^{M} (q^{\frac12}t^2;q)_{\infty}^M 
\oint \prod_{i=N+1}^{N+M} \frac{ds_{i}}{2\pi is_{i}}
\prod_{i\neq j}\left( \frac{s_{i}}{s_{j}};q \right)_{\infty}
\left( q^{\frac12}t^2 \frac{s_{i}}{s_{j}};q \right)_{\infty}
}_{\mathbb{II}_{\mathcal{N}'}^{\textrm{3d $U(M)$}}}
\nonumber\\
&\times 
\prod_{i=1}^{N}\prod_{j=N+1}^{N+M}
\underbrace{
\frac{1}{(q^{\frac14}t s_{i}^{\pm} s_{j}^{\mp};q)_{\infty}}
}_{\mathbb{II}_{N'}^{\textrm{3d HM}} \left(\frac{s_{i}}{s_{j}} \right) }
\cdot 
\underbrace{
(q^{\frac12}\prod_{i=1}^{N}s_{i}^{\pm} \prod_{j=N+1}^{N+M}s_{j}^{\mp}z^{\pm};q)_{\infty}
}_{F\left( q^{\frac12} \frac{\prod_{i=1}^{N}s_{i}}{\prod_{j=N+1}^{N+M}s_{j}}z \right)}. 
\end{align}

On the other hand, the dual junction has no gauge symmetry. 
For $N=M$, 3d $\mathcal{N}=4$ twisted hypermultiplet transforming as fundamental representation 
under the $U(N)$ gauge symmetry should arise from D3-D5$'$ string. 
It would obey the Nahm pole boundary condition 
with and embedding $\rho:$ $\mathfrak{su}(2)$ $\rightarrow$ $\mathfrak{u}(N)$. 
For $N\neq M$ we have no twisted hypermultiplet.

The quarter-index (\ref{4dun_3dumNN1}) would be equal to
\begin{align}
\label{4dun_3dumNN2}
\mathbb{IV}^{N|[M]|}_{\mathcal{D}\mathcal{D}'}
&=
\underbrace{
(q)_{\infty}
}_{\mathbb{IV}_{\mathcal{D}\mathcal{D}'}^{\textrm{4d $U(1)$}}} 
\cdot 
(q^{\frac34+\frac{N-1}{4}} t^{1+(N-1)} z^{\pm} ;q )_{\infty}^{ {\delta^{N}}_{M} }. 
\end{align}
The contributions include the quarter-index of 
4d $\mathcal{N}=4$ $U(1)$ gauge theory with a pair of Dirichlet b.c. $(\mathcal{D},\mathcal{D}')$ 
and the half-index of Nahm pole b.c. for 3d twisted hyper with and embedding 
$\rho:$ $\mathfrak{su}(2)$ $\rightarrow$ $\mathfrak{u}(N)$ \cite{Gaiotto:2019jvo}.

%%%%%%%%%%%%%%%%%%%%%%%%%%%%%%%%%%%%%%%%%%%%%
%%%%%%%%%%%%%%%%%%%%%%%%%%%%%%%%%%%%%%%%%%%%%
\subsection{4d $U(N)|$3d $(M_{1})- \cdots -(M_{k-1})$ with $\mathcal{N}'$}
\label{sec_4d3dNN}
\begin{figure}
\begin{center}
\includegraphics[width=14cm]{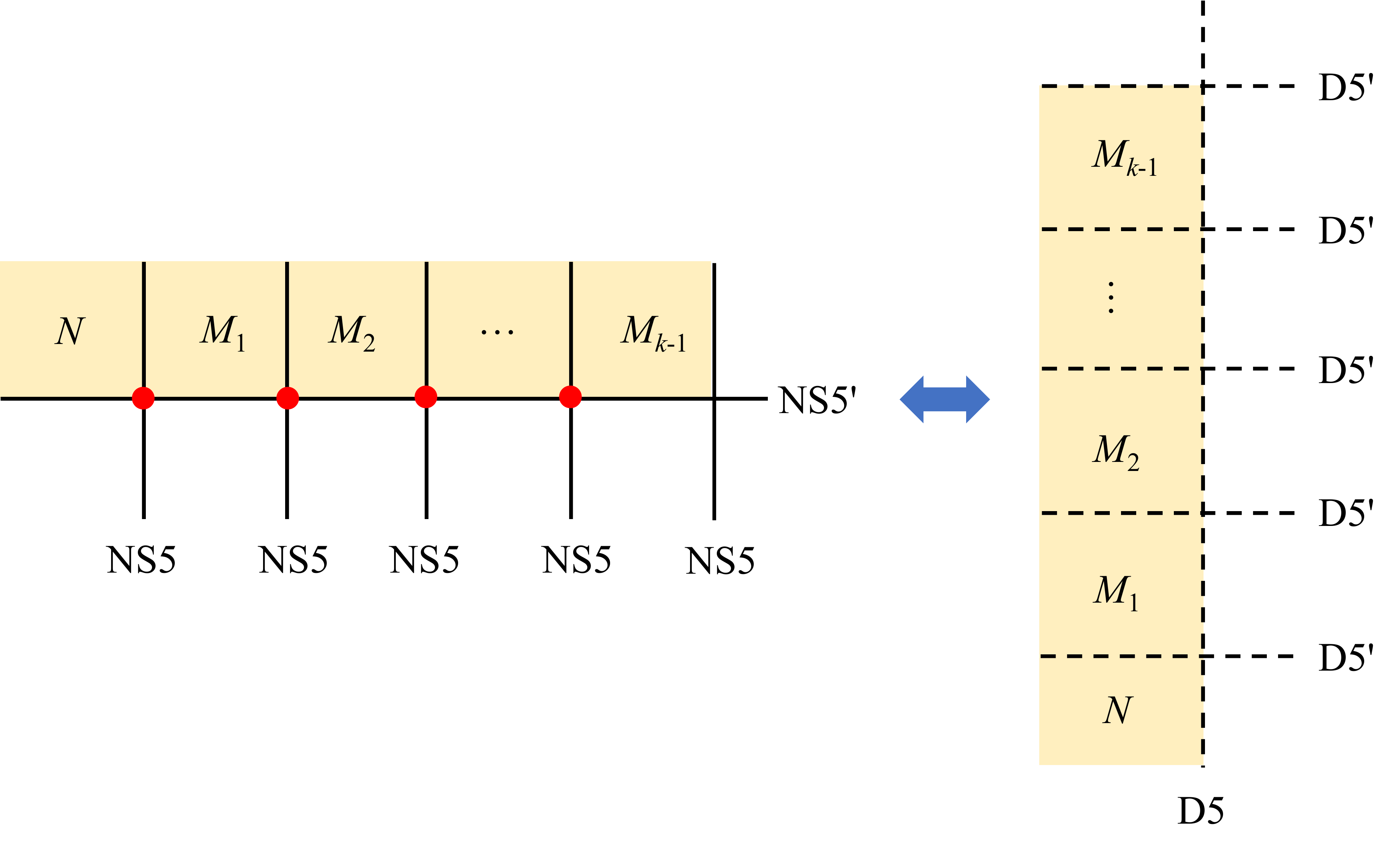}
\caption{
The brane constructions of the 4d $U(N)|$3d $(M_{1})- \cdots -(M_{k-1})$ with Neumann b.c. $\mathcal{N}'$ 
and its mirror corner configuration that has no gauge symmetry. 
}
\label{fig4dun3dumx}
\end{center}
\end{figure}

%%%%%%%%%%%%%%%%%%%%%%%%%%%%%%%%%%%%%%%%%%%%%
%%%%%%%%%%%%%%%%%%%%%%%%%%%%%%%%%%%%%%%%%%%%%
Let us take the quarter-BPS corner of 4d $\mathcal{N}=4$ $U(N)$ gauge theory 
with a pair of Neumann b.c. $(\mathcal{N}, \mathcal{N}')$ 
and the half-BPS boundary of 3d $\mathcal{N}=4$ linear quiver gauge theory 
$(M_{1})-\cdots-(M_{k-1})$ obeying Neumann b.c. $(\mathcal{N}',N')$. 
The corresponding brane setup is illustrated in Figure \ref{fig4dun3dumx}. 
At the NS5-NS5$'$ junction 
for which the NS5-brane has both-sided D3-branes,  
there appears a cross-determinant Fermi multiplet.

%%%%%%%%%%%%%%%%%%%%%%%%%%%%%%%%%%%%%%%%%%%%%
\subsubsection{4d $U(1)|$3d $U(1)\times U(1)$ with $\mathcal{N}'$}
\label{sec_4du13du1u1NN}
%%%%%%%%%%%%%%%%%%%%%%%%%%%%%%%%%%%%%%%%%%%%%
For the junction 4d $U(1)|$3d $U(1)\times U(1)$ with $\mathcal{N}'$, 
we have 4d $U(1)$ gauge theory and 3d $U(1)\times U(1)$ gauge symmetry. 

The quarter-index is 
\begin{align}
\label{4du1_3du1u1NN1}
&
\mathbb{IV}^{\textrm{4d $U(1)|(1)-(1)|$}}_{\mathcal{N}\mathcal{N}'}
\nonumber\\
&=
\underbrace{
(q)_{\infty}\oint \frac{ds_{1}}{2\pi is_{1}}
}_{\mathbb{IV}_{\mathcal{N}\mathcal{N}'}^{\textrm{4d $U(1)$}}}
\cdot 
\underbrace{
(q)_{\infty} (q^{\frac12}t^2;q)_{\infty}\oint \frac{ds_{2}}{2\pi is_{2}}
}_{\mathbb{II}_{\mathcal{N}'}^{\textrm{3d $U(1)$}}}
\cdot 
\underbrace{
(q)_{\infty}(q^{\frac12} t^2;q)_{\infty}\oint \frac{ds_{3}}{2\pi is_{3}}
}_{\mathbb{II}_{\mathcal{N}'}^{\textrm{3d $U(1)$}}}
\nonumber\\
&\times 
\underbrace{
\frac{1}{(q^{\frac14}ts_{1}^{\pm}s_{2}^{\mp};q)_{\infty}}
}_{\mathbb{II}_{N'}^{\textrm{3d HM}} \left(\frac{s_{1}}{s_{2}} \right)}
\cdot 
\underbrace{
\frac{1}{(q^{\frac14}t s_{2}^{\pm} s_{3}^{\mp};q)_{\infty}}
}_{\mathbb{II}_{N'}^{\textrm{3d HM}} \left(\frac{s_{2}}{s_{3}} \right)}
\cdot 
\underbrace{
(q^{\frac12}s_{1}^{\pm}s_{2}^{\mp}z_{1}^{\pm};q)_{\infty}
}_{F \left(q^{\frac12}\frac{s_{1}}{s_{2}} \right)}
\cdot 
\underbrace{
(q^{\frac12}s_{2}^{\pm}s_{3}^{\mp}z_{2}^{\pm};q)_{\infty}
}_{F \left( q^{\frac12}\frac{s_{2}}{s_{3}} \right)}. 
\end{align}
In the last line there are contributions 
from two bi-fundamental hypermultiplets 
as well as two cross-determinant Fermi multiplets.

The S-dual configuration includes two D5$'$-branes 
which have a single D3-brane in their both sides. 
They should give rise to two fundamental twisted hypermultiplets. 
In addition, they should obey the Dirichlet b.c. $D$ 
as the D3-brane terminates on a single D5-brane. 

We have checked that he quarter-index (\ref{4du1_3du1u1NN1}) agrees with
\begin{align}
\label{4du1_3du1u1NN2}
\mathbb{IV}^{1|[1]-[1]|}_{\mathcal{D}\mathcal{D}'}
&=
\underbrace{
(q)_{\infty}
}_{\mathbb{IV}_{\mathcal{D}\mathcal{D}'}^{\textrm{4d $U(1)$}}}
\cdot 
\underbrace{
(q^{\frac34}tz_{1}^{\pm};q)_{\infty}
}_{\mathbb{II}_{D}^{\textrm{3d tHM}} (z_{1})}
\cdot 
\underbrace{
(q^{\frac34}tz_{2}^{\pm};q)_{\infty}
}_{\mathbb{II}_{D}^{\textrm{3d tHM}} (z_{2})}. 
\end{align}

%%%%%%%%%%%%%%%%%%%%%%%%%%%%%%%%%%%%%%%%%%%%%
\subsubsection{4d $U(2)|$3d $U(1)\times U(1)$ with $\mathcal{N}'$}
\label{sec_4du23du1u1NN}
%%%%%%%%%%%%%%%%%%%%%%%%%%%%%%%%%%%%%%%%%%%%%
In the configuration 4d $U(2)|$3d $U(1)\times U(1)$ with $\mathcal{N}'$, 
4d $U(2)$ gauge theory satisfies a pair of Neumann b.c. $(\mathcal{N}, \mathcal{N}')$  
and 3d $U(1)\times U(1)$ gauge theory obeys Neumann b.c. $\mathcal{N}'$. 
It has two 3d bi-fundamental hyper multiplets with Neumann b.c. $N'$ and two 2d cross-determinant Fermi multiplets. 

The quarter-index is evaluated as
\begin{align}
\label{4du2_3du1u1NN1}
&
\mathbb{IV}^{\textrm{4d $U(2)|(1)-(1)|$}}_{\mathcal{N}\mathcal{N}'}
\nonumber\\
&=
\underbrace{
\frac12 (q)_{\infty}^2 \oint \prod_{i=1}^{2} \frac{ds_{i}}{2\pi is_{i}}
\prod_{i\neq j}\left( \frac{s_{i}}{s_{j}};q \right)_{\infty}
}_{\mathbb{IV}_{\mathcal{N}\mathcal{N}'}^{\textrm{4d $U(2)$}}}
\cdot 
\underbrace{
(q)_{\infty} (q^{\frac12}t^2;q)_{\infty}
\oint \frac{ds_{3}}{2\pi is_{3}}
}_{\mathbb{II}_{\mathcal{N}'}^{\textrm{3d $U(1)$}}}
\cdot 
\cdot 
\underbrace{
(q)_{\infty} (q^{\frac12}t^2;q)_{\infty}
\oint \frac{ds_{4}}{2\pi is_{4}}
}_{\mathbb{II}_{\mathcal{N}'}^{\textrm{3d $U(1)$}}}
\nonumber\\
&\times 
\prod_{i=1}^{2}
\underbrace{
\frac{1}{(q^{\frac14}t s_{i}^{\pm}s_{3}^{\mp};q)_{\infty}}
}_{\mathbb{II}_{N'}^{\textrm{3d HM}} \left(\frac{s_{i}}{s_{3}} \right)}
\cdot 
\underbrace{
\frac{1}{(q^{\frac14}t s_{3}^{\pm}s_{4}^{\mp};q)_{\infty}}
}_{\mathbb{II}_{N'}^{\textrm{3d HM}} \left(\frac{s_{3}}{s_{4}} \right)}
\cdot 
\underbrace{
(q^{\frac12}s_{1}^{\pm}s_{2}^{\pm}s_{3}^{\mp}z_{1}^{\pm};q)_{\infty}
}_{F\left(\frac{s_{1}s_{2}}{s_{3}}z_{1} \right)}
\cdot 
\underbrace{
(q^{\frac12}s_{3}^{\pm}s_{4}^{\mp}z_{2}^{\pm};q)_{\infty}
}_{F\left(q^{\frac12}\frac{s_{3}}{s_{4}} z_{2} \right)}.
\end{align}

In the S-dual junction, the gauge symmetry is completely broken. 
As one of the D5$'$ branes has a single D3-brane in its both sides, 
there would be a 3d charged twisted hypermultiplet. 
Due to the presence of D5-brane, it should obey the Dirichlet b.c. $D$. 

The quarter-index (\ref{4du2_3du1u1NN1}) agrees with
\begin{align}
\label{4du2_3du1u1NN2}
\mathbb{IV}^{2|[1]-[1]|}_{\mathcal{D}\mathcal{D}'}
&=
\underbrace{
(q)_{\infty}
}_{\mathbb{IV}_{\mathcal{D}\mathcal{D}'}^{\textrm{4d $U(1)$}}}
\cdot 
\underbrace{
(q^{\frac34}t z_{2}^{\pm};q)_{\infty}
}_{\mathbb{II}_{D}^{\textrm{3d tHM}}(z_{2})}. 
\end{align}

%%%%%%%%%%%%%%%%%%%%%%%%%%%%%%%%%%%%%%%%%%%%%
\subsubsection{4d $U(2)|$3d $U(2)\times U(1)$ with $\mathcal{N}'$}
\label{sec_4du23du2u1NN}
%%%%%%%%%%%%%%%%%%%%%%%%%%%%%%%%%%%%%%%%%%%%%
Consider the junction 4d $U(2)|$3d $U(2)\times U(1)$ with $\mathcal{N}'$. 
The configuration has 4d $U(2)$ gauge symmetry and 3d $U(2)\times U(1)$ gauge symmetry. 
There are two 3d bi-fundamental  hypermultiplets with Neumann b.c. $N'$ 
and two 2d cross-determinant Fermi multiplets.  

The quarter-index is 
\begin{align}
\label{4du2_3du2u1NN1}
&
\mathbb{IV}^{\textrm{4d $U(2)|(2)-(1)|$}}_{\mathcal{N}\mathcal{N}'}
\nonumber\\
&=
\underbrace{
\frac12 (q)_{\infty}^2 
\oint \prod_{i=1}^{2} \frac{ds_{i}}{2\pi is_{i}} 
\prod_{i\neq j}\left(\frac{s_{i}}{s_{j}};q \right)_{\infty}
}_{\mathbb{IV}_{\mathcal{N}\mathcal{N}'}^{\textrm{4d $U(2)$}}}
\cdot 
\underbrace{
\frac12 (q)_{\infty}^2 (q^{\frac12} t^2;q)_{\infty}^2 
\oint \prod_{i=3}^{4}\frac{ds_{i}}{2\pi is_{i}} 
\prod_{i\neq j}\left(\frac{s_{i}}{s_{j}};q \right)_{\infty}
\left(q^{\frac12} t^2 \frac{s_{i}}{s_{j}};q \right)_{\infty}
}_{\mathbb{II}_{\mathcal{N}'}^{\textrm{3d $U(2)$}}}
\nonumber\\
&\times 
\underbrace{
(q)_{\infty} (q^{\frac12}t^2;q)_{\infty} 
\oint \frac{ds_{5}}{2\pi is_{5}}
}_{\mathbb{II}_{\mathcal{N}'}^{\textrm{3d $U(1)$}}}
\nonumber\\
&\times 
\prod_{i=1}^{2}
\prod_{j=3}^{4}
\underbrace{
\frac{1}{(q^{\frac14}t s_{i}^{\pm}s_{j}^{\mp};q)_{\infty}}
}_{\mathbb{II}_{N'}^{\textrm{3d HM}}\left(\frac{s_{i}}{s_{j}} \right) }
\cdot 
\prod_{i=3}^{4}
\underbrace{
\frac{1}{(q^{\frac14}ts_{i}^{\pm}s_{5}^{\mp};q)_{\infty}}
}_{\mathbb{II}_{N'}^{\textrm{3d HM}} \left(\frac{s_{i}}{s_{5}} \right)}
\cdot 
\underbrace{
(q^{\frac12}s_{1}^{\pm}s_{2}^{\pm}s_{3}^{\mp}s_{4}^{\mp}z_{1}^{\pm};q)_{\infty}
}_{F\left( q^{\frac12} \frac{s_{1}s_{2}}{s_{3}s_{4}}z_{1} \right)}
\cdot 
\underbrace{
(q^{\frac12}s_{3}^{\pm}s_{4}^{\pm}s_{5}^{\mp}z_{2}^{\pm};q)_{\infty}
}_{F\left( q^{\frac12}\frac{s_{3}s_{4}}{s_{5}} \right)}. 
\end{align}

The quarter-index (\ref{4du2_3du2u1NN1}) agrees with
\begin{align}
\label{4du2_3du2u1NN2}
\mathbb{IV}^{2|[2]-[1]|}_{\mathcal{D}\mathcal{D}'}
&=
\underbrace{
(q)_{\infty}
}_{\mathbb{IV}_{\mathcal{D}\mathcal{D}'}^{\textrm{4d $U(1)$}}}
\cdot 
(q t^2 z_{1}^{\pm};q)_{\infty}. 
\end{align}
This involves the contribution from 
3d fundamental twisted hypermultiplet satisfying 
the Nahm pole boundary condition with an embedding $\rho:$ $\mathfrak{su}(2)$ $\rightarrow$ $\mathfrak{u}(2)$.

%%%%%%%%%%%%%%%%%%%%%%%%%%%%%%%%%%%%%%%%%%%%%
\subsubsection{4d $U(3)|$3d $U(2)\times U(1)$ with $\mathcal{N}'$}
\label{sec_4du33du2u1NN}
%%%%%%%%%%%%%%%%%%%%%%%%%%%%%%%%%%%%%%%%%%%%%
Next consider the configuration 4d $U(3)|$3d $U(2)\times U(1)$ with $\mathcal{N}'$. 
It keeps 4d $U(3)$ gauge symmetry and 3d $U(2)\times U(1)$ gauge symmetry. 
There are two 3d bi-fundamental hypermultiplets obeying Neumann b.c. $N'$ 
and two 2d cross-determinant Fermi multiplets.

The quarter-index reads 
\begin{align}
\label{4du3_3du2u1NN1}
&
\mathbb{IV}^{\textrm{4d $U(3)|(2)-(1)|$}}_{\mathcal{N}\mathcal{N}'}
\nonumber\\
&=
\underbrace{
\frac{1}{3!}(q)_{\infty}^3 
\oint \prod_{i=1}^{3}\frac{ds_{i}}{2\pi is_{i}} 
\prod_{i\neq j}
\left( \frac{s_{i}}{s_{j}};q \right)_{\infty}
}_{\mathbb{IV}_{\mathcal{N}\mathcal{N}'}^{\textrm{4d $U(3)$}}}
\cdot 
\underbrace{
\frac12 (q)_{\infty}^2 (q^{\frac12}t^2;q)_{\infty}^2 
\oint \prod_{i=4}^{5} \frac{ds_{i}}{2\pi i s_{i}} 
\left(\frac{s_{i}}{s_{j}};q \right)_{\infty}
\left(q^{\frac12}t^2 \frac{s_{i}}{s_{j}};q \right)_{\infty}
}_{\mathbb{II}_{\mathcal{N}'}^{\textrm{3d $U(2)$}}}
\nonumber\\
&\times 
\underbrace{
(q)_{\infty}(q^{\frac12}t^2;q)_{\infty}\oint \frac{ds_{6}}{2\pi is_{6}}
}_{\mathbb{II}_{N'}^{\textrm{3d $U(1)$}}}
\nonumber\\
&\times 
\prod_{i=1}^{3}\prod_{j=4}^{5}
\underbrace{
\frac{1}{(q^{\frac14}t s_{i}^{\pm}s_{j}^{\mp};q)_{\infty}}
}_{\mathbb{II}_{N'}^{\textrm{3d HM}} \left(\frac{s_{i}}{s_{j}}\right)}
\cdot 
\prod_{i=4}^{5}
\underbrace{
\frac{1}{(q^{\frac14}t s_{i}^{\pm}s_{6}^{\mp};q)_{\infty}}
}_{\mathbb{II}_{N'}^{\textrm{3d HM}} \left(\frac{s_{i}}{s_{6}} \right)}
\cdot 
\underbrace{
(q^{\frac12}s_{1}^{\pm}s_{2}^{\pm}s_{3}^{\pm}s_{4}^{\mp}s_{5}^{\mp}z_{1}^{\pm};q)_{\infty}
}_{F\left( q^{\frac12} \frac{s_{1}s_{2}s_{3}}{s_{4}s_{5}}z_{1} \right)}
\cdot 
\underbrace{
(q^{\frac12}s_{4}^{\pm}s_{5}^{\pm}s_{6}^{\mp}z_{2}^{\pm};q)_{\infty}
}_{F\left( q^{\frac12}\frac{s_{4}s_{5}}{s_{6}}z_{2} \right)}. 
\end{align}

The S-dual corner configuration is viewed as a corner of 4d $U(3)$ gauge theory. 
It is obtained from the Dirichlet b.c. $\mathcal{D}'$ for 4d $U(3)$ gauge theory 
further satisfying boundary condition $\mathcal{D}$ corresponding to the D5-brane. 

The quarter-index (\ref{4du3_3du2u1NN1}) agrees with 
the quarter-index of 4d $U(1)$ gauge theory with a pair of Dirichlet b.c. $(\mathcal{D},\mathcal{D}')$
\begin{align}
\label{4du3_3du2u1NN2}
\mathbb{IV}^{3|[2]-[1]|}_{\mathcal{D}\mathcal{D}'}
&=
\underbrace{
(q)_{\infty}
}_{\mathbb{IV}_{\mathcal{D}\mathcal{D}'}^{\textrm{4d $U(1)$}}}. 
\end{align}
This reflects the fact that 
the unequal numbers of D3-branes across D5-branes do not admit the 3d twisted hypermultiplet.

%%%%%%%%%%%%%%%%%%%%%%%%%%%%%%%%%%%%%%%%%%%%%
\subsubsection{4d $U(2)|$3d $U(2)\times U(2)$ with $\mathcal{N}'$}
\label{sec_4du33du2u2NN}
%%%%%%%%%%%%%%%%%%%%%%%%%%%%%%%%%%%%%%%%%%%%%
Let us consider the junction 4d $U(2)|$3d $U(2)\times U(2)$ with $\mathcal{N}'$. 
This includes a corner of 4d $U(2)$ gauge theory with a pair of Neumann b.c. $(\mathcal{N},\mathcal{N}')$ 
and 3d $U(2)\times U(2)$ gauge theory with Neumann b.c. $\mathcal{N}'$. 
It also has two 3d bi-fundamental hypermultiplets transforming as 
$({\bf 2}, \overline{\bf 2}, {\bf 1})$ $\oplus$ $(\overline{\bf 2}, {\bf 2}, {\bf 1})$ and 
$({\bf 1}, {\bf 2}, \overline{\bf 2})$ $\oplus$ $({\bf 1}, \overline{\bf 2}, {\bf 2})$ 
under the $U(2)\times U(2)\times U(2)$ gauge symmetry 
and 2d two cross-determinant Fermi multiplets 
transforming as $(\det, \det^{-1},{\bf 1})$ and $({\bf 1}, \det, \det^{-1})$. 

The quarter-index takes the form
\begin{align}
\label{4du2_3du2u2NN1}
&
\mathbb{IV}^{\textrm{4d $U(2)|(2)-(2)|$}}_{\mathcal{N}\mathcal{N}'}
\nonumber\\
&=
\underbrace{
\frac12 (q)_{\infty}^2 \oint \prod_{i=1}^{2}
\frac{ds_{i}}{2\pi is_{i}} \prod_{i\neq j}
\left( \frac{s_{i}}{s_{j}};q \right)_{\infty}
}_{\mathbb{IV}_{\mathcal{N}\mathcal{N}'}^{\textrm{4d $U(2)$}}}
\cdot 
\underbrace{
\frac12 (q)_{\infty}^2 (q^{\frac12}t^2;q)_{\infty}^2 
\oint \prod_{i=3}^{4} \frac{ds_{i}}{2\pi is_{i}}\prod_{i\neq j}
\left( \frac{s_{i}}{s_{j}};q \right)_{\infty}
\left( q^{\frac12} t^2 \frac{s_{i}}{s_{j}};q \right)_{\infty}
}_{\mathbb{II}_{\mathcal{N}'}^{\textrm{3d $U(2)$}}}
\nonumber\\
&\times 
\underbrace{
\frac12 (q)_{\infty}^2 (q^{\frac12}t^2;q)_{\infty}^2 
\oint \prod_{i=5}^{6}
\frac{ds_{i}}{2\pi is_{i}} \prod_{i\neq j}
\left( \frac{s_{i}}{s_{j}};q \right)_{\infty}
\left( q^{\frac12} t^2 \frac{s_{i}}{s_{j}};q \right)_{\infty}
}_{\mathbb{II}_{\mathcal{N}'}^{\textrm{3d $U(2)$}}}
\nonumber\\
&\times 
\prod_{i=1}^{2}\prod_{j=3}^{4}
\underbrace{
\frac{1}{(q^{\frac14}t s_{i}^{\pm} s_{j}^{\mp};q)_{\infty}}
}_{\mathbb{II}_{N'}^{\textrm{3d HM}} \left(\frac{s_{i}}{s_{j}} \right)} 
\cdot 
\prod_{i=3}^{4}\prod_{j=5}^{6}
\underbrace{
\frac{1}{(q^{\frac14}ts_{i}^{\pm}s_{j}^{\mp};q)_{\infty}}
}_{\mathbb{II}_{N'}^{\textrm{3d HM}} \left(\frac{s_{i}}{s_{j}} \right)}
\cdot 
\underbrace{
(q^{\frac12} s_{1}^{\pm} s_{2}^{\pm}s_{3}^{\mp}s_{4}^{\mp}z_{1}^{\pm};q)_{\infty}
}_{F\left( q^{\frac12}\frac{s_{1}s_{2}}{s_{3}s_{4}} z_{1} \right)}
\cdot 
\underbrace{
(q^{\frac12}s_{3}^{\pm}s_{4}^{\pm}s_{5}^{\mp}s_{6}^{\mp}z_{2}^{\pm};q)_{\infty}
}_{F\left( q^{\frac12}\frac{s_{3}s_{4}}{s_{5}s_{6}} z_{2} \right)}. 
\end{align}

The S-dual configuration has no gauge symmetry. 
As the numbers of D3-branes are the same across the two D5$'$-branes, 
the junction includes the contributions from two 3d fundamental twisted hypers. 
According to the D5-brane, 
they should obey the Nahm pole boundary condition characterized by an embedding 
$\rho:$ $\mathfrak{su}(2)$ $\rightarrow$ $\mathfrak{u}(2)$. 

We find that the quarter-index (\ref{4du2_3du2u2NN1}) coincides with
\begin{align}
\label{4du2_3du2u2NN2}
\mathbb{IV}^{2|[2]-[2]|}_{\mathcal{D}\mathcal{D}'}
&=
\underbrace{
(q)_{\infty}
}_{\mathbb{IV}_{\mathcal{D}\mathcal{D}'}^{\textrm{4d $U(1)$}}}
\cdot 
(q t^2 z_{1}^{\pm};q)_{\infty}
\cdot 
(q t^2 z_{2}^{\pm};q)_{\infty}.
\end{align}
As expected, it involves the two contributions 
which capture the two 3d fundamental twisted hypers 
obeying Nahm pole boundary condition with 
an embedding $\rho:$ $\mathfrak{su}(2)$ $\rightarrow$ $\mathfrak{u}(2)$.

%%%%%%%%%%%%%%%%%%%%%%%%%%%%%%%%%%%%%%%%%%%%%
\subsubsection{4d $U(3)|$3d $U(3)\times U(1)$ with $\mathcal{N}'$}
\label{sec_4du33du3u1NN}
%%%%%%%%%%%%%%%%%%%%%%%%%%%%%%%%%%%%%%%%%%%%%
To gain more insight, let us examine the junction 4d $U(3)|$3d $U(3)\times U(1)$ with $\mathcal{N}'$. 
The surviving gauge symmetry is 4d $U(3)$ and 3d $U(3)\times U(1)$. 
We have two 3d bi-fundamental hypers with Neumann b.c. $N'$ and two 2d cross-determinant Fermi multiplets. 

The quarter-index is given by
\begin{align}
\label{4du3_3du3u1NN1}
&
\mathbb{IV}^{\textrm{4d $U(3)|(3)-(1)|$}}_{\mathcal{N}\mathcal{N}'}
\nonumber\\
&=
\underbrace{
\frac{1}{3!}(q)_{\infty}^3 
\oint \prod_{i=1}^{3}\frac{ds_{i}}{2\pi is_{i}} 
\prod_{i\neq j}\left(\frac{s_{i}}{s_{j}};q \right)_{\infty}
}_{\mathbb{IV}_{\mathcal{N}\mathcal{N}'}^{\textrm{4d $U(3)$}}}
\cdot 
\underbrace{
\frac{1}{3!}
(q)_{\infty}^3 (q^{\frac12}t^2;q)_{\infty}^3 
\oint \prod_{i=4}^{6}\frac{ds_{i}}{2\pi is_{i}}
\prod_{i\neq j}
\left(\frac{s_{i}}{s_{j}};q \right)_{\infty}
\left(q^{\frac12} t^2 \frac{s_{i}}{s_{j}};q \right)_{\infty}
}_{\mathbb{II}_{\mathcal{N}'}^{\textrm{3d $U(3)$}}}
\nonumber\\
&\times 
\underbrace{
(q)_{\infty} (q^{\frac12}t^2;q)_{\infty}\oint \frac{ds_{7}}{2\pi is_{7}}
}_{\mathbb{II}_{\mathcal{N}'}^{\textrm{3d $U(1)$}}}
\nonumber\\
&\times 
\prod_{i=1}^{3}\prod_{j=4}^{6}
\underbrace{
\frac{1}{(q^{\frac14}ts_{i}^{\pm}s_{j}^{\mp};q)_{\infty}}
}_{\mathbb{II}_{N'}^{\textrm{3d HM}} \left(\frac{s_{i}}{s_{j}} \right)}
\cdot 
\prod_{i=4}^{6}
\underbrace{
\frac{1}{(q^{\frac14}t s_{i}^{\pm}s_{7}^{\mp};q)_{\infty}}
}_{\mathbb{II}_{N'}^{\textrm{3d HM}} \left(\frac{s_{i}}{s_{7}} \right)}
\cdot 
\underbrace{
(q^{\frac12}s_{1}^{\pm}s_{2}^{\pm}s_{3}^{\pm}s_{4}^{\mp}s_{5}^{\mp}s_{6}^{\mp}z_{1}^{\pm};q)_{\infty}
}_{F\left(q^{\frac12} \frac{s_{1}s_{2}s_{3}}{s_{4}s_{5}s_{6}}z_{1} \right)}
\cdot 
\underbrace{
(q^{\frac12}s_{4}^{\pm}s_{5}^{\pm}s_{6}^{\pm}s_{7}^{\mp}z_{2}^{\pm};q)_{\infty}
}_{F\left( q^{\frac12}\frac{s_{4}s_{5}s_{6}}{s_{7}} z_{2} \right)}.
\end{align}

In the S-dual junction 
three D3-branes live in each side of one of the D5$'$-branes 
so that one obtains a 3d fundamental twisted hypermultiplet 
obeying the Nahm pole boundary condition specified by an embedding 
$\rho:$ $\mathfrak{su}(2)$ $\rightarrow$ $\mathfrak{u}(3)$ corresponding to the D5-brane. 

We have checked that 
the quarter-index (\ref{4du3_3du3u1NN1}) agrees with
\begin{align}
\label{4du3_3du3u1NN2}
\mathbb{IV}^{3|[3]-[1]|}_{\mathcal{D}\mathcal{D}'}
&=
\underbrace{
(q)_{\infty}
}_{\mathbb{IV}_{\mathcal{D}\mathcal{D}'}^{\textrm{4d $U(1)$}}}
\cdot 
(q^{\frac54} t^3 z_{1}^{\pm};q)_{\infty}. 
\end{align}
The factors $(q^{\frac54}t^3 z_{1}^{\pm};q)_{\infty}$ are now familiar 
contributions of 3d fundamental twisted hypermultiplet obeying 
Nahm pole boundary condition associated to an embedding 
$\rho:$ $\mathfrak{su}(2)$ $\rightarrow$ $\mathfrak{u}(3)$.

%%%%%%%%%%%%%%%%%%%%%%%%%%%%%%%%%%%%%%%%%%%%%
\subsubsection{4d $U(3)|$3d $U(3)\times U(2)$ with $\mathcal{N}'$}
\label{sec_4du33du3u2NN}
%%%%%%%%%%%%%%%%%%%%%%%%%%%%%%%%%%%%%%%%%%%%%
As a further check, consider the junction 
4d $U(3)|$3d $U(3)\times U(2)$ with $\mathcal{N}'$. 

Similarly, we can compute the quarter-index as
\begin{align}
\label{4du3_3du3u2NN1}
&
\mathbb{IV}^{\textrm{4d $U(3)|(3)-(2)|$}}_{\mathcal{N}\mathcal{N}'}
\nonumber\\
&=
\underbrace{
\frac{1}{3!}(q)_{\infty}^3 \oint \prod_{i=1}^3 
\frac{ds_{i}}{2\pi is_{i}} \prod_{i\neq j}
\left(\frac{s_{i}}{s_{j}};q \right)_{\infty}
}_{\mathbb{IV}_{\mathcal{N}\mathcal{N}'}^{\textrm{4d $U(3)$}}}
\cdot 
\underbrace{
\frac{1}{3!}(q)_{\infty}^3 (q^{\frac12}t^2;q)_{\infty}^3 
\oint \prod_{i=4}^{6}\frac{ds_{i}}{2\pi is_{i}}
\prod_{i\neq j}
\left( \frac{s_{i}}{s_{j}};q \right)_{\infty}
\left( q^{\frac12} t^2 \frac{s_{i}}{s_{j}};q \right)_{\infty}
}_{\mathbb{II}_{\mathcal{N}'}^{\textrm{3d $U(3)$}}}
\nonumber\\
&\times 
\underbrace{
\frac12 (q)_{\infty}^2 (q^{\frac12} t^2;q)_{\infty}^2 
\oint \prod_{i=7}^{8}\frac{ds_{i}}{2\pi i s_{i}}
\prod_{i\neq j}
\left(\frac{s_{i}}{s_{j}};q \right)_{\infty}
\left( q^{\frac12} t^2 \frac{s_{i}}{s_{j}};q \right)_{\infty}
}_{\mathbb{II}_{\mathcal{N}'}^{\textrm{3d $U(2)$}}}
\nonumber\\
&\times 
\prod_{i=1}^{3}\prod_{j=4}^{6}
\underbrace{
\frac{1}{(q^{\frac14}t s_{i}^{\pm} s_{j}^{\mp};q)_{\infty}}
}_{\mathbb{II}_{N'}^{\textrm{3d HM}} \left(\frac{s_{i}}{s_{j}} \right)}
\cdot 
\prod_{i=4}^{6}\prod_{j=7}^{8}
\underbrace{
\frac{1}{(q^{\frac14}t s_{i}^{\pm}s_{j}^{\mp};q)_{\infty}}
}_{\mathbb{II}_{N'}^{\textrm{3d HM}} \left(\frac{s_{i}}{s_{j}} \right)}
\nonumber\\
&\times 
\underbrace{
(q^{\frac12}s_{1}^{\pm}s_{2}^{\pm}s_{3}^{\pm} s_{4}^{\mp}s_{5}^{\mp}s_{6}^{\mp}z_{1}^{\pm};q)_{\infty}
}_{F\left(\frac{s_{1}s_{2}s_{3}}{s_{4}s_{5}s_{6}}z_{1} \right)}
\cdot 
\underbrace{
(q^{\frac12} s_{4}^{\pm}s_{5}^{\pm}s_{6}^{\pm}s_{7}^{\mp}s_{8}^{\mp}z_{2}^{\pm};q)_{\infty}
}_{F\left(q^{\frac12}\frac{s_{4}s_{5}s_{6}}{s_{7}s_{8}}z_{2} \right)}. 
\end{align}

For the dual junction, gauge symmetry is completely broken. 
Again three D3-branes in each side of one of the D5$'$-branes 
yield a 3d fundamental twisted hypermultiplet 
obeying the Nahm pole boundary condition with an embedding 
$\rho:$ $\mathfrak{su}(2)$ $\rightarrow$ $\mathfrak{u}(3)$ due to the D5-brane. 

In fact, the quarter-index (\ref{4du3_3du3u2NN1}) agrees with
\begin{align}
\label{4du3_3du3u2NN2}
\mathbb{IV}^{3|[3]-[2]|}_{\mathcal{D}\mathcal{D}'}
&=
\underbrace{
(q)_{\infty}
}_{\mathbb{IV}_{\mathcal{D}\mathcal{D}'}^{\textrm{4d $U(1)$}}}
\cdot 
(q^{\frac54} t^3 z_{1}^{\pm};q)_{\infty}.
\end{align}

%%%%%%%%%%%%%%%%%%%%%%%%%%%%%%%%%%%%%%%%%%%%%
\subsubsection{4d $U(N)|$3d $U(N)^{k-1}$ with $\mathcal{N}'$}
\label{sec_4du33du3u2NN}
%%%%%%%%%%%%%%%%%%%%%%%%%%%%%%%%%%%%%%%%%%%%%
Let us consider the junction 4d $U(N)|$3d $U(N)^{k-1}$ with $\mathcal{N}'$, 
consisting of a corner of 4d $U(N)$ gauge theory with a pair of Neumann b.c. $(\mathcal{N}, \mathcal{N}')$ 
and a boundary of 3d quiver gauge theory of gauge group $U(N)^{k-1}$ with $(k-2)$ bi-fundamental hypermultiplets. 
The 3d vector and hypers obey Neumann b.c. $\mathcal{N}'$ and $N'$ respectively. 
We have an additional 3d bi-fundamental hypermultiplet that couples to 4d $U(N)$ and 3d $U(N)$ vector multiplets 
and that has Neumann b.c. $N'$. 
Besides we have $(k-1)$ 2d cross-determinant Fermi multiplets which cancel the Abelian part of boundary gauge anomaly. 
In the brane construction, 
this junction is realized by $N$ D3-branes intersecting with 
$k$ NS5-branes and a single NS5$'$-brane (see Figure \ref{fig4dun3dumx}). 

For the junction 4d $U(N)|$3d $U(N)^{k-1}$ with Neumann b.c. $\mathcal{N}'$, 
we can evaluate the quarter-index as
\begin{align}
\label{4dun_3dunxkNN1}
&
\mathbb{IV}^{\textrm{4d $U(N)|(N)^{k-1}|$}}_{\mathcal{N}\mathcal{N}'}
\nonumber\\
&=
\underbrace{
\frac{1}{N!}(q)_{\infty}^{N}\oint \prod_{i=1}^{N}\frac{ds_{i}}{2\pi is_{i}}
\prod_{i\neq j}\left(\frac{s_{i}}{s_{j}};q \right)_{\infty}
}_{\mathbb{IV}_{\mathcal{N}\mathcal{N}'}^{\textrm{4d $U(N)$}}}
\nonumber\\
&\times 
\underbrace{
(q)_{\infty}^{(k-1)N}
(q^{\frac12}t^2;q)_{\infty}^{(k-1)N}
\prod_{l=1}^{k-1}\oint \prod_{i=1}^{N}
\frac{ds_{i}^{(l)}}{2\pi is_{i}^{(l)}} 
\prod_{i\neq j}
\left( \frac{s_{i}^{(l)}}{s_{j}^{(l)}};q \right)_{\infty}
\left(q^{\frac12}t^2 \frac{s_{i}^{(l)}}{s_{j}^{(l)}};q \right)_{\infty}
}_{\mathbb{II}_{\mathcal{N}'}^{\textrm{3d $U(N)^{k-1}$}}}
\nonumber\\
&\times 
\prod_{i=1}^{N}\prod_{j=1}^{N}
\underbrace{
\frac{1}{(q^{\frac14}ts_{i}^{\pm}s_{j}^{(1)\mp};q)_{\infty}}
}_{\mathbb{II}_{N'}^{\textrm{3d HM}}\left(\frac{s_{i}}{s_{j}^{(1)}} \right)}
\cdot 
\prod_{l=1}^{k-2}\prod_{i=1}^{N}\prod_{j=1}^{N}
\underbrace{
\frac{1}{(q^{\frac14}t s_{i}^{(l)\pm}s_{j}^{(l+1)\mp};q)_{\infty}}
}_{\mathbb{II}^{\textrm{3d HM}}_{N'}\left(\frac{s^{(l)}_{i}}{s^{(l+1)}_{j}} \right)}
\nonumber\\
&\times 
\underbrace{
(q^{\frac12}\prod_{i=1}^{N}s_{i}^{\pm}\prod_{j=1}^{N}s_{j}^{(1)\mp}z_{1}^{\pm};q)_{\infty}
}_{F\left( q^{\frac12}\frac{\prod_{i=1}^{N}s_{i}}{\prod_{j=1}^{N}s_{j}^{(1)}} z_{1} \right)}
\cdot 
\prod_{l=1}^{k-2}
\underbrace{
(q^{\frac12}\prod_{i=1}^{N}s_{i}^{(l)\pm}\prod_{j=1}^{N}s_{i}^{(l+1)\mp}z_{l+1}^{\pm};q)_{\infty}
}_{F\left( q^{\frac12} \frac{\prod_{i=1}^{N}s_{i}^{(l)}}{\prod_{j=1}^{N}s_{j}^{(l+1)}} z_{l+1};q \right)_{\infty}}.
\end{align}

The gauge symmetry is completely broken in the S-dual junction. 
As the numbers of D3-branes remain the same across the $k$ D5$'$-branes, 
the junction would include the contributions from $k$ 3d fundamental twisted hypermultiplets. 
Since the number of D3-branes jumps from $N$ to $0$ across the D5-brane, 
they should obey the Nahm pole boundary condition characterized by an embedding 
$\rho:$ $\mathfrak{su}(2)$ $\rightarrow$ $\mathfrak{u}(N)$. 

We expect that the quarter-index (\ref{4dun_3dunxkNN1}) would be equal to
\begin{align}
\label{4dun_3dunxkNN2}
\mathbb{IV}^{N|[N]^{k-1}|}_{\mathcal{D}\mathcal{D}'}
&=
\underbrace{
(q)_{\infty}
}_{\mathbb{IV}_{\mathcal{D}\mathcal{D}'}^{\textrm{4d $U(1)$}}}
\cdot 
\prod_{l=1}^{k} 
(q^{\frac34+\frac{N-1}{4}} t^{1+(N-1)}z_{l}^{\pm};q)_{\infty},
\end{align}
which contains the quarter-index of 4d $U(1)$ gauge theory with a pair of Dirichlet b.c. $(\mathcal{D}, \mathcal{D}')$ 
and the contributions from $k$ 3d fundamental twisted hypermultiplets with the Nahm pole boundary condition with an embedding 
$\rho:$ $\mathfrak{su}(2)$ $\rightarrow$ $\mathfrak{u}(N)$.

%%%%%%%%%%%%%%%%%%%%%%%%%%%%%%%%%%%%%%%%%%%%%
\subsubsection{4d $U(2)|$3d $U(2)\times U(1)\times U(1)$ with $\mathcal{N}'$}
\label{sec_4du23du2u1u1NN}
%%%%%%%%%%%%%%%%%%%%%%%%%%%%%%%%%%%%%%%%%%%%%
Let us examine the case where several same numbers of D3-branes are arranged in both sides of D5$'$-branes. 
For the junction 4d $U(2)|$3d $U(2)\times U(1)\times U(1)$ with $\mathcal{N}'$, 
we have 4d $U(2)$ gauge symmetry and 3d $U(2)\times U(1)\times U(1)$ gauge symmetry. 
There are three 3d bi-fundamental hypermultiplets and three 2d cross-determinant Fermi multiplets.

The quarter-index takes the form
\begin{align}
\label{4du2_3d211NN1}
&
\mathbb{IV}^{\textrm{4d $U(2)|(2)-(1)-(1)|$}}_{\mathcal{N}\mathcal{N}'}
\nonumber\\
&=
\underbrace{
\frac12 (q)_{\infty}^2 
\oint \prod_{i=1}^{2}\frac{ds_{i}}{2\pi is_{i}}
\prod_{i\neq j}\left(\frac{s_{i}}{s_{j}};q \right)_{\infty}
}_{\mathbb{IV}_{\mathcal{N}\mathcal{N}'}^{\textrm{4d $U(2)$}}}
\cdot 
\underbrace{
\frac12 (q)_{\infty}^2 (q^{\frac12} t^2;q)_{\infty}^2 
\oint \prod_{i=3}^{4}\frac{ds_{i}}{2\pi is_{i}}
\prod_{i\neq j}
\left(\frac{s_{i}}{s_{j}};q \right)_{\infty}
\left(q^{\frac12} t^2 \frac{s_{i}}{s_{j}};q \right)_{\infty}
}_{\mathbb{II}_{\mathcal{N}'}^{\textrm{3d $U(2)$}}}
\nonumber\\
&\times 
\underbrace{
(q)_{\infty} (q^{\frac12}t^2;q)_{\infty}\oint \frac{ds_{5}}{2\pi is_{5}}
}_{\mathbb{II}_{\mathcal{N}'}^{\textrm{3d $U(1)$}}}
\cdot 
\underbrace{
(q)_{\infty} (q^{\frac12} t^2;q)_{\infty}\oint \frac{ds_{6}}{2\pi is_{6}}
}_{\mathbb{II}_{\mathcal{N}'}^{\textrm{3d $U(1)$}}}
\nonumber\\
&\times 
\prod_{i=1}^{2}\prod_{j=3}^{4}
\underbrace{
\frac{1}{(q^{\frac14}t s_{i}^{\pm}s_{j}^{\mp};q)_{\infty}}
}_{\mathbb{II}_{N'}^{\textrm{3d HM}} \left(\frac{s_{i}}{s_{j}} \right)}
\cdot 
\prod_{i=3}^{4}
\underbrace{
\frac{1}{(q^{\frac14}t s_{i}^{\pm}s_{5}^{\mp};q)_{\infty}}
}_{\mathbb{II}_{N'}^{\textrm{3d HM}} \left(\frac{s_{i}}{s_{5}} \right)}
\cdot 
\underbrace{
\frac{1}{(q^{\frac14}t s_{5}^{\pm} s_{6}^{\mp};q)_{\infty}}
}_{\mathbb{II}_{N'}^{\textrm{3d HM}}\left(\frac{s_{5}}{s_{6}} \right)}
\nonumber\\
&\times 
\underbrace{
(q^{\frac12}s_{1}^{\pm}s_{2}^{\pm}s_{3}^{\mp}s_{4}^{\mp}z_{1}^{\pm};q)_{\infty}
}_{F\left( q^{\frac12}\frac{s_{1}s_{2}}{s_{3}s_{4}}z_{1} \right)}
\cdot 
\underbrace{
(q^{\frac12}s_{3}^{\pm}s_{4}^{\pm}s_{5}^{\mp}z_{2}^{\pm};q)_{\infty}
}_{F\left(q^{\frac12} \frac{s_{3}s_{4}}{s_{5}} z_{2} \right)}
\cdot 
\underbrace{
(q^{\frac12} s_{5}^{\pm}s_{6}^{\mp}z_{3}^{\pm};q)_{\infty}
}_{F\left( q^{\frac12}\frac{s_{5}}{s_{6}}z_{3} \right)}
\end{align}

In the dual junction, there are four D5$'$-branes. 
For each side of the first D5$'$-brane, there are two D3-branes. 
Across the second one, the numbers of D3-branes jumps from two to one. 
In both sides of the third one, there is a single D3-brane. 
Across the last one, the number of D3-branes jumps from one to zero. 
Therefore we expect that there are two kinds of 3d fundamental twisted hypermultiplets 
associated to the first and third D5$'$-branes.

We find that the quarter-index (\ref{4du2_3d211NN1}) agrees with
\begin{align}
\label{4du2_3d211NN2}
\mathbb{IV}^{2|[2]-[1]-[1]|}_{\mathcal{D}\mathcal{D}'}
&=
\underbrace{
(q)_{\infty}
}_{\mathbb{IV}_{\mathcal{D}\mathcal{D}'}^{\textrm{4d $U(1)$}}}
\cdot 
(q t^2 z_{1}^{\pm};q)_{\infty}
\cdot 
\underbrace{
(q^{\frac34}t z_{3}^{\pm};q)_{\infty}
}_{\mathbb{II}_{D}^{\textrm{3d tHM}}(z_{3})}. 
\end{align}
This includes two factors; $(q t^2 z_{1}^{\pm};q)_{\infty}$ and $(q^{\frac34}tz_{3}^{\pm};q)_{\infty}$
as well as the quarter-index of 4d $U(1)$ gauge theory with a pair of Dirichlet b.c. $(\mathcal{D}, \mathcal{D}')$. 
The first factor is the contribution from 
the 3d fundamental twisted hyper with Nahm pole b.c. specified by an embedding $\rho:$ $\mathfrak{su}(2)$ 
$\rightarrow$ $\mathfrak{u}(2)$ corresponding to the first D5$'$-brane, 
and the second describes the contribution from the 3d fundamental twisted hyper with Dirichlet b.c. $D$ 
corresponding to the third D5$'$-brane.

%%%%%%%%%%%%%%%%%%%%%%%%%%%%%%%%%%%%%%%%%%%%%
\subsubsection{4d $U(3)|$3d $U(3)\times U(1)\times U(1)$ with $\mathcal{N}'$}
\label{sec_4du33du3u1u1NN}
%%%%%%%%%%%%%%%%%%%%%%%%%%%%%%%%%%%%%%%%%%%%%
For the configuration 4d $U(3)|$3d $U(3)\times U(1)\times U(1)$ with $\mathcal{N}'$, 
there are 4d $U(3)$ gauge symmetry and 3d $U(3)\times U(1)\times U(1)$ gauge symmetry. 
Again there are three 3d bi-fundamental hypermultiplets with Neumann b.c. $N'$ 
and three 2d cross-determinant Fermi multiplets 
which cancel the Abelian part of gauge anomaly. 

The quarter-index is evaluated as 
\begin{align}
\label{4du3_3d311NN1}
&
\mathbb{IV}^{\textrm{4d $U(3)|(3)-(1)-(1)|$}}_{\mathcal{N}\mathcal{N}'}
\nonumber\\
&=
\underbrace{
\frac{1}{3!}(q)_{\infty}^3 \oint \prod_{i=1}^{3}\frac{ds_{i}}{2\pi is_{i}} 
\prod_{i\neq j}\left(\frac{s_{i}}{s_{j}};q \right)_{\infty}
}_{\mathbb{IV}_{\mathcal{N}\mathcal{N}'}^{\textrm{4d $U(3)$}}}
\cdot 
\underbrace{
\frac{1}{3!}(q)_{\infty}^3 (q^{\frac12} t^2;q)_{\infty}^3 
\oint \prod_{i=4}^{6}\frac{ds_{i}}{2\pi is_{i}}
\prod_{i\neq j}
\left(\frac{s_{i}}{s_{j}};q \right)_{\infty}
\left(q^{\frac12} t^2 \frac{s_{i}}{s_{j}};q \right)_{\infty}
}_{\mathbb{II}_{\mathcal{N}'}^{\textrm{3d $U(3)$}}}
\nonumber\\
&\times 
\underbrace{
(q)_{\infty} (q^{\frac12} t^2;q)_{\infty} \oint \frac{ds_{7}}{2\pi is_{7}}
}_{\mathbb{II}_{\mathcal{N}'}^{\textrm{3d $U(1)$}}}
\cdot 
\underbrace{
(q)_{\infty} (q^{\frac12} t^2;q)_{\infty} \oint \frac{ds_{8}}{2\pi is_{8}}
}_{\mathbb{II}_{\mathcal{N}'}^{\textrm{3d $U(1)$}}}
\nonumber\\
&\times 
\prod_{i=1}^{3}\prod_{j=4}^{6}
\underbrace{
\frac{1}
{(q^{\frac14} t s_{i}^{\pm} s_{j}^{\mp};q )_{\infty}}
}_{\mathbb{II}_{N'}^{\textrm{3d HM}} \left(\frac{s_{i}}{s_{j}} \right)}
\cdot 
\prod_{i=4}^{6}
\underbrace{
\frac{1}{(q^{\frac14} t s_{i}^{\pm}s_{7}^{\mp};q)_{\infty}}
}_{\mathbb{II}_{N'}^{\textrm{3d HM}}\left(\frac{s_{i}}{s_{7}} \right) }
\cdot 
\underbrace{
\frac{1}{(q^{\frac14}t s_{7}^{\pm}s_{8}^{\mp};q)_{\infty}}
}_{\mathbb{II}_{N'}^{\textrm{3d HM}}\left(\frac{s_{7}}{s_{8}} \right) }
\nonumber\\
&\times 
\underbrace{
(q^{\frac12}s_{1}^{\pm}s_{2}^{\pm}s_{3}^{\pm} s_{4}^{\mp}s_{5}^{\mp}s_{6}^{\mp}z_{1}^{\pm};q)_{\infty}
}_{F\left(q^{\frac12}\frac{s_{1}s_{2}s_{3}}{s_{4}s_{5}s_{6}}z_{1} \right)}
\cdot 
\underbrace{
(q^{\frac12} s_{4}^{\pm}s_{5}^{\pm}s_{6}^{\pm} s_{7}^{\mp}z_{2}^{\pm};q)_{\infty}
}_{F\left(q^{\frac12}\frac{s_{4}s_{5}s_{6}}{s_{7}}z_{2} \right)}
\cdot 
\underbrace{
(q^{\frac12}s_{7}^{\pm}s_{8}^{\mp}z_{3}^{\pm};q)_{\infty}
}_{F\left( q^{\frac12} \frac{s_{7}}{s_{8}}z_{3} \right)}. 
\end{align}

The S-dual junction would involve 
the two types of 3d fundamental twisted hypermultiplets. 
One should obey the Nahm pole b.c. with a homomorphism $\rho:$ $\mathfrak{su}(2)$ $\rightarrow$ $\mathfrak{u}(3)$ 
and the other should satisfy the Dirichlet b.c. $D$. 

In fact, we have confirmed that the quarter-index (\ref{4du3_3d311NN1}) coincides with
\begin{align}
\label{4du3_3d311NN2}
\mathbb{IV}^{3|[3]-[1]-[1]|}_{\mathcal{D}\mathcal{D}'}
&=
\underbrace{
(q)_{\infty}
}_{\mathbb{IV}_{\mathcal{D}\mathcal{D}'}^{\textrm{4d $U(1)$}}}
\cdot 
(q^{\frac54}t^3 z_{1}^{\pm};q)_{\infty} 
\cdot 
\underbrace{
(q^{\frac34}t z_{3}^{\pm};q)_{\infty}
}_{\mathbb{II}_{N'}^{\textrm{3d tHM}} (z_{3})}.
\end{align}

%%%%%%%%%%%%%%%%%%%%%%%%%%%%%%%%%%%%%%%%%%%%%
\subsubsection{4d $U(N)|$3d $(M_{1})- \cdots -(M_{k-1})$ with $\mathcal{N}'$}
\label{sec_4dun3dumx}
%%%%%%%%%%%%%%%%%%%%%%%%%%%%%%%%%%%%%%%%%%%%%
Now we would like to propose the generalization. 
We consider the junction 4d $U(N)|$3d $(M_{1})- \cdots -(M_{k-1})$ with $\mathcal{N}'$, 
which is constructed by 
the quarter-BPS corner of 4d $\mathcal{N}=4$ $U(N)$ SYM theory with a pair of Neumann b.c. $(\mathcal{N}, \mathcal{N}')$ 
and the half-BPS boundary of 3d linear quiver gauge theory $(M_{1})-\cdots-(M_{k-1})$ 
whose vector and $(k-2)$ hypers obey Neumann b.c. $\mathcal{N}'$ and $N'$ respectively. 
Furthermore the junction contains a 3d bi-fundamental hypermultiplet that 
couples to 4d $U(N)$ and 3d $U(M_{1})$ gauge fields 
and $(k-1)$ 2d cross-determinant Fermi multiplets. 

The quarter-index takes the form
\begin{align}
\label{4duN_3duMxNN1}
&
\mathbb{IV}^{\textrm{4d $U(N)|(M_{1})-(M_{2})-\cdots-(M_{k-1})|$}}_{\mathcal{N}\mathcal{N}'}
\nonumber\\
&=
\underbrace{
\frac{1}{N!}(q)_{\infty}^N \oint \prod_{i=1}^{N} \frac{ds_{i}}{2\pi is_{i}}
\prod_{i\neq j}\left( \frac{s_{i}}{s_{j}};q \right)_{\infty}
}_{\mathbb{IV}_{\mathcal{N}\mathcal{N}'}^{\textrm{4d $U(N)$}}}
\nonumber\\
&\times 
\prod_{l=1}^{k-1}
\underbrace{
\frac{1}{M_{l}!} 
(q)_{\infty}^{M_{l}} (q^{\frac12} t^2;q)_{\infty}^{M_{l}} 
\oint \prod_{i=1}^{M_{l}} 
\frac{ds_{i}^{(l)}}{2\pi is_{i}^{(l)}} 
\prod_{i\neq j}
\left( \frac{s_{i}^{(l)}}{s_{j}^{(l)}};q \right)_{\infty}
\left( q^{\frac12} t^2 \frac{s_{i}^{(l)}}{s_{j}^{(l)}};q \right)_{\infty}
}_{\mathbb{II}_{\mathcal{N}'}^{\textrm{3d $U(M_{l})$}}}
\nonumber\\
&\times 
\prod_{i=1}^{N}\prod_{j=1}^{M_{1}}
\underbrace{
\frac{1}{(q^{\frac14}ts_{i}^{\pm}s_{j}^{(1)\mp};q)_{\infty}}
}_{\mathbb{II}_{N'}^{\textrm{3d HM}} \left(\frac{s_{i}}{s_{j}^{(1)}} \right)}
\cdot 
\prod_{l=1}^{k-2}\prod_{i=1}^{M_{l}} \prod_{j=1}^{M_{l+1}}
\underbrace{
\frac{1}{(q^{\frac14}t s_{i}^{(l)\pm} s_{j}^{(l+1)\mp};q)_{\infty}}
}_{\mathbb{II}_{N'}^{\textrm{3d HM}}\left(\frac{s_{i}^{(l)}}{s_{j}^{(l+1)}} \right)}
\nonumber\\
&\times 
\underbrace{
(q^{\frac12}\prod_{i=1}^{N}s_{i}^{\pm} \prod_{j=1}^{M_{1}}s_{j}^{(1)\mp}z_{1}^{\pm};q)_{\infty}
}_{F\left( q^{\frac12} \frac{\prod_{i=1}^{N}s_{i}}{\prod_{j=1}^{M_{1}}s_{j}^{(1)}} z_{1} \right)}
\cdot 
\prod_{l=1}^{k-2}
\underbrace{
(q^{\frac12} \prod_{i=1}^{M_{l}} s_{i}^{(l)\pm} \prod_{j=1}^{M_{l+1}} s_{j}^{(l+1)\mp}z_{l+1}^{\pm};q)_{\infty}
}_{F\left( q^{\frac12} \frac{\prod_{i=1}^{M_{l}}s_{i}^{(l)} }{\prod_{j=1}^{M_{l+1}} s_{j}^{(l+1)}} z_{l+1} \right)}.
\end{align}

Under the action of S-duality, we obtain the corner of 4d $\mathcal{N}=4$ gauge theory 
with a pair of Dirichlet b.c. $(\mathcal{D}, \mathcal{D}')$ 
as the gauge symmetry is completely broken according to the D5- and D5$'$-branes. 
The junction may contain the 3d fundamental twisted hypermultiplets 
which are associated to the D5$'$-branes with equal numbers of D3-branes in their both sides. 
Due to the D5-brane on which D3-branes end, 
these twisted hypers should receive the Dirichlet/Nahm pole boundary conditions. 

We expect that 
the quarter-index (\ref{4duN_3duMxNN1}) agrees with
\begin{align}
\label{4duN_3duMxNN2}
&
\mathbb{IV}^{N|[M_{1}]-[M_{2}]-\cdots -[M_{k-1}]|}_{\mathcal{D}\mathcal{D}'}
\nonumber\\
&=
\underbrace{
(q)_{\infty}
}_{\mathbb{IV}_{\mathcal{D}\mathcal{D}'}^{\textrm{4d $U(1)$}}}
\cdot 
(q^{\frac34+\frac{N-1}{4}} t^{1+(N-1)} z_{1}^{\pm};q)_{\infty}^{{\delta^{N}}_{M_{1}} }
\cdot 
\prod_{l=1}^{k-2}
(q^{\frac34+\frac{M_{l}-1}{4}} t^{1+(M_{l}-1)} z_{l+1};q)_{\infty}^{{\delta^{M_{l} }}_{M_{l+1}}}
.
\end{align}
This contains the quarter-index of 4d $U(1)$ gauge theory with a pair of Dirichlet b.c. $(\mathcal{D}, \mathcal{D}')$ 
and the contributions from 3d fundamental twisted hypers with Dirichlet/Nahm pole b.c. 
specified by an embedding $\rho:$ $\mathfrak{su}(2)$ $\rightarrow$ $\mathfrak{u}(N)$.

%%%%%%%%%%%%%%%%%%%%%%%%%%%%%%%%%%%%%%%%%%%%%
%%%%%%%%%%%%%%%%%%%%%%%%%%%%%%%%%%%%%%%%%%%%%
\subsection{4d $U(N)|$3d $(M)-[1]$ with $(\mathcal{N}',N')$}
\label{sec_4dN3dMhm1}
\begin{figure}
\begin{center}
\includegraphics[width=11cm]{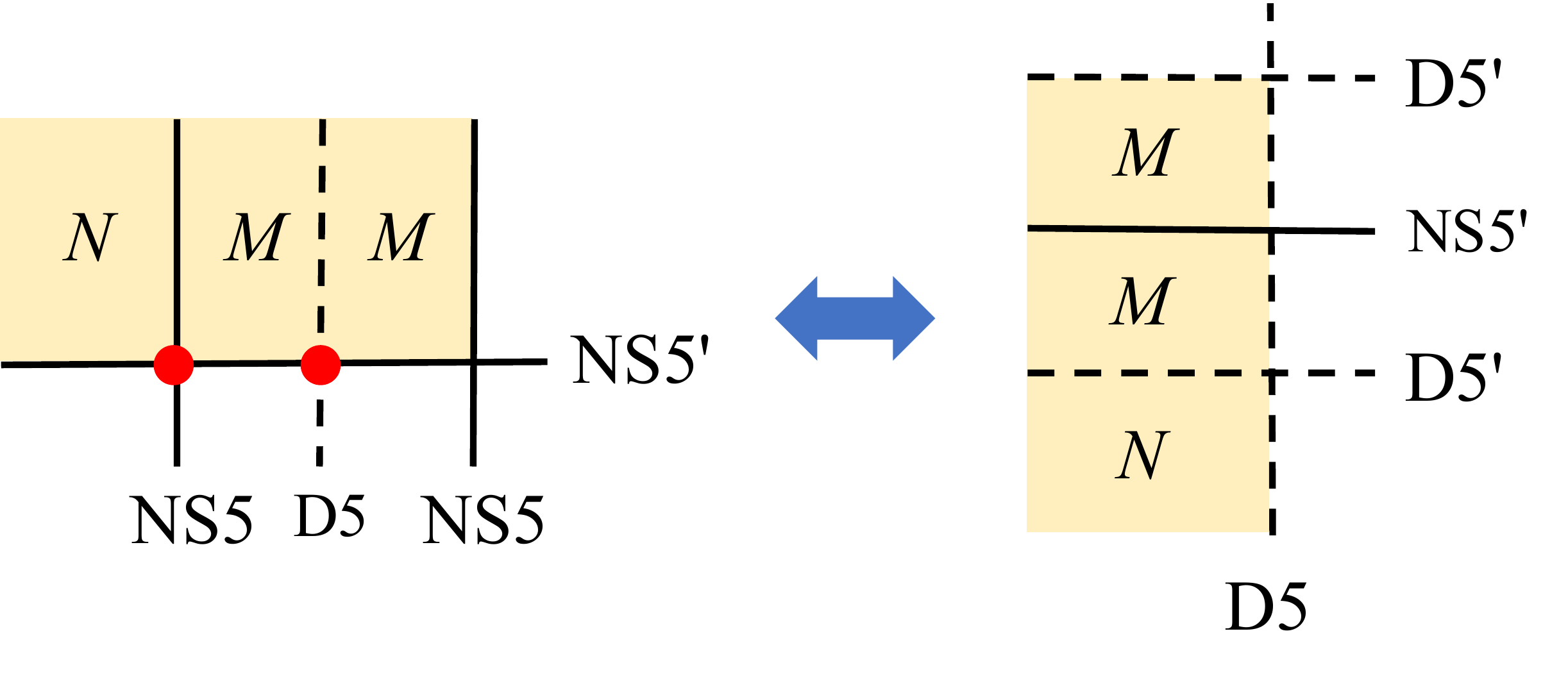}
\caption{
The brane constructions of the configuration 4d $U(N)|$3d $(M)-[1]$ with Neumann b.c. $(\mathcal{N}',N')$ 
and its mirror corner configuration that has no gauge symmetry. 
}
\label{fig4dun3dumhm1}
\end{center}
\end{figure}

%%%%%%%%%%%%%%%%%%%%%%%%%%%%%%%%%%%%%%%%%%%%%
%%%%%%%%%%%%%%%%%%%%%%%%%%%%%%%%%%%%%%%%%%%%%
Now consider the junctions which 
are constructed from a corner of 4d $\mathcal{N}=4$ $U(N)$ gauge theory with a pair of Neumann b.c. $(\mathcal{N}, \mathcal{N}')$ 
and a boundary of 3d $\mathcal{N}=4$ $U(M)$ gauge theory with a fundamental hypermultiplet obeying Neumann b.c. $(\mathcal{N}', N')$. 
The corresponding brane configuration is illustrated in Figure \ref{fig4dun3dumhm1}. 
The junction contains 3d bi-fundamental hypermultiplet which couples to 
4d $U(N)$ and 3d $U(M)$ gauge fields, 
2d cross-determinant Fermi multiplet at the NS5-NS5$'$ junction, 
and 2d fundamental Fermi multiplet corresponding to the D5-NS5$'$ intersection. 
We denote this junction by 4d $U(N)|$3d $(M)-[1]$ with Neumann b.c. $(\mathcal{N}',N')$.  

By applying S-duality, we can find the dual junction in Figure \ref{fig4dun3dumhm1}. 
It has a corner of 4d gauge theory with a pair of boundary conditions $(\mathcal{N}', \mathcal{D})$. 
It may include two types of 3d twisted hypermultiplets associated to the 
D3-D5$'$ string and D3-D3 string across the NS5$'$-brane. 
They should obey the Dirichlet/Nahm pole boundary conditions due to the D5-brane.

%%%%%%%%%%%%%%%%%%%%%%%%%%%%%%%%%%%%%%%%%%%%%
\subsubsection{4d $U(1)|$3d $(1)-[1]$ with $(\mathcal{N}',N')$}
\label{sec_4du13du1hm1}
%%%%%%%%%%%%%%%%%%%%%%%%%%%%%%%%%%%%%%%%%%%%%
Our first example is the junction 4d $U(1)|$3d $(1)-[1]$ with $(\mathcal{N}',N')$. 
It has a corner of 4d $U(1)$ gauge theory with a pair of Neumann b.c. $(\mathcal{N}, \mathcal{N}')$ 
and 3d $U(1)$ gauge theory with a charged hypermultiplet obeying Neumann b.c. $(\mathcal{N}', N')$. 
The 4d $U(1)$ gauge theory and 3d $U(1)$ gauge theory are coupled through 3d bi-fundamental hypermultiplet 
satisfying Neumann b.c. $N'$. 
There are two 2d Fermi multiplets. 
One is the determinant Fermi multiplet with charge $+$ and $-$  
under the 4d $U(1)$ and 3d $U(1)$ gauge symmetries. 
The another is the fundamental Fermi multiplet which carries charge under the 3d gauge symmetry.

The quarter-index is 
\begin{align}
\label{4du1_3du1hm1NN1}
&
\mathbb{IV}^{\textrm{4d $U(1)|(1)-[1]|$}}_{\mathcal{N}\mathcal{N}'}
\nonumber\\
&=
\underbrace{
(q)_{\infty} \oint \frac{ds_{1}}{2\pi is_{1}}
}_{\mathbb{IV}_{\mathcal{N}\mathcal{N}'}^{\textrm{4d $U(1)$}}}
\cdot 
\underbrace{
(q)_{\infty} (q^{\frac12} t^2;q)_{\infty}\oint \frac{ds_{2}}{2\pi is_{2}}
}_{\mathbb{II}_{\mathcal{N}'}^{\textrm{3d $U(1)$}}}
\nonumber\\
&\times 
\underbrace{
\frac{1}{(q^{\frac14}t s_{1}^{\pm}s_{2}^{\mp};q)_{\infty}}
}_{\mathbb{II}_{N'}^{\textrm{3d HM}} \left(\frac{s_{1}}{s_{2}} \right)}
\cdot 
\underbrace{
\frac{1}{(q^{\frac14}t s_{2}^{\pm};q)_{\infty}}
}_{\mathbb{II}_{N'}^{\textrm{3d HM}} (s_{2})}
\cdot 
\underbrace{
(q^{\frac12}s_{1}^{\pm}s_{2}^{\mp}z^{\pm};q)_{\infty}
}_{F\left(\frac{s_{1}}{s_{2}}z \right)}
\cdot 
\underbrace{
(q^{\frac12}s_{2}^{\pm}x^{\pm};q)_{\infty}
}_{F\left(q^{\frac12} s_{2}x \right)}
\end{align}

In the dual junction we have a corner of 4d $U(1)$ gauge theory 
with a pair of boundary condition $(\mathcal{N}', \mathcal{D})$. 
There would be two kinds of 3d twisted hypermultiplets with Dirichlet b.c. $D$. 
As there is a single D3-brane in both sides of one of the D5$'$-branes, 
3d fundamental twisted hyper appears from D3-D5$'$ string. 
The another twisted hyper arises from D3-D3 string across the NS5$'$-brane. 

As expected, we find that the quarter-index (\ref{4du1_3du1hm1NN1}) agrees with
\begin{align}
\label{4du1_3du1hm1NN2}
\mathbb{IV}^{1|[1]|}_{\mathcal{N}'\mathcal{D}'\mathcal{D}}
&=
\underbrace{
\frac{1}{(q^{\frac12}t^2;q)_{\infty}}
}_{\mathbb{IV}_{\mathcal{N}'\mathcal{D}}^{\textrm{4d $U(1)$}}}
\cdot 
\underbrace{
(q^{\frac34}tz^{\pm};q)_{\infty}
}_{\mathbb{II}_{D}^{\textrm{3d tHM}} (z)}
\cdot 
\underbrace{
(q^{\frac34}tx^{\pm};q)_{\infty}
}_{\mathbb{II}_{D}^{\textrm{3d tHM}} (x)}.
\end{align}

%%%%%%%%%%%%%%%%%%%%%%%%%%%%%%%%%%%%%%%%%%%%%
\subsubsection{4d $U(2)|$3d $(1)-[1]$ with $(\mathcal{N}',N')$}
\label{sec_4du23du1hm1}
%%%%%%%%%%%%%%%%%%%%%%%%%%%%%%%%%%%%%%%%%%%%%
Next consider the junction 4d $U(2)|$3d $(1)-[1]$ with $(\mathcal{N}',N')$. 
It keeps 4d $U(2)$ gauge symmetry and 3d $U(1)$ gauge symmetry. 
There are two 3d hypermultiplets; 
one transforms as the bi-fundamental representation under the $U(2)\times U(1)$ gauge symmetry, 
while the other transforms as the fundamental representation under the 3d $U(1)$ gauge symmetry. 
The configuration also includes two 2d Fermi multiplets; 
one is the cross-determinant Fermi at the NS5-NS5$'$ junction 
and the other is the fundamental Fermi at the D5-NS5$'$ junction. 

The quarter-index is written as
\begin{align}
\label{4du2_3du1hm1NN1}
&
\mathbb{IV}^{\textrm{4d $U(2)|(1)-[1]|$}}_{\mathcal{N}\mathcal{N}'}
\nonumber\\
&=
\underbrace{
\frac12 (q)_{\infty}^2 \oint \frac{ds_{1}}{2\pi is_{1}} \frac{ds_{2}}{2\pi is_{2}}
\left(\frac{s_{1}}{s_{2}};q \right)_{\infty}
\left( \frac{s_{2}}{s_{1}};q \right)_{\infty}
}_{\mathbb{IV}_{\mathcal{N}\mathcal{N}'}^{\textrm{4d $U(2)$}}}
\cdot 
\underbrace{
(q)_{\infty} (q^{\frac12} t^2;q)_{\infty}\oint \frac{ds_{3}}{2\pi is_{3}}
}_{\mathbb{II}_{\mathcal{N}'}^{\textrm{3d $U(1)$}}}
\nonumber\\
&\times 
\prod_{i=1}^{2}
\underbrace{
\frac{1}{(q^{\frac14}t s_{i}^{\pm}s_{3}^{\mp};q)_{\infty}}
}_{\mathbb{II}_{N'}^{\textrm{3d HM}} \left(\frac{s_{i}}{s_{3}} \right)}
\cdot 
\underbrace{
\frac{1}{(q^{\frac14}ts_{3}^{\pm};q)_{\infty}}
}_{\mathbb{II}_{N'}^{\textrm{3d HM}} (s_{3})}
\cdot 
\underbrace{
(q^{\frac12}s_{1}^{\pm}s_{2}^{\pm}s_{3}^{\mp} z^{\pm};q)_{\infty}
}_{F\left( q^{\frac12}\frac{s_{1}s_{2}}{s_{3}}z \right)}
\cdot 
\underbrace{
(q^{\frac12}s_{3}^{\pm}x^{\pm};q)_{\infty}
}_{F(q^{\frac12}s_{3}x)}. 
\end{align}

The dual junction has no gauge symmetry. 
Unequal numbers of D3-branes across the D5$'$-brane 
would not admit the 3d fundamental twisted hypermultiplet. 
However, there still exists a 3d twisted hypermultiplet 
arising from D3-D3 string across the NS5$'$-brane. 
It should receive Dirichlet b.c. $D$ due to the D5-brane. 

Actually we have checked that the quarter-index (\ref{4du2_3du1hm1NN1}) coincides with
\begin{align}
\label{4du2_3du1hm1NN2}
\mathbb{IV}^{2|[1]|}_{\mathcal{N}'\mathcal{D}'\mathcal{D}}
&=
\underbrace{
\frac{1}{(q^{\frac12}t^2;q)_{\infty}}
}_{\mathbb{IV}_{\mathcal{N}'\mathcal{D}}^{\textrm{4d $U(1)$}}}
\cdot 
\underbrace{
(q^{\frac34}t x^{\pm};q)_{\infty}
}_{\mathbb{II}_{D}^{\textrm{3d tHM}}(x)}. 
\end{align}

%%%%%%%%%%%%%%%%%%%%%%%%%%%%%%%%%%%%%%%%%%%%%
\subsubsection{4d $U(2)|$3d $(2)-[1]$ with $(\mathcal{N}',N')$}
\label{sec_4du23du2hm1}
%%%%%%%%%%%%%%%%%%%%%%%%%%%%%%%%%%%%%%%%%%%%%
Let us investigate the case where the 3d gauge symmetry is non-Abelian. 
The simplest example is the configuration 
4d $U(2)|$3d $(2)-[1]$ with $(\mathcal{N}',N')$, 
in which we have a corner of 4d $U(2)$ gauge theory with a pair of Neumann b.c. $(\mathcal{N},\mathcal{N}')$ 
and a boundary of 3d $U(2)$ gauge theory with a fundamental hyper subject to Neumann b.c. $(\mathcal{N}', N')$. 
At the NS5-NS5$'$ junction there appears a 2d cross-determinant Fermi multiplet 
and at the NS5$'$-D5 junction one finds a 2d fundamental Fermi multiplet. 

The quarter-index is 
\begin{align}
\label{4du2_3du2hm1NN1}
&
\mathbb{IV}^{\textrm{4d $U(2)|(2)-[1]|$}}_{\mathcal{N}\mathcal{N}'}
\nonumber\\
&=
\underbrace{
\frac12 (q)_{\infty}^2 \oint \frac{ds_{1}}{2\pi is_{1}} \frac{ds_{2}}{2\pi is_{2}}
\left( \frac{s_{1}}{s_{2}};q \right)_{\infty}
\left( \frac{s_{2}}{s_{1}};q \right)_{\infty}
}_{\mathbb{IV}_{\mathcal{N}\mathcal{N}'}^{\textrm{4d $U(2)$}}}
\nonumber\\
&\times 
\underbrace{
\frac12 (q)_{\infty}^2 (q^{\frac12}t^2;q)_{\infty}^2 
\oint \frac{ds_{3}}{2\pi is_{3}} \frac{ds_{4}}{2\pi is_{4}} 
\prod_{i\neq j}
\left( \frac{s_{i}}{s_{j}};q \right)_{\infty}
\left( q^{\frac12} t^2 \frac{s_{i}}{s_{j}};q \right)_{\infty}
}_{\mathbb{II}_{\mathcal{N}'}^{\textrm{3d $U(2)$}}}
\nonumber\\
&\times 
\prod_{i=1}^{2}\prod_{j=3}^{4}
\underbrace{
\frac{1}{(q^{\frac14}t s_{i}^{\pm}s_{j}^{\mp};q)_{\infty}}
}_{\mathbb{II}_{N'}^{\textrm{3d HM}} \left(\frac{s_{i}}{s_{j}} \right)}
\cdot 
\prod_{i=3}^{4}
\underbrace{
\frac{1}{(q^{\frac14}ts_{i}^{\pm};q)_{\infty}}
}_{\mathbb{II}_{N'}^{\textrm{3d HM}} (s_{i})}
\cdot 
\underbrace{
(q^{\frac12}s_{1}^{\pm}s_{2}^{\pm}s_{3}^{\mp}s_{4}^{\mp}z^{\pm};q)_{\infty}
}_{F\left(q^{\frac12}\frac{s_{1}s_{2}}{s_{3}s_{4}} z \right)}
\cdot 
\prod_{i=3}^{4} 
\underbrace{
(q^{\frac12} s_{i}^{\pm}x^{\pm};q)_{\infty}
}_{F(q^{\frac12}s_{i}x)}.
\end{align}

In the S-dual junction the gauge symmetry is completely broken. 
The junction would involve two kinds of 3d twisted hypermultiplets. 
One is the 3d fundamental twisted hyper with Nahm pole b.c. 
with an embedding $\rho:$ $\mathfrak{su}(2)$ $\rightarrow$ $\mathfrak{u}(2)$. 
The other is the 3d bi-fundamental hyper associated to the string across the NS5$'$-brane. 

The quarter-index (\ref{4du2_3du2hm1NN1}) turns out to coincide with
\begin{align}
\label{4du2_3du2hm1NN2}
\mathbb{IV}^{2|[2]|}_{\mathcal{N}'\mathcal{D}'\mathcal{D}}
&=
\underbrace{
\frac{1}{(q^{\frac12}t^2;q)_{\infty}}
}_{\mathbb{IV}_{\mathcal{N}'\mathcal{D}}^{\textrm{4d $U(1)$}}}
\cdot 
(q t^2 z^{\pm};q)_{\infty}
\cdot 
\underbrace{
(q^{\frac34}tx^{\pm};q)_{\infty}
}_{\mathbb{II}_{D}^{\textrm{3d tHM}}(x)}.
\end{align}
The first term is the quarter-index of 4d $U(1)$ gauge theory with a pair of 
boundary conditions $(\mathcal{N}', \mathcal{D})$. 
The second term is identified with the contributions 
from 3d fundamental twisted hyper satisfying the Nahm pole b.c. of rank $2$. 
The last term is the contribution from the 3d bi-fundamental hyper. 
Unlike the 3d fundamental twisted hyper, 
it is identifies with the Dirichlet half-index of 3d twisted hyper. 
The fact that 
we get the quarter index $\mathbb{IV}_{\mathcal{N}'\mathcal{D}}^{\textrm{4d $U(1)$}}$ 
and the Dirichlet half-index $\mathbb{II}_{D}^{\textrm{3d tHM}}$ of 3d twisted hyper 
would be caused by the D5$'$-brane on which two D3-branes terminate.

%%%%%%%%%%%%%%%%%%%%%%%%%%%%%%%%%%%%%%%%%%%%%
\subsubsection{4d $U(3)|$3d $(3)-[1]$ with $(\mathcal{N}',N')$}
\label{sec_4du33du3hm1}
%%%%%%%%%%%%%%%%%%%%%%%%%%%%%%%%%%%%%%%%%%%%%
Consider the junction 4d $U(3)|$3d $(3)-[1]$ with $(\mathcal{N}',N')$ 
in which 4d $U(3)$ gauge symmetry and 3d $U(3)$ gauge symmetry are preserved.  

The quarter-index is 
\begin{align}
\label{4du3_3du3hm1NN1}
&
\mathbb{IV}^{\textrm{4d $U(3)|(3)-[1]|$}}_{\mathcal{N}\mathcal{N}'}
\nonumber\\
&=
\underbrace{
\frac{1}{3!}(q)_{\infty}^3 
\oint \prod_{i=1}^{3}\frac{ds_{i}}{2\pi is_{i}}
\prod_{i\neq j}
\left(\frac{s_{i}}{s_{j}};q \right)_{\infty}
}_{\mathbb{IV}_{\mathcal{N}\mathcal{N}'}^{\textrm{4d $U(3)$}}}
\cdot 
\underbrace{
\frac{1}{3!}(q)_{\infty}^3 (q^{\frac12}t^2;q)_{\infty}^3 
\prod_{i\neq j}
\left(\frac{s_{i}}{s_{j}};q \right)_{\infty}
\left(q^{\frac12} t^2 \frac{s_{i}}{s_{j}};q \right)_{\infty}
}_{\mathbb{II}_{\mathcal{N}'}^{\textrm{3d $U(3)$}}}
\nonumber\\
&\times 
\prod_{i=1}^{3}\prod_{j=4}^{6}
\underbrace{
\frac{1}{(q^{\frac14}t s_{i}^{\pm}s_{j}^{\mp};q)_{\infty}}
}_{\mathbb{II}_{N'}^{\textrm{3d HM}}\left(\frac{s_{i}}{s_{j}} \right)}
\cdot 
\prod_{i=4}^{6}
\underbrace{
\frac{1}{(q^{\frac14}t s_{i}^{\pm};q)_{\infty}}
}_{\mathbb{II}_{N'}^{\textrm{3d HM}} (s_{i})}
\nonumber\\
&\times 
\underbrace{
(q^{\frac12}s_{1}^{\pm}s_{2}^{\pm}s_{3}^{\pm}s_{4}^{\mp}s_{5}^{\mp}s_{6}^{\mp}z^{\pm};q)_{\infty}
}_{F\left(q^{\frac12} \frac{s_{1}s_{2}s_{3}}{s_{4}s_{5}s_{6}}z \right)}
\cdot 
\prod_{i=4}^{6}
\underbrace{
(q^{\frac12} s_{i}^{\pm}x^{\pm};q)_{\infty}
}_{F(q^{\frac12}s_{i}x)}. 
\end{align}

We have confirmed that the quarter-index (\ref{4du3_3du3hm1NN1}) agrees with 
\begin{align}
\label{4du3_3du3hm1NN2}
\mathbb{IV}^{3|[3]|}_{\mathcal{N}'\mathcal{D}'\mathcal{D}}
&=
\underbrace{
\frac{1}{(q^{\frac12}t^2;q)_{\infty}}
}_{\mathbb{IV}_{\mathcal{N}'\mathcal{D}}^{\textrm{4d $U(1)$}}}
\cdot 
(q^{\frac54}t^3 z^{\pm};q)_{\infty}
\cdot 
\underbrace{
(q^{\frac34}t x^{\pm};q)_{\infty}
}_{\mathbb{II}_{D}^{\textrm{3d tHM}} (x)}.
\end{align}
Again it includes the quarter-index of 4d $U(1)$ gauge theory with a pair of 
boundary conditions $(\mathcal{N}', \mathcal{D})$. 
The second term is the contributions 
from 3d fundamental twisted hyper with the Nahm pole b.c. of rank $3$. 
The last term is the Dirichlet half-index of 3d twisted hyper.

%%%%%%%%%%%%%%%%%%%%%%%%%%%%%%%%%%%%%%%%%%%%%
\subsubsection{4d $U(N)|$3d $(M)-[1]$ with $(\mathcal{N}',N')$}
\label{sec_4duN3duMhm1}
%%%%%%%%%%%%%%%%%%%%%%%%%%%%%%%%%%%%%%%%%%%%%
Let us discuss the general junction 4d $U(N)|$3d $(M)-[1]$ with $(\mathcal{N}',N')$, 
which is constructed in the brane setup in Figure \ref{fig4dun3dumhm1}. 

It has a corner of 4d $\mathcal{N}=4$ $U(N)$ SYM theory subject to 
a pair of Neumann b.c. $(\mathcal{N}, \mathcal{N}')$ 
and a boundary of 3d $\mathcal{N}=4$ $U(M)$ gauge theory with a fundamental hypermultiplet 
obeying Neumann b.c. $(\mathcal{N}', N')$. 
The 4d $U(N)$ and 3d $U(M)$ vector multiplets 
are coupled through the 3d bi-fundamnetal hypermultiplet. 
In addition, there are 2d cross-determinant Fermi multiplet appearing at the NS5-NS5$'$ junction 
and fundamental Fermi multiplet at the NS5$'$-D5 junction. 
These Fermi multiplets are required to cancel the boundary gauge anomaly.

The quarter-index is given by
\begin{align}
\label{4duN_3duMhm1NN1}
&
\mathbb{IV}^{\textrm{4d $U(N)|(M)-[1]|$}}_{\mathcal{N}\mathcal{N}'}
\nonumber\\
&=
\underbrace{
\frac{1}{N!}(q)_{\infty}^{N} 
\oint \prod_{i=1}^{N}
\frac{ds_{i}}{2\pi is_{i}}
\prod_{i\neq j}\left( \frac{s_{i}}{s_{j}};q \right)_{\infty}
}_{\mathbb{IV}_{\mathcal{N}\mathcal{N}}^{\textrm{4d $U(N)$}}}
\nonumber\\
&\times 
\underbrace{
\frac{1}{M!}
(q)_{\infty}^{M} (q^{\frac12}t^2;q)_{\infty}^{M}
\oint \prod_{i=N+1}^{N+M}
\frac{ds_{i}}{2\pi is_{i}}
\prod_{i\neq j}
\left( \frac{s_{i}}{s_{j}};q \right)_{\infty}
\left(q^{\frac12} t^2 \frac{s_{i}}{s_{j}};q \right)_{\infty}
}_{\mathbb{II}_{\mathcal{N}'}^{\textrm{3d $U(M)$}}}
\nonumber\\
&\times 
\prod_{i=1}^{N}\prod_{j=N+1}^{N+M}
\underbrace{
\frac{1}{(q^{\frac14}t s_{i}^{\pm}s_{j}^{\mp};q)_{\infty}}
}_{\mathbb{II}_{N'}^{\textrm{3d HM}} \left(\frac{s_{i}}{s_{j}} \right)}
\cdot 
\prod_{i=N+1}^{N+M}
\underbrace{
\frac{1}{(q^{\frac14}t s_{i}^{\pm};q)_{\infty}}
}_{\mathbb{II}_{N'}^{\textrm{3d HM}}(s_{i})}
\nonumber\\
&\times 
\underbrace{
(q^{\frac12}\prod_{i=1}^{N}s_{i}^{\pm}\prod_{j=N+1}^{N+M}s_{j}^{\mp}z^{\pm};q)_{\infty}
}_{F\left( q^{\frac12} \frac{\prod_{i=1}^{N}s_{i}}{\prod_{j=N+1}^{N+M}s_{j}}z \right)}
\cdot 
\underbrace{
\prod_{i=N+1}^{N+M}(q^{\frac12}s_{i}^{\pm}x^{\pm};q)_{\infty}
}_{F (q^{\frac12}s_{i}x)}. 
\end{align}

The dual junction has no gauge symmetry. 
It contains a corner of 4d gauge theory with a pair of boundary conditions $(\mathcal{N}',\mathcal{D})$. 
There would be the 3d twisted hypermultiplet arising from D3-D3 string across the NS5$'$-brane 
obeying the boundary condition due to the D5-brane. 
For $N=M$, the junction would involve 3d fundamental twisted hypermultiplet arising from D3-D5$'$ string. 
Due to the D5-brane, it should satisfy the Dirichlet/Nahm pole boundary condition 
specified by a homomorphism $\rho:$ $\mathfrak{su}(2)$ $\rightarrow$ $\mathfrak{u}(N)$.

The quarter-index (\ref{4duN_3duMhm1NN1}) would lead to the same answer as 
\begin{align}
\label{4duN_3duMhm1NN2}
\mathbb{IV}^{N|[M]|}_{\mathcal{N}'\mathcal{D}'\mathcal{D}}
&=
\underbrace{
\frac{1}{(q^{\frac12}t^2;q)_{\infty}}
}_{\mathbb{IV}_{\mathcal{N}'\mathcal{D}}^{\textrm{4d $U(1)$}}}
\cdot 
(q^{\frac34+\frac{N-1}{4}} t^{1+(N-1)} z^{\pm};q)_{\infty}^{{\delta^{N}}_{M} }
\cdot 
\underbrace{
(q^{\frac34}t x^{\pm};q)_{\infty}
}_{\mathbb{II}_{D}^{\textrm{3d tHM}} (x)}.
\end{align}
The first term is the quarter-index of 4d $U(1)$ gauge theory 
with a pair of boundary conditions $(\mathcal{N}',\mathcal{D})$. 
The second is a sort of half-index of Dirichlet/Nahm pole b.c. for 3d fundamental twisted hypermultiplet 
which only appears for $N=M$. 
The last is the Dirichlet half-index for 3d twisted hypermultiplet 
which is associated to the 3d twisted hypermultiplet arising from D3-D3 string across the NS5$'$-brane.

%%%%%%%%%%%%%%%%%%%%%%%%%%%%%%%%%%%%%%%%%%%%%
%%%%%%%%%%%%%%%%%%%%%%%%%%%%%%%%%%%%%%%%%%%%%
\subsection{4d $U(N)|$3d $(M)-[2]$ with $(\mathcal{N}',N')$}
\label{sec_4dN3dMhm2}
\begin{figure}
\begin{center}
\includegraphics[width=12.5cm]{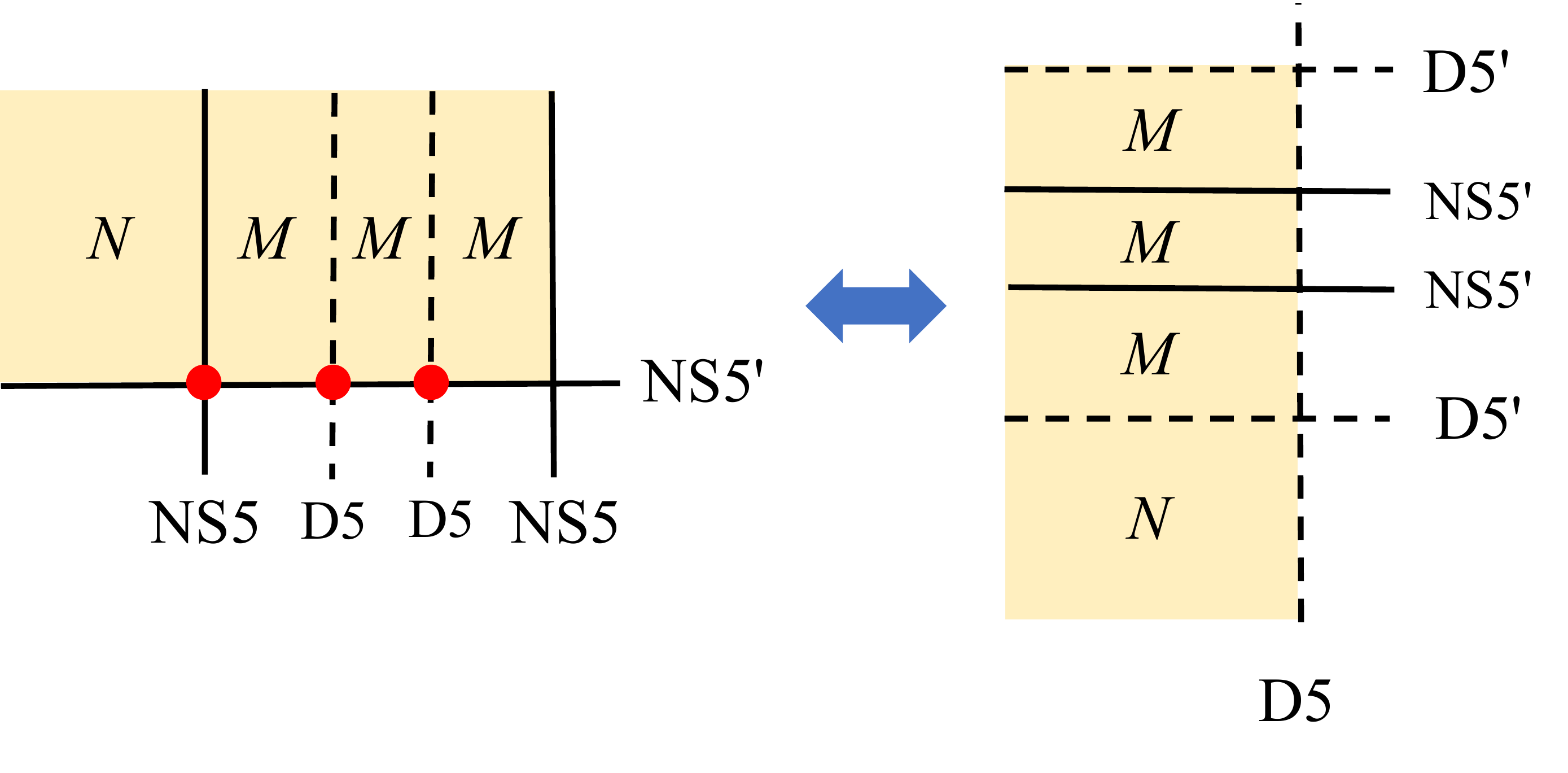}
\caption{
The brane constructions of the configuration 4d $U(N)|$3d $(M)-[2]$ with Neumann b.c. $(\mathcal{N}',N')$ 
and its mirror corner configuration which involves Dirichlet b.c. $\mathcal{D}$ or Nahm pole b.c. for a three-dimensional $U(M)$ gauge theory. 
}
\label{fig4dun3dumhm2}
\end{center}
\end{figure}

%%%%%%%%%%%%%%%%%%%%%%%%%%%%%%%%%%%%%%%%%%%%%
%%%%%%%%%%%%%%%%%%%%%%%%%%%%%%%%%%%%%%%%%%%%%
Let us consider the junction 4d $U(N)|$3d $(M)-[2]$ with $(\mathcal{N}',N')$. 
It can be obtained by further adding a fundamental hypermultiplet to 
the junction 4d $U(N)|$3d $(M)-[1]$ with $(\mathcal{N}',N')$ in section \ref{sec_4dN3dMhm1} 
and the corresponding brane construction is shown in Figure \ref{fig4dun3dumhm2}. 

Unlike the previous cases, 
the S-dual junction has 3d $U(M)$ gauge theory subject to Dirichlet/Nahm pole b.c. 
Hence the evaluation of the quarter-indices for the dual junctions may include the non-perturbative monopole sum.

%%%%%%%%%%%%%%%%%%%%%%%%%%%%%%%%%%%%%%%%%%%%%
\subsubsection{4d $U(1)|$3d $(1)-[2]$ with $(\mathcal{N}',N')$}
\label{sec_4d13d1hm2}
%%%%%%%%%%%%%%%%%%%%%%%%%%%%%%%%%%%%%%%%%%%%%
We begin with the simplest example, 
that is the junction 4d $U(1)|$3d $(1)-[2]$ with $(\mathcal{N}',N')$. 
It has 4d $U(1)$ gauge symmetry and 3d $U(1)$ gauge symmetry. 
There are two 3d fundamental hypermultiplets with charges under the 3d $U(1)$ gauge symmetry 
and a 3d bi-fundamental hypermultiplet coupled to 4d $U(1)$ and 3d $U(1)$ gauge fields. 
In addition, there are a 2d cross-determinant Fermi multiplet at the NS5-NS5$'$ junction 
and two fundamental Fermi multiplet at the NS5$'$-D5 junction. 

The quarter-index is calculated as
\begin{align}
\label{4du1_3du1hm2NN1}
&
\mathbb{IV}^{\textrm{4d $U(1)|(1)-[2]|$}}_{\mathcal{N}\mathcal{N}'}
\nonumber\\
&=
\underbrace{
(q)_{\infty}\oint \frac{ds_{1}}{2\pi is_{1}}
}_{\mathbb{IV}_{\mathcal{N}\mathcal{N}'}^{\textrm{4d $U(1)$}}}
\cdot 
\underbrace{
(q)_{\infty} (q^{\frac12}t^2;q)_{\infty} \oint \frac{ds_{2}}{2\pi is_{2}}
}_{\mathbb{II}_{\mathcal{N}'}^{\textrm{3d $U(1)$}}}
\nonumber\\
&\times 
\underbrace{
\frac{1}{(q^{\frac14}t s_{1}^{\pm}s_{2}^{\mp};q)_{\infty}}
}_{\mathbb{II}_{N'}^{\textrm{3d HM}} \left(\frac{s_{1}}{s_{2}} \right)}
\cdot 
\prod_{\alpha=1}^{2}
\underbrace{
\frac{1}{(q^{\frac14}t s_{2}^{\pm}x_{\alpha}^{\pm};q)_{\infty}}
}_{\mathbb{II}_{N'}^{\textrm{3d HM}} (s_{2}x_{\alpha})}
\nonumber\\
&\times 
\underbrace{
(q^{\frac12}s_{1}^{\pm}s_{2}^{\mp}z^{\pm};q)_{\infty}
}_{F\left( q^{\frac12}\frac{s_{1}}{s_{2}}z \right)}
\cdot 
\underbrace{
(q^{\frac12}s_{2}^{\pm}x_{1}^{\pm}u^{\pm};q)_{\infty}
}_{F(q^{\frac12}s_{2} x_{1} u )}
\cdot 
\underbrace{
(q^{\frac12}s_{2}^{\pm}x_{2}^{\pm}u^{\mp};q)_{\infty}
}_{F(q^{\frac12} s_{2}x_{2}u^{-1})}
\end{align}
where $x_{1}x_{2}=1$. 

The S-dual configuration contains a corner of 
4d $U(1)$ gauge theory with a pair of boundary conditions $(\mathcal{N}', \mathcal{D})$. 
This 4d gauge theory has a defect of D5$'$-brane that couples 3d fundamental twisted hyper to 4d gauge theory. 
It should obey the Dirichlet b.c. $D$ required from the D5-brane. 
The junction should have a 3d $\widetilde{U(1)}$ twisted vector multiplet arising from 
a single D3-brane suspended between NS5$'$-branes 
and two 3d bi-fundamental twisted hypers arising strings across the two NS5$'$-branes. 
The D5-brane again requires the twisted vector and twisted hypers to obey Dirichlet b.c. $\mathcal{D}$ and $D$ respectively. 

We then obtain the quarter-index for the dual junction
\begin{align}
\label{4du1_3du1hm2NN2}
&
\mathbb{IV}^{\textrm{4d $U(1)$}|(1)-[1]|}_{\mathcal{N}'\mathcal{D}}
\nonumber\\
&=
\underbrace{
\frac{1}{(q^{\frac12}t^2;q)_{\infty}}
}_{\mathbb{IV}_{\mathcal{N}'\mathcal{D}}^{\textrm{4d $U(1)$}}}
\cdot 
\frac{1}{(q)_{\infty} (q^{\frac12}t^2;q)_{\infty}} \sum_{m\in \mathbb{Z}}
\nonumber\\
&\times 
\underbrace{
(q^{\frac34}t z^{\pm};q)_{\infty} 
}_{\mathbb{II}_{D}^{\textrm{3d tHM}} (z)}
\cdot 
\underbrace{
(q^{\frac34\pm m}t u^{\pm};q)_{\infty}
}_{\mathbb{II}_{D}^{\textrm{3d tHM}} (q^{m}u)}
\cdot 
\underbrace{
(q^{\frac34\pm m}t u^{\pm};q)_{\infty}
}_{\mathbb{II}_{D}^{\textrm{3d tHM}}(q^{m}u)}
\nonumber\\
&\times 
q^{m^2} u^{2m} x_{1}^{m} x_{2}^{-m}.
\end{align}
In fact, we have confirmed that 
this matches with the quarter-index (\ref{4du1_3du1hm2NN1}).

%%%%%%%%%%%%%%%%%%%%%%%%%%%%%%%%%%%%%%%%%%%%%
\subsubsection{4d $U(2)|$3d $(1)-[2]$ with $(\mathcal{N}',N')$}
\label{sec_4d23d1hm2}
%%%%%%%%%%%%%%%%%%%%%%%%%%%%%%%%%%%%%%%%%%%%%
Let us turn to the junction 4d $U(2)|$3d $(1)-[2]$ with $(\mathcal{N}',N')$. 
We have 4d $U(2)$ gauge symmetry and 3d $U(1)$ gauge symmetry.

Similarly, we can compute the quarter-index as 
\begin{align}
\label{4du2_3du1hm2NN1}
&
\mathbb{IV}^{\textrm{4d $U(2)|(1)-[2]|$}}_{\mathcal{N}\mathcal{N}'}
\nonumber\\
&=
\underbrace{
\frac12 (q)_{\infty}^2 \oint \frac{ds_{1}}{2\pi is_{1}} \frac{ds_{2}}{2\pi is_{2}} 
\left(\frac{s_{1}}{s_{2}};q \right)_{\infty}
\left(\frac{s_{2}}{s_{1}};q \right)_{\infty}
}_{\mathbb{IV}_{\mathcal{N}\mathcal{N}'}^{\textrm{4d $U(2)$}}}
\cdot 
\underbrace{
(q)_{\infty}(q^{\frac12}t^2;q)_{\infty}\oint \frac{ds_{3}}{2\pi is_{3}}
}_{\mathbb{II}_{\mathcal{N}'}^{\textrm{3d $U(1)$}}}
\nonumber\\
&\times 
\prod_{i=1}^{2}
\underbrace{
\frac{1}{(q^{\frac14}t s_{i}^{\pm}s_{3}^{\mp};q)_{\infty}}
}_{\mathbb{II}_{N'}^{\textrm{3d HM}}\left(\frac{s_{i}}{s_{3}} \right)}
\prod_{\alpha=1}^{2}
\underbrace{
\frac{1}{(q^{\frac14}t s_{3}^{\pm}x_{\alpha}^{\pm};q)_{\infty}}
}_{\mathbb{II}_{N'}^{\textrm{3d HM}}(s_{3}x_{\alpha})}
\nonumber\\
&\times 
\underbrace{
(q^{\frac12}s_{1}^{\pm}s_{2}^{\pm}s_{3}^{\mp}z_{1}^{\pm};q)_{\infty}
}_{F\left(q^{\frac12}\frac{s_{1}s_{2}}{s_{3}}z_{1} \right)}
\cdot 
\underbrace{
(q^{\frac12}s_{3}^{\pm}x_{1}^{\pm}u^{\pm};q)_{\infty}
}_{F(q^{\frac12}s_{3}x_{1}u)}
\cdot 
\underbrace{
(q^{\frac12}s_{3}^{\pm}x_{2}^{\pm}u^{\mp};q)_{\infty}
}_{F(q^{\frac12}s_{3}x_{2}u^{-1})}
\end{align}
with $x_{1}x_{2}=1$. 

In the S-dual configuration 
4d $U(2)\times U(1)$ gauge symmetry is firstly broken down to $U(1)$ due to the D5$'$-brane and 
we have a corner of 4d $U(1)$ gauge theory with a pair of boundary conditions $(\mathcal{N}',\mathcal{D})$. 
The 4d gauge theory includes a defect of D5$'$-brane, however, 
the unequal numbers of D3-branes across the D5$'$-brane should not lead to the 3d fundamental twisted hypermultiplet. 
As a single D3-brane is stretched between the NS5$'$-branes, 
there are 3d $\widetilde{U(1)}$ twisted vector multiplet and two twisted hypers arising strings across the NS5$'$-branes. 
They should obey the Dirichlet b.c. $\mathcal{D}$ and $D$. 

We have checked that the quarter-index (\ref{4du2_3du1hm2NN1}) agrees with
\begin{align}
\label{4du2_3du1hm2NN2}
&
\mathbb{IV}^{\textrm{4d $U(2)\rightarrow U(1)$}|(1)-[1]|}_{\mathcal{N}'\mathcal{D}}
\nonumber\\
&=
\underbrace{
\frac{1}{(q^{\frac12}t^2;q)_{\infty}}
}_{\mathbb{IV}_{\mathcal{N}'\mathcal{D}}^{\textrm{4d $U(1)$}}}
\cdot 
\frac{1}{(q)_{\infty}(q^{\frac12}t^2;q)_{\infty}}\sum_{m\in \mathbb{Z}}
\nonumber\\
&\times 
\underbrace{
(q^{\frac34\pm m}t u^{\pm};q)_{\infty}
}_{\mathbb{II}_{D}^{\textrm{3d tHM}} (q^{m}u)}
\cdot 
\underbrace{
(q^{\frac34\pm m}t u^{\pm};q)_{\infty}
}_{\mathbb{II}_{D}^{\textrm{3d HM}} (q^{m}u)}
\cdot 
q^{m^2} \cdot 
u^{2m} \cdot 
x_{1}^{m} x_{2}^{-m}.
\end{align}

More challenging classes are
the junction 4d $U(N)|$3d $(M)-[2]$ with $(\mathcal{N}',N')$ where $M$ is larger than one. 
To study them, one would need to figure out the Nahm pole half-indices for 3d $\mathcal{N}=4$ gauge theories. 
We leave this for future work.

%%%%%%%%%%%%%%%%%%%%%%%%%%%%%%%%%%%%%%%%%%%%%
%%%%%%%%%%%%%%%%%%%%%%%%%%%%%%%%%%%%%%%%%%%%%
\subsection{4d $U(L)|$3d $U(M)|$4d $U(N)$ with $\mathcal{N}'$}
\label{sec_4dL3dM4dN}
%%%%%%%%%%%%%%%%%%%%%%%%%%%%%%%%%%%%%%%%%%%%%
%%%%%%%%%%%%%%%%%%%%%%%%%%%%%%%%%%%%%%%%%%%%%
Now we would like to examine the $\mathcal{N}=(0,4)$ junction  
which involves a pair of corners of 4d $\mathcal{N}=4$ gauge theories 
and a boundary of 3d $\mathcal{N}=4$ gauge theory. 

\begin{figure}
\begin{center}
\includegraphics[width=12.5cm]{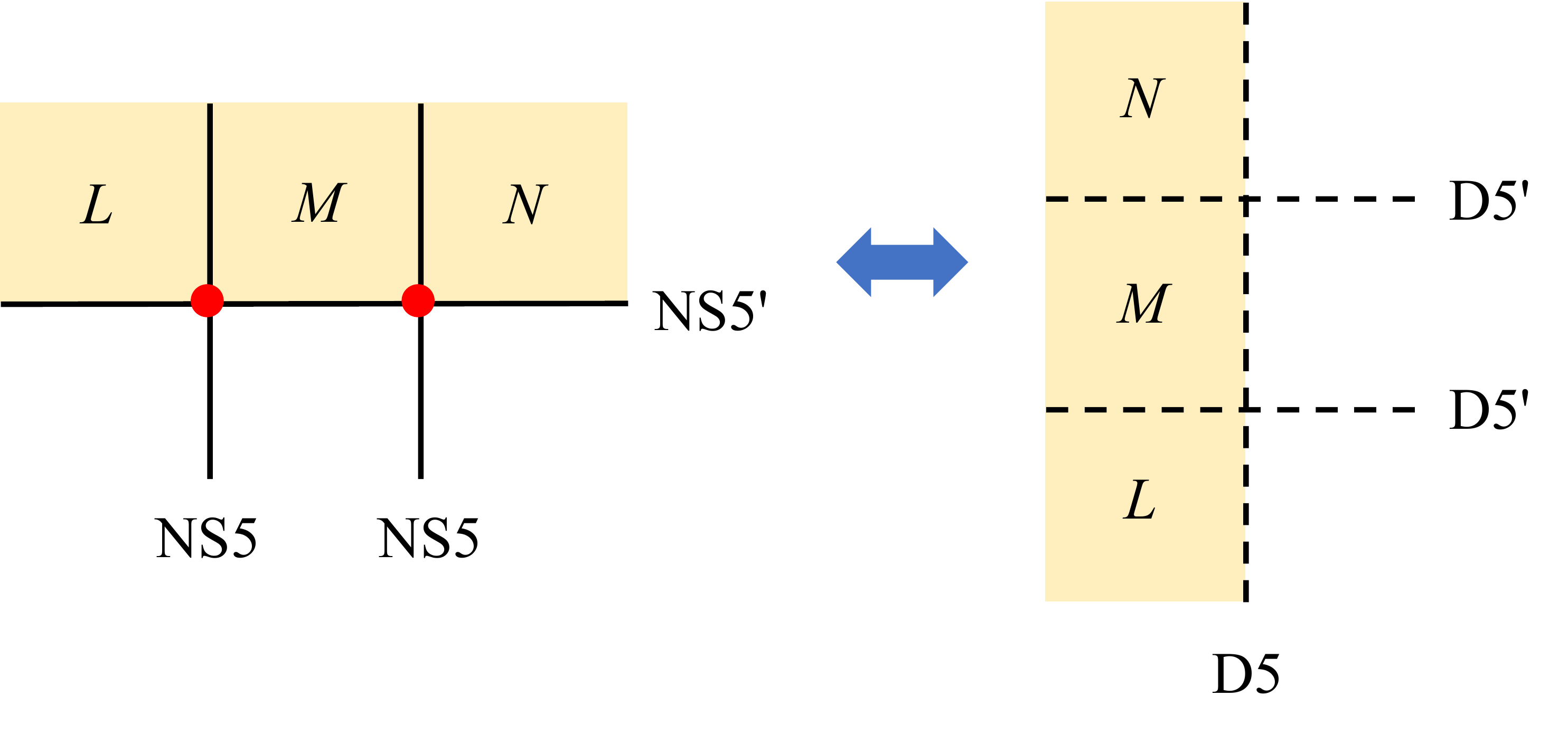}
\caption{
The brane constructions of the configuration 4d $U(L)|$3d $U(M)|$4d $U(N)$ with Neumann b.c. $\mathcal{N}'$ 
and its mirror corner configuration that has no gauge symmetry. 
}
\label{fig4dul3dum4dun}
\end{center}
\end{figure}

We begin with the junction 4d $U(L)|$3d $U(M)|$4d $U(N)$ with $\mathcal{N}'$, 
which has a pair of 4d $U(L)$ and $U(N)$ gauge theories obeying 
a pair of Neumann b.c. $(\mathcal{N}, \mathcal{N}')$ 
and 3d $U(M)$ gauge theory obeying Neumann b.c. $\mathcal{N}'$. 
There are two 3d hypermultiplets. 
Under $U(L)\times U(M)\times U(N)$ 
one transforms as $({\bf L}, \overline{\bf M},{\bf 1})$ $\oplus$ $(\overline{\bf L}, {\bf M},{\bf 1})$ 
and the other transforming as $({\bf 1}, {\bf M}, \overline{\bf N})$ $\oplus$ $({\bf 1}, \overline{\bf M}, {\bf N})$. 
They obey the Neumann b.c. $N'$. 
In addition, we have two 2d cross-determinant Fermi multiplets. 
One transforms as $(\det, \det^{-1},{\bf 1})$ and the other transforms as $({\bf 1},\det, \det^{-1})$. 
It is constructed from the brane configuration in Figure \ref{fig4dul3dum4dun}. 

By applying S-duality, we can address the dual junction. 
Due to the D5$'$-branes, 4d $U(L)\times U(M)\times U(N)$ gauge symmetry should be broken down to $U(\min \{L,M,N\})$. 
Furthermore, the D5-brane breaks down the gauge symmetry $U(\min \{L,M,N\})$ 
by requiring the Dirichlet/Nahm pole boundary condition specified by an embedding 
$\rho:$ $\mathfrak{su}(2)$ $\rightarrow$ $\mathfrak{u}(\min \{L,M,N\})$. 
We check the dualities of these junctions by evaluating the quarter-indices.

%%%%%%%%%%%%%%%%%%%%%%%%%%%%%%%%%%%%%%%%%%%%%
\subsubsection{4d $U(1)|$3d $U(1)|$4d $U(1)$ with $\mathcal{N}'$}
\label{sec_4du13du14du1NN}
%%%%%%%%%%%%%%%%%%%%%%%%%%%%%%%%%%%%%%%%%%%%%
Let us start with the junction 4d $U(1)|$3d $U(1)|$4d $U(1)$ with $\mathcal{N}'$. 
This has two corners of 4d $U(1)$ gauge theories with a pair of Neumann b.c. 
$(\mathcal{N}, \mathcal{N}')$ and 3d $U(1)$ gauge theory between them obeying Neumann b.c. $\mathcal{N}'$. 
We have two 3d bi-fundamental hypermultiplets with Neumann b.c. $N'$ 
and two 2d determinant Fermi multiplets cancelling the gauge anomaly. 

The quarter-index reads
\begin{align}
\label{111Na}
&
\mathbb{IV}^{\textrm{4d $U(1)|(1)|$4d $U(1)$}}_{\mathcal{N}\mathcal{N}'}
\nonumber\\
&=
\underbrace{
(q)_{\infty}\oint \frac{ds_{1}}{2\pi is_{1}}
}_{\mathbb{IV}_{\mathcal{N}\mathcal{N}'}^{\textrm{4d $U(1)$}}}
\cdot 
\underbrace{
(q)_{\infty} (q^{\frac12}t^2;q)_{\infty}
\oint \frac{ds_{2}}{2\pi is_{2}}
}_{\mathbb{II}_{\mathcal{N}'}^{\textrm{3d $U(1)$}}}
\cdot 
\underbrace{
(q)_{\infty} \oint \frac{ds_{3}}{2\pi is_{3}}
}_{\mathbb{IV}_{\mathcal{N}\mathcal{N}'}^{\textrm{4d $U(1)$}}}
\nonumber\\
&\times 
\underbrace{
\frac{1}{(q^{\frac14}ts_{1}^{\pm}s_{2}^{\mp};q)_{\infty}}
}_{\mathbb{II}_{N'}^{\textrm{3d HM}} \left( \frac{s_{1}}{s_{2}} \right)}
\cdot 
\underbrace{
\frac{1}{(q^{\frac14}ts_{3}^{\pm}s_{4}^{\mp};q)_{\infty}}
}_{\mathbb{II}_{N'}^{\textrm{3d HM}}\left(\frac{s_{3}}{s_{4}} \right) }
\cdot 
\underbrace{
(q^{\frac12} s_{1}^{\pm} s_{2}^{\mp} z_{1}^{\pm};q)_{\infty}
}_{F\left(q^{\frac12}\frac{s_{1}}{s_{2}}z_{1} \right)}
\cdot 
\underbrace{
(q^{\frac12} s_{2}^{\pm} s_{3}^{\mp} z_{2}^{\pm};q)_{\infty}
}_{F\left( q^{\frac12} \frac{s_{2}}{s_{3}}z_{2} \right) }.
\end{align}

The dual configuration has 4d $U(1)$ gauge theory with Dirichlet b.c. $\mathcal{D}$. 
It also contains two defects of D5$'$-branes. 
As the numbers of D3-branes are the same across them, 
they couple two 3d fundamental twisted hypermultiplets to the 4d $U(1)$ gauge theory satisfying Dirichlet b.c. $D$. 

Then the quarter-index would take the form
\begin{align}
\label{111Nb}
\mathbb{IV}^{1|[1]|1}_{\mathcal{D}\mathcal{D}'}
&=
\underbrace{
\frac{(q)_{\infty}}{(q^{\frac12}t^2;q)_{\infty}}
}_{\mathbb{II}_{\mathcal{D}}^{\textrm{4d $U(1)$}}}
\cdot 
\underbrace{
(q^{\frac34}tz_{1}^{\pm};q)_{\infty}
}_{\mathbb{II}_{D}^{\textrm{3d tHM}} (z_{1})}
\cdot 
\underbrace{
(q^{\frac34}tz_{2}^{\pm};q)_{\infty}
}_{\mathbb{II}_{D}^{\textrm{3d tHM}} (z_{2})}.
\end{align}
In fact, the quarter-indices (\ref{111Na}) and (\ref{111Nb}) agree with each other.

%%%%%%%%%%%%%%%%%%%%%%%%%%%%%%%%%%%%%%%%%%%%%
\subsubsection{4d $U(2)|$3d $U(1)|$4d $U(1)$ with $\mathcal{N}'$}
\label{sec_4du23du14du1NN}
%%%%%%%%%%%%%%%%%%%%%%%%%%%%%%%%%%%%%%%%%%%%%
For the configuration 4d $U(2)|$3d $U(1)|$4d $U(1)$ with $\mathcal{N}'$, 
we have two corners of 4d $U(2)$ and $U(1)$ gauge theories with Neumann boundary conditions $(\mathcal{N}, \mathcal{N}')$ 
and 3d $U(1)$ gauge theory with Neumann b.c. $\mathcal{N}'$. 
There are two 3d bi-fundamental hypermultiplets subject to Neumann b.c. $N'$ 
together with two 2d cross-determinant Fermi multiplets. 

The quarter-index reads
\begin{align}
\label{211Na}
&
\mathbb{IV}^{\textrm{4d $U(2)|(1)|$4d $U(1)$}}_{\mathcal{N}\mathcal{N}'}
\nonumber\\
&=
\underbrace{
\frac12 (q)_{\infty}^2 \oint \frac{ds_{1}}{2\pi is_{1}}\frac{ds_{2}}{2\pi is_{2}}
\left(\frac{s_{1}}{s_{2}};q \right)_{\infty}
\left(\frac{s_{2}}{s_{1}};q \right)_{\infty}
}_{\mathbb{IV}_{\mathcal{N}\mathcal{N}'}^{\textrm{4d $U(2)$}}}
\cdot 
\underbrace{
(q)_{\infty} (q^{\frac12}t^2;q)_{\infty}
\oint \frac{ds_{3}}{2\pi is_{3}}
}_{\mathbb{II}_{\mathcal{N}'}^{\textrm{3d $U(1)$}}}
\cdot 
\underbrace{
(q)_{\infty} \oint \frac{ds_{4}}{2\pi is_{4}}
}_{\mathbb{IV}_{\mathcal{N}\mathcal{N}'}^{\textrm{4d $U(1)$}}}
\nonumber\\
&\times 
\prod_{i=1}^{2}
\underbrace{
\frac{1}{(q^{\frac14}t s_{i}^{\pm}s_{3}^{\mp};q)_{\infty}}
}_{\mathbb{II}_{N'}^{\textrm{3d HM}} \left( \frac{s_{i}}{s_{3}} \right)}
\cdot 
\underbrace{
\frac{1}
{(q^{\frac14}t s_{3}^{\pm}s_{4}^{\mp};q)_{\infty}}
}_{\mathbb{II}_{N'}^{\textrm{3d HM}} \left(\frac{s_{3}}{s_{4}} \right)}
\cdot 
\underbrace{
(q^{\frac12} s_{1}^{\pm} s_{2}^{\pm} s_{3}^{\mp} z_{1}^{\pm};q)_{\infty}
}_{F\left(q^{\frac12}\frac{s_{1}s_{2}}{s_{3}}z_{1} \right)}
\cdot
\underbrace{
(q^{\frac12} s_{3}^{\pm} s_{4}^{\mp} z_{2}^{\pm};q)_{\infty}
}_{F\left(q^{\frac12}\frac{s_{3}}{s_{4}}z_{2} \right)}. 
\end{align}

In the dual configuration, 
the 4d $U(2)\times U(1)\times U(1)$ gauge symmetry is broken down to $U(1)$ by the D5$'$-branes, 
which is further broken down by the D5-brane. 
As the number of D3-branes jumps across one of the D5$'$-brane, there would be just a single twisted hypermultiplet 
obeying Dirichlet b.c. $D$. 

The quarter-index for the dual junction then takes the form
\begin{align}
\label{211Nb}
\mathbb{IV}^{2|[1]|1}_{\mathcal{D}\mathcal{D}'}
&=
\underbrace{
\frac{(q)_{\infty}}{(q^{\frac12}t^2;q)_{\infty}}
}_{\mathbb{II}_{\mathcal{D}}^{\textrm{4d $U(1)$}}}
\cdot 
\underbrace{
(q^{\frac34} t z_{2}^{\pm};q)_{\infty}
}_{\mathbb{II}_{D}^{\textrm{3d tHM}}(z_{2})}.
\end{align}
As expected, the quarter-indices (\ref{211Na}) and (\ref{211Nb}) coincide to each other.

%%%%%%%%%%%%%%%%%%%%%%%%%%%%%%%%%%%%%%%%%%%%%
\subsubsection{4d $U(2)|$3d $U(1)|$4d $U(2)$ with $\mathcal{N}'$}
\label{sec_4du23du14du2NN}
%%%%%%%%%%%%%%%%%%%%%%%%%%%%%%%%%%%%%%%%%%%%%
Next consider the junction 4d $U(2)|$3d $U(1)|$4d $U(2)$ with $\mathcal{N}'$. 
In this case the 3d gauge symmetry is smaller than each of 4d gauge symmetries. 

We can similarly calculate the quarter-index as
\begin{align}
\label{212Na}
&
\mathbb{IV}^{\textrm{4d $U(2)|(1)|$4d $U(2)$}}_{\mathcal{N}\mathcal{N}'}
\nonumber\\
&=
\underbrace{
\frac12 (q)_{\infty}^2 
\oint \frac{ds_{1}}{2\pi is_{1}}
\frac{ds_{2}}{2\pi is_{2}}
\left(\frac{s_{1}}{s_{2}};q \right)_{\infty}
\left(\frac{s_{2}}{s_{1}};q \right)_{\infty}
}_{\mathbb{IV}_{\mathcal{N}\mathcal{N}'}^{\textrm{4d $U(2)$}}}
\cdot 
\underbrace{
(q)_{\infty} (q^{\frac12}t^2;q)_{\infty}
\oint \frac{ds_{3}}{2\pi is_{3}}
}_{\mathbb{II}_{\mathcal{N}'}^{\textrm{3d $U(1)$}}}
\nonumber\\
&\times 
\underbrace{
\frac12 (q)_{\infty}^2 
\oint \frac{ds_{4}}{2\pi is_{4}}
\frac{ds_{5}}{2\pi is_{5}}
\left(\frac{s_{4}}{s_{5}};q \right)_{\infty}
\left(\frac{s_{5}}{s_{4}};q \right)_{\infty}
}_{\mathbb{IV}_{\mathcal{N}\mathcal{N}'}^{\textrm{4d $U(2)$}}}
\nonumber\\
&\times 
\prod_{i=1}^{2}
\underbrace{
\frac{1}{(q^{\frac14}ts_{i}^{\pm}s_{3}^{\mp};q)_{\infty}}
}_{\mathbb{II}_{N'}^{\textrm{3d HM}} \left(\frac{s_{i}}{s_{3}} \right)}
\prod_{i=4}^{5}
\underbrace{
\frac{1}{(q^{\frac14}ts_{3}^{\pm}s_{i}^{\mp};q)_{\infty}}
}_{\mathbb{II}_{N'}^{\textrm{3d HM}}\left(\frac{s_{3}}{s_{i}} \right)}
\cdot 
\underbrace{
(q^{\frac12}s_{1}^{\pm}s_{2}^{\pm}s_{3}^{\mp}z_{1}^{\pm};q)_{\infty}
}_{F \left(q^{\frac12}\frac{s_{1}s_{2}}{s_{3}}z_{1} \right)}
\cdot 
\underbrace{
(q^{\frac12}s_{3}^{\pm}s_{4}^{\mp}s_{5}^{\mp}z_{2}^{\pm};q)_{\infty}
}_{F \left( q^{\frac12}\frac{s_{3}}{s_{4}s_{5}}z_{2} \right)}. 
\end{align}

In the S-dual configuration 
the D5$'$-branes break the $U(2)\times U(1)\times U(2)$ gauge symmetry down to $U(1)$ 
and they do not lead to the 3d fundamental twisted hypers. 
In addition, the $U(1)$ gauge symmetry is broken by the D5-brane 
as Dirichlet b.c. $\mathcal{D}$. 

The quarter-index for the dual configuration is then simply 
the Dirichlet half-index of 4d $U(1)$ gauge theory:
\begin{align}
\label{212Nb}
\mathbb{IV}^{2|[1]|2}_{\mathcal{D}\mathcal{D}'}
&=
\underbrace{
\frac{(q)_{\infty}}{(q^{\frac12}t^2;q)_{\infty}}
}_{\mathbb{II}_{D}^{\textrm{4d $U(1)$}}}.
\end{align}
In fact, this coincides with the quarter-index (\ref{212Na}).

%%%%%%%%%%%%%%%%%%%%%%%%%%%%%%%%%%%%%%%%%%%%%
\subsubsection{4d $U(3)|$3d $U(1)|$4d $U(1)$ with $\mathcal{N}'$}
\label{sec_4du33du14du1NN}
%%%%%%%%%%%%%%%%%%%%%%%%%%%%%%%%%%%%%%%%%%%%%
To get more insight, let us consider the junction 4d $U(3)|$3d $U(1)|$4d $U(1)$ with $\mathcal{N}'$. 
A pair of corners of 4d $U(3)$ and $U(1)$ gauge theories with boundary conditions $(\mathcal{N}, \mathcal{N}')$ 
are coupled to 3d $U(1)$ gauge theory through 3d bi-fundamental hypermultiplets which obey Neumann b.c. $N'$. 
The junction also has two 2d cross-determinant Fermi multiplets which cancel the boundary gauge anomaly. 

The quarter-index is
\begin{align}
\label{311Na}
&
\mathbb{IV}^{\textrm{4d $U(3)|(1)|$4d $U(1)$}}_{\mathcal{N}\mathcal{N}'}
\nonumber\\
&=
\underbrace{
\frac{1}{3!}(q)_{\infty}^3 
\oint \prod_{i=1}^{3} \frac{ds_{i}}{2\pi is_{i}}
\prod_{i\neq j} \left( \frac{s_{i}}{s_{j}};q \right)_{\infty}
}_{\mathbb{IV}_{\mathcal{N}\mathcal{N}'}^{\textrm{4d $U(3)$}}}
\cdot 
\underbrace{
(q)_{\infty} (q^{\frac12}t^2;q)_{\infty}
\oint \frac{ds_{4}}{2\pi is_{4}}
}_{\mathbb{II}_{\mathcal{N}'}^{\textrm{3d $U(1)$}}}
\cdot 
\underbrace{
(q)_{\infty} \oint \frac{ds_{5}}{2\pi is_{5}}
}_{\mathbb{IV}_{\mathcal{N}\mathcal{N}'}^{\textrm{4d $U(1)$}}}
\nonumber\\
&\times 
\prod_{i=1}^{3}
\underbrace{
\frac{1}{(q^{\frac14}ts_{i}^{\pm}s_{4}^{\mp};q)_{\infty}}
}_{\mathbb{II}_{N'}^{\textrm{3d HM}} \left(\frac{s_{i}}{s_{4}} \right)}
\cdot 
\underbrace{
\frac{1}{(q^{\frac14}t s_{4}^{\pm}s_{5}^{\mp};q)_{\infty}}
}_{\mathbb{II}_{N'}^{\textrm{3d HM}} \left(\frac{s_{4}}{s_{5}} \right)}
\cdot 
\underbrace{
(q^{\frac12}s_{1}^{\pm}s_{2}^{\pm}s_{3}^{\pm}s_{4}^{\mp}z_{1}^{\pm};q)_{\infty}
}_{F\left( q^{\frac12}\frac{s_{1}s_{2}s_{3}}{s_{4}}z_{1} \right)}
\cdot 
\underbrace{
(q^{\frac12}s_{4}^{\pm}s_{5}^{\mp}z_{2}^{\pm};q)_{\infty}
}_{F\left( q^{\frac12}\frac{s_{4}}{s_{5}}z_{2} \right)}. 
\end{align}

Under the action of S-duality, 
we find the configuration in which 
the D5$'$-branes break 4d $U(3)\times U(1)\times U(1)$ gauge symmetry down to $U(1)$ 
and the D5-brane further breaks the $U(1)$ gauge symmetry. 
One of the D5$'$-branes has a single D3-brane in its both sides, 
which gives rise to 3d fundamental twisted hypermultiplet obeying Dirichlet b.c. $D$. 

We have checked that 
the quarter-index (\ref{311Na}) agrees with
\begin{align}
\label{311Nb}
\mathbb{IV}^{3|[1]|1}_{\mathcal{D}\mathcal{D}'}
&=
\underbrace{
\frac{(q)_{\infty}}{(q^{\frac12}t^2;q)_{\infty}}
}_{\mathbb{II}_{\mathcal{D}}^{\textrm{4d $U(1)$}}}
\cdot 
\underbrace{
(q^{\frac34}t z_{2}^{\pm};q)_{\infty}
}_{\mathbb{II}_{D}^{\textrm{3d tHM}} (z_{2})},
\end{align}
which is the product of a half-index of Dirichlet b.c. $\mathcal{D}$ for 4d $U(1)$ gauge theory 
and a Dirichlet half-index of 3d twisted hyper.

%%%%%%%%%%%%%%%%%%%%%%%%%%%%%%%%%%%%%%%%%%%%%
\subsubsection{4d $U(1)|$3d $U(2)|$4d $U(2)$ with $\mathcal{N}'$}
\label{sec_4du13du24du2NN}
%%%%%%%%%%%%%%%%%%%%%%%%%%%%%%%%%%%%%%%%%%%%%
Now let us consider the junction whose 3d gauge symmetry is non-Abelian. 
For the junction 4d $U(1)|$3d $U(2)|$4d $U(2)$ with $\mathcal{N}'$, 
we have corners of 4d $U(1)$ and $U(2)$ gauge theories and a boundary of 3d $U(2)$ gauge theory between them.  
Again the 4d and 3d gauge theories are connected by 3d bi-fundamental hypers together with 
2d cross-determinant Fermi multiplets. 

The quarter-index takes the form
\begin{align}
\label{122Na}
&
\mathbb{IV}^{\textrm{4d $U(1)|(2)|$4d $U(2)$}}_{\mathcal{N}\mathcal{N}'}
\nonumber\\
&=
\underbrace{
(q)_{\infty} \oint \frac{ds_{1}}{2\pi is_{1}}
}_{\mathbb{IV}_{\mathcal{N}\mathcal{N}'}^{\textrm{4d $U(1)$}}}
\cdot 
\underbrace{
\frac12 (q)_{\infty}^2 (q^{\frac12}t^2;q)_{\infty}^2 
\oint \prod_{i=1}^{2}
\frac{ds_{i}}{2\pi is_{i}}
\prod_{i\neq j}
\left( \frac{s_{i}}{s_{j}};q \right)_{\infty}
\left(q^{\frac12}t^2 \frac{s_{i}}{s_{j}};q \right)_{\infty}
}_{\mathbb{II}_{\mathcal{N}'}^{\textrm{3d $U(2)$}}}
\nonumber\\
&\times 
\underbrace{
\frac12 (q)_{\infty}^2 \oint \prod_{i=4}^{5}\frac{ds_{i}}{2\pi is_{i}}
\prod_{i\neq j}\left( \frac{s_{i}}{s_{j}};q \right)_{\infty}
}_{\mathbb{IV}_{\mathcal{N}\mathcal{N}'}^{\textrm{4d $U(2)$}}}
\nonumber\\
&\times 
\prod_{i=2}^{3}
\underbrace{
\frac{1}{(q^{\frac14}t s_{1}^{\pm}s_{i}^{\mp};q)_{\infty}}
}_{\mathbb{II}_{N'}^{\textrm{3d HM}} \left(\frac{s_{1}}{s_{i}} \right)}
\cdot 
\prod_{i=2}^{3}\prod_{j=4}^{5}
\underbrace{
\frac{1}{(q^{\frac14}t s_{i}^{\pm}s_{j}^{\mp};q)_{\infty}}
}_{\mathbb{II}_{N'}^{\textrm{3d HM}} \left(\frac{s_{i}}{s_{j}} \right)}
\cdot 
\underbrace{
(q^{\frac12} s_{1}^{\pm}s_{2}^{\mp} s_{3}^{\mp}z_{1}^{\pm};q)_{\infty}
}_{F\left(q^{\frac12}\frac{s_{1}}{s_{2}s_{3}}z_{1} \right)}
\cdot 
\underbrace{
(q^{\frac12}s_{2}^{\pm}s_{3}^{\pm}s_{4}^{\mp}s_{5}^{\mp}z_{2}^{\pm};q)_{\infty}
}_{F\left( q^{\frac12} \frac{s_{2}s_{3}}{s_{4}s_{5}} z_{2} \right)}. 
\end{align}

In the dual configuration, 
the 4d $U(1)\times U(2)\times U(2)$ gauge symmetry is broken down to $U(1)$ according to the D5$'$-branes. 
The 4d $U(1)$ gauge theory further should satisfy Dirichlet b.c. $\mathcal{D}$ due to the D5-brane. 
In addition, we would have 3d fundamental twisted hypermultiplet as 
one of the D5$'$-branes has two D3-branes in its both sides, which should obey 
Nahm pole b.c. with an embedding $\rho:$ $\mathfrak{su}(2)$ $\rightarrow$ $\mathfrak{u}(2)$. 

The quarter-index (\ref{122Na}) agrees with
\begin{align}
\label{122Nb}
\mathbb{IV}^{1|[2]|2}_{\mathcal{D}\mathcal{D}'}
&=
\underbrace{
\frac{(q)_{\infty}}{(q^{\frac12}t^2;q)_{\infty}}
}_{\mathbb{II}_{D}^{\textrm{4d $U(1)$}}}
\cdot 
(q t^2 z_{2}^{\pm};q)_{\infty}.
\end{align}
This involves the half-index of 4d $U(1)$ gauge theory subject to Dirichlet b.c. 
and the half-index of 3d twisted hyper obeying Nahm pole b.c. with an embedding $\rho:$ $\mathfrak{su}(2)$ $\rightarrow$ $\mathfrak{u}(2)$.

%%%%%%%%%%%%%%%%%%%%%%%%%%%%%%%%%%%%%%%%%%%%%
\subsubsection{4d $U(2)|$3d $U(2)|$4d $U(2)$ with $\mathcal{N}'$}
\label{sec_4du23du24du2NN}
%%%%%%%%%%%%%%%%%%%%%%%%%%%%%%%%%%%%%%%%%%%%%
Consider the configuration 4d $U(2)|$3d $U(2)|$4d $U(2)$ with $\mathcal{N}'$. 
We have a pair of corners of 4d $U(2)$ gauge theories with Neumann b.c. $(\mathcal{N}, \mathcal{N}')$ 
coupled to a boundary of 3d $U(2)$ gauge theory through 
3d bi-fundamental hypers and 2d cross-determinant Fermi multiplets.

The quarter-index takes the form
\begin{align}
\label{222Na}
&
\mathbb{IV}^{\textrm{4d $U(2)|(2)|$4d $U(2)$}}_{\mathcal{N}\mathcal{N}'}
\nonumber\\
&=
\underbrace{
\frac12 (q)_{\infty}^2 
\oint \frac{ds_{1}}{2\pi is_{1}} \frac{ds_{2}}{2\pi is_{2}}
\left(\frac{s_{1}}{s_{2}};q \right)_{\infty}
\left(\frac{s_{2}}{s_{1}};q \right)_{\infty}
}_{\mathbb{IV}_{\mathcal{N}\mathcal{N}'}^{\textrm{4d $U(2)$}}}
\nonumber\\
&\times 
\underbrace{
\frac12 (q)_{\infty}^2 (q^{\frac12} t^2;q)_{\infty}^2 
\oint \prod_{i=3}^{4} 
\frac{ds_{i}}{2\pi is_{i}}
\prod_{i\neq j}
\left(\frac{s_{i}}{s_{j}};q \right)_{\infty}
\left(q^{\frac12} t^2 \frac{s_{i}}{s_{j}};q \right)_{\infty}
}_{\mathbb{II}_{\mathcal{N}'}^{\textrm{3d $U(2)$}}}
\nonumber\\
&\times 
\underbrace{
\frac12 (q)_{\infty}^2 \oint \frac{ds_{5}}{2\pi is_{5}} \frac{ds_{6}}{2\pi is_{6}}
\left( \frac{s_{5}}{s_{6}};q \right)_{\infty}
\left( \frac{s_{6}}{s_{5}};q \right)_{\infty}
}_{\mathbb{IV}_{\mathcal{N}\mathcal{N}'}^{\textrm{4d $U(2)$}}}
\nonumber\\
&\times 
\prod_{i=1}^{2}\prod_{j=3}^{4}
\underbrace{
\frac{1}{(q^{\frac14}t s_{i}^{\pm} s_{j}^{\mp};q)_{\infty}}
}_{\mathbb{II}_{N'}^{\textrm{3d HM}} \left( \frac{s_{i}}{s_{j}} \right)}
\cdot 
\prod_{i=3}^{4}\prod_{j=5}^{6}
\underbrace{
\frac{1}{(q^{\frac14}t s_{i}^{\pm}s_{j}^{\mp};q)_{\infty}}
}_{\mathbb{II}_{N'}^{\textrm{3d HM}}\left(\frac{s_{i}}{s_{j}} \right)}
\cdot 
\underbrace{
(q^{\frac12}s_{1}^{\pm}s_{2}^{\pm}s_{3}^{\mp}s_{4}^{\mp}z_{1}^{\pm};q)_{\infty}
}_{F\left(q^{\frac12} \frac{s_{1}s_{2}}{s_{3}s_{4}} z_{1} \right)}
\cdot 
\underbrace{
(q^{\frac12}s_{3}^{\pm}s_{4}^{\pm}s_{5}^{\mp}s_{6}^{\mp}z_{2}^{\pm};q)_{\infty}
}_{F\left( q^{\frac12}\frac{s_{3}s_{4}}{s_{5}s_{6}} \right)}. 
\end{align}

The $U(2)\times U(2)\times U(2)$ gauge symmetry in the dual configuration is broken down to $U(2)$ due to the D5$'$-branes. 
The $U(2)$ gauge symmetry is further broken because of Nahm pole boundary condition of rank $2$. 
As the numbers of D3-branes do not change across the D5$'$-branes, 
the dual junction admits two 3d fundamental twisted hypermultiplets 
obeying Nahm pole b.c. specified by a homomorphism $\rho:$ $\mathfrak{su}(2)$ $\rightarrow$ $\mathfrak{u}(2)$. 

The quarter-index (\ref{222Na}) agrees with
\begin{align}
\label{222Nb}
\mathbb{IV}^{2|[2]|2}_{\mathcal{D}\mathcal{D}'}
&=
\underbrace{
\frac{
(q)_{\infty} (q^{\frac32}t^2;q)_{\infty}
}
{(q^{\frac12}t^2;q)_{\infty} (q t^4;q)_{\infty}}
}_{\mathbb{II}_{\textrm{Nahm}}^{\textrm{4d $U(2)$}}}
\cdot 
(q t^2 z_{1}^{\pm};q)_{\infty}
\cdot 
(q t^2 z_{2}^{\pm};q)_{\infty}. 
\end{align}
The first terms are the half-index of Nahm pole boundary condition for 4d $U(2)$ gauge theory 
and the other two terms are the two contributions from 3d twisted hypers obeying the Nahm pole b.c.

%%%%%%%%%%%%%%%%%%%%%%%%%%%%%%%%%%%%%%%%%%%%%
\subsubsection{4d $U(3)|$3d $U(3)|$4d $U(3)$ with $\mathcal{N}'$}
\label{sec_4du33du34du3NN}
%%%%%%%%%%%%%%%%%%%%%%%%%%%%%%%%%%%%%%%%%%%%%
As a further example of non-Abelian cases, 
let us examine the junction 4d $U(3)|$3d $U(3)|$4d $U(3)$ with $\mathcal{N}'$. 
The junction preserves 4d $U(3)$ and $U(3)$ gauge symmetries and 3d $U(3)$ gauge symmetry. 
The 4d $U(3)$ gauge theories are coupled to 3d $U(3)$ gauge theory through 
3d bi-fundamental hypers and 2d cross-determinant Fermi multiplets. 

The quarter-index takes the form
\begin{align}
\label{333Na}
&
\mathbb{IV}^{\textrm{4d $U(3)|(3)|$4d $U(3)$}}_{\mathcal{N}\mathcal{N}'}
\nonumber\\
&=
\underbrace{
\frac{1}{3!}(q)_{\infty}^3 
\oint \prod_{i=1}^{3}
\frac{ds_{i}}{2\pi is_{i}}
\prod_{i\neq j}
\left(\frac{s_{i}}{s_{j}};q \right)_{\infty}
}_{\mathbb{IV}_{\mathcal{N}\mathcal{N}'}^{\textrm{4d $U(3)$}}}
\nonumber\\
&\times 
\underbrace{
\frac{1}{3!}
(q)_{\infty}^3 (q^{\frac12}t^2;q)_{\infty}^3 
\oint \prod_{i=4}^{6}\frac{ds_{i}}{2\pi is_{i}}
\prod_{i\neq j}
\left(\frac{s_{i}}{s_{j}};q \right)_{\infty}
\left(q^{\frac12} t^2 \frac{s_{i}}{s_{j}};q \right)_{\infty}
}_{\mathbb{II}_{\mathcal{N}'}^{\textrm{3d $U(3)$}}}
\nonumber\\
&\times 
\underbrace{
\frac{1}{3!}(q)_{\infty}^3 
\oint \prod_{i=7}^{9}\frac{ds_{i}}{2\pi is_{i}}
\prod_{i\neq j}\left( \frac{s_{i}}{s_{j}};q \right)_{\infty}
}_{\mathbb{IV}_{\mathcal{N}\mathcal{N}'}^{\textrm{4d $U(3)$}}}
\nonumber\\
&\times 
\prod_{i=1}^{3}\prod_{j=4}^{6}
\underbrace{
\frac{1}{(q^{\frac14}ts_{i}^{\pm}s_{j}^{\mp};q)_{\infty}}
}_{\mathbb{II}_{N'}^{\textrm{3d HM}} \left( \frac{s_{i}}{s_{j}} \right)}
\cdot 
\prod_{i=4}^{6}\prod_{j=7}^{9}
\underbrace{
\frac{1}{(q^{\frac14}t s_{i}^{\pm} s_{j}^{\mp};q)_{\infty}}
}_{\mathbb{II}_{N'}^{\textrm{3d HM}} \left(\frac{s_{i}}{s_{j}} \right)}
\nonumber\\
&\times 
\underbrace{
(q^{\frac12}s_{1}^{\pm}s_{2}^{\pm}s_{3}^{\pm}s_{4}^{\mp}s_{5}^{\mp}s_{6}^{\mp}z_{1}^{\pm};q)_{\infty}
}_{F\left( q^{\frac12}\frac{s_{1}s_{2}s_{3}}{s_{4}s_{5}s_{6}}z_{1} \right)}
\cdot 
\underbrace{
(q^{\frac12}s_{4}^{\pm}s_{5}^{\pm}s_{6}^{\pm}s_{7}^{\mp}s_{8}^{\mp}s_{9}^{\mp}z_{2}^{\pm};q)_{\infty}
}_{F(q^{\frac12} \frac{s_{4}s_{5}s_{6}}{s_{7}s_{8}s_{9}}z_{2})}. 
\end{align}

According to the D5$'$-branes, the dual configuration keeps 4d $U(3)$ gauge theory in a half-space. 
It should satisfy Nahm pole boundary condition of rank $3$ required from D5-brane. 
In addition, there would be two 3d twisted hypermultiplets with Nahm pole b.c. 
with an embedding $\rho:$ $\mathfrak{su}(2)$ $\rightarrow$ $\mathfrak{u}(3)$. 

The quarter-index (\ref{333Na}) agrees with
\begin{align}
\label{333Nb}
\mathbb{IV}^{3|[3]|3}_{\mathcal{D}\mathcal{D}'}
&=
\underbrace{
\frac{(q)_{\infty}(q^{\frac32}t^2;q)_{\infty}(q^2 t^4;q)_{\infty}}
{(q^{\frac12}t^2;q)_{\infty}(qt^{4};q)_{\infty}(q^{\frac32}t^6;q)_{\infty}}
}_{\mathbb{II}_{\textrm{Nahm}}^{\textrm{4d $U(3)$}}}
\cdot 
(q^{\frac54}t^3 z_{1}^{\pm};q)_{\infty}
\cdot 
(q^{\frac54}t^3 z_{2}^{\pm};q)_{\infty}. 
\end{align}
Again the first terms are identified with the half-index of Nahm pole boundary condition for 4d $U(3)$ gauge theory 
and the other two are the two contributions from 3d twisted hypers subject to the Nahm pole b.c.

%%%%%%%%%%%%%%%%%%%%%%%%%%%%%%%%%%%%%%%%%%%%%
\subsubsection{4d $U(L)|$3d $U(M)|$4d $U(N)$ with $\mathcal{N}'$}
\label{sec_4duL3duM4duNNN}
%%%%%%%%%%%%%%%%%%%%%%%%%%%%%%%%%%%%%%%%%%%%%
Let us discuss the generalization of the dualities for 
the junction 4d $U(L)|$3d $U(M)|$4d $U(N)$ with $\mathcal{N}'$. 
The junction is composed of two corners of 4d $\mathcal{N}=4$ $U(L)$ and $U(N)$ gauge theories 
with a pair of Neumann  b.c. $(\mathcal{N}, \mathcal{N}')$ 
and a boundary of 3d $\mathcal{N}=4$ $U(M)$ gauge theory with Neumann b.c. $\mathcal{N}'$. 
There are two 3d bi-fundamental hypermultiplets with Neumann b.c. $N'$ 
and two 2d cross-determinant Fermi multiplets.

The quarter-index takes the form
\begin{align}
\label{lmnNa}
&
\mathbb{IV}^{\textrm{4d $U(L)|(M)|$4d $U(N)$}}_{\mathcal{N}\mathcal{N}'}
\nonumber\\
&=
\underbrace{
\frac{1}{L!}(q)_{\infty}^{L}
\oint \prod_{i=1}^{L}\frac{ds_{i}}{2\pi is_{i}}
\prod_{i\neq j}
\left( \frac{s_{i}}{s_{j}};q \right)_{\infty}
}_{\mathbb{IV}_{\mathcal{N}\mathcal{N}'}^{\textrm{4d $U(L)$}}}
\nonumber\\
&\times 
\underbrace{
\frac{1}{M!}(q)_{\infty}^{M}(q^{\frac12}t^2;q)_{\infty}^{M}
\oint \prod_{i=L+1}^{L+M}
\frac{ds_{i}}{2\pi is_{i}}\prod_{i\neq j}
\left( \frac{s_{i}}{s_{j}};q \right)_{\infty}
\left(q^{\frac12} t^2 \frac{s_{i}}{s_{j}};q \right)_{\infty}
}_{\mathbb{II}_{\mathcal{N}'}^{\textrm{3d $U(M)$}}}
\nonumber\\
&\times 
\underbrace{
\frac{1}{N!}(q)_{\infty}^{N}\oint \prod_{i=L+M+1}^{L+M+N}
\frac{ds_{i}}{2\pi is_{i}}\prod_{i\neq j}
\left(\frac{s_{i}}{s_{j}};q \right)_{\infty}
}_{\mathbb{IV}_{\mathcal{N}\mathcal{N}'}^{\textrm{4d $U(N)$}}}
\nonumber\\
&\times 
\prod_{i=1}^{L}\prod_{j=L+1}^{L+M}
\underbrace{
\frac{1}{(q^{\frac14}ts_{i}^{\pm}s_{j}^{\mp};q)_{\infty}}
}_{\mathbb{II}_{N'}^{\textrm{3d HM}}\left(\frac{s_{i}}{s_{j}} \right)}
\cdot 
\prod_{i=L+1}^{L+M}\prod_{j=L+M+1}^{L+M+N}
\underbrace{
\frac{1}{(q^{\frac14}ts_{i}^{\pm}s_{j}^{\mp};q)_{\infty}}
}_{\mathbb{II}_{N'}^{\textrm{3d HM}}\left(\frac{s_{i}}{s_{j}} \right)}
\nonumber\\
&\times 
(q^{\frac12}\prod_{i=1}^{L}s_{i}^{\pm}\prod_{j=L+1}^{L+M}s_{j}^{\mp}z_{1}^{\pm};q)_{\infty}
\cdot 
(q^{\frac12}\prod_{i=L+1}^{L+M}s_{i}^{\pm}\prod_{j=L+M+1}^{L+M+N}s_{j}^{\mp}z_{2}^{\pm};q)_{\infty}. 
\end{align}

In the dual junction 
4d $U(L)\times U(M)\times U(N)$ gauge symmetry is broken down 
to $U(\min\{L,M,N\})$ due to the D5$'$-branes 
so that there is a 4d $U(\min \{L,M,N \})$ gauge theory in a half-space. 
According to the D5-brane, it should obey Dirichlet/Nahm pole b.c. 
specified by an  embedding $\rho:$ $\mathfrak{su}(2)$ $\rightarrow$ $\mathfrak{u}(\min \{L,M,N\})$. 
Besides, the junction may include 3d twisted hypermultiplets. 
For $L=M$, there would be 3d twisted hypermultiplet obeying the Dirichlet/Nahm pole b.c. 
with an embedding $\rho:$ $\mathfrak{su}(2)$ $\rightarrow$ $\mathfrak{u}(L)$. 
Analogously, for $N=M$, we would have 3d twisted hypermultiplet satisfying the Dirichlet/Nahm pole b.c. 
with an embedding $\rho:$ $\mathfrak{su}(2)$ $\rightarrow$ $\mathfrak{u}(N)$. 

Then the quarter-indices for the dual junction is written as 
\begin{align}
\label{lmnNb}
\mathbb{IV}^{L|[M]|N}_{\mathcal{D}\mathcal{D}'}
&=
\underbrace{
\prod_{k=1}^{\min\{L,M,N\}} 
\frac{(q^{\frac{k+1}{2}} t^{2(k-1)} ;q)_{\infty}}
{(q^{\frac{k}{2}} t^{2k};q)_{\infty}}
}_{\mathbb{II}_{\textrm{Nahm}}^{\textrm{4d $U(\min\{L,M,N\})$}} }
\nonumber\\
&\times 
(q^{\frac34+\frac{L-1}{4}} t^{1+(L-1)}z_{1}^{\pm};q )_{\infty}^{{\delta^{L}}_{M} }
\cdot 
(q^{\frac34+\frac{M-1}{4}} t^{1+(M-1)}z_{2}^{\pm};q )_{\infty}^{{\delta^{M}}_{N} }.
\end{align}
We expect that 
the quarter-indices (\ref{lmnNa}) and (\ref{lmnNb}) agree with each other.

%%%%%%%%%%%%%%%%%%%%%%%%%%%%%%%%%%%%%%%%%%%%%
%%%%%%%%%%%%%%%%%%%%%%%%%%%%%%%%%%%%%%%%%%%%%
\subsection{4d $U(L)|$3d $(M_{1})-(M_{2})-\cdots-(M_{k-1})|$4d $U(N)$ with $\mathcal{N}'$}
\label{sec_4dL3dMx4dN}
\begin{figure}
\begin{center}
\includegraphics[width=14cm]{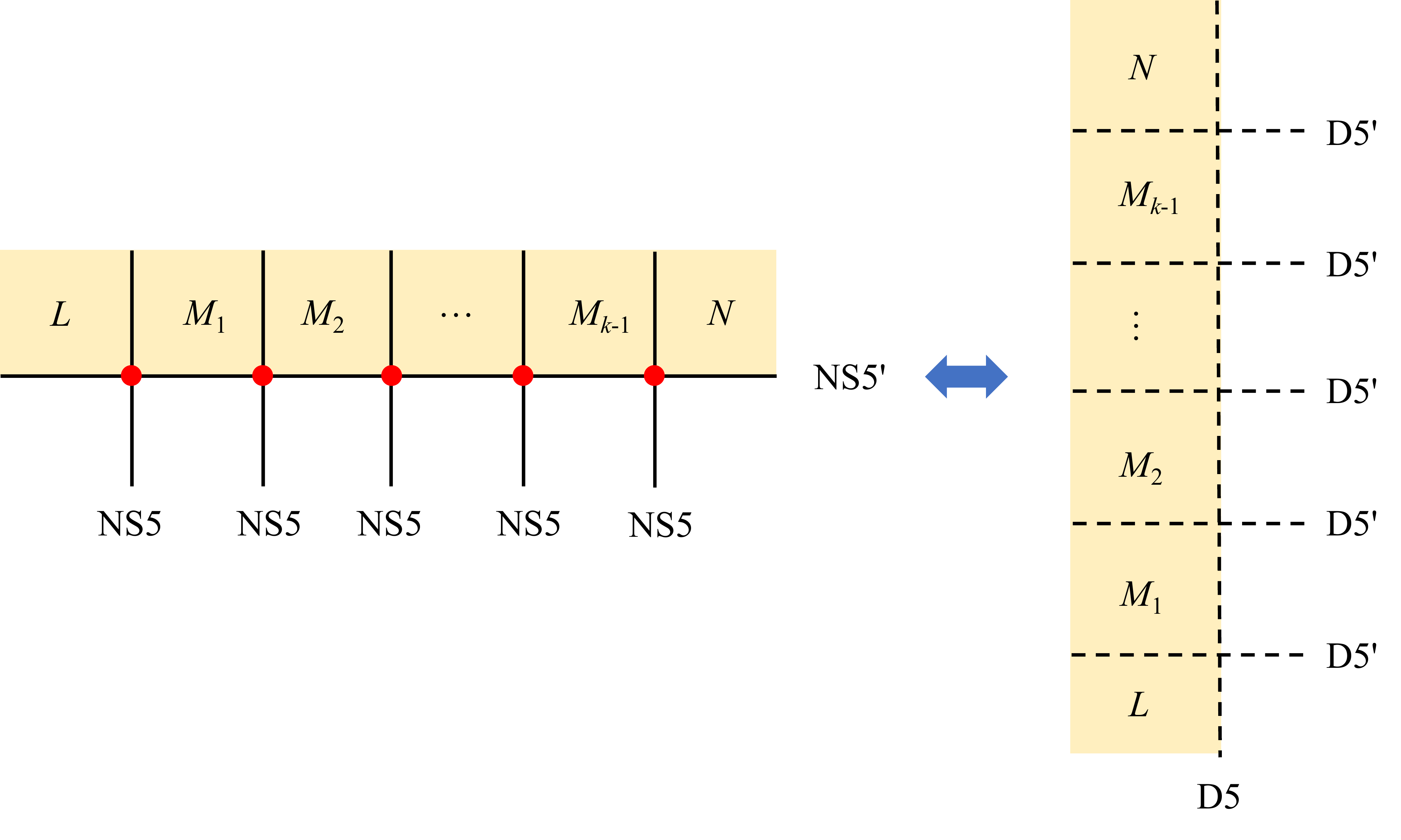}
\caption{
The brane constructions of the configuration 4d $U(L)|$3d $(M_{1})-(M_{2})-\cdots-(M_{k-1})|$4d $U(N)$ with Neumann b.c. $\mathcal{N}'$ 
and its mirror configuration. 
}
\label{fig4dul3dumx4dun}
\end{center}
\end{figure}

%%%%%%%%%%%%%%%%%%%%%%%%%%%%%%%%%%%%%%%%%%%%%
%%%%%%%%%%%%%%%%%%%%%%%%%%%%%%%%%%%%%%%%%%%%%
Let us generalize the junction in section \ref{sec_4dL3dM4dN} by including 3d linear quiver gauge theory 
$(M_{1})-\cdots-(M_{k-1})$ with a gauge group $U(M_{1})\times \cdots \times U(M_{k-1})$ 
and $(k-2)$ bi-fundamental hypermultiplets. 
The brane construction is shown in Figure \ref{fig4dul3dumx4dun}. 
The junction has 4d $U(L)$ and $U(N)$ gauge symmetry 
ans 3d $U(M_{1})\times U(M_{k-1})$ gauge symmetry. 
In addition to the $(k-2)$ bi-fundamental hypers in the 3d quiver gauge theory, 
there are two bi-fundamental hypers; one couples to 4d $U(L)$ and 3d $U(M_{1})$, 
the other couples to 4d $U(N)$ and 3d $U(M_{k-1})$. 
Furthermore, there are $k$ cross-determinant Fermi multiplets 
corresponding to the $k$ NS5-NS5$'$-junction. 

Contrast to the original junction, 
the dual junction is rather simple 
in that it can be described in terms of free fields as it has no gauge symmetry. 
In particular, it has the 4d gauge theory with Dirichlet/Nahm pole boundary conditions. 
We confirm the dualities of these junctions by showing the matching of indices.

%%%%%%%%%%%%%%%%%%%%%%%%%%%%%%%%%%%%%%%%%%%%%
\subsubsection{4d $U(1)|$3d $(1)-(1)|$4d $U(1)$ with $\mathcal{N}'$}
\label{sec_4du13d114du1NN}
%%%%%%%%%%%%%%%%%%%%%%%%%%%%%%%%%%%%%%%%%%%%%
For the junction 4d $U(1)|$3d $(1)-(1)|$4d $U(1)$ with $\mathcal{N}'$, 
we have a pair of corners of 4d $U(1)$ gauge theories obeying the boundary conditions $(\mathcal{N}, \mathcal{N}')$ 
and a boundary of 3d $U(1)\times U(1)$ gauge theory with Neumann b.c. $\mathcal{N}'$. 
In total, we have three 3d bi-fundamental hypermultiplets and three 2d cross-determinant Fermi multiplets. 

The quarter-index takes the form
\begin{align}
\label{1111Na}
&
\mathbb{IV}^{\textrm{4d $U(1)|(1)-(1)|$4d $U(1)$}}_{\mathcal{N}\mathcal{N}'}
\nonumber\\
&=
\underbrace{
(q)_{\infty}\oint \frac{ds_{1}}{2\pi is_{1}}
}_{\mathbb{IV}_{\mathcal{N}\mathcal{N}'}^{\textrm{4d $U(1)$}}}
\cdot 
\underbrace{
(q)_{\infty}(q^{\frac12}t^2;q)_{\infty}\oint \frac{ds_{2}}{2\pi is_{2}}
}_{\mathbb{II}_{\mathcal{N}'}^{\textrm{3d $U(1)$}}}
\cdot 
\underbrace{
(q)_{\infty}(q^{\frac12}t^2;q)_{\infty}\oint \frac{ds_{3}}{2\pi is_{3}}
}_{\mathbb{II}_{\mathcal{N}'}^{\textrm{3d $U(1)$}}}
\cdot 
\underbrace{
(q)_{\infty}\oint \frac{ds_{4}}{2\pi is_{4}}
}_{\mathbb{IV}_{\mathcal{N}\mathcal{N}'}^{\textrm{4d $U(1)$}}}
\nonumber\\
&\times 
\underbrace{
\frac{1}{(q^{\frac14}t s_{1}^{\pm}s_{2}^{\mp};q)_{\infty}}
}_{\mathbb{II}_{N'}^{\textrm{3d HM}} \left( \frac{s_{1}}{s_{2}} \right)}
\cdot 
\underbrace{
\frac{1}{(q^{\frac14}t s_{2}^{\pm}s_{3}^{\mp};q)_{\infty}}
}_{\mathbb{II}_{N'}^{\textrm{3d HM}} \left(\frac{s_{2}}{s_{3}} \right)}
\cdot 
\underbrace{
\frac{1}{(q^{\frac14}t s_{3}^{\pm}s_{4}^{\mp};q)_{\infty}}
}_{\mathbb{II}_{N'}^{\textrm{3d HM}} \left( \frac{s_{3}}{s_{4}} \right)}
\nonumber\\
&\times 
\underbrace{
(q^{\frac12}s_{1}^{\pm}s_{2}^{\mp}z_{1}^{\pm};q)_{\infty}
}_{F\left( q^{\frac12} \frac{s_{1}}{s_{2}}z_{1} \right)}
\cdot 
\underbrace{
(q^{\frac12} s_{2}^{\pm} s_{3}^{\mp}z_{2}^{\pm};q)_{\infty}
}_{F\left( q^{\frac12}\frac{s_{2}}{s_{3}}z_{2} \right)}
\cdot 
\underbrace{
(q^{\frac12}s_{3}^{\pm}s_{4}^{\mp}z_{3}^{\pm};q)_{\infty}
}_{F\left(\frac{s_{3}}{s_{4}}z_{3} \right)}. 
\end{align}

The dual junction has three defects of D5$'$-branes, 
which break down $U(1)\times U(1)\times U(1)\times U(1)$ gauge symmetry to $U(1)$. 
The 4d $U(1)$ gauge theory in a half-space should be subject to Dirichlet b.c. $\mathcal{D}$. 
The equal numbers of D3-branes across three D5$'$-branes yield three 3d twisted hypermultiplets, 
which should satisfy Dirichlet b.c. $D$. 

We then obtain the quarter-index for the dual junction
\begin{align}
\label{1111Nb}
\mathbb{IV}^{1|[1]-[1]|1}_{\mathcal{D}\mathcal{D}'}
&=
\underbrace{
\frac{(q)_{\infty}}{(q^{\frac12}t^2;q)_{\infty}}
}_{\mathbb{II}_{\mathcal{D}}^{\textrm{4d $U(1)$}}}
\cdot 
\prod_{\alpha=1}^{3}
\underbrace{
(q^{\frac34} t z_{\alpha}^{\pm};q)_{\infty}
}_{\mathbb{II}_{D}^{\textrm{3d tHM}} (z_{\alpha})}. 
\end{align}
We have checked that the quarter-indices (\ref{1111Na}) and (\ref{1111Nb}) coincide to each other.

%%%%%%%%%%%%%%%%%%%%%%%%%%%%%%%%%%%%%%%%%%%%%
\subsubsection{4d $U(2)|$3d $(1)-(1)|$4d $U(1)$ with $\mathcal{N}'$}
\label{sec_4du23d114du1NN}
%%%%%%%%%%%%%%%%%%%%%%%%%%%%%%%%%%%%%%%%%%%%%
Next consider the junction 4d $U(2)|$3d $(1)-(1)|$4d $U(1)$ with $\mathcal{N}'$. 
This is obtained from the previous example by replacing one of the 4d $U(1)$ gauge theories with 4d $U(2)$ gauge theory. 

In a similar manner, the quarter-index takes the form
\begin{align}
\label{2111Na}
&
\mathbb{IV}^{\textrm{4d $U(2)|(1)-(1)|$4d $U(1)$}}_{\mathcal{N}\mathcal{N}'}
\nonumber\\
&=
\underbrace{
\frac12 (q)_{\infty}^2 \oint \prod_{i=1}^{2}\frac{ds_{i}}{2\pi is_{i}}
\prod_{i\neq j}\left(\frac{s_{i}}{s_{j}};q \right)_{\infty}
}_{\mathbb{IV}_{\mathcal{N}\mathcal{N}'}^{\textrm{4d $U(2)$}}}
\cdot 
\underbrace{
(q)_{\infty}(q^{\frac12}t^2;q)_{\infty}
\oint \frac{ds_{3}}{2\pi is_{3}}
}_{\mathbb{II}_{\mathcal{N}'}^{\textrm{3d $U(1)$}}}
\nonumber\\
&\times 
\underbrace{
(q)_{\infty}(q^{\frac12}t^2;q)_{\infty}\oint \frac{ds_{4}}{2\pi is_{4}}
}_{\mathbb{II}_{\mathcal{N}'}^{\textrm{3d $U(1)$}}}
\cdot 
\underbrace{
(q)_{\infty}\oint \frac{ds_{5}}{2\pi is_{5}}
}_{\mathbb{IV}_{\mathcal{N}\mathcal{N}'}^{\textrm{4d $U(1)$}}}
\nonumber\\
&\times 
\prod_{i=1}^{2}
\underbrace{
\frac{1}{(q^{\frac14}t s_{i}^{\pm}s_{3}^{\mp};q)_{\infty}}
}_{\mathbb{II}_{N'}^{\textrm{3d HM}} \left(\frac{s_{i}}{s_{3}} \right)}
\cdot 
\underbrace{
\frac{1}{(q^{\frac14}t s_{3}^{\pm}s_{4}^{\mp};q)_{\infty}}
}_{\mathbb{II}_{N'}^{\textrm{3d HM}}\left( \frac{s_{3}}{s_{4}} \right) }
\cdot 
\underbrace{
\frac{1}{(q^{\frac14}t s_{4}^{\pm}s_{5}^{\mp};q)_{\infty}}
}_{\mathbb{II}_{N'}^{\textrm{3d HM}} \left(\frac{s_{4}}{s_{5}} \right)}
\nonumber\\
&\times 
\underbrace{
(q^{\frac12}s_{1}^{\pm}s_{2}^{\pm}s_{3}^{\mp}z_{1}^{\pm};q)_{\infty}
}_{F\left(\frac{s_{1}s_{2}}{s_{3}}z_{1} \right)}
\cdot 
\underbrace{
(q^{\frac12}s_{3}^{\pm}s_{4}^{\mp}z_{2}^{\pm};q)_{\infty}
}_{F\left( \frac{s_{3}}{s_{4}}z_{2} \right)}
\cdot 
\underbrace{
(q^{\frac12}s_{4}^{\pm}s_{5}^{\mp}z_{3}^{\pm};q)_{\infty}
}_{F\left( \frac{s_{4}}{s_{5}} z_{3} \right)}.
\end{align}

In the dual configuration 
$U(1)\times U(1)\times U(1)\times U(1)$ gauge symmetry is broken down to $U(1)$ by D5$'$-branes, 
which gives rise to 4d $U(1)$ gauge theory with Dirichlet b.c. $\mathcal{D}$. 
The equal numbers of D3-branes across two D5$'$-branes yield two 3d twisted hypermultiplets, 
which should satisfy Dirichlet b.c. $D$. 

In fact, we find that the quarter-index (\ref{2111Na}) agrees with
\begin{align}
\label{2111Nb}
\mathbb{IV}^{2|[1]-[1]|1}_{\mathcal{D}\mathcal{D}'}
&=
\underbrace{
\frac{(q)_{\infty}}{(q^{\frac12}t^2;q)_{\infty}}
}_{\mathbb{II}_{\mathcal{D}}^{\textrm{4d $U(1)$}}}
\cdot 
\underbrace{
(q^{\frac34}t z_{2}^{\pm};q)_{\infty}
}_{\mathbb{II}_{D}^{\textrm{3d tHM}} (z_{2})}
\cdot 
\underbrace{
(q^{\frac34}tz_{3}^{\pm};q)_{\infty}
}_{\mathbb{II}_{D}^{\textrm{3d tHM}} (z_{3})}.
\end{align}

%%%%%%%%%%%%%%%%%%%%%%%%%%%%%%%%%%%%%%%%%%%%%
\subsubsection{4d $U(2)|$3d $(2)-(1)|$4d $U(1)$ with $\mathcal{N}'$}
\label{sec_4du23d214du1NN}
%%%%%%%%%%%%%%%%%%%%%%%%%%%%%%%%%%%%%%%%%%%%%
Consider the junction 4d $U(2)|$3d $(2)-(1)|$4d $U(1)$ with $\mathcal{N}'$. 
In this case the junction has a pair of $U(2)$ gauge symmetries and a pair of $U(1)$ gauge symmetries. 

The quarter-index takes the form
\begin{align}
\label{2211Na}
&
\mathbb{IV}^{\textrm{4d $U(2)|(2)-(1)|$4d $U(1)$}}_{\mathcal{N}\mathcal{N}'}
\nonumber\\
&=
\underbrace{
\frac12 (q)_{\infty}^2 \oint \prod_{i=1}^{2}
\frac{ds_{i}}{2\pi is_{i}}
\prod_{i\neq j}\left(\frac{s_{i}}{s_{j}};q \right)_{\infty}
}_{\mathbb{IV}_{\mathcal{N}\mathcal{N}'}^{\textrm{4d $U(2)$}}}
\cdot 
\underbrace{
\frac12 (q)_{\infty}^2 (q^{\frac12}t^2;q)_{\infty}^2
\oint \prod_{i=3}^{4}
\frac{ds_{i}}{2\pi is_{i}}
\prod_{i\neq j}
\left(\frac{s_{i}}{s_{j}};q \right)_{\infty}
\left(q^{\frac12}t^2\frac{s_{i}}{s_{j}};q \right)_{\infty}
}_{\mathbb{II}_{\mathcal{N}'}^{\textrm{3d $U(2)$}}}
\nonumber\\
&\times 
\underbrace{
(q)_{\infty}(q^{\frac12}t^2;q)_{\infty}\oint \frac{ds_{5}}{2\pi is_{5}}
}_{\mathbb{II}_{\mathcal{N}'}^{\textrm{3d $U(1)$}}}
\cdot 
\underbrace{
(q)_{\infty}\oint \frac{ds_{6}}{2\pi is_{6}}
}_{\mathbb{IV}_{\mathcal{N}\mathcal{N}'}^{\textrm{4d $U(1)$}}}
\nonumber\\
&\times 
\prod_{i=1}^{2}\prod_{j=3}^{4}
\underbrace{
\frac{1}{(q^{\frac14}ts_{i}^{\pm}s_{j}^{\mp};q)_{\infty}}
}_{\mathbb{II}_{N'}^{\textrm{3d HM}} \left( \frac{s_{i}}{s_{j}} \right)}
\cdot 
\prod_{i=3}^{4}
\underbrace{
\frac{1}{(q^{\frac14}t s_{i}^{\pm} s_{5}^{\mp};q)_{\infty}}
}_{\mathbb{II}_{N'}^{\textrm{3d HM}} \left(\frac{s_{i}}{s_{5}} \right)}
\cdot 
\underbrace{
\frac{1}{(q^{\frac14}t s_{5}^{\pm}s_{6}^{\mp};q)_{\infty}}
}_{\mathbb{II}_{N'}^{\textrm{3d HM}}\left(\frac{s_{5}}{s_{6}} \right)}
\nonumber\\
&\times 
\underbrace{
(q^{\frac12}s_{1}^{\pm}s_{2}^{\pm} s_{3}^{\mp}s_{4}^{\mp}z_{1}^{\pm};q)_{\infty}
}_{F\left(q^{\frac12} \frac{s_{1}s_{2}}{s_{3}s_{4}} \right)}
\cdot 
\underbrace{
(q^{\frac12}s_{3}^{\pm}s_{4}^{\pm} s_{5}^{\mp}z_{2}^{\pm};q)_{\infty}
}_{F\left(q^{\frac12} \frac{s_{3}s_{4}}{s_{5}}z_{2} \right)}
\cdot 
\underbrace{
(q^{\frac12}s_{5}^{\pm}s_{6}^{\mp}z_{3}^{\pm};q)_{\infty}
}_{F\left(q^{\frac12} \frac{s_{5}}{s_{6}} z_{3} \right)}.
\end{align}

From S-duality, 
one find the configuration 
in which the three D5$'$-branes break down $U(2)\times U(2)\times U(1)\times U(1)$ to $U(1)$ 
and the D5-brane imposes Dirichlet b.c. $\mathcal{D}$ for 4d $U(1)$ gauge theory. 
As there are two D5$'$-branes with equal numbers of D3-branes in their each side, 
there would be two 3d twisted hypermultiplets. 
Corresponding to the numbers of D3-branes, 
one has Nahm pole of rank $2$ and the other receives Dirichlet b.c. $D$.

We get the quarter-index 
\begin{align}
\label{2211Nb}
\mathbb{IV}^{2|[2]-[1]|1}_{\mathcal{D}\mathcal{D}'}
&=
\underbrace{
\frac{(q)_{\infty}}{(q^{\frac12}t^2;q)_{\infty}}
}_{\mathbb{II}_{\mathcal{D}}^{\textrm{4d $U(1)$}}}
\cdot 
(qt^2 z_{1}^{\pm};q)_{\infty}
\cdot 
\underbrace{
(q^{\frac34}t z_{3}^{\pm};q)_{\infty}
}_{\mathbb{II}_{D}^{\textrm{3d HM}} (z_{3})}.
\end{align}
In fact, the quarter-indiceds (\ref{2211Na}) and (\ref{2211Nb}) agree with each other.

%%%%%%%%%%%%%%%%%%%%%%%%%%%%%%%%%%%%%%%%%%%%%
\subsubsection{4d $U(2)|$3d $(2)-(2)|$4d $U(2)$ with $\mathcal{N}'$}
\label{sec_4du23d224du2NN}
%%%%%%%%%%%%%%%%%%%%%%%%%%%%%%%%%%%%%%%%%%%%%
Let us consider the junction in which both 3d and 4d gauge symmetries are non-Abelian. 
For the junction 4d $U(2)|$3d $(2)-(2)|$4d $U(2)$ with $\mathcal{N}'$ 
we have four factors of $U(2)$ gauge symmetry, 
three 3d bi-fundamental hypermultiplets 
and three 2d cross-determinant Fermi multiplets. 

The quarter-index is
\begin{align}
\label{2222Na}
&
\mathbb{IV}^{\textrm{4d $U(2)|(2)-(2)|$4d $U(2)$}}_{\mathcal{N}\mathcal{N}'}
\nonumber\\
&=
\underbrace{
\frac12 (q)_{\infty}^2 \oint \prod_{i=1}^{2}
\frac{ds_{i}}{2\pi is_{i}} \prod_{i\neq j}\left(\frac{s_{i}}{s_{j}};q \right)_{\infty}
}_{\mathbb{IV}_{\mathcal{N}\mathcal{N}'}^{\textrm{4d $U(2)$}}}
\cdot 
%\nonumber\\
%&\times 
\underbrace{
\frac12 (q)_{\infty}^2 (q^{\frac12}t^2;q)_{\infty}^2 
\oint \prod_{i=3}^{4}\frac{ds_{i}}{2\pi is_{i}} 
\prod_{i\neq j}
\left( \frac{s_{i}}{s_{j}};q \right)_{\infty}
\left( q^{\frac12} t^2 \frac{s_{i}}{s_{j}};q \right)_{\infty}
}_{\mathbb{II}_{\mathcal{N}'}^{\textrm{3d $U(2)$}}}
\nonumber\\
&\times 
\underbrace{
\frac12 (q)_{\infty}^2 (q^{\frac12} t^2;q)_{\infty}^2 
\oint \prod_{i=5}^{6} \frac{ds_{i}}{2\pi is_{i}}
\prod_{i\neq j}
\left( \frac{s_{i}}{s_{j}};q \right)_{\infty}
\left( q^{\frac12} t^2 \frac{s_{i}}{s_{j}};q \right)_{\infty}
}_{\mathbb{II}_{\mathcal{N}'}^{\textrm{3d $U(2)$}}}
\cdot 
\underbrace{
\frac12 (q)_{\infty}^2 
\oint \prod_{i=7}^{8}
\frac{ds_{i}}{2\pi is_{i}}
\prod_{i\neq j}
\left( \frac{s_{i}}{s_{j}};q \right)_{\infty}
}_{\mathbb{IV}_{\mathcal{N}\mathcal{N}'}^{\textrm{4d $U(2)$}}}
\nonumber\\
&\times 
\prod_{i=1}^{2}\prod_{j=3}^{4}
\underbrace{
\frac{1}{(q^{\frac14} t s_{i}^{\pm} s_{j}^{\mp};q)_{\infty}}
}_{\mathbb{II}_{N'}^{\textrm{3d HM}} \left(\frac{s_{i}}{s_{j}} \right)}
\cdot 
\prod_{i=3}^{4}\prod_{j=5}^{6}
\underbrace{
\frac{1}{(q^{\frac14}ts_{i}^{\pm}s_{j}^{\mp};q)_{\infty}}
}_{\mathbb{II}_{N'}^{\textrm{3d HM}} \left(\frac{s_{i}}{s_{j}} \right)}
\cdot 
\prod_{i=5}^{6}\prod_{j=7}^{8}
\underbrace{
\frac{1}{(q^{\frac14}ts_{i}^{\pm}s_{j}^{\mp};q)_{\infty}}
}_{\mathbb{II}_{N'}^{\textrm{3d HM}} \left(\frac{s_{i}}{s_{j}} \right)}
\nonumber\\
&\times 
\underbrace{
(q^{\frac12}s_{1}^{\pm}s_{2}^{\pm}s_{3}^{\mp}s_{4}^{\mp}z_{1}^{\pm};q)_{\infty}
}_{F\left(q^{\frac12} \frac{s_{1}s_{2}}{s_{3}s_{4}} z_{1} \right)}
\cdot 
\underbrace{
(q^{\frac12}s_{3}^{\pm}s_{4}^{\pm}s_{5}^{\mp}s_{6}^{\mp}z_{2}^{\pm};q)_{\infty}
}_{F\left(q^{\frac12} \frac{s_{3}s_{4}}{s_{5}s_{6}} z_{2} \right)}
\cdot 
\underbrace{
(q^{\frac12}s_{5}^{\pm}s_{6}^{\pm}s_{7}^{\mp}s_{8}^{\mp}z_{3}^{\pm};q)_{\infty}
}_{F\left( q^{\frac12} \frac{s_{5}s_{6}}{s_{7}s_{8}} z_{3} \right)}.
\end{align}

In the dual configuration 
there remains 4d $U(2)$ gauge theory in a half-space 
with Nahm pole boundary condition 
with a homomorphism $\rho:$ $\mathfrak{su}(2)$ $\rightarrow$ $\mathfrak{u}(2)$. 
As the numbers of D3-branes are constant across the three D5$'$-branes, 
we find three 3d twisted hypermultiplets. 
Due to the D5-brane, they should satisfy 
Nahm pole boundary condition characterized by the same homomorphism. 

Then we have the quarter-index for the dual junction 
\begin{align}
\label{2222Nb}
\mathbb{IV}^{2|[2]-[2]|2}_{\mathcal{D}\mathcal{D}'}
&=
\underbrace{
\frac{(q)_{\infty} (q^{\frac32}t^2;q)_{\infty}}
{(q^{\frac12}t^2;q)_{\infty} (qt^{4};q)_{\infty}}
}_{\mathbb{II}_{\textrm{Nahm}}^{\textrm{4d $U(2)$}}}
\cdot 
\prod_{\alpha=1}^{3}
(q t^2 z_{\alpha}^{\pm}:q)_{\infty}.
\end{align}
We have confirmed that 
the quarter-indices (\ref{2222Na}) and (\ref{2222Nb}) coincides with each other.

%%%%%%%%%%%%%%%%%%%%%%%%%%%%%%%%%%%%%%%%%%%%%
\subsubsection{4d $U(L)|$3d $(M_{1})-(M_{2})-\cdots-(M_{k-1})|$4d $U(N)$ with $\mathcal{N}'$}
\label{sec_4dul3dumx4dunNN}
%%%%%%%%%%%%%%%%%%%%%%%%%%%%%%%%%%%%%%%%%%%%%
Now we would like to propose the generalization. 
For the junction 
4d $U(L)|$3d $(M_{1})-(M_{2})-\cdots-(M_{k-1})|$4d $U(N)$ with $\mathcal{N}'$, 
we have two corners of 4d $\mathcal{N}=4$ $U(L)$ and $U(N)$ gauge theories 
with a pair of Neumann b.c. $(\mathcal{N}, \mathcal{N}')$. 
Between these 4d gauge theories 
we have 3d $\mathcal{N}=4$ linear quiver gauge theory 
$(M_{1})-(M_{2})-\cdots-(M_{k-1})$ 
which has a gauge group $U(M_{1})\times \cdots U(M_{k-1})$ 
and $(k-2)$ bi-fundamental hypermultiplets. 
The 4d and 3d gauge theories are coupled by additional two 3d bi-fundamental hypermultiplets. 
In total, there are $k$ 3d bi-fundamental hypermultiplets. 
Besides, there are $k$ cross-determinant Fermi multiplets.

The quarter-index can be computed as 
\begin{align}
\label{LMxNNa}
&
\mathbb{IV}^{\textrm{4d $U(L)|(M_{1})-(M_{2})-\cdots-(M_{k-1})|$4d $U(N)$}}_{\mathcal{N}\mathcal{N}'}
\nonumber\\
&=
\underbrace{
\frac{1}{L!}(q)_{\infty}^{L}
\oint \prod_{i=1}^{L}\frac{ds_{i}}{2\pi is_{i}}
\prod_{i\neq j}\left( \frac{s_{i}}{s_{j}};q \right)_{\infty}
}_{\mathbb{IV}_{\mathcal{N}\mathcal{N}'}^{\textrm{4d $U(L)$}}}
\nonumber\\
&\times 
\prod_{l=1}^{k-1}
\underbrace{
\frac{1}{M_{l}!}(q)_{\infty}^{M_{l}}(q^{\frac12}t^2;q)_{\infty}^{M_{l}}
\oint \prod_{i=1}^{M_{l}} \frac{ds_{i}^{(l)}}{2\pi is_{i}^{(l)}}
\prod_{i\neq j}
\left( \frac{s_{i}^{(l)}}{s_{j}^{(l)}};q \right)_{\infty}
\left( q^{\frac12} t^2 \frac{s_{i}^{(l)}}{s_{j}^{(l)}};q \right)_{\infty}
}_{\mathbb{II}_{\mathcal{N}'}^{\textrm{3d $U(M_{l})$}}}
\nonumber\\
&\times 
\underbrace{
\frac{1}{N!}(q)_{\infty}^{N}
\oint \prod_{i=L+1}^{L+N}\frac{ds_{i}}{2\pi is_{i}}
\prod_{i\neq j}\left( \frac{s_{i}}{s_{j}};q \right)_{\infty}
}_{\mathbb{IV}_{\mathcal{N}\mathcal{N}'}^{\textrm{4d $U(N)$}}}
\nonumber\\
&\times 
\prod_{i=1}^{L}\prod_{j=1}^{M_{1}}
\underbrace{
\frac{1}{(q^{\frac14}t s_{i}^{\pm}s_{j}^{(1)\mp};q)_{\infty}}
}_{\mathbb{II}_{N'}^{\textrm{3d HM}} \left(\frac{s_{i}}{s_{j}^{(1)}} \right)}
\cdot 
\prod_{l=1}^{k-2}
\prod_{i=1}^{M_{l}}
\prod_{j=1}^{M_{l+1}}
\underbrace{
\frac{1}{(q^{\frac14}t s_{i}^{(l)\pm} s_{j}^{(l+1)\mp};q)_{\infty}}
}_{\mathbb{II}_{N'}^{\textrm{3d HM}} \left( \frac{s_{i}^{(l)}}{s_{j}^{(l+1)}} \right)}
\cdot 
\prod_{i=1}^{M_{k-1}}\prod_{j=L+1}^{L+N}
\underbrace{
\frac{1}{(q^{\frac14}t s_{i}^{(k-1)\pm} s_{j}^{\mp};q)_{\infty}}
}_{\mathbb{II}_{N'}^{\textrm{3d HM}} \left(\frac{s_{i}^{(k-1)}}{s_{j}} \right)}
\nonumber\\
&\times 
\underbrace{
\left(q^{\frac12}\prod_{i=1}^{L}s_{i}^{\pm} \prod_{j=1}^{M_{1}}s_{j}^{(1)\mp}z_{1}^{\pm};q\right)_{\infty}
}_{F\left(q^{\frac12} \frac{\prod_{i=1}^{L}s_{i}}{\prod_{j=1}^{M_{1}} s_{j}^{(1)}}z_{1} \right)}
\cdot 
\prod_{l=1}^{k-2}
\underbrace{
\left(q^{\frac12}\prod_{i=1}^{M_{l}} s_{i}^{(l)\pm} \prod_{j=1}^{M_{l+1}} s_{j}^{(l+1)\mp} z_{l+1}^{\pm};q\right)_{\infty}
}_{F\left( q^{\frac12} \frac{\prod_{i=1}^{M_{l}} s_{i}^{(l)}}{\prod_{j=1}^{M_{l+1}} s_{j}^{(l+1)}} z_{l+1} \right)}
\cdot 
\underbrace{
\left(q^{\frac12}\prod_{i=1}^{M_{k-1}} s_{i}^{(k-1)\pm} \prod_{j=L+1}^{L+N} s_{j}^{\mp}z_{k}^{\pm};q\right)_{\infty}
}_{F\left(q^{\frac12} \frac{\prod_{i=1}^{M_{k-1}} s_{i}^{(k-1)}}{\prod_{j=L+1}^{L+N}s_{j}} z_{k} \right)}. 
\end{align}

Under the action of S-duality, 
we find the dual junction 
in which the $k$ defects of D5$'$-branes 
break the gauge symmetry down to $U(\min\{L,M_{i},N\})$ 
and the D5-brane requires the $U(\min\{L,M_{i},N\})$ gauge theory to satisfy 
Dirichlet/Nahm pole boundary condition characterized by an embedding 
$\rho:$ $\mathfrak{su}(2)$ $\rightarrow$ $\mathfrak{u}(\min\{L,M_{i},N\})$. 
For D5$'$-branes which have equal numbers of D3-branes in their both sides, 
we find 3d twisted hypermultiplets. 
Due to the D5-brane, they should obey 
Dirichlet/Nahm pole boundary condition with an embedding specified by 
the corresponding numbers of D3-branes.

The quarter-index would be given by
\begin{align}
\label{LMxNNb}
&
\mathbb{IV}^{L|[M_{1}]-[M_{2}]-\cdots-[M_{k-1}]|N}_{\mathcal{D}\mathcal{D}'}
\nonumber\\
&=
\underbrace{
\prod_{k=1}^{\min \{L,M_{i},N\}}
\frac{(q^{\frac{k+1}{2}} t^{2(k-1)} ;q)_{\infty}}
{(q^{\frac{k}{2}} t^{2k};q)_{\infty}}
}_{\mathbb{II}_{\textrm{Nahm}}^{U(\min\{L,M_{i},N \})}}
\nonumber\\
&\times 
(q^{\frac34+\frac{L-1}{4}} t^{1+(L-1)}z_{1}^{\pm};q)_{\infty}^{{\delta^{L}}_{M_{1}}}
\cdot 
\prod_{l=1}^{k-2}
(q^{\frac34+\frac{M_{l}-1}{4}} t^{1+(M_{l}-1)} z_{l+1}^{\pm};q)_{\infty}^{{\delta^{M_{l}}}_{M_{l+1}}}
\cdot 
(q^{\frac34+\frac{N-1}{4}} t^{1+(N-1)}z_{k}^{\pm};q)_{\infty}^{{\delta^{N}}_{M_{k-1}}}.
\end{align}
We expect that 
the quarter-indices (\ref{LMxNNa}) and (\ref{LMxNNb}) give the same answer.

%%%%%%%%%%%%%%%%%%%%%%%%%%%%%%%%%%%%%%%%%%%%%
%%%%%%%%%%%%%%%%%%%%%%%%%%%%%%%%%%%%%%%%%%%%%
\subsection{4d $U(L)|$3d $(M)-[1]|$4d $U(N)$ with $\mathcal{N}'$}
\label{sec_4dL3dhm1M4dN}
\begin{figure}
\begin{center}
\includegraphics[width=12.5cm]{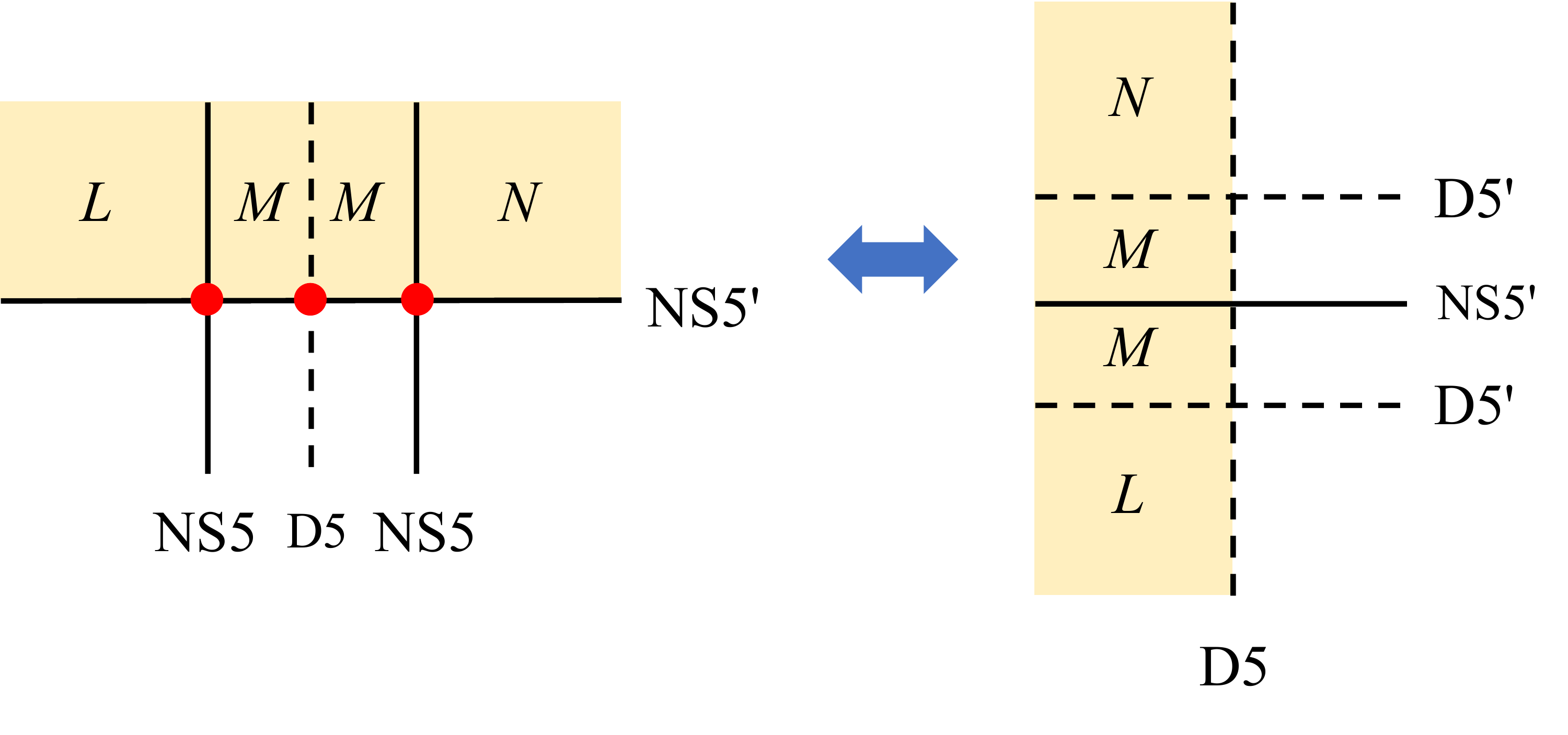}
\caption{
The brane constructions of the configuration 4d $U(L)|$3d $(M)-[1]|$4d $U(N)$ with Neumann b.c. $\mathcal{N}'$ 
and its mirror corner configuration that has no gauge symmetry. 
}
\label{fig4dul3dumhm14dun}
\end{center}
\end{figure}

%%%%%%%%%%%%%%%%%%%%%%%%%%%%%%%%%%%%%%%%%%%%%
%%%%%%%%%%%%%%%%%%%%%%%%%%%%%%%%%%%%%%%%%%%%%
Now turn to the junction 4d $U(L)|$3d $(M)-[1]|$4d $U(N)$ with $\mathcal{N}'$, 
which has a 3d fundamental hypermultiplet. 
This is a generalization of the junction in section \ref{sec_4dN3dMhm1} by adding another corner of 4d theory. 
The 4d $U(L)$ and $U(N)$ gauge theories are coupled 
to 3d $U(M)$ gauge theory through two 3d bi-fundamental hypermultiplets. 
There are two cross-determinant Fermi multiplets and a fundamental Fermi multiplet. 
The brane configuration is illustrated in Figure \ref{fig4dul3dumhm14dun}. 

By applying S-duality we can obtain the dual junction. 
Unlike the junction in section \ref{sec_4dN3dMhm1}, the D3-branes do not terminate on D5$'$-brane. 
This fact admits non-trivial Nahm pole boundary conditions 
for 3d bi-fundamental twisted hypermultiplets as well as 
a pair of boundary conditions $(\mathcal{N}', \mathcal{D}/\textrm{Nahm})$ for 4d gauge theory.

%%%%%%%%%%%%%%%%%%%%%%%%%%%%%%%%%%%%%%%%%%%%%
\subsubsection{4d $U(1)|$3d $(1)-[1]|$4d $U(1)$ with $\mathcal{N}'$}
\label{sec_4du13du1hm14du1NN}
%%%%%%%%%%%%%%%%%%%%%%%%%%%%%%%%%%%%%%%%%%%%%
Let us start with the junction 4d $U(1)|$3d $(1)-[1]|$4d $U(1)$ with $\mathcal{N}'$. 
This junction is obtained from the junction in section \ref{sec_4du13du14du1NN} 
by adding a fundamental hypermultiplet with Neumann b.c. $N'$ and a 2d fundamental Fermi multiplet. 
There is a pair of corners of 4d $U(1)$ gauge theories with Neumann b.c. $(\mathcal{N}, \mathcal{N}')$
and a boundary of 3d $U(1)$ gauge theory with Neumann b.c. $\mathcal{N}'$. 
Again there are two 3d bi-fundamental hypers and two cross-determinant Fermi multiplets.

The quarter-index takes the form
\begin{align}
\label{111hm1Na}
&
\mathbb{IV}^{\textrm{4d $U(1)|(1)-[1]|$4d $U(1)$}}_{\mathcal{N}\mathcal{N}'}
\nonumber\\
&=
\underbrace{
(q)_{\infty} \oint \frac{ds_{1}}{2\pi is_{1}}
}_{\mathbb{IV}_{\mathcal{N}\mathcal{N}'}^{\textrm{4d $U(1)$}}}
\cdot 
\underbrace{
(q)_{\infty}(q^{\frac12}t^2;q)_{\infty}\oint \frac{ds_{2}}{2\pi is_{2}}
}_{\mathbb{II}_{\mathcal{N}'}^{\textrm{3d $U(1)$}}}
\cdot 
\underbrace{
(q)_{\infty}\oint \frac{ds_{3}}{2\pi is_{3}}
}_{\mathbb{IV}_{\mathcal{N}\mathcal{N}'}^{\textrm{4d $U(1)$}}}
\nonumber\\
&\times 
\underbrace{
\frac{1}{(q^{\frac14}t s_{1}^{\pm}s_{2}^{\mp};q)_{\infty}}
}_{\mathbb{II}_{N'}^{\textrm{3d HM}} \left(\frac{s_{1}}{s_{2}} \right)}
\cdot 
\underbrace{
\frac{1}{(q^{\frac14}t s_{2}^{\pm};q)_{\infty}}
}_{\mathbb{II}_{N'}^{\textrm{3d HM}} \left( s_{2}\right)}
\cdot 
\underbrace{
\frac{1}{(q^{\frac14}t s_{2}^{\pm}s_{3}^{\mp};q)_{\infty}}
}_{\mathbb{II}_{N'}^{\textrm{3d HM}}\left(\frac{s_{2}}{s_{3}} \right)}
\nonumber\\
&\times 
\underbrace{
(q^{\frac12}s_{1}^{\pm}s_{2}^{\mp}z_{1}^{\pm};q)_{\infty}
}_{F\left(q^{\frac12}\frac{s_{1}}{s_{2}}z_{1} \right)}
\cdot 
\underbrace{
(q^{\frac12}s_{2}^{\pm}x^{\pm};q)_{\infty}
}_{F(q^{\frac12}s_{2}x)}
\cdot 
\underbrace{
(q^{\frac12}s_{2}^{\pm}s_{3}^{\mp}z_{2}^{\pm};q)_{\infty}
}_{F\left(q^{\frac12}\frac{s_{2}}{s_{3}} z_{2} \right)}. 
\end{align}

The dual junction has a pair of corners of 
4d $U(1)$ gauge theories with sets of boundary conditions $(\mathcal{N}', \mathcal{D})$. 
From the D3-D5$'$ strings, we have two 3d fundamental hypermultiplets with Dirichlet b.c. $D$. 
In addition, from the D3-D3 strings across the NS5$'$-brane, 
we have a 3d bi-fundamental twisted hypermultiplet, which again  obeys Dirichlet b.c. $D$. 

Then one obtains the quarter-index for the dual junction
\begin{align}
\label{111hm1Nb}
&
\mathbb{IV}^{1|[1]|1}_{\mathcal{N}'\mathcal{D}}
\nonumber\\
&=
\underbrace{
\frac{1}{(q^{\frac12} t^2;q)_{\infty}}
}_{\mathbb{IV}_{\mathcal{N}'\mathcal{D}}^{\textrm{4d $U(1)$}}}
\cdot 
\underbrace{
\frac{1}{(q^{\frac12} t^2;q)_{\infty}}
}_{\mathbb{IV}_{\mathcal{N}'\mathcal{D}}^{\textrm{4d $U(1)$}}}
\cdot 
\underbrace{
(q^{\frac34}t z_{1}^{\pm};q)_{\infty}
}_{\mathbb{II}_{D}^{\textrm{3d tHM}}(z_{1})}
\cdot 
\underbrace{
(q^{\frac34}t x^{\pm};q)_{\infty}
}_{\mathbb{II}_{D}^{\textrm{3d tHM}}(x)}
\cdot 
\underbrace{
(q^{\frac34}t z_{2}^{\pm};q)_{\infty}
}_{\mathbb{II}_{D}^{\textrm{3d tHM}}(z_{2})}.
\end{align}
In fact, we find that the quarter-indices (\ref{111hm1Na}) 
and (\ref{111hm1Nb}) coincide with each other.

%%%%%%%%%%%%%%%%%%%%%%%%%%%%%%%%%%%%%%%%%%%%%
\subsubsection{4d $U(2)|$3d $(1)-[1]|$4d $U(1)$ with $\mathcal{N}'$}
\label{sec_4du23du1hm14du1NN}
%%%%%%%%%%%%%%%%%%%%%%%%%%%%%%%%%%%%%%%%%%%%%
Consider the junction 4d $U(2)|$3d $(1)-[1]|$4d $U(1)$ with $\mathcal{N}'$, 
which is obtained from the junction in section \ref{sec_4du23du14du1NN} 
by adding a 3d charged hypermultiplet with Neumann b.c. $N'$ 
together with 2d charged Fermi multiplet in the 3d gauge theory.

The quarter-index takes the form
\begin{align}
\label{211hm1Na}
&
\mathbb{IV}^{\textrm{4d $U(2)|(1)-[1]|$4d $U(1)$}}_{\mathcal{N}\mathcal{N}'}
\nonumber\\
&=
\underbrace{
\frac12 (q)_{\infty}^2 \oint \prod_{i=1}^{2}
\frac{ds_{i}}{2\pi is_{i}}
\prod_{i\neq j}
\left(\frac{s_{i}}{s_{j}};q \right)_{\infty}
}_{\mathbb{IV}_{\mathcal{N}\mathcal{N}'}^{\textrm{4d $U(2)$}}}
\cdot 
\underbrace{
(q)_{\infty} (q^{\frac12}t^2;q)_{\infty}\oint \frac{ds_{3}}{2\pi is_{3}}
}_{\mathbb{II}_{\mathcal{N}'}^{\textrm{3d $U(1)$}}}
\cdot 
\underbrace{
(q)_{\infty}\oint \frac{ds_{4}}{2\pi is_{4}}
}_{\mathbb{IV}_{\mathcal{N}\mathcal{N}'}^{\textrm{4d $U(1)$}}}
\nonumber\\
&\times 
\prod_{i=1}^{2}
\underbrace{
\frac{1}{(q^{\frac14}t s_{i}^{\pm}s_{3}^{\mp};q)_{\infty}}
}_{\mathbb{II}_{N'}^{\textrm{3d HM}}\left(\frac{s_{i}}{s_{3}} \right) }
\cdot 
\underbrace{
\frac{1}{(q^{\frac14}t s_{3}^{\pm};q)_{\infty}}
}_{\mathbb{II}_{N'}^{\textrm{3d HM}} (s_{3}) }
\cdot 
\underbrace{
\frac{1}{(q^{\frac14}t s_{3}^{\pm}s_{4}^{\mp};q)_{\infty}}
}_{\mathbb{II}_{N'}^{\textrm{3d HM}} \left(\frac{s_{3}}{s_{4}} \right)}
\nonumber\\
&\times 
\underbrace{
(q^{\frac12}s_{1}^{\pm}s_{2}^{\pm}s_{3}^{\mp}z_{1}^{\pm};q)_{\infty}
}_{F\left(q^{\frac12} \frac{s_{1}s_{2}}{s_{3}}z_{1} \right)}
\cdot 
\underbrace{
(q^{\frac12}s_{3}^{\pm}x^{\pm};q)_{\infty}
}_{F\left(q^{\frac12} s_{3}x \right)}
\cdot 
\underbrace{
(q^{\frac12}s_{3}^{\pm}s_{4}^{\mp}z_{2}^{\pm};q)_{\infty}
}_{F\left(q^{\frac12} \frac{s_{3}}{s_{4}}z_{2} \right)}.
\end{align}

In the dual configuration 
there is a pair of corners of 4d gauge theories. 
Due to the D5$'$-brane, 
the $U(2)$ gauge symmetry should be broken down to $U(1)$ 
so that one finds a pair of corners of 
4d $U(1)$ gauge theories with two kinds of boundary conditions $(\mathcal{N}', \mathcal{D})$. 
For the D5$'$-brane with equal numbers of D3-branes in its both sides, 
we have 3d fundamental hypermultiplet with Dirichlet b.c. $D$. 
In addition, from the D3-D3 strings across the NS5$'$-brane, 
we have a 3d bi-fundamental twisted hypermultiplet, which again  obeys Dirichlet b.c. $D$. 

As expected, we have checked that 
the quarter-index (\ref{211hm1Na}) matches with 
\begin{align}
\label{211hm1Nb}
&
\mathbb{IV}^{2|[1]|1}_{\mathcal{N}'\mathcal{D}}
\nonumber\\
&=
\underbrace{
\frac{1}{(q^{\frac12} t^2;q)_{\infty}}
}_{\mathbb{IV}_{\mathcal{N}'\mathcal{D}}^{\textrm{4d $U(1)$}}}
\cdot 
\underbrace{
\frac{1}{(q^{\frac12} t^2;q)_{\infty}}
}_{\mathbb{IV}_{\mathcal{N}'\mathcal{D}}^{\textrm{4d $U(1)$}}}
\cdot 
\underbrace{
(q^{\frac34}t z_{2}^{\pm};q)_{\infty}
}_{\mathbb{II}_{D}^{\textrm{3d tHM}} (z_{2})}
\cdot 
\underbrace{
(q^{\frac34}t x^{\pm};q)_{\infty}
}_{\mathbb{II}_{D}^{\textrm{3d tHM}} (x)}. 
\end{align}

%%%%%%%%%%%%%%%%%%%%%%%%%%%%%%%%%%%%%%%%%%%%%
\subsubsection{4d $U(2)|$3d $(1)-[1]|$4d $U(2)$ with $\mathcal{N}'$}
\label{sec_4du23du1hm14du2NN}
%%%%%%%%%%%%%%%%%%%%%%%%%%%%%%%%%%%%%%%%%%%%%
As a further example, let us consider the junction 4d $U(2)|$3d $(1)-[1]|$4d $U(2)$ with $\mathcal{N}'$. 
This is obtained by adding a 3d fundamental hypermultiplet with Neumann b.c. $N'$ 
and a 2d fundamental Fermi multiplet.

Analogously, we can compute the quarter-index as
\begin{align}
\label{212hm1Na}
&
\mathbb{IV}^{\textrm{4d $U(2)|(1)-[1]|$4d $U(2)$}}_{\mathcal{N}\mathcal{N}'}
\nonumber\\
&=
\underbrace{
\frac12 (q)_{\infty}^2 \oint \prod_{i=1}^{2}\frac{ds_{i}}{2\pi is_{i}}
\prod_{i\neq j}\left(\frac{s_{i}}{s_{j}};q \right)_{\infty}
}_{\mathbb{IV}_{\mathcal{N}\mathcal{N}'}^{\textrm{4d $U(2)$}}}
\cdot 
\underbrace{
(q)_{\infty} (q^{\frac12}t^2;q)_{\infty}\oint \frac{ds_{3}}{2\pi is_{3}}
}_{\mathbb{II}_{\mathcal{N}'}^{\textrm{3d $U(1)$}}}
\cdot 
\underbrace{
\frac12 (q)_{\infty}^{2}\oint \prod_{i=4}^{5}
\frac{ds_{i}}{2\pi is_{i}}
\prod_{i\neq j}
\left(\frac{s_{i}}{s_{j}};q \right)_{\infty}
}_{\mathbb{IV}_{\mathcal{N}\mathcal{N}'}^{\textrm{4d $U(2)$}}}
\nonumber\\
&\times 
\prod_{i=1}^{2}
\underbrace{
\frac{1}{(q^{\frac14}t s_{i}^{\pm}s_{3}^{\mp};q)_{\infty}}
}_{\mathbb{II}_{N'}^{\textrm{3d HM}}\left(\frac{s_{i}}{s_{j}} \right)}
\cdot 
\underbrace{
\frac{1}{(q^{\frac14}t s_{3}^{\pm};q)_{\infty}}
}_{\mathbb{II}_{N'}^{\textrm{3d HM}} (s_{i})}
\cdot 
\prod_{i=4}^{5}
\underbrace{
\frac{1}{(q^{\frac14}t s_{3}^{\pm}s_{i}^{\mp};q)_{\infty}}
}_{\mathbb{II}_{N'}^{\textrm{3d HM}} \left(\frac{s_{i}}{s_{j}} \right)}
\nonumber\\
&\times 
\underbrace{
(q^{\frac12} s_{1}^{\pm}s_{2}^{\pm}s_{3}^{\pm}z_{1}^{\pm};q)_{\infty}
}_{F\left(q^{\frac12}\frac{s_{1}s_{2}}{s_{3}}z_{1} \right)}
\cdot 
\underbrace{
(q^{\frac12}s_{3}^{\pm}x^{\pm};q)_{\infty}
}_{F(q^{\frac12}s_{3}x)}
\cdot 
\underbrace{
(q^{\frac12}s_{3}^{\pm}s_{4}^{\mp}s_{5}^{\mp}z_{2}^{\pm};q)_{\infty}
}_{F\left(q^{\frac12} \frac{s_{3}}{s_{4}s_{5}}z_{2} \right)}. 
\end{align}

There are two corners of 4d gauge theories in the dual configuration. 
According to the D5$'$-branes, each $U(2)$ gauge symmetry in the two corners is broken down to $U(1)$. 
As a consequence, there remains a pair of corners of 4d $U(1)$ gauge theories. 
Since the numbers of D3-branes jump across the D5-branes, 
the junction would not include 3d fundamental twisted hypermultiplet. 
Instead, there would be 3d bi-fundamental twisted hypermultiplet arising from D3-D3 string across the NS5$'$-brane. 

The quarter-index for the dual junction gets simplified as
\begin{align}
\label{212hm1Nb}
&
\mathbb{IV}^{2|[1]|2}_{\mathcal{N}'\mathcal{D}}
\nonumber\\
&=
\underbrace{
\frac{1}{(q^{\frac12}t^2;q)_{\infty}}
}_{\mathbb{IV}_{\mathcal{N}'\mathcal{D}}^{\textrm{4d $U(1)$}}}
\cdot 
\underbrace{
\frac{1}{(q^{\frac12}t^2;q)_{\infty}}
}_{\mathbb{IV}_{\mathcal{N}'\mathcal{D}}^{\textrm{4d $U(1)$}}}
\cdot 
(q^{\frac34}t x^{\pm};q)_{\infty}. 
\end{align}
In fact, we have confirmed that 
the quarter-indices (\ref{212hm1Na}) and (\ref{212hm1Nb}) coincide to each other.

%%%%%%%%%%%%%%%%%%%%%%%%%%%%%%%%%%%%%%%%%%%%%
\subsubsection{4d $U(2)|$3d $(2)-[1]|$4d $U(2)$ with $\mathcal{N}'$}
\label{sec_4du23du2hm14du2NN}
%%%%%%%%%%%%%%%%%%%%%%%%%%%%%%%%%%%%%%%%%%%%%
Let us turn to the non-Abelian case. 
For the junction 4d $U(2)|$3d $(2)-[1]|$4d $U(2)$ with $\mathcal{N}'$, 
we have three factors of $U(2)$ gauge symmetry. 
The 4d and 3d gauge theories are coupled 
through two 3d bi-fundamental hypers and two 2d cross-determinant Fermi multiplets. 
The junction further contains 
a 3d hypermultiplet with Neumann b.c. $N'$ and a 2d fundamental Fermi multiplet 
transforming as the fundamental representation under the 3d $U(2)$ gauge symmetry.

Then the quarter-index is evaluated as
\begin{align}
\label{222hm1Na}
&
\mathbb{IV}^{\textrm{4d $U(2)|(2)-[1]|$4d $U(2)$}}_{\mathcal{N}\mathcal{N}'}
\nonumber\\
&=
\underbrace{
\frac12 (q)_{\infty}^2 
\oint \prod_{i=1}^{2}
\frac{ds_{i}}{2\pi is_{i}}
\prod_{i\neq j}
\left(\frac{s_{i}}{s_{j}};q \right)_{\infty}
}_{\mathbb{IV}_{\mathcal{N}\mathcal{N}'}^{\textrm{4d $U(2)$}}}
\cdot 
\underbrace{
\frac12 (q)_{\infty}^2 (q^{\frac12} t^2;q)_{\infty}^2 
\oint \prod_{i=3}^{4} \frac{ds_{i}}{2\pi is_{i}}
\left(\frac{s_{i}}{s_{j}}:q \right)_{\infty}
\left(q^{\frac12} t^2 \frac{s_{i}}{s_{j}}:q \right)_{\infty}
}_{\mathbb{II}_{\mathcal{N}'}^{\textrm{3d $U(2)$}}}
\nonumber\\
&\times 
\underbrace{
\frac12 (q)_{\infty}^2 \oint \prod_{i=5}^{6}
\frac{ds_{i}}{2\pi is_{i}}\prod_{i\neq j}
\left(\frac{s_{i}}{s_{j}};q \right)_{\infty}
}_{\mathbb{IV}_{\mathcal{N}\mathcal{N}'}^{\textrm{4d $U(2)$}}}
\nonumber\\
&\times 
\prod_{i=1}^{2}\prod_{j=3}^{4}
\underbrace{
\frac{1}{(q^{\frac14}ts_{i}^{\pm}s_{j}^{\mp};q)_{\infty}}
}_{\mathbb{II}_{N'}^{\textrm{3d HM}} \left(\frac{s_{i}}{s_{j}} \right)}
\cdot 
\prod_{i=3}^{4}
\underbrace{
\frac{1}{(q^{\frac34}t s_{i}^{\pm};q)_{\infty}}
}_{\mathbb{II}_{N'}^{\textrm{3d HM}} (s_{i})}
\cdot 
\prod_{i=3}^{4}\prod_{j=5}^{6}
\underbrace{
\frac{1}{(q^{\frac14}t s_{i}^{\pm}s_{j}^{\mp};q)_{\infty}}
}_{\mathbb{II}_{N'}^{\textrm{3d HM}} \left(\frac{s_{i}}{s_{j}} \right)}
\nonumber\\
&\times 
\underbrace{
(q^{\frac12}s_{1}^{\pm}s_{2}^{\pm}s_{3}^{\mp}s_{4}^{\mp}z_{1}^{\pm};q)_{\infty}
}_{F\left( q^{\frac12} \frac{s_{1}s_{2}}{s_{3}s_{4}}z_{1} \right)}
\cdot 
\prod_{i=3}^{4}
\underbrace{
(q^{\frac12}s_{i}^{\pm}x^{\pm};q)_{\infty}
}_{F(q^{\frac12}s_{i}x)}
\cdot 
\underbrace{
(q^{\frac12}s_{3}^{\pm}s_{4}^{\pm}s_{5}^{\mp}s_{6}^{\mp}z_{2}^{\pm};q)_{\infty}
}_{F\left( q^{\frac12} \frac{s_{3}s_{4}}{s_{5}s_{6}} z_{2}\right)}. 
\end{align}

Taking S-duality, 
we get the junction which has two corners of 
4d $U(2)$ gauge theories with a pair of boundary conditions $(\mathcal{N}', \textrm{Nahm})$. 
As the numbers of D3-branes do not change across the two D5$'$-branes, 
we would get two 3d fundamental twisted hypers obeying Nahm pole boundary condition 
with an embedding $\rho:$ $\mathfrak{su}(2)$ $\rightarrow$ $\mathfrak{u}(2)$. 
In addition, there would be a 3d bi-fundamental twisted hyper 
satisfying Nahm pole boundary condition with an embedding $\rho:$ $\mathfrak{su}(2)$ $\rightarrow$ $\mathfrak{u}(2)$. 

We find that 
the quarter-index (\ref{222hm1Na}) coincides with
\begin{align}
\label{222hm1Nb}
&
\mathbb{IV}^{2|[2]|2}_{\mathcal{N}'\mathcal{D}}
\nonumber\\
&=
\underbrace{
\frac{1}{(q^{\frac12}t^2;q)_{\infty} (qt^{4};q)_{\infty}}
}_{\mathbb{IV}_{\mathcal{N}'\mathcal{D}}^{\textrm{4d $U(2)$}} }
\cdot 
\underbrace{
\frac{1}{(q^{\frac12}t^2;q)_{\infty} (qt^{4};q)_{\infty}}
}_{\mathbb{IV}_{\mathcal{N}'\mathcal{D}}^{\textrm{4d $U(2)$}} }
\nonumber\\
&\times 
(q t^2 z_{1}^{\pm};q)_{\infty}
\cdot 
(q t^2 z_{2}^{\pm};q)_{\infty}
\cdot 
(q^{\frac34}t x^{\pm};q)_{\infty}
\cdot 
(q^{\frac54}t^3 x^{\pm};q)_{\infty}.
\end{align}
It contains two quarter-indices of 4d $U(2)$ gauge theory 
with a pair of boundary conditions $(\mathcal{N}', \mathcal{D})$. 
The factors $(qt^2 z_{1}^{\pm};q)_{\infty}$ and $(qt^2 z_{2};q)_{\infty}$ 
are now familiar contributions from the two 3d fundamental twisted hypers obeying Nahm pole boundary condition. 
We interprete the factors $(q^{\frac34}tx^{\pm};q)_{\infty}$ $(q^{\frac54}t^3 x^{\pm};q)_{\infty}$ 
as the contributions from the 3d bi-fundamental twisted hyper obeying Nahm pole boundary condition.

%%%%%%%%%%%%%%%%%%%%%%%%%%%%%%%%%%%%%%%%%%%%%
\subsubsection{4d $U(3)|$3d $(3)-[1]|$4d $U(3)$ with $\mathcal{N}'$}
\label{sec_4du33du3hm14du3NN}
%%%%%%%%%%%%%%%%%%%%%%%%%%%%%%%%%%%%%%%%%%%%%
In order to gain more insight, 
let us examine the configuration 4d $U(3)|$3d $(3)-[1]|$4d $U(3)$ with $\mathcal{N}'$. 
The junction has a pair of corners of 4d $U(3)$ gauge theories with Neumann b.c. 
$(\mathcal{N}, \mathcal{N}')$ 
coupled to a boundary of 3d $U(3)$ gauge theory with Neumann b.c. $\mathcal{N}'$. 
There are two 3d bi-fundamental hypermultiplets which couple to both 4d and 3d gauge theories 
and a 3d fundamental hypermultiplet that couples to 3d gauge theory. 
In addition, there are two 2d cross-determinant Fermi multiplets and a 2d fundamental Fermi multiplet.

The quarter-index is given by
\begin{align}
\label{333hm1Na}
&
\mathbb{IV}^{\textrm{4d $U(3)|(3)-[1]|$4d $U(3)$}}_{\mathcal{N}\mathcal{N}'}
\nonumber\\
&=
\underbrace{
\frac{1}{3!} (q)_{\infty}^{3} 
\oint \prod_{i=1}^{3}\frac{ds_{i}}{2\pi is_{i}}
\prod_{i\neq j}\left(\frac{s_{i}}{s_{j}};q \right)_{\infty}
}_{\mathbb{IV}_{\mathcal{N}\mathcal{N}'}^{\textrm{4d $U(3)$}}}
\nonumber\\
&\times 
\underbrace{
\frac{1}{3!}(q)_{\infty}^3 (q^{\frac12}t^2;q)_{\infty}^3 
\oint \prod_{i=4}^{6}
\frac{ds_{i}}{2\pi is_{i}}\prod_{i\neq j}
\left(\frac{s_{i}}{s_{j}};q \right)_{\infty}
\left(q^{\frac12} t^2 \frac{s_{i}}{s_{j}};q \right)_{\infty}
}_{\mathbb{II}_{\mathcal{N}'}^{\textrm{3d $U(3)$}}}
\nonumber\\
&\times 
\underbrace{
\frac{1}{3!}(q)_{\infty}^3 \oint \prod_{i=7}^{9}
\frac{ds_{i}}{2\pi is_{i}}
\prod_{i\neq j}\left( \frac{s_{i}}{s_{j}};q \right)_{\infty}
}_{\mathbb{IV}_{\mathcal{N}\mathcal{N}'}^{\textrm{4d $U(3)$}}}
\nonumber\\
&\times 
\prod_{i=1}^{3}\prod_{j=4}^{6}
\underbrace{
\frac{1}{(q^{\frac14}ts_{i}^{\pm}s_{j}^{\mp};q)_{\infty}}
}_{\mathbb{II}_{N'}^{\textrm{3d HM}} \left( \frac{s_{i}}{s_{j}} \right)}
\cdot 
\prod_{i=4}^{6}
\underbrace{
\frac{1}{(q^{\frac14}ts_{i}^{\pm};q)_{\infty}}
}_{\mathbb{II}_{N'}^{\textrm{3d HM}}(s_{i})}
\cdot 
\prod_{i=4}^{6}\prod_{j=7}^{9}
\underbrace{
\frac{1}{(q^{\frac14}t s_{i}^{\pm}s_{j}^{\mp};q)_{\infty}}
}_{\mathbb{II}_{N'}^{\textrm{3d HM}} \left(\frac{s_{i}}{s_{j}} \right)}
\nonumber\\
&\times 
\underbrace{
(q^{\frac12}s_{1}^{\pm}s_{2}^{\pm}s_{3}^{\pm}s_{4}^{\mp}s_{5}^{\mp}s_{6}^{\mp}z_{1}^{\pm};q)_{\infty}
}_{F\left( q^{\frac12} \frac{s_{1}s_{2}s_{3}}{s_{4}s_{5}s_{6}}z_{1} \right)}
\cdot 
\prod_{i=4}^{6}
\underbrace{
(q^{\frac12}s_{i}^{\pm}x^{\pm};q)_{\infty}
}_{F(q^{\frac12}s_{i}x)}
\cdot 
\underbrace{
(q^{\frac12}s_{4}^{\pm}s_{5}^{\pm}s_{6}^{\pm}s_{7}^{\mp}s_{8}^{\mp}s_{9}^{\mp}z_{2}^{\pm};q)_{\infty}
}_{F\left(\frac{s_{4}s_{5}s_{6}}{s_{7}s_{8}s_{9}} z_{2} \right)}. 
\end{align}

Similarly, the dual junction has two corners of 
4d $U(3)$ gauge theories with a pair of boundary conditions $(\mathcal{N}', \textrm{Nahm})$. 
The equal numbers of D3-branes across the two D5$'$-branes 
give rise to two 3d fundamental twisted hypers obeying Nahm pole boundary condition 
with a homomorphism $\rho:$ $\mathfrak{su}(2)$ $\rightarrow$ $\mathfrak{u}(3)$. 
In addition, there would exist a 3d bi-fundamental twisted hyper 
obeying Nahm pole boundary condition with an embedding $\rho:$ $\mathfrak{su}(2)$ $\rightarrow$ $\mathfrak{u}(3)$. 

We find that 
the quarter-index (\ref{333hm1Na}) agrees with
\begin{align}
\label{333hm1Nb}
&
\mathbb{IV}^{3|[3]|3}_{\mathcal{N}'\mathcal{D}}
\nonumber\\
&=
\underbrace{
\frac{1}{(q^{\frac12}t^2;q)_{\infty} (qt^4;q)_{\infty} (q^{\frac32}t^6;q)_{\infty}}
}_{\mathbb{IV}_{\mathcal{N}'\mathcal{D}}^{\textrm{4d $U(3)$}}}
\cdot 
\underbrace{
\frac{1}{(q^{\frac12}t^2;q)_{\infty} (qt^4;q)_{\infty} (q^{\frac32}t^6;q)_{\infty}}
}_{\mathbb{IV}_{\mathcal{N}'\mathcal{D}}^{\textrm{4d $U(3)$}}}
\nonumber\\
&\times 
(q^{\frac54}t^3 z_{1}^{\pm};q)_{\infty}
\cdot 
(q^{\frac54}t^3 z_{2}^{\pm};q)_{\infty}
\cdot 
(q^{\frac34}t x^{\pm};q)_{\infty}
\cdot 
(q^{\frac54}t^{3} x^{\pm};q)_{\infty}
\cdot 
(q^{\frac74}t^5 x^{\pm};q)_{\infty}.
\end{align}
As expected from brane picture, 
this has two quarter-indices of 4d $U(3)$ gauge theory 
with a pair of boundary conditions $(\mathcal{N}', \mathcal{D})$. 
The factors $(q^{\frac54}t^3 z_{1}^{\pm};q)_{\infty}$ and $(q^{\frac34}t^{3} z_{2}^{\pm};q)_{\infty}$ 
are contributions from the two 3d fundamental twisted hypers obeying Nahm pole boundary condition. 
We would like to identify the remaining factors 
$(q^{\frac34}tx^{\pm};q)_{\infty}$ 
$(q^{\frac54}t^3 x^{\pm};q)_{\infty}$ 
$(q^{\frac74}t^5 x^{\pm};q)_{\infty}$ 
with the contributions from the 3d bi-fundamental twisted hyper obeying Nahm pole boundary condition 
specified by an embedding $\rho:$ $\mathfrak{su}(2)$ $\rightarrow$ $\mathfrak{u}(3)$.

%%%%%%%%%%%%%%%%%%%%%%%%%%%%%%%%%%%%%%%%%%%%%
\subsubsection{4d $U(L)|$3d $(M)-[1]|$4d $U(N)$ with $\mathcal{N}'$}
\label{sec_4dul3dumhm14dunNN}
%%%%%%%%%%%%%%%%%%%%%%%%%%%%%%%%%%%%%%%%%%%%%
Now we would like to give the dualities for the junctions 4d $U(L)|$3d $(M)-[1]|$4d $U(N)$ with $\mathcal{N}'$ 
where we assume that $L\ge M$ and $N\ge M$. 
This junction has a pair of 4d $\mathcal{N}=4$ $U(L)$ and $U(N)$ gauge theories which obey a pair of Neumann b.c. $(\mathcal{N}, \mathcal{N}')$. 
Between these corners we have a boundary of 3d $\mathcal{N}=4$ $U(M)$ gauge theory with Neumann b.c. $\mathcal{N}'$. 

There are three 3d hypermultiplets. 
Under the $U(L)\times U(M)\times U(N)$ gauge symmetry, 
one transforms as $({\bf L}, \overline{\bf M}, \bf{1})$ $\oplus$ $(\overline{\bf L}, {\bf M}, \bf{1})$ 
and the other transforms as $({\bf 1}, {\bf M}, \overline{\bf{N}})$ $\oplus$ $({\bf 1}, \overline{\bf M}, \bf{N})$ 
and the last transforms as $({\bf 1}, {\bf M}, {\bf 1})$ $\oplus$ $({\bf 1}, \overline{\bf M}, {\bf 1})$. 
All of them obey the Neumann b.c. $N'$ due to the NS5$'$-brane. 

There are also three 3d Fermi multiplets. 
Under the $U(L)\times U(M)\times U(N)$ gauge symmetry, 
one transforms as $(\det, \det^{-1}, \bf{1})$ 
and the other transforms as $({\bf 1}, \det, \det^{-1})$ 
and the last transforms as $({\bf 1}, {\bf M}, {\bf 1})$.

The quarter-index is given by
\begin{align}
\label{LMNhm1Na}
&
\mathbb{IV}^{\textrm{4d $U(L)|(M)-[1]|$4d $U(N)$}}_{\mathcal{N}\mathcal{N}'}
\nonumber\\
&=
\underbrace{
\frac{1}{L!}(q)_{\infty}^{L}
\oint \prod_{i=1}^{L}
\frac{ds_{i}}{2\pi is_{i}}
\prod_{i\neq j}
\left( \frac{s_{i}}{s_{j}};q \right)_{\infty}
}_{\mathbb{IV}_{\mathcal{N}\mathcal{N}'}^{\textrm{4d $U(L)$}}}
\cdot 
\underbrace{
\frac{1}{M!}
(q)_{\infty}^{M} (q^{\frac12} t^2;q)_{\infty}^{M}
\oint \prod_{i=L+1}^{L+M}
\frac{ds_{i}}{2\pi is_{i}}
\prod_{i\neq j}
\left(\frac{s_{i}}{s_{j}};q \right)_{\infty}
\left( q^{\frac12} t^2 \frac{s_{i}}{s_{j}};q \right)_{\infty}
}_{\mathbb{II}_{\mathcal{N}'}^{\textrm{3d $U(M)$}}}
\nonumber\\
&\times 
\underbrace{
\frac{1}{N!}(q)_{\infty}^{N}
\oint \prod_{i=L+M+1}^{L+M+N}
\frac{ds_{i}}{2\pi is_{i}}
\prod_{i\neq j}
\left( \frac{s_{i}}{s_{j}};q \right)_{\infty}
}_{\mathbb{IV}_{\mathcal{N}\mathcal{N}'}^{\textrm{4d $U(N)$}}}
\nonumber\\
&\times 
\prod_{i=1}^{L}\prod_{j=L+1}^{L+M}
\underbrace{
\frac{1}{(q^{\frac14}t s_{i}^{\pm}s_{j}^{\mp};q)_{\infty}}
}_{\mathbb{II}_{N'}^{\textrm{3d HM}} \left(\frac{s_{i}}{s_{j}} \right)}
\cdot 
\prod_{i=L+1}^{L+M}
\underbrace{
\frac{1}{(q^{\frac14} t s_{i}^{\pm};q)_{\infty}}
}_{\mathbb{II}_{N'}^{\textrm{3d HM}} (s_{i})}
\cdot 
\prod_{i=L+1}^{L+M}
\prod_{j=L+M+1}^{L+M+N}
\underbrace{
\frac{1}{(q^{\frac14}t s_{i}^{\pm}s_{j}^{\mp};q)_{\infty}}
}_{\mathbb{II}_{N'}^{\textrm{3d HM}} \left(\frac{s_{i}}{s_{j}} \right)}
\nonumber\\
&\times 
\underbrace{
\left( q^{\frac12} \prod_{i=1}^{L}s_{i}^{\pm} \prod_{j=L+1}^{L+M} s_{j}^{\mp}z_{1}^{\pm};q\right)_{\infty}
}_{F\left( q^{\frac12} \frac{\prod_{i=1}^{L}s_{i}}{\prod_{j=L+1}^{L+M}s_{j}} z_{1} \right)}
\cdot 
\prod_{i=L+1}^{L+M}
\underbrace{
\left(q^{\frac12} s_{i}^{\pm}x^{\pm};q \right)_{\infty}
}_{F(q^{\frac12}s_{i}x)}
\cdot 
\underbrace{
\left( q^{\frac12}\prod_{i=L+1}^{L+M}s_{i}^{\pm} \prod_{j=L+M+1}^{L+M+N}s_{j}^{\mp}z_{2}^{\pm};q \right)_{\infty}
}_{F\left( q^{\frac12} \frac{\prod_{i=L+1}^{L+M} s_{i}}{\prod_{j=L+M+1}^{L+M+N} s_{j}} z_{2} \right)}.
\end{align}

Under the action of S-duality, we obtain the dual configuration. 
It consists of a pair of corners of 4d gauge theories. 
According to the D5$'$-branes, 
the 4d $U(L)$ gauge symmetry in one corner is broken to $U(M)$ 
and the 4d $U(N)$ gauge symmetry in the other corner is broken to $U(M)$. 
As a result, we have a pair of corners of 4d $U(M)$ gauge theories 
with two boundary conditions $(\mathcal{N}', \textrm{Nahm})$. 

When $L=M$, we have a 3d fundamental twisted hyper with Dirichlet/Nahm b.c. 
specified by an embedding $\rho:$ $\mathfrak{su}(2)$ $\rightarrow$ $\mathfrak{u}(M)$. 
Likewise, for $N=M$, we have a 3d fundamental twisted hyper with Dirichlet/Nahm b.c. 
specified by an embedding $\rho:$ $\mathfrak{su}(2)$ $\rightarrow$ $\mathfrak{u}(M)$. 

The junction also would have a 3d twisted hypermultiplet 
arising from D3-D3 string across the NS5$'$-brane. 
Due to the D5-brane, it should be subject to 
Dirichlet/Nahm b.c. specified by an embedding $\rho:$ $\mathfrak{su}(2)$ $\rightarrow$ $\mathfrak{u}(M)$.

The quarter-index for the dual junction would take the form 
\begin{align}
\label{LMNhm1Nb}
&
\mathbb{IV}^{L|[M]|N}_{\mathcal{N}'\mathcal{D}}
\nonumber\\
&=
\underbrace{
\prod_{k=1}^{M}
\frac{1}{(q^{\frac{k}{2}} t^{2k};q)_{\infty}}
}_{\mathbb{IV}_{\mathcal{N}'\mathcal{D}}^{\textrm{4d $U(M)$}}}
\cdot 
\underbrace{
\prod_{k=1}^{M}
\frac{1}{(q^{\frac{k}{2}} t^{2k};q)_{\infty}}
}_{\mathbb{IV}_{\mathcal{N}'\mathcal{D}}^{\textrm{4d $U(M)$}}}
\nonumber\\
&\times 
\left(q^{\frac34+\frac{M-1}{4}} t^{1+(M-1)} z_{1}^{\pm};q \right)_{\infty}^{{\delta^{L}}_{M}}
\cdot 
\left(q^{\frac34+\frac{M-1}{4}} t^{1+(M-1)} z_{2}^{\pm};q \right)_{\infty}^{{\delta^{N}}_{M}}
\cdot 
\prod_{k=1}^{M}
\left( q^{\frac14+\frac{k}{2}} t^{-1+2k} x^{\pm};q \right)_{\infty}.
\end{align}
The expression includes 
two quarter-indices of 4d $U(M)$ gauge theories 
with a pair of boundary conditions $(\mathcal{N}',\mathcal{D}/\textrm{Nahm})$. 
The factors 
$(q^{\frac34+\frac{M-1}{4}} t^{1+(M-1)}z_{1}^{\pm};q)_{\infty}^{{\delta^{L}}_{M}}$ 
and 
$(q^{\frac34+\frac{M-1}{4}} t^{1+(M-1)}z_{1}^{\pm};q)_{\infty}^{{\delta^{N}}_{M}}$ 
describe the contributions 
from the possible 3d fundamental twisted hypers with Dirichlet/Nahm b.c. 
specified by an embedding $\rho:$ $\mathfrak{su}(2)$ $\rightarrow$ $\mathfrak{u}(M)$. 
The factors 
$\prod_{k=1}^{M}$ $(q^{\frac14+\frac{k}{2}} t^{-1+2k}x^{\pm};q)_{\infty}$ 
are the expected contributions from a 3d twisted hypermultiplet subject to 
Dirichlet/Nahm b.c. specified by an embedding $\rho:$ $\mathfrak{su}(2)$ $\rightarrow$ $\mathfrak{u}(M)$. 

We expect that 
the quarter-indices (\ref{LMNhm1Na}) and (\ref{LMNhm1Nb}) would give the same result.

%%%%%%%%%%%%%%%%%%%%%%%%%%%%%%%%%%%
\subsection*{Acknowledgements}
The author would like to thank Tudor Dimofte, Sergei Gukov, Miroslav Rapcak, Junya Yagi and Masahito Yamazaki for useful discussions and comments. 
Especially he is grateful to Davide Gaiotto for sharing many useful ideas, discussing and suggesting the problems. 
This work is supported in part by Perimeter Institute for Theoretical Physics and 
JSPS Overseas Research fellowships. Research at
Perimeter Institute is supported by the Government of Canada through the Department of
Innovation, Science and Economic Development and by the Province of Ontario through the
Ministry of Research, Innovation and Science.
%%%%%%

\bibliographystyle{utphys}
\bibliography{ref}

\end{document}